\documentclass[12pt,preprint,aps,showpacs,color,floatfix]{revtex4}
\usepackage{epsfig,colordvi}
\usepackage{color}

\begin{document}
\newcommand{\yl}[1]{\textcolor{red}{#1}}
\newcommand{\ylnote}[1]{[Note: \textcolor{magenta}{#1}]}
\newcommand{\npa}[1]{Nucl.~Phys.~{A#1}}
\newcommand{\etal}{\em{et al.,}}
\newcommand{\neglect}[1]{  }
\newcommand{\Label}[1]{\label{#1} \message{FIG \thefigure :label #1 fig.} }
\newcommand{\blabla}{{\em Dies \\ ist \\ nur \\ ein \\ sinnloser \\ Text \\ zum \\ Auff\"ullen  !}}
\newcommand{\pslash}{\mbox{$p\!\!\!/$}}
\newcommand{\rfl}{red full line}
\newcommand{\bdl}{blue dashed line}
\newcommand{\bpl}{black dotted line}
\newcommand{\gml}{green dash-dotted line}
\newcommand{\rml}{red dash-dotted line}
\newcommand{\gfl}{green full line}
\newcommand{\lhs}{left hand side}
\newcommand{\rhs}{right hand  side}
\newcommand{\Figref}[1]{Fig.\ \ref{#1}}
\newcommand{\figref}[1]{Fig.\ \ref{#1}}
\newcommand{\lhsref}[1]{left hand side of Fig.\ \ref{#1}}
\newcommand{\rhsref}[1]{right hand side of Fig.\ \ref{#1}}
\newcommand{\be}{\begin{equation}}
\newcommand{\ee}{\end{equation}}
\newcommand{\kp}{\mbox{K$^+$~}}
\newcommand{\km}{\mbox{K$^-$~}}
\newcommand{\Apart}[1][ ]{$A_{\rm part}${#1}}
\newcommand{\pip}[1][ ]{$\pi^{+}${#1}}
\newcommand{\pim}[1][ ]{$\pi^{-}${#1}}
\newcommand{\AGeV}[1][ ]{$A$~GeV{#1}}
\newcommand{\p}{\mbox{p}}
\newcommand{\K}{\mbox{K}}
\newcommand{\N}{\mbox{N}}
\newcommand{\Y}{\mbox{Y}}
\newcommand{\B}{\mbox{B}}
\renewcommand{\thefigure}{\roman{section}.\arabic{figure}}

\begin{center}
{\bf Strangeness Production close to Threshold\\ in Proton -
Nucleus and Heavy-Ion Collisions} \vspace{1truecm}

Christoph~Hartnack$^1$, Helmut Oeschler$^2$, Yvonne Leifels$^3$, \\ Elena L.~Bratkovskaya$^{4,5}$, J\"org Aichelin$^1$
\vspace{.5truecm}

$^1$ SUBATECH, Laboratoire de Physique Subatomique et des
Technologies Associ\'ees \\ UMR 6457, University of Nantes - IN2P3/CNRS -
Ecole des Mines
de Nantes \\
4 rue Alfred Kastler, F-44072 Nantes, Cedex 03, France  \\
$^2$ Institut f\"ur Kernphysik, Darmstadt University of Technology, 64289 Darmstadt, Germany\\
$^3$ GSI Helmholtzzentrum f\"ur Schwerionenforschung GmbH, 64291 Darmstadt, Germany \\
$^4$ Institute for Theoretical Physics, Frankfurt University,
         60438 Frankfurt-am-Main, Germany\\
$^5$ Frankfurt Institut for Advanced Studies,
         Frankfurt University, 60438 Frankfurt-am-Main, Germany
\end{center}


%

\date{\today}
%

We discuss strangeness production close to threshold in p+A and
A+A collision. Comparing the body of available \kp, K$^0$ , \km ,
and $\Lambda$ data with the IQMD transport code and for some key
observables as well  with the HSD transport code, we find good
agreement for the large majority of the observables. The
investigation of the reaction with help of these codes reveals the
complicated interaction of the strange particles with hadronic
matter which makes strangeness production in heavy-ion collisions
very different from that in elementary interactions. We show how
different strange particle observables can be used to study the
different facets of this interaction (production, rescattering and
potential interaction) which finally merge into  a comprehensive
understanding of these interactions. We identify those observables
which allow for studying (almost) exclusively one of these
processes to show how future high precision experiments can
improve our quantitative understanding. Finally, we discuss how
the \kp multiplicity can be used to study the hadronic equation of
state.





\newpage

\tableofcontents

\newpage

\section{Introduction}
\subsection{Introduction}
More than a quarter of a century ago, the first strange particles
have been observed in heavy-ion collisions at relativistic
energies \cite{Schnetzer:1989vy,Harris:1981tb} and since then,
strangeness production has become one of the major research fields
in this domain. In the beginning, the experiments had been
performed by different groups at the Bevalac at the Lawrence
Berkeley Laboratory (LBL). This search was limited to light ions
and only with the advent of the SIS accelerator at the
Gesellschaft f\"ur Schwerionenforschung (GSI) in Germany it became
possible to extend the experimental programs to heavy systems. Two
groups, the KaoS and the FOPI Collaborations, performed
experiments on different aspects of the physics of strange
particles over the last 15 years. Very recently, the HADES
Collaboration has joined the efforts and their first results on
strangeness production in heavy-ion collisions are appearing. In
parallel, at the COSY accelerator in J\"ulich, a research program
on close-to-threshold production of strange particles in
proton-proton and proton-nucleus collisions has been launched with
the intension to bridge the gap between the production of strange
particles in elementary (pp) and in heavy-ion collisions.
Heavy-ion collisions are of special interest as strange particles
can be produced below the corresponding threshold in NN
collisions, yet with quite small cross sections. Their yields and their
emission behavior test fundamental properties of hot and dense nuclear matter.
In the meantime the field has matured and a broad body of data
from the present generation of detectors is available.

In the \kp sector the start was difficult despite of many efforts
of different groups. The first measured  \kp spectrum from a
heavy-ion collisions below threshold (being at 1.58 $A$~GeV) was
far from theoretical predictions, and therefore the production
mechanism remained unclear. Later, after the original data had
been superseded by new data from the KaoS Collaboration which came
close to the original theoretical prediction, it became clear that
strange baryons and \kp are created in binary collisions of
baryons and mesons, even below threshold. The Fermi motion can
provide necessary additional energy in the center-of-mass system.
Thus, strange particle production tests the momentum distribution
of the nucleons inside the nucleus. In addition, the unexpectedly
large production yield of \kp mesons in heavy-ion reactions could
be explained by a new mechanism which is not available in
elementary collisions, the production via an intermediate nuclear
resonance, which acts as an energy reservoir. Thus production of
strange particles is sensitive to the behavior of nuclear
resonances inside the nucleus. Moreover, because \kp production
competes with the decay of the resonance, a shorter mean free
path, in other words a higher density, enhances the \kp production
via resonances and \kp production became a tool to study nuclear
matter properties, especially at the high compression which occurs
during the heavy-ion reaction.

Strange particles interact with the hadronic environment not only
by collisions but also by potential interaction. At finite
densities, this interaction has been investigated by extending the
chiral perturbation theory to the SU(3) sector. This procedure
allowed to predict how the \kp mesons are modified in matter but
failed for the description of the \km properties, where more
sophisticated methods are required, as will be discussed later.
Heavy-ion collisions are the only possibility to confront these
predictions with nature. This added another interesting aspect to
the strangeness production in heavy-ion collisions.

For the \km sector already the first measured spectrum in
heavy-ion reactions suggested that \km production in heavy-ion
collisions has little in common with \km production in elementary
or even in proton-heavy ion (pA) collisions. Indeed, the
interpretation of the data identified the strangeness-exchange
reaction Y$\pi\to\km$ N (Y = $\Lambda$, $\Sigma$) as the main
source of \km, a channel which is absent in pA collisions due to
the lack of $\pi$'s. The large cross section of the inverse
reaction and its energy dependence has the consequence that most
of the produced \km are, however, reabsorbed. Those which are
finally observed are created close to the surface as will be
demonstrated. The \km spectra carry a very complex information on
the system which to reproduce is a challenge for every theoretical
approach. What makes the situation even more complicated is the
very complex potential interaction with the hadronic environment.
In contradistinction to the \kp mesons, the \km can produce
baryonic resonances in the nuclear medium which render mean field
approaches invalid and require self-consistent multi-channel
Br\"uckner G-matrix calculations. These calculations have not
reached stable conclusions yet and the depth of the \km N
potential as a function of the nuclear density is still very much
debated. If the interaction were strong enough, a kaon condensate
could be created whose existence has also implications for the
stability of neutron stars. Also here heavy-ion reactions are the
only possibility to test the existence of such condensates because
kaonic atoms, the other source for the determination of the \km
-nucleon interaction probe this potential at very low densities
only. In Section V.F it is shown that the existence of a \km
condensate is not compatible with present heavy-ion data.

This complexity of the physics of strange particles in matter is
not easy to assess. The high-density zone, created in heavy-ion
reactions has a lifetime of a couple of fm/$c$. Then the system
expands rapidly. The time scale of this fast expansion  requires
to study the system with help of transport theories in which the
above discussed physics is embedded. Transport theories have been
developed in the eighties and early nineties and have been
continuously improved since then. They simulate the whole
reaction, from the initial separation of projectile and target
until the formation of the finally observed particles,
and use as input the elementary cross sections and the
interactions among the different particles. By comparing the
results for different input quantities (and it will turn out later
that uncertainties in the available input quantities limit in some
cases the possibility to draw definite conclusions) with
experimental data one is therefore able to test the different
theories. These transport theories have proven to be the essential
tool to interpret heavy-ion reaction data and many results of such
comparisons have been published. They are also a link
between many-body calculations, usually performed for infinite
matter, and experiment.

It is the purpose of this review
\begin{itemize}
\item to review the body of data on strange particle production at
threshold energies ($E_{\rm beam} \le 2$ \AGeV) which is available
by now and which will not increase much in near future.
\item to compare the body of data which the results of one version of one
selected transport theory, in order to assess whether the
present data can be reproduced consistently in terms of
the physics which is embedded in these transport approaches. For
this purpose we employ the Isospin Quantum Molecular Dynamics
(IQMD) approach \cite{Hartnack:1997ez}. For some key observables
we compare data as well with the most advanced version  of another
transport approach, the Hadron String Dynamics (HSD), which has
been developed over the years by Bratkovskaya and Cassing
\cite{Cassing:1999es}. This allows us to assess possible
uncertainties of the theoretical approaches and the consequences
of different ingredients. Transport theories have been developed
in parallel with the experiments. Therefore, the results published
in the literature are not always comparable among each other as
they are coming from different program versions. \item to study
systematically the dependence of the experimental observables on
the input of these transport approaches and to identify those
observables which carry the key information on debated physics
topics. It includes the \kp N and the \km N potential and the
(experimentally unknown) cross sections. \item to assess whether
the present body of data allows for conclusions on the underlying
physics like the in-medium modifications of strange meson
properties, rescattering of strange hadrons and creation of a
local statistical equilibrium. \item to study whether strangeness
production in heavy-ion reactions can contribute to the answer of
two of the most exciting questions debated presently in many-body
and heavy-ion physics:
\begin{itemize}
\item Will there be a \km condensate if matter is sufficiently compressed?
\item Can the nuclear equation of state be determined experimentally?
\end{itemize}
\end{itemize}

Recently two review articles on kaon production in this energy
domain have been published which are complementary to our
approach.

The review by Fuchs \cite{Fuchs:2003pc} puts the emphasis on
comparing published results of different transport models employed to
describe the heavy ion data. The study concentrates on the
possibility to explore the nuclear equation of state by \kp
production, a subject which has seen a large progress since then,
as we will discuss later.

Dohrmann \cite{Dohrmann:2006dx} investigates kaon production via different
entrance channels including leptons and photons and limits the
study of heavy-ion collisions to a comparison of the spectra with
theory without studying in detail how the surrounding matter
influences the production of strangeness.

Our report is organized as follows. In Chapter II we present
results from the different theoretical groups which study the
properties of kaons in infinite matter. This includes a discussion
on the role of the $\Lambda (1405)$ and on the information one may
obtain by studying kaonic atoms. This will allow to elucidate the
mean-field approach which we employ in the calculation.  In
Chapter III we outline the IQMD transport theory which is employed
to interpret the data. We discuss the input quantities which are
experimentally known like the measured cross sections, and comment
the different parameterizations for the yet unknown ones (which
can be found in the literature). We explain as well the virtual
particle concept which allow to study strangeness production
despite of the very small production cross section in this energy
domain. Section IV is devoted to the discussion of \kp (K$^0$) and
$\Lambda$ production in matter. We begin with the excitation
function of the multiplicity followed by a discussion of the
spectra and the information obtained from their slopes. A study of
the information which one may obtain from polar, in-plane and
azimuthal distributions continues this section. We conclude this
section by a discussion of the information on the nuclear reaction
mechanism which one may obtain from the $\Lambda (\Sigma^0)$
observables. In Chapter V we discuss the \km observables in the
same sequence as the \kp observables in Chapter IV and in Chapter
VI we confront the mechanism of \kp and of \km production and
outline similarities and differences. We demonstrate that the two
strange mesons are sensitive to different stages of the heavy-ion
reaction and discuss the consequences for the study of the KN
potential in heavy-ion reactions. In the last part of this
section, we study to which extent measurements of \kp production
allow to extract information on the hadronic equation of state. We
conclude the report by a summary.

\section{Theory}
\label{theory}
\subsection{Experimental information on Kaon properties in matter}

There are several ways to obtain information on the
properties of kaons in matter. First of all one can scatter kaons
elastically with nucleons to obtain experimentally the scattering length.
Its knowledge allows --- as we will see --- to determine unambiguously the change of
the mass of the \kp  even for moderate densities.
However, for \km the knowledge of the scattering length
alone is not sufficient to determine the in medium properties
because in \km reactions strange baryonic resonances
can be created whose properties are density dependent.
Using inelastic \km N data in coupled-channel
calculations allows for exploring the properties of \km in matter.
This analysis may be complemented by that of kaonic atoms, which
are bound \km-nucleus systems in which the low-density \km
-nucleon interaction is tested. The binding energy of kaonic atoms
gives a direct information on the \km potential at low densities.
Finally, \km nucleus cluster states have been suggested to
determine the depth of the \km N potential. We will discuss the
information obtained from these approaches in turn. However, we
first show which properties can be directly inferred from
experiments in a model-independent way.

Knowing the kaon-nucleon scattering amplitude, the KN potential for low densities
of nuclear matter, $\rho$, can be obtained in a simple $t\rho$
approximation
\cite{Kaiser:1995cy,Kaiser:1996js,Oller:2000ma,SchaffnerBielich:2000jy,Tolos:2005jg}.
Employing
\begin{equation}
\Pi_{K} = -4\pi t_{{\rm KN}} \cdot \rho \label{pitrho}
\end{equation}
where $t_{{\rm KN}}$ is the (known) free elementary ${\rm KN}$
scattering amplitude, and $\Pi_{K}= 2EU_{opt}$ is the self energy
of the $K$ meson. Using the scattering length given below, we
obtain for the \kp a repulsive optical potential and a
corresponding mass shift of
\begin {equation}
U_{\rm opt}^{{\rm K}^+}(\bf{q}=0,\rho) = + 30 \ \mbox{ MeV } \cdot
\rho/\rho_0. \label{ress}
\end{equation}
$\rho_0$ is the saturation density of nuclear matter. This value
is compatible with the experimental results
\cite{Nekipelov:2002sd}.

Later it has been realized that at low density $\rho< \rho_0$ the
in-medium kaon self-energies can be obtained by a systematic
expansion of the chiral Lagrangian. All
phenomenological approaches should therefore be consistent with this result.
In this approach a low-density theorem \cite{Dover:1971hr} can be
formulated which relates the change of the kaon mass squared to
the kaon-nucleon scattering amplitude $a_{{\rm KN}}$
\cite{Lutz:1997wt,Lutz:1994cf} as
\begin{eqnarray}
\Delta m_{K}^2 &=& -\pi \left( 1+x \right) \left( a^{(I=0)}_{{\rm
KN}}+3\,a^{(I=1)}_{{\rm KN}} \right) \,\rho
\nonumber\\
&+& \alpha \left( \left(a_{{\rm KN}}^{(I=0)}\right)^2
+3\,\left(a_{{\rm KN}}^{(I=1)}\right)^2 \right)
 k_F^4 +{\cal O}\left( k_F^5 \right) \,,
\label{trho}
\end{eqnarray}
where
\begin{eqnarray}
\alpha &=&\frac{1-x^2+x^2\,\log \left(x^2\right)}{\pi^2\,(1-x)^2}
\simeq 0.166 \,, \qquad x= \frac{m_{\rm K}}{m_{\rm N}}\,,\qquad
\rho = \frac{2\,k_F^3}{3\,\pi^2 }\,. \label{lowdensity}
\end{eqnarray}
The KN scattering lengths, $a^{(I=0)}_{{\rm KN}}$ and $a^{(I=1)}_{{\rm KN}}$, have been measured in the eighties and
the results of the different experiments are summarized in
Refs.~\cite{Martin:1980qe,Dover:1982zh}. They are $a^{(I=0)}_{{\rm
K}^+{\rm N}} \approx 0.02$ fm and $a^{(I=1)}_{{\rm K}^+{\rm N}}
\approx -0.32$ fm for the K$^+$ and $a^{(I=0)}_{{\rm K}^-{\rm N}}
\approx (-1.70 + i0.68)$ fm and $a^{(I=1)}_{{\rm K}^-{\rm N}}
\approx (0.37 + i0.60)$ fm for the K$^-$. With these values the
result of Eq.~(\ref{pitrho}) is compatible with
Eq.~(\ref{ress}).

For the K$^+$ meson, the term $\propto k_F^4$ modifies only slightly the
leading term (the mass shift changes from 28 MeV to 35 MeV
\cite{luhab} at normal nuclear matter density). Thus this
expansion in $\rho$ or $k_F$  seems to converges quite well even
at moderate densities. Consequently, the increase of the K$^+$
mass with density is theoretically well under control and a simple
$t\rho$ approximation gives already quantitatively good results.

For the K$^-$ meson, on the contrary, the situation is
considerably more complicated. At saturation density $\rho_0$, the
$k_F^4$ term is larger than the term proportional to $k_F^3$ and
hence the expansion does not converge. Thus already at moderate densities
the low-density theorem - although still valid at very low
densities - is of little use. In addition, according to
Eq.~(\ref{trho}), the interaction is repulsive whereas the
results in kaonic atoms \cite{Baca:2000ic} indicate that already
at quite low densities the interaction becomes attractive. Koch
\cite{Koch:1994mj} and later Waas {\em et
al.} ~\cite{Waas:1996xh,Waas:1996fy,Waas:1997pe} found that this
change of sign of the potential can be explained by the influence
of the $\Lambda (1405)$ state, a resonance 27 MeV below the KN
threshold. They showed that the Pauli blocking of intermediate KN
states at finite density shifts the mass of this resonance above
threshold and renders the potential attractive. Later, it has been
realized that the situation becomes even more complex: When the
K$^-$ potential is calculated self-consistently the
attractive self-energy of the K$^-$ counteracts the Pauli blocking
and pushes the $\Lambda (1405)$ mass again to lower values but leaving it
moderately attractive, as we will see.

\subsection{Strange particles in infinite matter - theoretical
approaches}

The theoretical study of kaon properties in dense matter goes
back to Kaplan and Nelson \cite{Kaplan:1986yq,Nelson:1987dg}. The degrees of freedom are the
baryon octet $B$ with a degenerate mass $m_B$ and the pseudoscalar meson octet
giving rise to the non-linear Sigma field $\Sigma$
\begin{equation}
\Sigma=\exp (2i\pi/f)
\end{equation}
with $\pi = \pi_\alpha \tau_\alpha (Tr \tau_\alpha \tau_\beta = \frac{1}{2} \delta_{\alpha \beta })$.
The pseudoscalar meson
decay constants f are all equal in the SU(3)$_V$ limit and are denoted by
$f=f_\pi \simeq 93$ MeV.
Using
\begin{equation}
\xi=\sqrt\Sigma=\exp(i\pi/f)
\end{equation}
the meson vector $V_\mu$ and axial vector
$A_\mu$ currents are defined as
\begin{equation}
V_\mu={1\over 2}(\xi^+\partial_\mu\xi+\xi\partial_\mu\xi^+)~~{\rm and}~~
A_\mu={i\over 2}(\xi^+\partial_\mu\xi-\xi\partial_\mu\xi^+),
\end{equation}
respectively. In chiral perturbation theory the meson - baryon
Lagrangian is formulated in terms of operators which are invariant
under SU(3)$_L\times$SU(3)$_R$. Expanding this Lagrangian in powers
of $\frac{m_\pi}{\Lambda}$ and $\frac{\partial }{\Lambda}$ is the
chiral symmetry breaking scale which is of the order of 1 GeV and
neglecting terms with four or more baryon fields as well as terms of
the order $\frac{m_\pi}{\Lambda}$ and $\frac{\partial }{\Lambda}$ and higher one obtains
the nonlinear chiral Lagrangian
\begin{eqnarray}
{\cal L}&=&{1\over 4}f^2{\rm Tr}\partial^\mu\Sigma\partial_\mu\Sigma^+
+{1\over 2}f^2\Lambda[{\rm Tr}M_q(\Sigma-1)+{\rm h.c.}]
+{\rm Tr}{\bar B}(i\gamma^\mu\partial_\mu-m_B)B\nonumber\\
&+&i{\rm Tr}{\bar B}\gamma^\mu[V_\mu, B]
+D{\rm Tr}{\bar B}\gamma^\mu\gamma^5\{A_\mu, B\}
+F{\rm Tr}{\bar B}\gamma^\mu\gamma^5[A_\mu, B]\nonumber\\
&+&a_1{\rm Tr}{\bar B}(\xi M_q\xi+{\rm h.c.})B
+a_2{\rm Tr}{\bar B}B(\xi M_q\xi+{\rm h.c.})\nonumber\\
&+&a_3[{\rm Tr}M_q\Sigma+{\rm h.c.}]{\rm Tr}{\bar B}B + {\cal O}(\frac{m_\pi}{\Lambda})+ {\cal O}(\frac{\partial}{\Lambda}). \label{LAG}
\end{eqnarray}
The current quark mass matrix is given by $M_q={\rm
diag}\{m_q,m_q,m_s\}$, neglecting the small difference between
the up and down quark masses. For the reduction of the different
terms to the kaon-nucleon problem we follow here the very clear
and straightforward presentation by Ko and Li \cite{Ko:1996yy}. For a more detailed discussion
of the effective field theory treatment of the kaon nucleon interaction we refer to
\cite{Kaiser:1995eg,Lee:1994my}. Terms
involving the axial vector current can be ignored as they have no
effects on the kaon mass. Expanding $\Sigma$ to order of $1/f^2$
and keeping explicitly only the kaon field, the first two terms in
Eq.~(\ref{LAG}) can be written as
\begin{equation}
\partial^\mu{\rm \bar K}\partial_\mu {\rm K}-\Lambda(m_q+m_s){\rm \bar KK},
\end{equation}
where
\begin{equation}
K=\left(\matrix{{\rm K}^+ \cr
                {\rm K}^0 \cr}\right)~~
{\rm and} ~~\bar K=({\rm K}^- ~~\bar {{\rm K}^0}).
\end{equation}

Keeping explicitly only the nucleon and kaon, the third and fourth
terms in Eq.~(\ref{LAG}) become
\begin{equation}
{\bar N}(i\gamma^\mu\partial_\mu-m_B)N
-{3i\over 8f^2}{\bar N}\gamma^0 N
\bar K \buildrel \leftrightarrow\over \partial_t K,
\end{equation}
where
\begin{equation}
N=\left(\matrix{{\rm p} \cr
                {\rm n} \cr}\right)~~{\rm and}~~\bar N=({\rm \bar p}~~ {\rm \bar
                n}).
\end{equation}
The last three terms in Eq.~(\ref{LAG}) can be similarly worked
out, and the results are
\begin{eqnarray}
{\rm Tr}{\bar B}(\xi M_q\xi+{\rm h.c.})B&=&2m_q{\bar N}N-{{\bar N}N\over 2f^2}
(m_q+m_s){\bar K}K,\nonumber\\
{\rm Tr}{\bar B}B(\xi M_q\xi+{\rm h.c.})&=&2m_s{\bar N}N-{{\bar N}N\over f^2}
(m_q+m_s){\bar K}K,\nonumber\\
\,[{\rm Tr}M_q\Sigma+{\rm h.c.}]{\rm Tr}{\bar B}B&=&2(2m_q+m_s){\bar N}N
-{2{\bar N}N\over f^2}(m_q+m_s){\bar K}K.
\end{eqnarray}
Combining the above expressions, one arrives at the Lagrangian,
\begin{eqnarray}
{\cal L}&=&{\bar N}(i\gamma^\mu\partial_\mu-m_B)N
+\partial^\mu{\bar K}\partial_\mu K-\Lambda(m_q+m_s){\bar K}K\nonumber\\
&-&{3i\over 8f^2_{\rm K}}{\bar N}\gamma^0 N \bar K \buildrel
\leftrightarrow\over \partial_t K
+[2a_1m_q+2a_2m_s+2a_3(2m_q+m_s)]{\bar N}N\nonumber\\
&-&{{\bar N}N{\bar K}K\over 2f^2_{\rm K}}(m_q+m_s)(a_1+2a_2+4a_3)
\label{LAG1}
\end{eqnarray}
where one can identify the kaon mass as
\begin{equation}
m_{\rm K}^2=\Lambda(m_q+m_s)
\end{equation}
and the nucleon mass as
\begin{eqnarray}\label{mn}
m_{\rm N}=m_B-2[a_1m_q+a_2m_s+a_3(2m_q+m_s)].
\end{eqnarray}
The ${\rm KN}$ sigma term is given as
\begin{eqnarray}
\Sigma_{{\rm KN}}&\equiv&{1\over 2}(m_q+m_s)\langle N|{\bar
u}u+{\bar s}s|
N\rangle\nonumber\\
&=&{1\over 2}(m_q+m_s)\Big[{1\over 2}\frac{\partial m_{\rm
N}}{\partial m_q}+
\frac{\partial m_{\rm N}}{\partial m_s}\Big]\nonumber\\
&=&-{1\over 2}(m_q+m_s)(a_1+2a_2+4a_3).
\label{sigkn}
\end{eqnarray}
The second line in the above equation follows from explicit chiral
symmetry breaking in the QCD Lagrangian, and the last step is
obtained using Eq.~(\ref{mn}).   Then Eq.~(\ref{LAG1}) can be
rewritten as
\begin{eqnarray}
{\cal L}&=&{\bar N}(i\gamma^\mu\partial_\mu-m_{\rm N})N
+\partial^\mu{\bar K}\partial_\mu K
-(m_{\rm K}^2-{\Sigma_{{\rm KN}}\over f^2_{\rm K}}{\bar N}N){\bar K}K\nonumber\\
&-&{3i\over 8f^2_{\rm K}}{\bar N}\gamma^0 N \bar K \buildrel
\leftrightarrow\over \partial_t K .\label{LAG2}
\end{eqnarray}
From the Euler-Lagrange equation and using the mean-field
approximation for the nucleon field, i.e.~ the factorization of products of four or
more field operators into products of bilinear terms over which one averages
independently and the replacement of  $\langle{\bar
N}\gamma ^0 N\rangle$ by the nuclear density $\rho_N$ and of
$\langle{\bar N}N\rangle$ by the scalar density $\rho_S$, one
obtains the following Klein-Gordon equation for a kaon in the
nuclear medium,
\begin{eqnarray}
\Big[\partial _\mu\partial ^\mu+{3i\over 4f^2_{\rm K}} \rho_N
\partial _t +\big(m_{\rm K}^2-{\Sigma _{{\rm KN}}\over f^2_{\rm K}}\rho
_S\big)\Big]K=0.\label{KG}
\end{eqnarray}
 The kaon dispersion relation in nuclear matter is then given by
\begin{eqnarray}
\omega^2({\bf k},\rho _N) =m_{\rm K}^2+{\bf k}^2 -{\Sigma_{{\rm
KN}}\over f^2_{\rm K}} \rho_S \pm {3\over 4}{\omega\over f^2_{\rm
K}}\rho_N,\label{DIS}
\end{eqnarray}
where {\bf k} is the three-momentum of the kaon and the
upper(lower) sign refers to K$^+$(K$^-$). The third term
in the above equation results from the attractive scalar interaction
due to explicit chiral symmetry breaking and depends on the
kaon-nucleon sigma term $\Sigma _{{\rm KN}}$. With a strangeness
content of the nucleon, normally taken as $y=2\langle N|\bar
ss|N\rangle /\langle N|\bar uu+\bar dd|N\rangle \approx 0.1-0.2$,
its value is $350<\Sigma_{{\rm KN}}<405$ MeV taking the
strange to light quark mass ratio to be $m_s/m_q\approx 29$. On
the other hand, recent lattice gauge calculations
\cite{Dong:1994zs,Fukugita:1994ba} show that $y\approx 0.33$ which
would give $\Sigma_{{\rm KN}}\approx 450$ MeV.  The last term in
Eq.~(\ref{DIS}) is due to the repulsive vector interaction and is
proportional to the nuclear density $\rho _N$. For a K$^-$ meson,
this term becomes attractive due to G parity.

From the dispersion relation, the in-medium kaon energy  can be
obtained
\begin{eqnarray}\label{omek}
E({\bf k}, \rho_N)=\left[m_{\rm K}^2+{\bf k}^2-{\Sigma_{{\rm
KN}}\over f^2_{\rm K}}\rho_S +\left({3\over 8}{\rho_N\over
f^2_{\rm K}}\right)^2\right]^{1/2} \pm {3\over 8}{\rho_N\over
f^2_{\rm K}}.\label{DIS1}
\end{eqnarray}
The in medium mass, defined as
\be
E({\bf k}\to 0, \rho_N),
\ee
is the central quantity for the discussion of the experimental observables which follows in the next chapters.
Using the KFSR relation ($m_\rho=2\sqrt{2}fg_\rho$)
and the SU(3) relation ($g_\omega=3g_\rho$) \cite{bhaduri}, the
kaon vector potential in the last term can be written as
$(1/3)(g_\omega/m_\omega)^2\rho_N$ which is just 1/3 of the
nucleon vector potential. This can be understood in the
constituent quark model since the kaon contains only one light
quark as compared to three in a nucleon. Since an antikaon has one
light antiquark, its vector potential becomes attractive, leading
to a reduction of its in-medium energy.

In their original papers, Kaplan and Nelson
\cite{Kaplan:1986yq,Nelson:1987dg}  obtained for both, K$^+$ and
K$^-$, only the mass correction term
\begin{equation}
\Delta m_{\rm K}^2(\rho) = - \frac{\Sigma_{{\rm KN}}}{f_{\rm K}^2}
\rho_S \label{kane}
\end{equation}
with $\Sigma_{{\rm KN}}$ as defined in Eq.~(\ref{sigkn}),
which reflects the Pauli blocking of the light quarks in matter.
With this approximation, with $f_{\rm K}=106$ MeV and
$\Sigma_{{\rm KN}} \approx 350 $ MeV and with the assumption that
the only medium modifications are those of Eq.~(\ref{kane}),
Eq.~(\ref{DIS}) predicts a kaon condensation around 3 $\rho_0$
which would have enormous consequences for supernovae explosions
as well as for heavy-ion collisions. Employing a Nambu
Jona-Lasinio (NJL) Lagrangian, Lutz {\em et al.}
\cite{Lutz:1994cf} showed later that there are additional
correction terms
\begin{equation}
\Delta m_{\rm K}^2(\rho) = - \frac{\Sigma_{{\rm KN}}}{f_{\rm K}^2}
\rho_S - \frac{m_{\rm K}^2\Delta f_{\rm K}^2}{f_{\rm K}^2} \pm
\frac{m_{\rm K}(\rho_u-\rho_s)}{4 f_{\rm K}^2} +
{\cal{O}}(m^0_u,s). \label{lut}
\end{equation}
which are not at all negligible. $\Delta f_{\rm K}^2$ is the
change of the kaon decay constant in matter which reflects the
change of the intrinsic wave function and $\rho_u(\rho_s)$ are the
{\it u} and {\it s} quark densities, respectively. Calculations in
the framework of the NJL model showed that these two terms dominate the
first term and that therefore the mass of the K$^+$ increases in
matter as predicted by the low-density theorem. These correction
terms are very close to the terms of the chiral Lagrangian which
have been neglected by Kaplan and Nelson. Assuming that $\omega
\approx m_{\rm K}$ and $\rho_s$ = 0, the last terms of
Eqs.~(\ref{DIS}) and (\ref{lut}) differ only by the decay
constants $f_{\rm K}$ and $f = f_\pi$. Their numerical values
differ  by 10 \% only.

\subsubsection{K$^+$ mesons} 
\begin{figure}[htb]
     \includegraphics[width=0.6\textwidth]{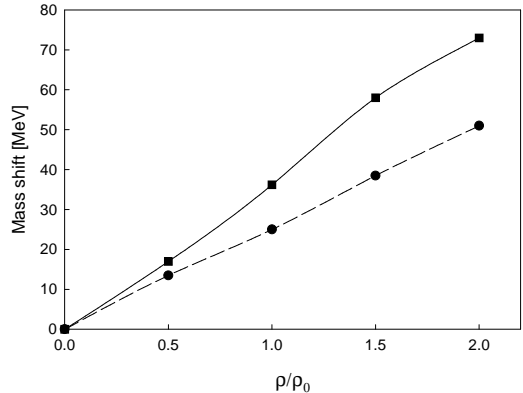}
\vspace*{-1.8cm} \caption[]{The K$^+$ mass shift as a function of
the nucleon density, with $\rho_0$ denoting the saturation
density. The circles show the results based on the vacuum
kaon-nucleon scattering amplitude, while the squares correspond to
the self-consistent result
\cite{Korpa:2004ae}. } \label{fig4}
\end{figure}

\begin{figure}[htb]
\vspace*{-0.2cm}
     \includegraphics[width=0.6\textwidth]{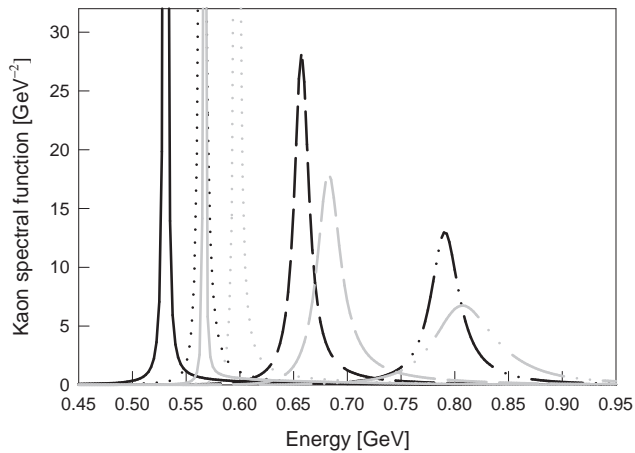}
\vspace*{-1.cm}
\caption[]{The kaon spectral function for different momenta:
0 (solid line), $200\,$MeV (dotted line), $400\,$MeV (dashed line),
and $600\,$MeV (dash-dot-dot line). Black lines show the result
at saturation density, gray lines, at twice
saturation density. \cite{Korpa:2004ae}.
}
\label{fig3}
\end{figure}

The results for the in-medium
properties of the K$^+$ obtained in the chiral perturbation
theory, in the NJL approach or in self-consistent calculations
\cite{Korpa:2004ae} agree among each other and at low density also
with those obtained by analyzing the scattering length.
Figure~\ref{fig4} shows the mass shift calculated by
Korpa and Lutz \cite{Korpa:2004ae} in comparison with that from
the scattering length analysis. Figure \ref{fig3} shows the spectral function of the
K$^+$ at $\rho_0$ and at $2\rho_0$ \cite{Korpa:2004ae}
demonstrating that the K$^+$ is a good quasi-particle. Therefore
it is a good approximation to propagate it as stable particle in
the simulation programs.

\subsubsection{K$^-$ mesons}
As discussed above, the K$^-$ mesons can easily form resonances
with baryons and its coupling to these resonances modifies its
medium properties. The repulsive free-space scattering of K$^-$
mesons is dominated by the presence of the zero isospin
$\Lambda(1405)$ resonance. At zero density the interaction is
repulsive but already at very moderate densities the potential
becomes attractive. The rapid change in sign cannot be understood
in the framework of a simple $t\rho$ approximation
(Eq.~(\ref{trho})). Also an expansion in the chiral parameter
$\Lambda$ is not sufficient to understand the properties of \km
mesons in matter.

Motivated by the K$^-$ data obtained at the SIS facility at GSI in
Darmstadt, many efforts have been advanced to understand the K$^-$
properties in matter as we will discuss now. All present-day
calculations use a self-consistent approach for the calculation
the K$^-$ properties. The starting point of these renewed efforts
is the on-shell scattering amplitude
\begin{eqnarray}
\langle M^{i}(\bar q)\,B^k(\bar p)|\,T\,| M^{j}(q)\,B^l(p) \rangle
&=&(2\pi)^4\,\delta^4(p+q-\bar p-\bar q)\,
\nonumber\\
&& \! \times\,\bar u_B^k(\bar p)\,
T^{ij}_{M^i \ B^k \rightarrow M^j B^l}(\bar q,\bar p ; q,p)\,u_B^l(p)
\label{on-shell-scattering}
\end{eqnarray}
where  $M$ and $B$ represent the different mesons (K$^-, \pi,
\eta$) and baryons (N,$\Lambda,\Sigma, \Xi$), respectively,
including their corresponding quantum numbers. The term
$\delta^4(...)$ guarantees energy-momentum conservation and
$u_B(p)$ is the nucleon isospin-doublet spinor. In quantum field
theory the free space scattering amplitude $T_{M B \rightarrow M
B}$ is obtained by solving the Bethe-Salpeter matrix equations
\begin{eqnarray}
T(K ,k ;w ) &=& K(K ,k ;w ) +\int \frac{d^4l}{(2\pi)^4}\,K(K , l;w
)\, G(l;w)\,T(l,k;w )\;,
\nonumber\\
G(l;w)&=&-i\,S_B({\textstyle {1\over 2}}\,w+l)\,D_M({\textstyle
{1\over 2}}\,w-l)\,, \label{BS-eq} \label{mb}
\end{eqnarray}
defined in terms of the elementary MB $\to$ MB scattering kernel $K(K,k;w
)$, the baryon propagator $ S_N(p)=1/(\pslash - m_B +
i\,\epsilon)$, and the pion propagator
$D_\pi(q)=1/(q^2-m_M^2+i\,\epsilon)$. In Eq.~(\ref{BS-eq})
convenient kinematical variables are used:
\begin{eqnarray}
w = p+q = \bar p+\bar q\,, \quad k= {\textstyle{1\over
2}}\,(p-q)\,,\quad K ={\textstyle{1\over 2}}\,(\bar p-\bar q)\,.
\label{def-moment}
\end{eqnarray}

The K$^-$N channel has isospin $I=0$ or $I=1$. Neglecting the
$\eta $ and $\Xi$ (which due to their mass do not play an
important role but have to be taken into account for a
quantitative comparison with data \cite{Oset:1997it}), it can only couple
to the $\pi \Sigma$ channel and the corresponding I=0 matrix has
the following structure

\[\left(\begin{array}{cc}
T_{{\bar{K}}N\rightarrow{\bar{K}}N} &
T_{{\bar{K}}N\rightarrow{\pi}{\Sigma}} \\
T_{{\pi}{\Sigma}\rightarrow{\bar{K}}N} &
T_{{\pi}{\Sigma}\rightarrow{\pi}{\Sigma}}
        \end{array}
\right) \ ,\]
whereas for $I=1$ it can couple to both, the $\pi \Sigma$ and the $\pi \Lambda$
channel,

\[\left(\begin{array}{ccc}
T_{{\bar{K}}N\rightarrow{\bar{K}}N} &
T_{{\bar{K}}N\rightarrow{\pi}{\Sigma}} &
T_{{\bar{K}}N\rightarrow{\pi}{\Lambda}}  \\
T_{{\pi}{\Sigma}\rightarrow{\bar{K}}N} &
T_{{\pi}{\Sigma}\rightarrow{\pi}{\Sigma}} &
T_{{\pi}{\Sigma}\rightarrow{\pi}{\Lambda}}  \\
T_{{\pi}{\Lambda}\rightarrow{\bar{K}}N} &
T_{{\pi}{\Lambda}\rightarrow{\pi}{\Sigma}} &
T_{{\pi}{\Lambda}\rightarrow{\pi}{\Lambda}}
        \end{array}
  \right) \ .\]

The free scattering amplitude, Eq.~(\ref{on-shell-scattering}), is
used to determine the elementary interaction vertices K by comparing the
calculation with the available KN $\rightarrow$ MB differential
cross sections in a partial wave expansion.

In the medium, the Bethe-Salpeter equation takes the form
\begin{equation}
{\mathcal T}={\mathcal K}+{\mathcal K}\cdot{\mathcal
G}\cdot{\mathcal T} \label{hatt}
\end{equation}

with ${\mathcal K}$ and ${\mathcal G}$ being the in-medium interaction
and propagator, respectively.

This equation is only formal because ${\mathcal
K}$ is not known. Working with effective
meson-baryon theories, in which ${\mathcal K}$ cannot be derived
from a more fundamental QCD approach, approximations are required.
Making the assumption that the interaction does not change in a
hadronic environment by identifying ${\mathcal K}$ with K one can
proceed to quantitative results.

\begin{figure}

\includegraphics[width=0.4\textwidth]{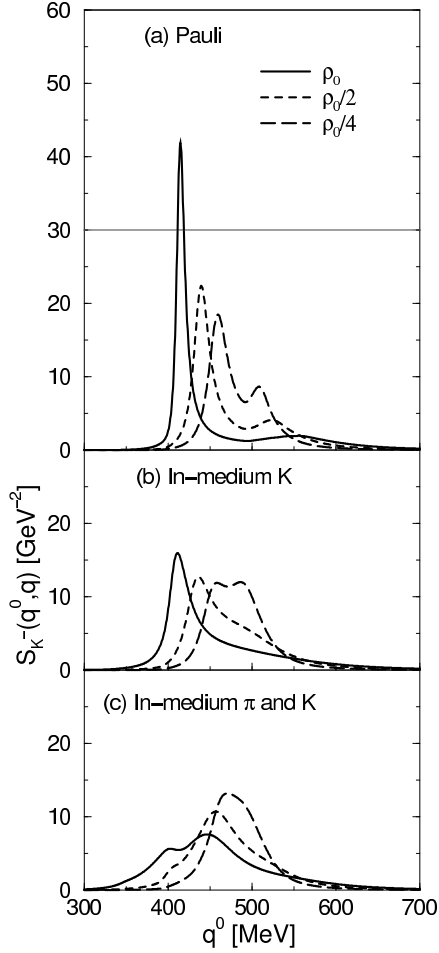}
\caption{ K$^-$ spectral density for zero momentum as a function
of energy at several densities: $\rho_0$ (solid line), $\rho_0/2$
(short-dashed line) and $\rho_0/4$ (long-dashed line). Results are
shown for the three approximations discussed in the text: a) Pauli blocking
(top panel), b) Pauli blocking and in-medium kaons (middle panel) and c) Pauli
blocking and in-medium pions and kaons (bottom panel). The figure is taken from
Ramos et al. (Ref.~\cite{Ramos:1999ku}). \label{fig:specka}}
\end{figure}

Particles feel the hadronic environment in different ways as
studied in Refs.~\cite{Ramos:1999ku,Lutz:1997wt}. First of all,
the Pauli principle prevents scattering into intermediate nucleon
states below the Fermi momentum. Thus, the free nucleon propagator
$S_B^0$ is replaced by an in-medium propagator which projects on
the available states. The consequences of the Pauli blocking of
the states below the Fermi surface is shown in
Fig.~\ref{fig:specka} (top), taken from  Ref.~\cite{Ramos:1999ku}.
At a moderate density a double-hump structure is seen. The left
hump is the kaon pole whereas the right one corresponds to the
$\Lambda (1405)$ - hole state which at this density is located
above the K$^-$ mass. With increasing density the right hump
disappears and at $\rho_0$ the strength of the K$^-$ is
concentrated around a peak value which is 80 MeV lower than the
mass of the free K$^-$ and hence the potential is attractive.
There remains, however, a long tail toward higher energies. The
phase-space restriction of the Pauli principle has  shifted the
center of the $\Lambda (1405)$ resonance above the K$^-$N
threshold.

The second medium modification of the free propagators is due to
the interaction of the K$^-$ with the hadronic environment which
is encoded in the self-energies $\Pi_{i}(\omega,{\bf q},\rho) $.
For the K$^-$ the free propagator $D^{-1}_0$ has to be replaced
by the dressed propagator $D^{-1}$
\begin{equation}
D^{-1}_0 \propto \omega^2 - {\bf q^2} - m_{\rm K}^2 \rightarrow
D^{-1}\propto \omega^2 - {\bf q^2} - m_{\rm K}^2 -
\Pi_{{\rm K}^-}(\omega,{\bf q},\rho) \ . \label{eq:disp1}
\end{equation}

The middle part of Fig.~\ref{fig:specka}  shows the spectral
function if we use both, the Pauli principle and the dressed K$^-$
propagator. The spectral function broadens and the energy of the
$\Lambda (1405)$ is lowered because the attractive K$^-$N
interaction counteracts the increase of the energy of the $\Lambda
(1405)$ - hole state \cite{Lutz:1997wt} due to the Pauli
principle. Consequently, the two separate peaks merge into one. In
addition, the strength in the K$^-$ channel is lowered.

Finally one also has to dress the pion propagator. The result of
this is shown in the bottom part of Fig.~\ref{fig:specka}. For the
choice of the dressing of Ref.~\cite{Ramos:1999ku} the spectral
function becomes very broad which raises the question whether the
K$^-$ can still be treated as a quasi particle, i.e. with a fixed
relation $\omega^2 = m^{*2} + \bf{q}^2$. The way how the pion
dressing has to be done is currently very much debated and
different suggestions gave different results for the spectral
function \cite{Lutz:1997wt,Tolos:2002ud,Friedman:1993cu}.

The other particles which appear in Eq.~(\ref{mb}) have large
masses. Therefore, their dressing does not influence the results
in the K$^-$ sector.

Limiting the discussion to those theories in which the in-medium
modification of the K$^-$ propagators are calculated
self-consistently, one finds essentially two different approaches,
that of the group of Ramos, Oset and Tolos
\cite{Ramos:1999ku,Oset:2000eg,Rios:2005ps,Tolos:2002ud,Tolos:2000fj}
and that of Lutz and Korpa \cite{Lutz:1997wt,luhab,Lutz:2001dq}.
Both of them became more and more refined over the years.
Sometimes the calculations are limited to s-wave scattering
whereas others include p-waves in order to have a wider dynamical
range for constraining the parameters. Both use the same strategy
to determine the parameters of the K$^-$N interaction in fitting
the available free scattering data and both use this interaction
subsequently in a Bethe-Salpeter equation. The models differ in
the form of the interaction as well as in the dressing of the
propagators and in the regularization of the loop integrals of
Eq.~(\ref{BS-eq}).

\subsubsection{Approach by Ramos et al.}
Ramos, Oset and Tolos \cite{Oset:2000eg,Tolos:2002ud} start from
the chiral Lagrangian in lowest order in momentum and with the
complete set of the pseudoscalar mesons and the baryon $1/2^+$
octet including s- and p- waves. Later Tolos {\em et
al.}~\cite{Tolos:2006ny} added self-consistency in the K$^-$
propagator. Projecting onto positive-energy states in the Dirac
equation for the nucleons they obtained a three-dimensional
equation of the Br\"uckner-Goldstone type

\begin{eqnarray}
\langle M_1 B_1 \mid T(\Omega) \mid M_2 B_2 \rangle &&= \langle M_1 B_1
\mid V({\sqrt s}) \mid M_2 B_2 \rangle   \nonumber \\
&& \hspace*{-2cm}+\sum_{M_3 B_3} \langle M_1 B_1 \mid V({\sqrt s }) \mid
M_3 B_3 \rangle
\frac {Q_{M_3 B_3}}{\Omega-E_{M_3} -E_{B_3}+i\eta} \langle M_3 B_3 \mid
T(\Omega)
\mid M_2 B_2 \rangle
\label{eq:spen1}
\end{eqnarray}
with
\begin{equation}
 E_{M_i(B_i)}(k,\omega)=\sqrt{k^2 +m_{M_i(B_i)}^2} + \Pi_{M_i(B_i)}
(k,\omega) \ .
\label{eq:spen}
\end{equation}
$Q_{M_3 B_3}$ is the Pauli operator. Due to the possible decay
into the $\pi$-hyperon channel the self-energies
$\Pi_{M_i(B_i)}(k,\omega) $ become complex. The optical K$^-$N
potential $U_{\rm opt}(k,\rho)$  is obtained  via the relation
\begin{equation}
U_{\rm opt}(k,\rho)= \frac{\Pi_{{\rm
K}^-}(k,\omega=\sqrt{k^2+m_k^2})}{2\sqrt{k^2+m_k^2}}. \label{vopt}
\end{equation}
The baryon potential is parameterized in different ways
with values of
about -70 MeV for nucleons and -30 MeV for hyperons  at $\rho_0$.
The K$^-$ self-energy is calculated self-consistently on the
quasi-particle level using
\begin{equation}
 \Pi_{{\rm K}^-}(k,\omega)= \sum_{N \leq F} \langle {\rm K}^- N \mid
 T_{{\rm K}^- N\rightarrow
{\rm K}^- N} (\Omega = E_{\rm N}+E_{{\rm K}^-}) \mid {\rm K}^- N
\rangle, \label{eq:self}
\end{equation}
where T is defined in Eq.~(\ref{eq:spen1}). An extension to
self-consistency on the level of the spectral function has been
given in Ref.~\cite{Oset:2000eg}. The self-energy of the pion is
taken from Ref.~\cite{Tolos:2002ud}.

\begin{figure}

\includegraphics[width=0.4\textwidth]{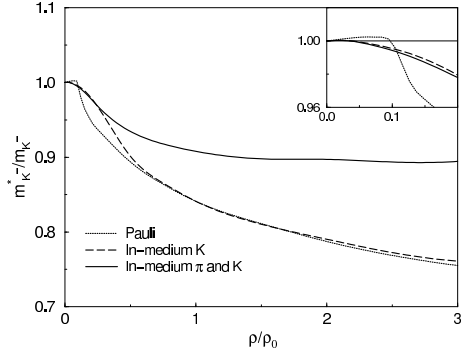}
\caption{ Effective mass of the K$^-$ meson in nuclear matter as a
function of density for the three approximations shown in Fig.~\ref{fig:specka} and discussed in the
text: {\it Pauli blocking} (dotted lines), {\it Pauli blocking and in-medium kaons} (dashed
lines) and {\it Pauli blocking and in-medium pions and kaons} (solid lines). The inset shows details at low densities.
The figure is taken from Ref.~\cite{Ramos:1999ku}
\label{fig:emass}}
\end{figure}

Given the self-energies of the K$^-$ and of the $\pi $, we can
write their propagators as
\begin{equation}
D_i(q^0,{\bf q},\rho) = \frac{1}{(q^0)^2-{\bf q\,}^2 - m_i^2 -
\Pi_i(q^0,\bf{q},\rho)}
\end{equation}
($i= {\rm K}^-, \pi$) with the corresponding spectral densities
\begin{equation}
S_i(\omega,{\bf q},\rho)= -\frac{1}{\pi} {\rm Im}\,
D_i(\omega,{\bf q},\rho) = -\frac{1}{\pi}\frac{{\rm Im}
\Pi_i(\omega,{\bf q},\rho)} {\mid \omega^2-{\bf q}\,^2-m_i^2-
\Pi_i(\omega,{\bf q},\rho) \mid^2} \ .
\end{equation}
The spectral density obtained in this approach for the K$^-$ has
already been shown in Fig.~\ref{fig:specka}. The effective
mass of the K$^-$, the solution of the dispersion relation
\begin{equation}
\omega^2 = {\bf q}\,^2 + m_{\rm K}^2 + {\rm Re\, } \Pi_{{\rm
K}^-}(\omega,{\bf q},\rho) \  \label{eq:disp}
\end{equation}
for $\bf{q} =0 $, is displayed as a function of density in
Fig.~\ref{fig:emass}. The effective mass of the K$^-$ is reduced
but by far not as much as shown in the original paper of Kaplan
and Nelson.

\begin{figure}[htb]
\centerline{
     \includegraphics[width=0.4\textwidth]{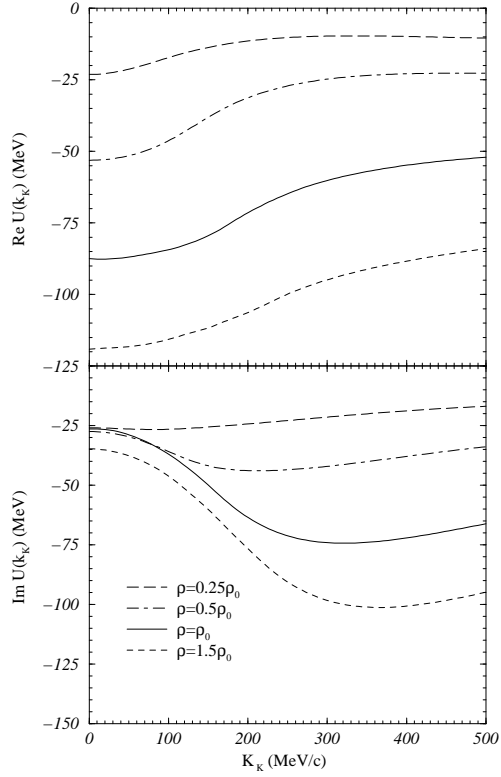}
}
      \caption{\small
Real and imaginary parts of the K$^-$ optical potential as
functions of the K$^-$ momentum for various densities. The figure is taken from
Ref.\cite{Tolos:2000fj}.}
        \label{fig:kaonramos}
\end{figure}

\begin{figure}[htb]
\centerline{
     \includegraphics[width=0.6\textwidth,angle=-90]{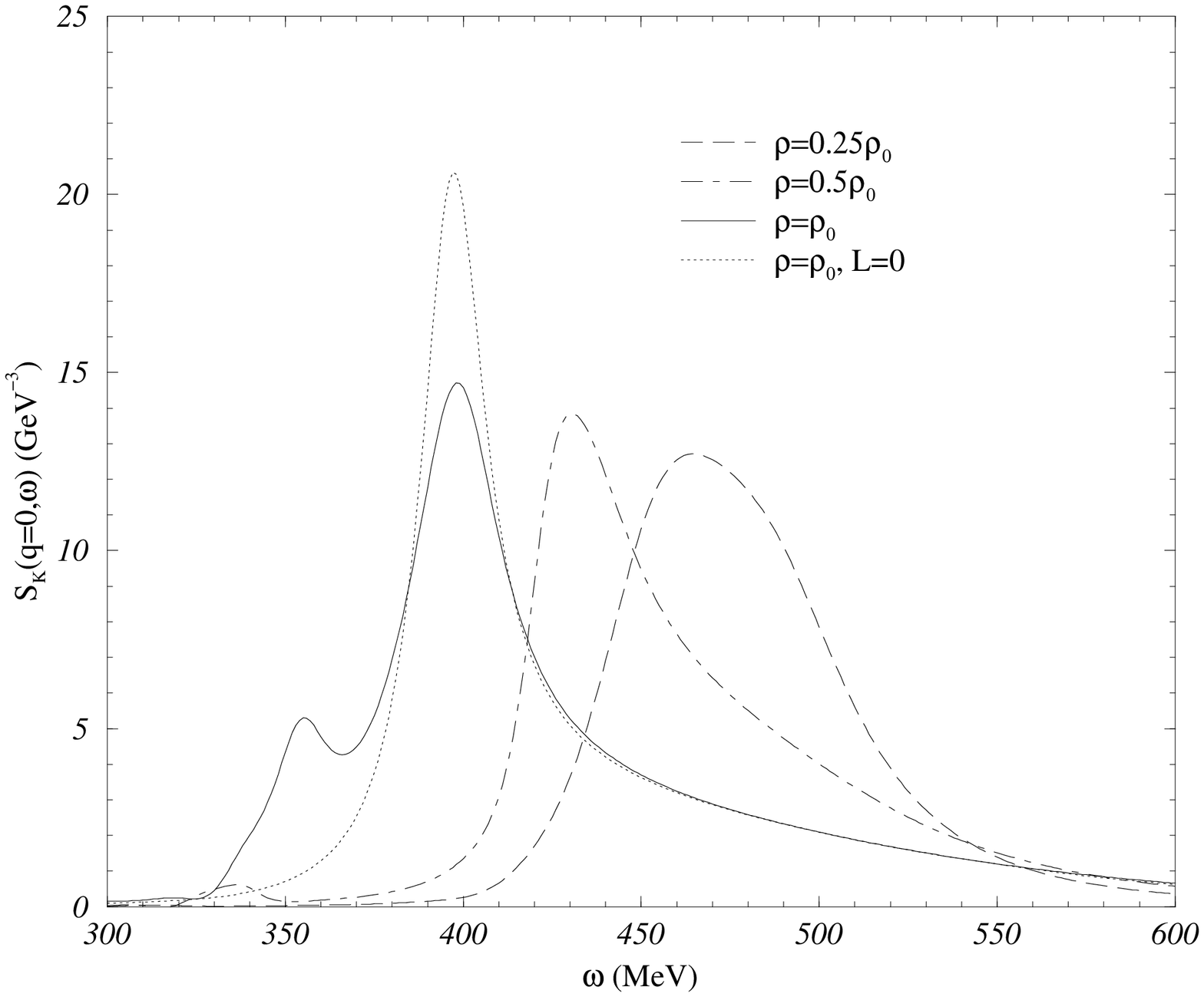}
}
      \caption{\small
$\bar{K}$ spectral density at $k_{\bar{K}}=0$ as a function of energy
 for various densities. The curve L=0 includes s-wave only.
 The figure is taken from Ref.\cite{Tolos:2000fj}.}
        \label{fig:kaon7}
\end{figure}

\begin{figure}[htb]
\centerline{
     \includegraphics[width=0.6\textwidth,angle=-90]{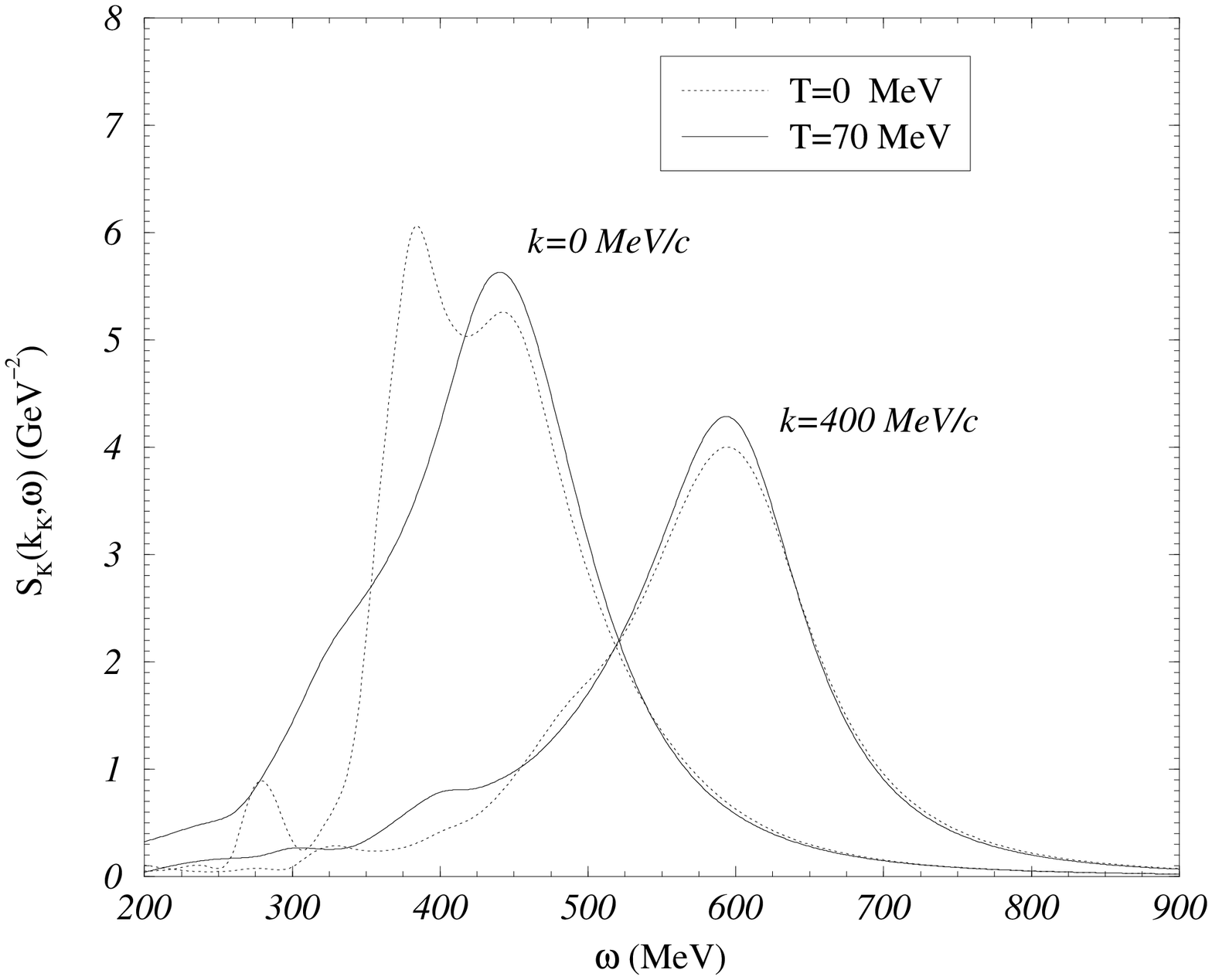}
}
      \caption{\small Spectral density of the K$^-$ at $\rho_0=0.17 \ \rm{fm^{-3}}$
      as a function of energy for K$^-$ momenta $k_{\bar{K}}=0$ and
      $k_{\bar{K}}=400 \ \rm{MeV/c}$, and for $T=0 \ \rm{MeV}$ and
      $T=70 \ \rm{MeV}$. The figure is taken from Tolos et al. ~\cite{Tolos:2002ud}.}
        \label{fig:temp8}
\end{figure}

Tolos {\em et al.}~\cite{Tolos:2000fj} use the same equation
(Eq.~\ref{eq:spen1}) but with two important changes: For the
interaction $V(\sqrt{s})$ they use the K$^-$N potential of the
J\"ulich group \cite{MuellerGroeling:1990cw} which is based on
meson-exchange  and which describes the available scattering data
also quite reasonably. To simplify the calculation they limit the
self-consistent calculation of the self-energy $\Pi_{{\rm
K}^-}(\omega,{\bf q},\rho)$ to the quasi-particle branch
$\Pi_{{\rm K}^-}^{qp}(\omega = \sqrt{m_k^2+k^2},{\bf q},\rho)$. In
addition they use a different parametrization of the baryon
potentials
\begin{equation}
U_{\Lambda,\Sigma}=-30\frac{\rho}{\rho_{0}},
\end{equation}
and
\begin{equation}
U_{N}(\rho,k)
=\alpha(\rho)+\frac{\beta(\rho)}{1+\left[
\displaystyle\frac{k}{\gamma(\rho)}\right]^2}
\
\label{eq:param}
\end{equation}
with density-dependent functions $\alpha,\beta,\gamma$. From the
self-consistent on-shell self-energy $\Pi_{{\rm K}^-}(k_{{\rm
K}^-},E_{{\rm K}^-}^{qp})$ one obtains via dispersion relation the
complete energy dependence of the self-energy, $\Pi_{{\rm
K}^-}(k_{{\rm K}^-},\omega)$, which can be used to determine the
 K$^-$ single-particle propagator in the medium,
\begin{equation}
D_{{\rm K}^-}(k_{{\rm K}^-},\omega) = \frac {1}{\omega^2 -k_{{\rm
K}^-}^2 -m_{{\rm K}^-}^2 -2 m_{{\rm K}^-} \Pi_{{\rm K}^-}(k_{{\rm
K}^-},\omega)} \ , \label{eq:prop}
\end{equation}
and the corresponding spectral density
\begin{equation}
S_{{\rm K}^-}(k_{{\rm K}^-},\omega) = - \frac {1}{\pi} {\mathrm
Im\,} D_{ {\rm K}^-}(k_{{\rm K}^-},\omega) . \label{eq:spec}
\end{equation}
The ${\rm K}^-$N potential of this approach is displayed in
Fig.~\ref{fig:kaonramos} and the spectral function in
Fig.~\ref{fig:kaon7}. The influence of the $K^-$ momentum
dependence of the optical potential on the observables in
heavy-ion collisions will be discussed in the $K^-chapter$.
%
Later this model has been extended to finite temperatures
\cite{Tolos:2002ud}. The results, displayed in
Fig.~\ref{fig:temp8}, show that at $\rho_0$ and a temperature of
70 MeV, which we encounter in reactions with beam energies around
$E_{\rm beam} = 1-2 A$ GeV, the general behavior does not change
but the double-hump structure of the spectral function has
disappeared. In heavy-ion collisions the K$^-$ therefore behaves
as a quasi particle albeit with a large width of about 150 MeV.

\subsubsection{Approach by Lutz et al.}
The work of Lutz {\em et al.}~\cite{luhab,Korpa:2004ae} starts out
as well from the Bethe-Salpeter equation (Eq.(~\ref{hatt})). It
uses, however, a different reduction to the three-dimensional
Br\"uckner-Goldstone equation. Assuming also that the
interaction vertices do not change in the medium, ${\mathcal K}=K$,
the in-medium reaction amplitude ${\mathcal T}$ can be expressed
in terms of the on-shell $T$ matrices and propagator modifications
$\Delta G \equiv {\mathcal G} - G$. In the case of the $I=0$
channel \cite{Korpa:2004ae} (denoting the channel
${\rm K}^- N$ by the index ``1'' and $\pi\Sigma$ by ``2'') we
write out the [11]-component of the matrix Bethe-Salpeter
equation:
\begin{equation}
{\mathcal T}_{11}=T_{11}+T_{11}\,\Delta G_1\,{\mathcal T}_{11}
+T_{12}\,\Delta G_2\,{\mathcal T}_{21}.
\label{comp11}
\end{equation}
Similarly, the [21]-component allows to solve for the ${\mathcal T}_{21}$ in
terms of ${\mathcal T}_{11}$ and ${\mathcal T}_{22}$:

\begin{equation}
{\mathcal T}_{21}=(1-T_{22}\,\Delta G_2)^{-1}\left[
T_{21}+T_{21}\,\Delta G_1\,{\mathcal T}_{11}\right].
\label{comp21}
\end{equation}
Substituting Eq.~(\ref{comp21}) into (\ref{comp11}) we obtain:
\begin{equation}
{\mathcal T}_{11}=\hat T_{11}+\hat T_{11}\,\Delta G_1\,{\mathcal T}_{11},
\label{comp11n}
\end{equation}
where we introduced
\begin{equation}
\hat T_{11}\equiv T_{11}+T_{12}\,\Delta G_2\,(1-T_{22}\Delta G_2)^{-1}\,T_{21}.
\label{that11}
\end{equation}

\begin{figure}[htb]
\centerline{
     \includegraphics[width=0.6\textwidth]{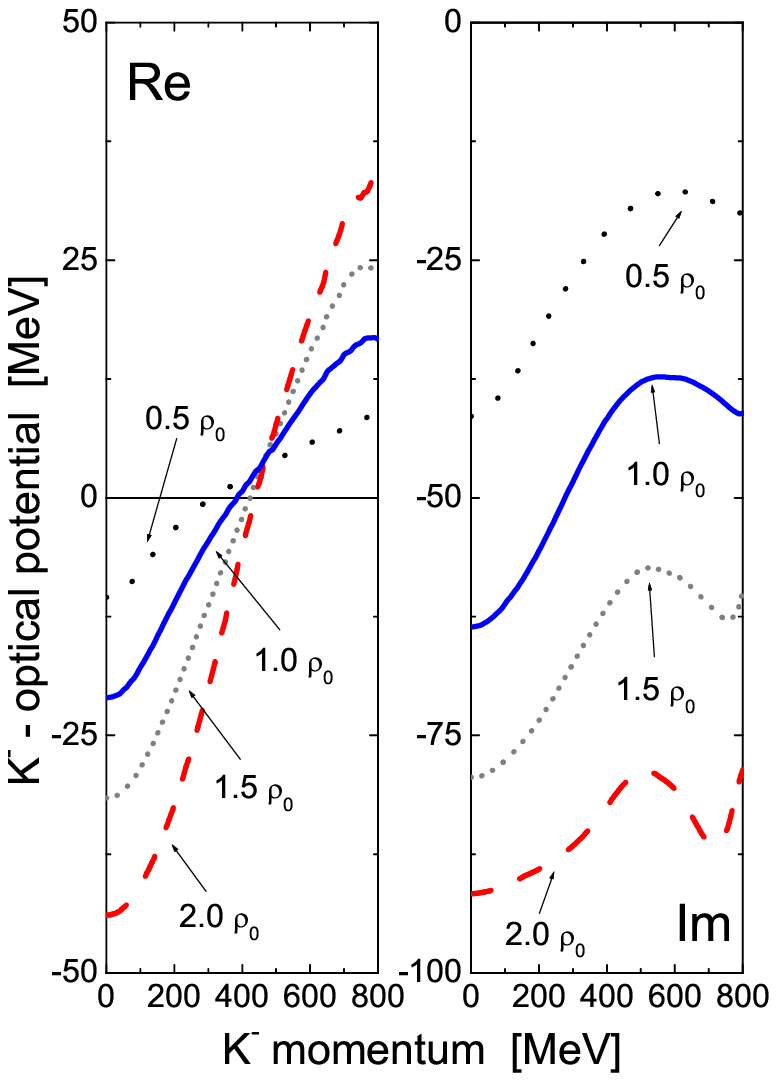}
}
      \caption{\small
Real and imaginary parts of the ${{\rm K}^-}$ optical potential as
functions of the K$^-$ momentum for various densities from
Ref.~\cite{Lutz:1994cf}.}
        \label{fig:kaonlutz}
\end{figure}

\begin{figure}[htb]
\vspace*{-0.2cm}
      \includegraphics[width=0.6\textwidth]{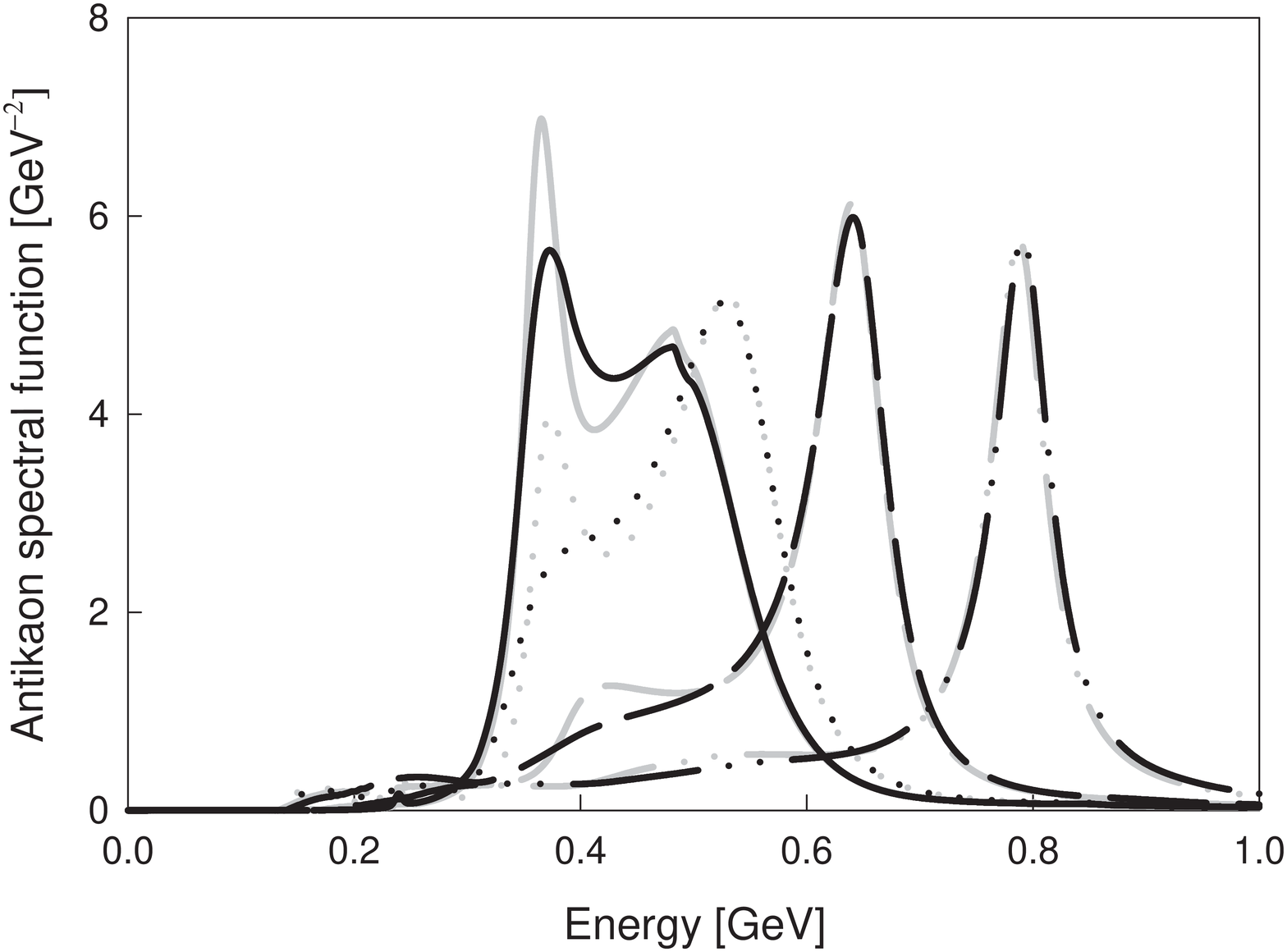}

\caption[]{The K$^-$ spectral function for different momenta: 0
(solid line), $200\,$MeV (dotted line), $400\,$MeV (dashed line),
and $600\,$MeV (dash-dot-dot line). Black lines show the result if
both, pion and K$^-$, are dressed \cite{Korpa:2004ae}, gray lines previous results
without pion dressing \cite{Lutz:2001dq}, in the nuclear matter at saturation density. } \label{fig2}
\end{figure}

In the approach of Lutz {\em et al.}~\cite{Lutz:2001dq} neither
nucleons nor $\pi$ are dressed. The form of the Bethe-Salpeter
equation with and without self-energy of the pions is formally identical
but the propagators are different. These pion propagators are input quantities
and not self-consistently obtained by the solution of the Bethe-Salpeter equation.
The interaction vertices are determined by an extensive comparison
of the available KN scattering data with a relativistic chiral
Lagrangian in the large $N_c$ limit which allows to reduce the
number of parameters to be adjusted. This comparison is made
using s- and p-waves in the partial-wave expansion. The K$^-$
nucleon potential which is obtained in this approach is displayed
in Fig.~\ref{fig:kaonlutz}, the spectral function in
Fig.~\ref{fig2}.


Comparing the three spectral functions (Figs.~\ref{fig:specka},
\ref{fig:kaon7} and \ref{fig2}) one finds that their general
structure is quite similar in the three approaches.
In absolute
magnitude as well as in detail they differ by factors up to two whereas the
models agree in the description of the \kp in dense matter.
Therefore, one has to conclude that the \km properties in matter
are much less known than those of the \kp. More important for
our numerical simulations is the fact that the extracted optical
potentials (Figs.~\ref{fig:kaonramos} and  \ref{fig:kaonlutz})
differ by about a factor of three at normal nuclear matter
density. Consequently, in the case of the K$^-$, the predictive
power of the results of the simulation programs is strongly
limited by the uncertainty of the K$^-$ properties the nuclear
matter.

\subsubsection{Kaonic Atoms}
Another source of information of the behavior of K$^-$ mesons in
matter are kaonic atoms. In kaonic atoms an electron is replaced
by a K$^-$ which, due to its larger mass, has a finite probability
to be in the tail region of the nucleus. Therefore, these systems
test the ${\rm K}^-$N potential at very low densities. Both, the
fits of phenomenological density-dependent optical potential to
the kaonic atom data \cite{Friedman:1993cu,Friedman:1994hx} and
relativistic mean field (RMF) calculations by Friedman {\em et
al.} \cite{Friedman:1998xa}, which describe the data
quantitatively very well, lead to  a deeply attractive potential
($-$Re $V_{\rm opt}(\rho_0) \approx 150-200$~MeV). Thus, the
potential values fitted to kaonic atoms are in conflict with the
results of the self-consistent approaches mentioned above which
give values of less or about 80 MeV at $\rho_0$ (see
Figs.~\ref{fig:kaonramos} and \ref{fig:kaonlutz}). Cieply {\em et
al.} \cite{Cieply:2001yg} were the first who tried to fit kaonic
atom data, low-energy K$^-$N and \km nucleus scattering data
simultaneously using a self-consistent coupled-channels approach,
similar to the one of Ref.~\cite{Ramos:1999ku}, for K$^-$ in
matter and assuming a Klein-Gordon equation of the form
\begin{equation}
\label{eq:KG1} \left[ \nabla^2 - 2{\mu}(B+V_{\rm opt} + V_{\rm
Coul}) + (V_{\rm Coul}+B)^2\right] \psi = 0~~ \;\;\;
\end{equation}
for the kaonic atoms. $B$ is the complex binding energy and
$V_{\rm Coul}$ is the Coulomb interaction of the
hadron with the finite-size nucleus, including vacuum-polarization terms.
Equation~(\ref{eq:KG1}) assumes that the optical potential
$V_{{\rm opt}}$ and also $V_c$ behave as a Lorentz scalars.  $V_{{\rm opt}}$ is given by a
$t(\rho)\rho$ form (see Eq.~\ref{trho})
\begin{equation}
\label{eq:Vopt}
2\mu V_{{\rm opt}}(r) =
 -{4\pi}(1+\frac{\mu}{M})[a_{{\rm K}^-{\rm p}}(\rho) \rho_{\rm p} (r)
+a_{{\rm K}^-{\rm n}}(\rho) \rho_{\rm n} (r)],
\end{equation}
where $M$ is the nucleon mass,
$\mu$ is the reduced
mass of the \km and the nucleus,
$a_{{\rm K}^-{\rm p}}$ and $a_{{\rm K}^-{\rm n}}$ are the
${\rm K}^-{\rm p}$ and ${\rm K}^-{\rm n}$ threshold scattering
amplitudes evaluated at nuclear matter density $\rho = \rho_{\rm
p} + \rho_{\rm n}$, and $\rho_{\rm p} (r)$ and $\rho_{\rm n} (r)$
are the proton and neutron density distributions.

These calculations confirmed that a slight modification of the
chiral meson-baryon interaction of
Ref.~\cite{Kaiser:1995cy,Kaiser:1996js} yields a relatively
shallow attractive real part of the potential ($\approx$ 55 MeV),
which is of about the same size as obtained from the ${\rm K}^-N$
scattering data if analyzed in the same chirally motivated coupled
channel approach. They confirm in addition that such a potential
can also fit well kaonic atoms with a $\chi^2$ value which is very
close to the value obtained for the deep attractive potential of
180 MeV of Refs.~\cite{Friedman:1998xa,Barnea:2006kv}. The authors
conclude that kaonic atom data cannot distinguish between shallow
and deep potentials and new experiments have to be designed to
clarify the depth of the ${\rm K}^-N$ potential. For the present
status of this discussion we refer to Ref.~\cite{Barnea:2006kv}.

\subsubsection{Deeply bound kaonic states in nuclei}
Motivated by the conjecture that a very strong \km  nucleon
potential could form a deeply bound ${\rm K}^-$ppn state
\cite{Akaishi:2002bg} a systematic experimental search has been
launched at the KEK facility. The first results by Suzuki {\em et
al.}~\cite{Suzuki:2005bj} showed a structure in the missing-mass
spectra in the ${}^4$He(${\rm K}^-_{\rm stopped},\thinspace {\rm
p}$) reaction which --- due to kinematic reasons --- cannot be due to
hypernucleus formation
 $({\rm K}^- +
 {}^4{\rm He})_{\rm atomic} \rightarrow
\pi^- + {}^4_{\Lambda}{\rm He}$, ${}^4_{\Lambda}{\rm He}
\rightarrow {\rm p} + {}^3_\Lambda{\rm H}$. This structure has
been identified with two deeply bound ${\rm \bar{K}NNN}$ states
\begin{equation}
({\rm K}^- + {}^4{\rm He})_{\rm atomic} \rightarrow {\rm
S}^0(3115) + {\rm p}
\end{equation}
and
\begin{equation}
({\rm K}^-  + {}^4{\rm He})_{\rm atomic} \rightarrow {\rm
S}^+(3140) + {\rm n}.
\end{equation}

\begin{figure}
     \includegraphics[width=9cm]{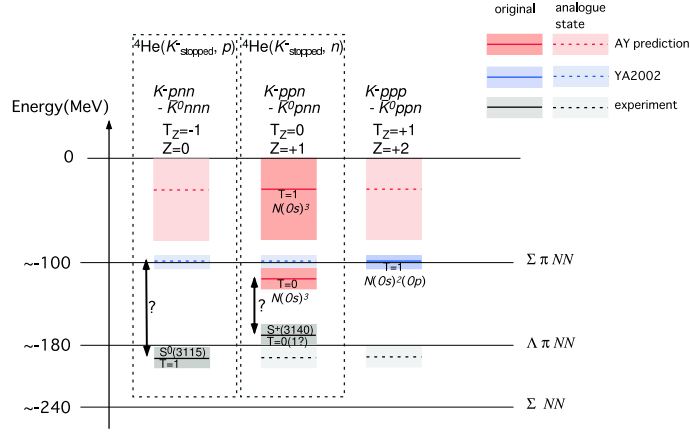}
  \caption{A comparison between non-relativistic calculations of deeply
bound kaonic nuclear states and experimental results. The
original \cite{Akaishi:2002bg},
 subsequent \cite{Yamazaki:2002uh,Dote:2003ac} predictions, and experimental results are
  represented by red, blue and grey, respectively.}\label{ex-th-comp}
\end{figure}

The properties of these states, however, and especially their binding energies
are far from the values predicted in Ref.~\cite{Akaishi:2002bg}
as seen in Fig.~\ref{ex-th-comp}.
More recently in the FINUDA experiment \cite{Agnello:2006jt} a
proton peak at a very similar proton momentum has been observed in
a reaction in which the $^4$He has been replaced by a $^6$Li. It
is not expected that for the two nuclei the binding energy of the
K$^-$ is identical. Hence this observation has reinforced the
interpretation of these structures as a K$^-$ absorption on a
quasi-deuteron. For the details of this interpretation of deeply
bound K$^-$ states we refer to Ref.~\cite{Oset:2005sn} which also
concludes that from deeply bound K$^-$ states there is presently
little experimental evidence for a strong $\bar{K}{\rm N}$
potential. Recently is has been argued
\cite{Ramos:2007vg,Magas:2008bp,Magas:2009qd} that there is
presently no experimental evidence for deeply bound kaonic
clusters.

\subsection{Strange baryons in matter}
The many-body scheme based on the Br\"uckner-Goldstone equation
which has been successfully applied to the description of kaons in
matter, was recently extended to the hyperonic sector
\cite{Schulze:1995jx,Schulze:1998jf,Vidana:1999jm}. It starts with
the construction of all the baryon-baryon (i.e., NN, YN and YY)
$G$-matrices, which describe in an effective way the interactions
between baryons in the presence of a surrounding baryonic medium
\begin{equation}
   G(\omega)_{B_1B_2,B_3B_4} =
   V_{B_1B_2,B_3B_4}+
  \displaystyle{ \sum_{B_5B_6}V_{B_1B_2,B_5B_6}
   \frac{Q_{B_5B_6}}{\omega-E_{B_5}-E_{B_6}+ i\eta}
   G(\omega)_{B_5B_6,B_3B_4} } \ .
   \label{eq:gmatrix} \end{equation}
In the above expression the first (last) two sub-indices indicate
the initial (final)  two-baryon states compatible with a given
value $S$ of the strangeness, namely NN for $S=0$, YN for
$S=-1,-2$, and YY for $S=-2,-3,-4$, $V$ is the bare baryon-baryon
interaction,
 $Q_{B_5B_6}$ is the Pauli operator which prevents the
intermediate baryons $B_5$ and $B_6$ from being scattered to
states below their respective Fermi momenta, and $\omega$, the
so-called starting energy, corresponds to the sum of
non-relativistic single-particle energies of the interacting
baryons (see Ref.~\cite{Vidana:1999jm} for details). The
single-particle energy of a baryon $B_i$ is given by
\begin{equation}
E_{B_i}=M_{B_i}+\frac{k^2}{2M_{B_i}}+{\mathrm Re}[U_{B_i}(k)] \ ,
\label{eq:spe}
\end{equation}
where $M_{B_i}$ denotes the rest mass of the baryon and the real
part of the single-particle potential $U_{B_i}(k)$ is the averaged
field ``felt'' by the baryon due to its interaction with the other
baryons of the medium. In the BHF approximation, $U_{B_i}(k)$ is
given by (see Eq.~\ref{eq:self}) \begin{equation}
       U_{B_i}(k) =
        \sum_{B_j}\sum_{k'}n_{B_j}(k')
       \left\langle \bf{k}\bf{k'}\right |
       G_{B_iB_j,B_iB_j}(\omega=E_{B_i}+E_{B_j})
       \left | \bf{k}\bf{k'} \right\rangle  \ ,
\label{eq:upot}
\end{equation}
where
\begin{equation}
n_{B_j}(k) =
\left\{
 \begin{array}{ll} 1 , \mbox{if $k \leq k_{F_{B_j}}$} \\ 0 ,
\mbox{otherwise} \end{array} \right.  \label{eq:ocnumb}
\end{equation} is the corresponding occupation number of the
species $B_j$. A sum over all the different baryon species is
performed and the matrix elements are anti-symmetrized when
baryons $B_i$ and $B_j$ belong to the same isomultiplet.

\begin{figure}[htp] \centering \includegraphics[width=9cm]{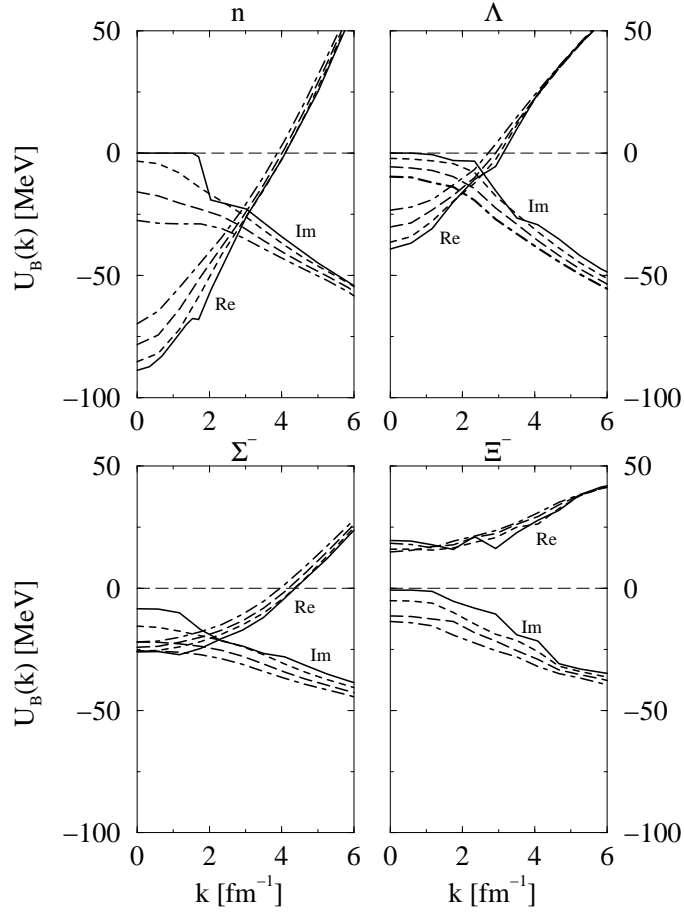}
\caption{Single-particle potential of the octet baryons at $\rho_0$ as
 a function of momentum for various temperatures: $T=0$ (solid lines),
$20$ MeV (short-dashed lines), $40$ MeV (long-dashed lines) and
$60$ MeV (dot-dashed lines). The figure is from
Ref.~\cite{Vidana:1999jm}.} \label{fig:udksym}
\end{figure}

The result of this calculation is shown in Fig.~\ref{fig:udksym}.
The potential of strange baryons is considerably shallower than
that of protons and neutrons and can even get repulsive for the
multi-strange baryons. These findings are in agreement with
measured $a_{\Lambda N} $ scattering length of about 2 fm
\cite{Rijken:1998yy}.

In the simulation of heavy-ion collisions, discussed in the
following, this reduction of the single-particle potential of strange baryons is
taken into account by assuming that the
strange quark is inert with respect to the hadronic mean field and
therefore $U_{\Lambda ,\Sigma} = 2/3 \ U_{{\rm p,n}}$.

\section{Microscopic Description within IQMD}
\label{micro}
\setcounter{figure}{0}
\newcommand{\chremark}[1]{}
The detailed analysis of strangeness production presented in this
article was performed using the Isospin Quantum Molecular Dynamics
model (IQMD)
\cite{Hartnack:1997ez,Hartnack:1989sd,hart,chhab,Bass:1995pj}, which is
a semi-classical model simulating heavy-ion collisions on an
event-by-event basis. We briefly sketch the model here and refer
for a detailed description to Refs.~\cite{hart,Hartnack:1997ez,chhab}.
Furthermore the derivation of the basic transport equations, the
description in terms of Skyrme potentials, and the considerations
for exploring the nuclear equation of state in heavy-ion
collisions can be found in Ref.~\cite{Stoecker:1986ci}. General
information on microscopic models of heavy-ion collisions and on
different numerical realizations are given in
Refs.~\cite{Cassing:1990dr,
Aichelin:1991xy,Bass:1998ca,Fuchs:2003pc}. In order to have a
comparison with other transport models, some of the results
presented in this report are as well calculated using the
Hadron-String-Dynamics model (HSD)
~\cite{Cassing:1996xx,Cassing:1999es,Cassing:2003vz} in the
version as presented in Ref.~\cite{Mishra:2004te}. This
microscopic model includes higher resonances in the baryonic and
mesonic sector and is therefore also applicable in the range of
several tens of GeV.
The results for antikaons \cite{Cassing:2003vz} in HSD have been
obtained in a version which includes off-shell propagation of
resonances.

\subsection{Nucleon-nucleon-potentials in IQMD}

In IQMD a particle is represented by the single-particle Wigner
density
\begin{equation} \Label{fdefinition}
 f_i ({\bf r, p},t) = \frac{1}{\pi^3 \hbar^3 }
 {\rm e}^{-\frac{2}{L} ({\bf r} - {\bf r_i} (t) )^2   }
 {\rm e}^{-\frac{L}{2\hbar^2} ({\bf p - p_i} (t) )^2 } \quad .
\end{equation}
The total one-body Wigner density is the sum of the Wigner densities of
all nucleons.
The particles move according to Hamiltonian equations of motion
\begin{equation}
\dot{r_i}=\frac{\partial\langle H \rangle}{\partial p_i} \qquad
\dot{p_i}=-\frac{\partial \langle H \rangle}{\partial r_i} \quad ,
\end{equation}

where the expectation value of the total Hamiltonian is
\begin{eqnarray}
\langle H \rangle &=& \langle T \rangle + \langle V \rangle
\nonumber \\ \Label{hamiltdef}
&=& \sum_i \frac{p_i^2}{2m_i} +
\sum_{i} \sum_{j>i}
 \int f_i({\bf r, p},t) \,
V({\bf r, r\,', p, p\,'})  f_j({\bf r\,', p\,'},t)\, \rm d{\bf r}\, \rm d{\bf r\,'}
\rm d{\bf p}\, \rm d{\bf p\,'} \quad.
\end{eqnarray}
The in-medium nucleon-nucleon potential interaction consists of
the real part of the Br\"uckner $G$-Matrix \cite{Jeukenne:1976uy}
with an additional Coulomb interaction between the charged
particles. The former is modelled here by a sum of several terms:
A contact interaction of Skyrme-type, a finite-range Yukawa term,
a momentum-dependent interaction, and a symmetry contact
interaction that distinguishes between protons and neutrons

%
\begin{eqnarray}
V({\bf r_i, r_j, p_i, p_j}) &=& G + V_{\rm Coul} \nonumber \\
       &=& V_{\rm Skyrme} + V_{\rm Yuk} + V_{\rm mdi} +
           + V_{sym} + V_{\rm Coul}  \nonumber \\
       &=& t_1 \delta ({\bf r_i} - {\bf r_j}) +
           t_2 \delta ({\bf r_i} - {\bf r_j}) \rho^{\gamma-1}({\bf r_i}) +
       \nonumber \\
       & &  t_3 \frac{\hbox{exp}\{-|{\bf r_i}-{\bf r_j}|/\mu\}}
               {|{\bf r_i}-{\bf r_j}|/\mu} + \Label{vijdef}  \\
       & & t_4\hbox{ln}^2 (1+t_5(\bf{p}_i-\bf{p}_j)^2)
               \delta ({\bf r_i} -{\bf r_j}) +\nonumber \\
       & & t_6 \frac{1}{\varrho_0}
                T_{3}^i T_{3}^j \delta({\bf r_i} - {\bf r_j}) + 
        \frac{Z_i Z_j e^2}{|{\bf r_i}-{\bf r_j}|} \nonumber.
\label{Vij}
\end{eqnarray}
The local Skyrme term contains an attractive term linear in the
baryonic density $\rho $ and a repulsive term  yielding a power law
dependence $\rho^{\gamma} $
which also simulates effectively many-particle correlations; for
more details see e.g.~Ref.~\cite{Stoecker:1986ci}. The
finite-range Yukawa term (with $t_3=-6.7$ MeV and $\mu=1.5$ fm) is
important to stabilize the surface of a finite nucleus.
The momentum dependence $V_{\rm mdi}$ of the NN interaction, which
may optionally be used in IQMD, is fitted to experimental data
\cite{Arnold:1982rf,pa67,Hama:1990vr} on the real part of the
nucleon optical potential
\cite{Aichelin:1987ti,Bertsch:1988ik,Hartnack:1994zz}. This yields
the parameters $t_4=1.57$ MeV and $t_5=5\cdot 10^{-4}$ MeV$^{-2}$.
The very simplified asymmetry term, where $T_{3}^i$ and $T_{3}^j$
denote the isospin projection $T_3$ of the nucleons $i$ and $j$
(i.e. 1/2 for protons and -1/2 for neutrons) with a strength of
$t_6 = 100$ MeV gives a symmetry energy which is linear in the
density difference of protons and neutrons.

For a nucleus in its ground state, the expectation value of the
total Hamiltonian has to correspond to its total binding energy.
When comparing to the Bethe-Weizs\"acker mass formula the kinetic
energy, the Skyrme interaction, and the momentum-dependent
interaction contribute to the volume energy, the Yukawa
interaction to the surface and the volume energy, and the symmetry
interactions to the volume symmetry energy. There is no term
corresponding to the pairing energy since it corresponds to a
global property of the nucleus which would be difficult to model
by microscopic local forces. Besides protons and neutrons the IQMD
model includes the $\Delta$ resonance as degree of freedom.
Concerning the Skyrme-, Yukawa- and momentum-dependent
interactions the $\Delta$'s are treated like ``heavy nucleons'',
i.e.~they interact with the same coupling constants as nucleons.
Coulomb interactions are performed with their real charges.
However, no symmetry interaction is assumed for the $\Delta$.

The nuclear equation of state (EoS), on the other hand, describes
the properties of infinite nuclear matter (without Coulomb
interactions) and is therefore given by the volume energy only.
The EoS describes the variation of the energy
$E(T=0,\rho/\rho_0)$
when changing the nuclear density to values different from the saturation
density $\rho_0$ for zero temperature. Often nuclear matter is assumed
to be isospin saturated, but we also consider asymmetric
nuclear matter, where the symmetry energy term contributes. In
fact, the density dependence of the symmetry energy has
recently been of great interest \cite{Li:2008gp}.

The single-particle potential resulting from the convolution of the distribution
functions $f_i$ and $f_j$ with the interactions
$V_{\rm Skyrme}+ V_{\rm Yuk}+V_{\rm mdi}$
(local interactions including their momentum dependence) is for symmetric nuclear matter
\begin{equation} \Label{eosinf}
U_i({\bf r_i},t) \,=\, \alpha
\left(\frac{\rho_{int}}{\rho_0}\right) +
        \beta \left(\frac{\rho_{int}}{\rho_0}\right)^{\gamma} +
        \delta \mbox{ln}^2 \left( \varepsilon
                \left( \Delta {\bf p} \right)^2 +1 \right)
                        \left(\frac{\rho_{int}}{\rho_0}\right) \quad ,
\label{Upot}
\end{equation}
where $\rho_{int}$ is the interaction density obtained by
convoluting the distribution function of a particle with the
distribution functions of all other particles of the surrounding
medium. $\Delta \bf{p}$ is the relative momentum of a particle
with respect to the surrounding medium.

The parameters $t_1 ... t_5$ in Eq.~(\ref{vijdef}) are uniquely
related to the coefficients $\alpha, \beta, \gamma, \delta$, and
$\epsilon$ in Eq.~(\ref{Upot}). Values of these parameters for the
different model choices can be found in Tab.~\ref{eostab}.

\begin{table}[hbt]
\begin{tabular}{lcccccc}
 &$\alpha$ (MeV)  &$\beta$ (MeV) & $\gamma$ & $\delta$ (MeV) &$\varepsilon \,
 \left(\frac{c^2}{\mbox{GeV}^2}\right) $ & $K$ (MeV) \\
\hline
 S  & -356 & 303 & 1.17 & ---  & ---  & 200   \\
 SM & -390 & 320 & 1.14 & 1.57 & 500  & 200 \\
 H  & -124 & 71  & 2.00 & ---  & ---  & 376  \\
 HM & -130 & 59  & 2.09 & 1.57 & 500  & 376 \\
\end{tabular}
\caption{Parameter sets for the nuclear equation of state used in
the IQMD model} \Label{eostab}
\end{table}

The parameters $\epsilon$ and $\delta$ are given by fits to the
experimentally measured optical potential. $E/A=-16$ MeV at the
ground-state density $\rho_0$ and $\rho_0=0.16$ fm$^{-3}$ are two
constants for the three parameters $\alpha, \beta, \gamma$. The
remaining degree of freedom is related to the compression modulus
$K$ of the nucleus, which corresponds to the curvature of the
volume energy at $\rho=\rho_0$ (for $T=0$) and is also given in
Tab. I
\begin{equation}
K  = -V\frac{{\rm d}p}{{\rm d}V}= 9 \rho^2
\frac{{\rm d}^2E/A(\rho)}{({\rm d}\rho)^2} |_{\rho=\rho_0} \qquad
.
\end{equation}

An equation of state with a rather low value
of the compression modulus $K$ yields a weak repulsion of
compressed nuclear matter and thus describes "soft" matter
(denoted by "S"). A high value of $K$ causes a strong repulsion of
nuclear matter under compression (called a "hard EoS", H). The
parameters of the momentum-dependent interactions $\delta$ and
$\epsilon$ can optionally be switched on (denoted "with mdi",
parametrization HM and SM) or off (denoted "no mdi").

In the calculations presented in this article the standard parametrization is SM.
It was shown to describe best many observables of intermediate
energy heavy-ion collisions, (see e.g.
\cite{Hartnack:1994bs,Stoicea:2004kp,Andronic:2004cp}) as will
also be seen for the kaon production in the later sections of this
article.

\subsection{Kaon-nucleon potentials}\label{knpot-desc}
The kaons and antikaons interact with the nuclear medium via a Schr\"odinger-type potential
of the form
\begin{equation}
U_{\rm opt} = \omega({\bf{k}},\rho) -\sqrt{{\bf{k}}^2+m^2}\label{uoptk}
\end{equation}
where $\omega$ and $\bf{k}$ are the energies and momenta of the
(anti)kaon.

For the calculation of the energy $\omega$ we consider the \km as stable quasi particles.
This allows us using
the relativistic mean field (RMF) calculations of Schaffner-Bielich et al. \cite{SchaffnerBielich:2000jy,Schaffner:1996kv}. For a comparison of this approach with chiral perturbation theory  we refer to \cite{Schaffner:1996kv}:
\begin{equation}
\omega({\bf{k}},\rho)= \sqrt{\left( {\bf{k}-\Sigma_v} \right)^2+m^2 + m
\Sigma_s} \pm \Sigma_v^0 \label{schaf}
\end{equation}
with a scalar self energy $\Sigma_s$ and a vector self energy ($\Sigma_v^0, {\bf \Sigma_v})$,
where the sign of the vector term $\pm\Sigma_v^0$ is positive for
kaons and negative for antikaons. This leads to different energies
of kaons and antikaons as it is seen in Fig.~\ref{opt-pot-33}.
The terms of the scalar potential $\Sigma_s$ are related to the
field of the $\sigma$ which itself is related in a non-linear way to the scalar
density  $\rho_s$
\begin{equation}
\Sigma_s= g_{\sigma {\rm K}}\ \sigma \ ; \qquad -g_{\sigma {\rm N}}
\rho_s = m_{\sigma} +b\sigma^2+c \sigma^3 .
\end{equation}
For the coupling constants we use the parametrization TM1 of
\cite{Schaffner:1996kv} :
\begin{equation}
g_{\sigma {\rm K}}=1.93 \qquad g_{\sigma {\rm N}}=10.0289
\end{equation}
and for the description of the $\sigma$-field we use
 \begin{equation}
m_{\sigma}=511.198\hbox{ MeV} \qquad b= -7.2325 {\rm fm}^{-1}
\qquad c=0.6183 .
\end{equation}

The vector potential is related to the baryon density $\rho_B$ by
\begin{equation}
\Sigma_v^0= g_{\omega {\rm K}}V_0 \qquad {\bf{\Sigma_v}}=g_{\omega
K}V_0 {\mathbf{\beta}}_{KN} \qquad V_0=\frac{g_{\omega N}}{m_{\omega}^2}
\rho_B
\end{equation}
where ${\mathbf \beta}_{KN}$ is the relative velocity of the kaon to
the nuclear medium. Again, the coupling constants are taken from
the TM1 set of \cite{Schaffner:1996kv} :
\begin{equation}
g_{\omega {\rm K}}=3.02 \qquad g_{\omega {\rm N}}=12.6139 \qquad
m_{\omega}=783\hbox{ MeV}.
\end{equation}

\begin{figure}[hbt]
\epsfig{file=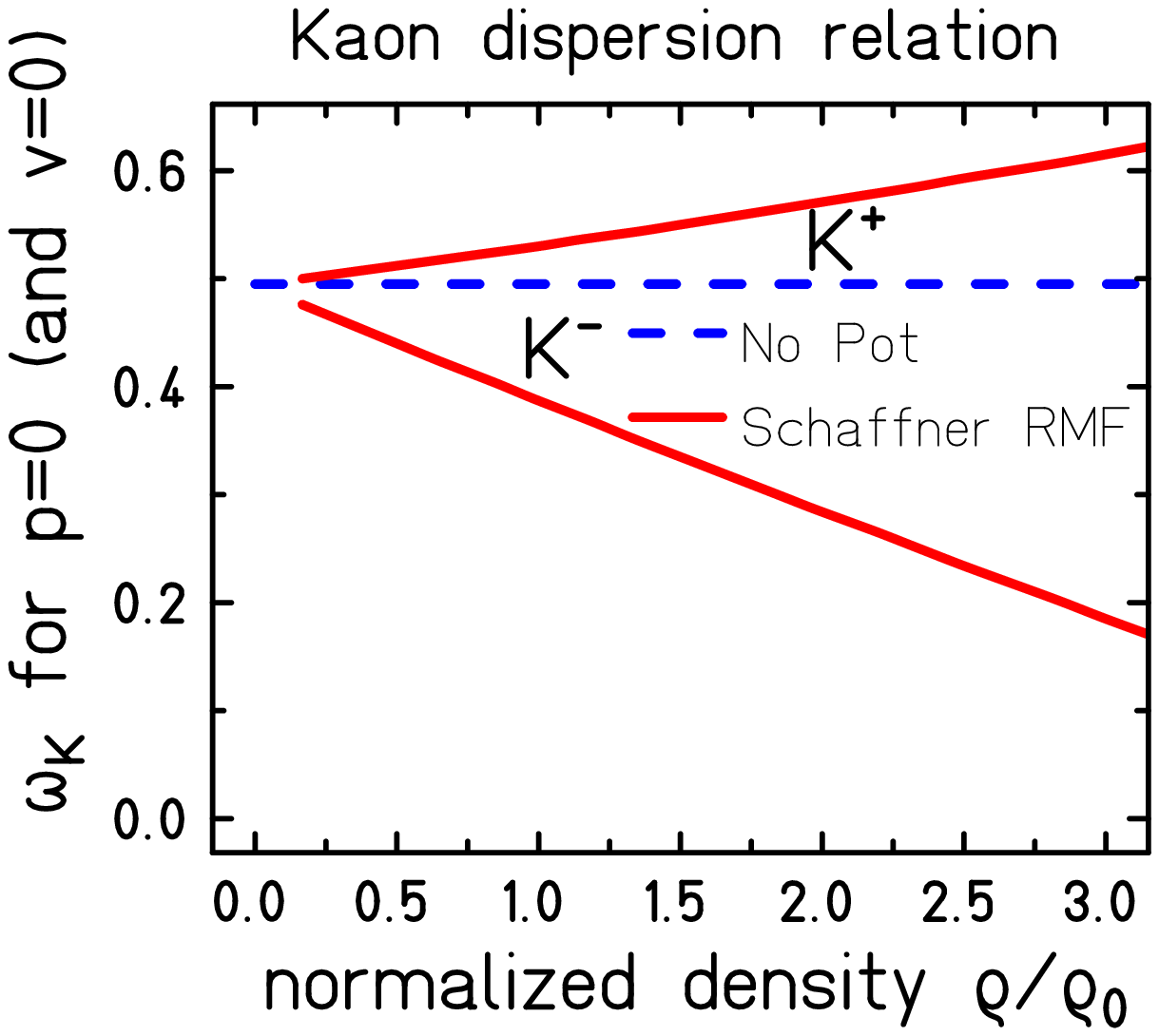,width=0.4\textwidth}\ \
\epsfig{file=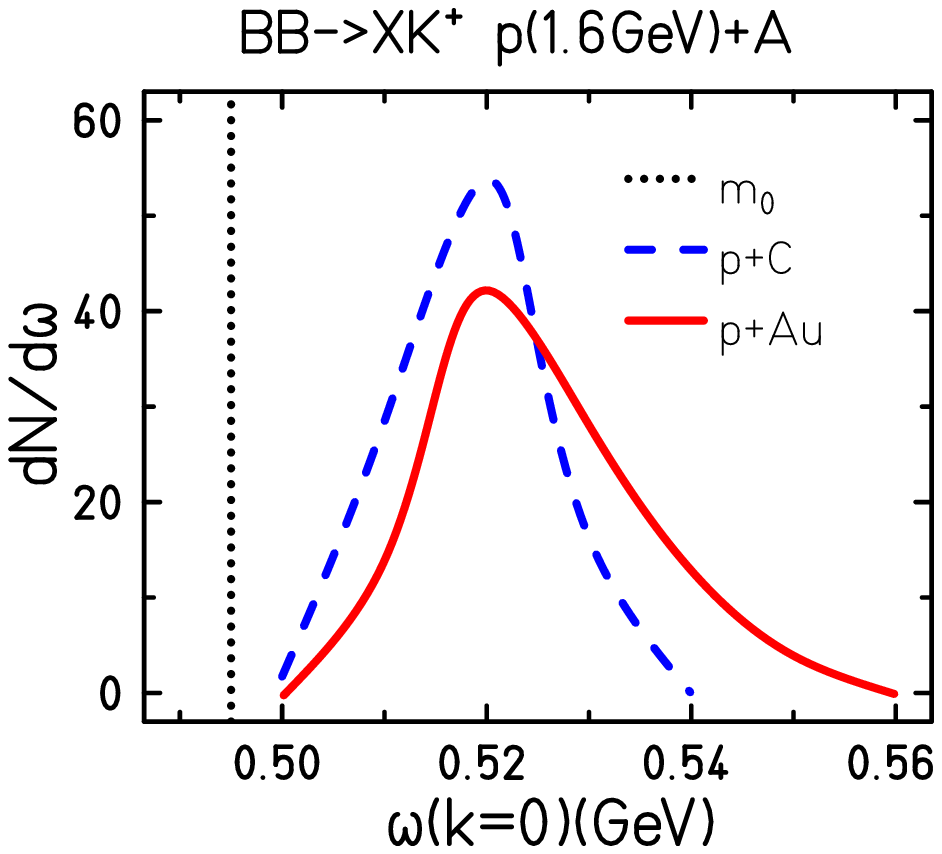,width=0.4\textwidth}
 \caption{Left: The function of the kaon rest energy $\omega(\bf{k}=0)$ as function of the
 baryonic density. Right: Distribution of the rest energies of \kp produced in
 BB-BYK collisions in a p+A reaction. }
\Label{opt-pot-33}
\end{figure}

In order to compare our approach with other models, we can
approximate $\omega({\bf{k}}=0,\rho)= m_{K^\pm}(\rho)$ by the linear relations
\begin{equation}
m_{\rm K^+}(\rho) = m_{\rm K^+}(\rho=0) \, (1+ \alpha_{\rm K^+}
\frac{\rho}{\rho_0}) \label{kpmass}
\end{equation}
\begin{equation}
m_{\rm K^-}(\rho) = m_{\rm K^-}(\rho=0) \, (1+ \alpha_{\rm K^-}
\frac{\rho}{\rho_0}) .
\label{kmmass}
\end{equation}
A value of  $\alpha_{\rm K^+} = 0.08$  reproduces the calculations
of~\cite{Korpa:2004ae}. The HSD model, advanced before detailed calculations
became available, uses $\alpha_{\rm K^+} = 0.04$. For $\alpha_{\rm K^-}$ we
obtain -0.21.

In later sections we may modify the strength of our potential in
order to study its significance. This will be done by applying a factor
$\alpha$ to the scalar and vector potentials:
\begin{equation}
\omega_{\alpha}({\bf{k}},\rho)=
\sqrt{\left( {\bf{k}-\alpha\Sigma_v} \right)^2+m^2 + m
\alpha\Sigma_s} \pm \alpha\Sigma_v^0 \label{kpotfactor}
\end{equation}
A value of $\alpha=0$ corresponds to a calculation without
KN-potentials, while $\alpha=1$ corresponds to our standard
potential described above. Note that we assume $\alpha=1$  as
default, if not explicitly stated to have a different value.

Figure~\ref{opt-pot-33}, left, shows the dependence of
$\omega(\bf{k}=0,\rho)$ of \kp and \km (eq.\ref{schaf})
as a function of the density of
the nuclear medium.
The nuclear medium is assumed to be at rest, otherwise an
additional contribution from the vector potential will show up.
The energy rises with density if we consider a kaon, but it falls
for the antikaons. Since this energy is the minimum energy needed
to create a (anti)kaon in the medium, it follows that the KN
potential enhances the threshold for kaon production but reduces
it for antikaons.

The right panel of Fig.~\ref{opt-pot-33} shows the effect of the
kaon potential on the distribution of the kaon rest energy
$\omega(\bf{k}=0,\rho)$ of \kp in p+A collisions at 1.6 GeV
incident energy. For this analysis we selected both production
channels, ${\rm NN}\to {\rm Y}{\rm N}\kp$ and ${\rm NN}\to {\rm
N}{\rm N}\kp \km$. The former is energetically more favorable (and
thus more probable). The kaons produced in p+Au collisions (red
full line) reach higher values of $\omega$ than the kaons produced
in p+C collisions (blue dashed line). The density profile of
carbon and gold nuclei are quite different and the gold nucleus
allows a better study of the effect of the nuclear medium.
For most of the kaons the rest
energy is enhanced by about 20-30 MeV with respect to the free
case (black dotted line) corresponding to a density at production
between 1/2$\rho_0$  and 3/4$\rho_0$.

\begin{figure}[hbt]
\epsfig{file=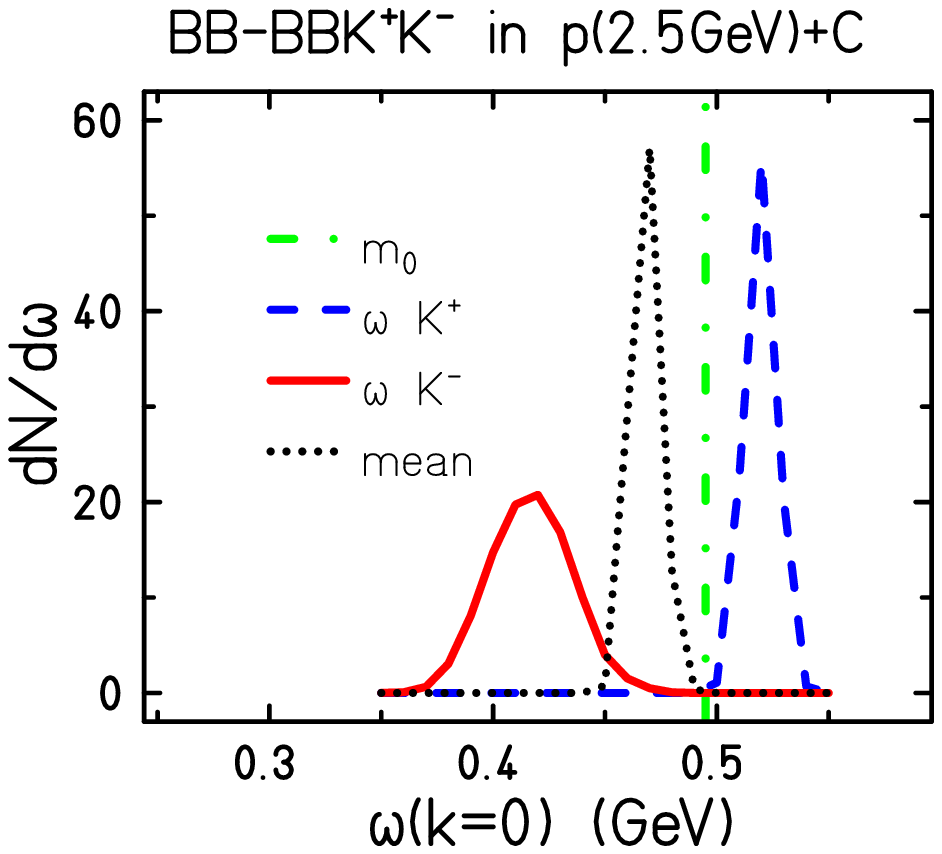,width=0.4\textwidth}\ \
\epsfig{file=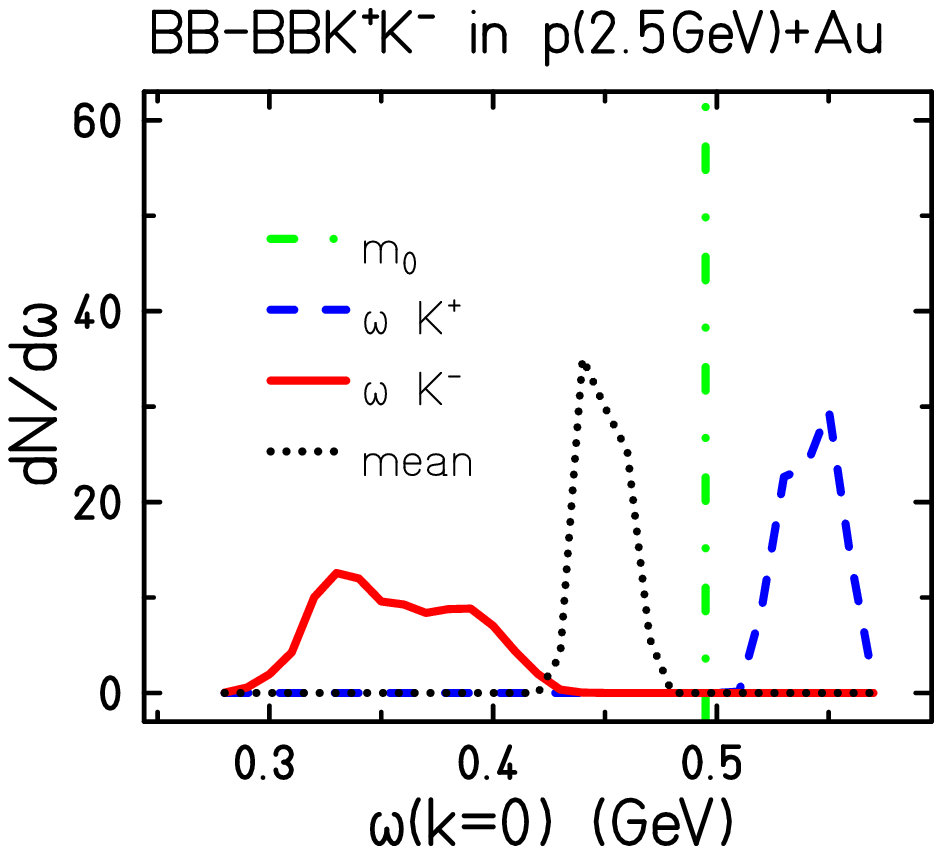,width=0.4\textwidth}
 \caption{Energies of kaons and antikaons produced in pair production reactions of
 p+C (left) and p+Au (right)}
\Label{opt-pot-km}
\end{figure}

For antikaons the situation is different. The optical potential
lowers their production threshold as shown in
Fig.~\ref{opt-pot-km} which displays the distribution of
the rest energies $\omega(\bf{k}=0,\rho)$ of K$^-$ (red full line) and \kp (dashed blue
line) produced in a p+C collisions (left) resp. p+Au collision
(right) at 2.5 GeV incident energy.  Again the effects are
stronger in the case of p+Au than for p+C. It should be
noted that due to strangeness conservation the K$^-$ have to be
produced together with a K$^+$. The decrease of the rest energy
of the antikaon is not compensated by the increase of the rest energy of the
kaon. On the average the change of the both rest energies (black dotted line) is negative
and therefore the threshold for the reaction NN $\to$ NN\kp \km
is lowered in the medium as compared to a production
in free space.

It should be noted that for the production of a kaon in matter,
the determination of the threshold has to include the potential
contributions of all particles (nucleons, hyperons, kaons) in the
ingoing and in the outgoing channels. For instance, the
momentum-dependent potential of an out-coming nucleon has to be
consistent with the total available energy. Similarly the momentum
dependence of the produced kaon depends on its relative momentum
with respect to the nuclear medium. This requires an iteration
procedure, since the available momentum of the nucleon and the
kaon are themselves dependent on the available energy. However, it
is found that the shift of the threshold (and thus also the
production yield of the kaon) depends on the application of a
momentum dependence of the nucleons but not on the application of
a momentum dependence of the kaons. This is due to the effect that
normally the center of mass of a high energy collision has a low
relative velocity with respect to the total nuclear medium. When a
collision takes place at an energy nearby the threshold, the kaon
will obtain a small relative velocity with respect to the center
of mass of the system and thus a small relative velocity with
respect to the nuclear medium. The incoming nucleons, however, had
to show a high relative velocity to the medium in order to obtain
the required high energy.

\subsection{Collisions}

In addition to the propagation of particles in the simulation,
collisions can take place. We use the common description that two
particles collide, if their minimum distance $d$ in their
c.m.-frame, i.e.\ the minimum distance of the centroids of the
Gaussians, fulfills the requirement
\begin{equation}
 d \le d_0 = \sqrt{ \frac { \sigma_{\rm tot} } {\pi}  }  , \qquad
 \sigma_{\rm tot} = \sigma(\sqrt{s},\hbox{ type} ),
\end{equation}
where "type" denotes the collision type considered (e.g. ${\rm
NN}$, $N\Delta$, \ldots). The total cross section is the sum of
the elastic and all inelastic cross sections
\begin{equation}
 \sigma_{\rm tot} = \sigma_{\rm el}+\sigma_{\rm inel}
 = \sigma_{\rm el}+\sum_{\rm channels} \sigma_i \quad .
\Label{tot-el-inel}
\end{equation}
For instance, for a pp collision the important contributions are
\begin{equation}
 \sigma_{\rm tot} = \sigma_{\rm el}+ \rm \sigma(pp \to p \Delta^+) +
                  \sigma(pp \to n \Delta^{++}) \quad .
\Label{pp-example}
\end{equation}
We use systematically free cross sections as given by experiments
with the exception of the NN $\to$ \kp $\Lambda$N channel, where
experiments have revealed a strong final state interaction which
is not present in matter. Different isospin channels are weighted
by isospin coefficients, e.g. for a pp collision we have
\begin{equation} \rm
\sigma(pp \to n \Delta^{++}) = 3 \sigma(pp \to p \Delta^+)
   =\frac{3}{4}\sigma_{\rm inelastic} .
\Label{pp-delta}
\end{equation}
Experimentally inaccessible cross sections like $\Delta {\rm N}\to {\rm
NN}$ are calculated from their reverse reactions  (here ${\rm NN}
\to\Delta N$) using detailed balance. For reactions involving
unstable particles with a finite width the form derived in
Ref.~\cite{Danielewicz:1991dh} is used.

The probability that a collision leads to a particular channel is given by the
contribution of this channel to the total cross section
\begin{equation}
 P_{\rm channel} = \frac{\sigma_{\rm channel}}{\sigma_{\rm tot}}\ ; \ \qquad {\rm e.g.}
 \quad P_{\rm pp \to p \Delta^+}
 = \frac{1}{4} \frac{\sigma_{\rm tot}-\sigma_{\rm el}}{\sigma_{\rm tot}}
\Label{proba}\quad .
\end{equation}
For the example we used
Eqs.~(\ref{tot-el-inel},\ref{pp-example},\ref{pp-delta}).
In the numerical simulation the channel is chosen randomly
according to the probability of the channel, e.g. in the above
case, there will be a 25\% chance to obtain a $\Delta^+$ in an
inelastic pp-collision.

\subsection{Virtual particles}
The production of kaons in this energy domain is a very rare
process with a production cross section of only a few nanobarns, in
comparison to the total ${\rm NN}$ cross section of about 20-40 mb.
The probability of producing strangeness is thus very small. Then,
the method presented above will lead to severe limitations in
statistics for the investigation of kaon production.
Nevertheless, simulation codes oriented toward higher collision energies,
like UrQMD \cite{Bass:1998ca,Bass:2001up} apply successfully this
method for reactions above the threshold. Results of such
calculations (without a mean field) on global kaon spectra and rapidity
distributions in Ni+Ni collisions at 1.93 \AGeV are quite
comparable to IQMD results \cite{bleicher}. However, detailed
investigations of triple differential cross sections and
differential flow patterns (like $v_2(p_t,y_{\rm c.m.}$)) cannot
be addressed in those approaches.

A way to overcome this problem is the method of ``perturbative
production'' of kaons, which was used in Ref.~\cite{Hartnack:1993bq}.
In this method one determines the probability for producing a kaon
in a collision from Eq.~(\ref{proba}). The production probability
$P$, the c.m.-momentum $\bf{p}$ and the invariant mass $\sqrt{s}$
of the collisions are stored but the strangeness production is not executed
(the momenta of the incoming particles are not changed). Instead,
the collision branches into another collision channel which is
then executed. At the end  the stored data are analyzed to
calculate the production of kaons in these elementary collisions
by summing up the probabilities.

However, this method has the disadvantage that further
interactions of the perturbatively produced particles can only roughly be estimated.
To overcome this we use the method
of virtual particles:
\begin{itemize}
\item With each particle $i$ we associate a probability $P_i$. Protons, neutrons, deltas
and pions initially have $P_i=1$.

\item After a collision which takes place with the probability
$P_r= \frac{\sigma(i,j\to k +X)}{\sigma_{tot}}$ according to Eq.~(\ref{proba}),
the particle k carries the probability
\begin{equation}
P_k=P_i\cdot P_j \cdot P_r.
\end{equation}
If this collision is not executed (because k is a virtual particle)
the parents continue undisturbed and after the collision have the
probabilities
$$P_i'({\rm undisturbed})=P_i(1-P_r\cdot P_j) \qquad  P_j'({\rm undisturbed})=P_j(1-P_r\cdot P_i). $$
\item
The interaction potential felt by particle $i$ is the sum of the
interaction potentials of the free particles multiplied by the
probabilities of the interacting particles.

\begin{equation}
V_{i}=\sum_k P_k V_{ik}
\end{equation}

\end{itemize}

\begin{figure}[hbt]
\epsfig{file=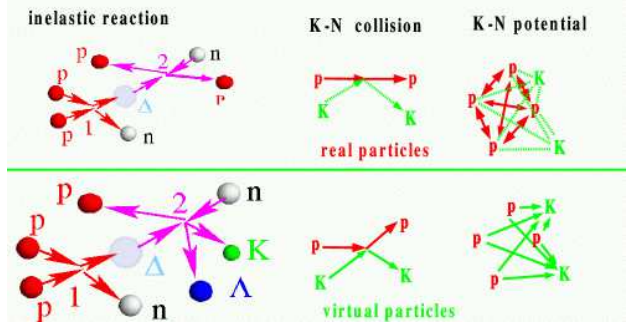,width=0.5\textwidth}
 \caption{Example for the differences in the treatment of ``real'' and ``virtual''
 particles (see text).}
\Label{virtual-particles}
\end{figure}

For the production of strangeness we have $P_r \ll 1$ and
therefore $ 1-P_r\cdot P_i \approx 1$ and $ 1-P_r\cdot P_j \approx 1$, we can simplify the above scheme
in the following way, which is illustrated in
Fig.~\ref{virtual-particles}:
\begin{itemize}

\item
Nucleons, deltas and pions are {\it real} particles with $P=1$. Strange
particles are {\it virtual} particles who have a very small probability $P_i$.

\item In a collision virtual particles are produced with a
reaction probability $P_r=\sigma({\rm BB \to K{\rm X}})/\sigma(tot)$. The
parent particles do not change their properties and follow another
channel of the collision according to its probability.
(Fig.~\ref{virtual-particles} , top-left, e.g. $\Delta^{++}\rm n \to pp$).

The produced strange particles proceed as if the production reaction had
taken place and carry a probability of $P_r$
(Fig.~\ref{virtual-particles} , bottom-left, e.g. $\Delta^{++}{\rm n\to p}
\Lambda K$).

\item In a collision of a real and a virtual particle, the momentum of the real
particle remains unchanged
(Fig.~\ref{virtual-particles} , top-middle). The virtual particle
acts, as if this collision had taken place
(Fig.~\ref{virtual-particles} , bottom-middle).

\item The real particles do not see a potential interaction with
the virtual particles  (Fig.~\ref{virtual-particles}, top-right).
The virtual particles feel a potential with the nuclear matter.
(Fig.~\ref{virtual-particles}, bottom-right).

\end{itemize}

This method has the advantage to allow for high-statistics
calculation of kaon one-body observables including all effects of
the medium like potential propagation, rescattering, absorption
etc. However, there are also some drawbacks of this method:

\begin{enumerate}
\item
The strange particles do not appear in the energy balance. Therefore
including strange particles violates the energy conservation, perfectly assured
for the real particles. The real particle do not 'feel' whether
strange particles are produced or not.

Questions like `{\it Do events with
kaons produce less highly energetic pions than other events}' (which would be
interesting for analogies between kaons and high energy pions)
cannot be addressed.
\item
KN- correlations cannot be calculated.

\item Higher-order processes might be described incorrectly. To
give an example:
\begin{enumerate}
\item An NN collision produces virtually N$\Lambda {\rm K}^+$. In
the `real world' it produces a N$\Delta$-pair.

\item The virtual $\Lambda$ rescatters with another nucleon while
the real $\Delta$ decays into N $\pi$.

\item The virtual $\Lambda$ and the real $\pi$ resulting from the
real $\Delta$-decay scatter and produce an N K$^-$-pair.
\end{enumerate}
The latter process should not be allowed because after a real production of the
$\Lambda$ the  $\Delta$-production could not take place and therefore the $\pi$
would not exist.
In our simulations such processes are explicitly forbidden (by
checking for different parents of the collision partners), but if
the $\pi$ in between has been reabsorbed by a nucleon $\pi {\rm
N}\to \Delta \to \pi$ N we cannot trace back its origin. However,
the probability of such a process is low.
\end{enumerate}

It should be noted that in IQMD kaons and antikaons are
propagating under the influence of an optical potential as described
in subsection \ref{knpot-desc}. Thus the particles have effective masses
as seen in Fig.~\ref{opt-pot-33}.
These effective masses enter in the time evolution equations.
Contrary to HSD \cite{Cassing:2003vz} there is no
propagation of the finite width of the spectral functions of antikaons in IQMD.

\subsection{Elementary production of  K$^+$ }

\begin{figure}[hbt]
\begin{tabular}{cc}
\epsfig{file=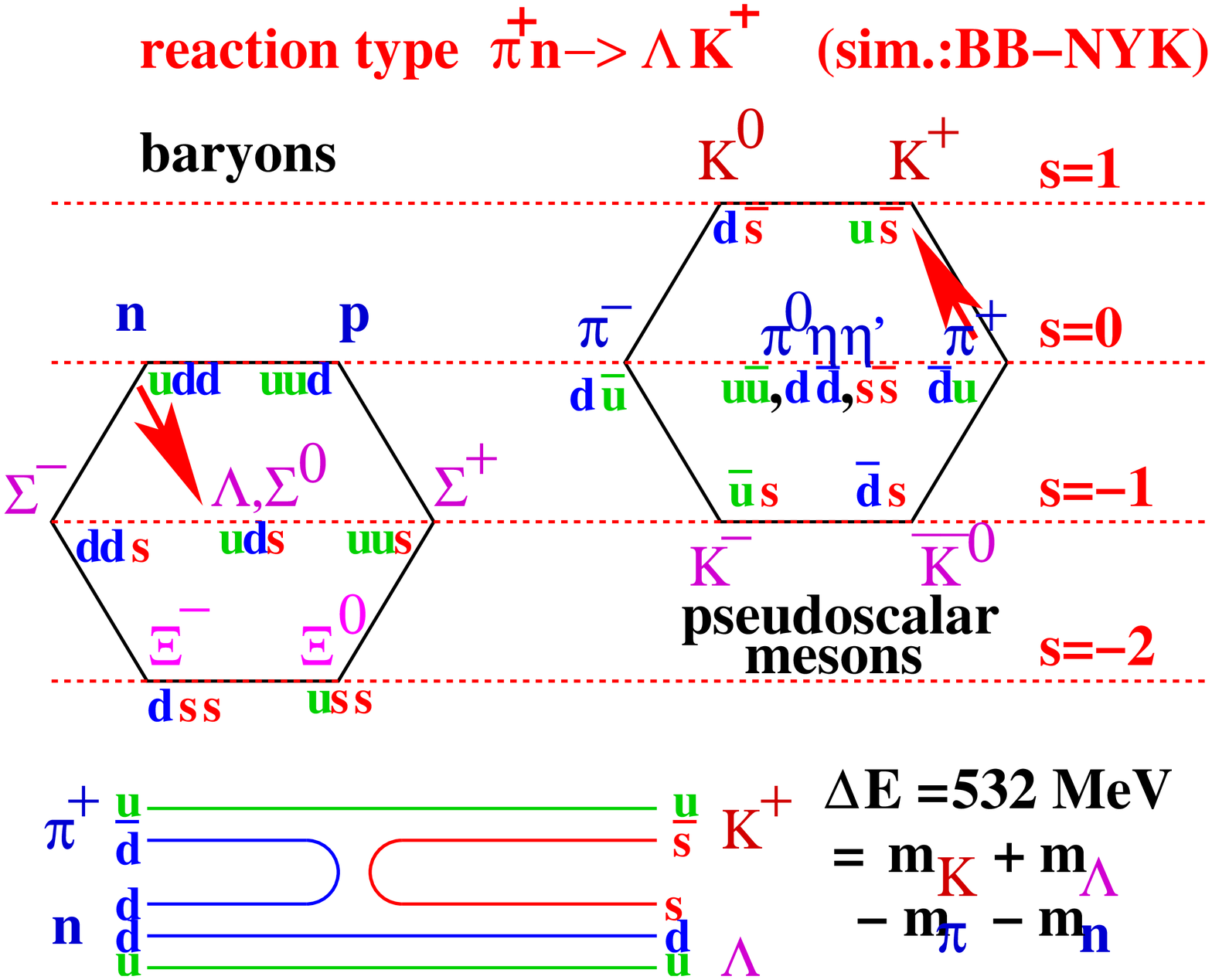,width=0.45\textwidth} &
\epsfig{file=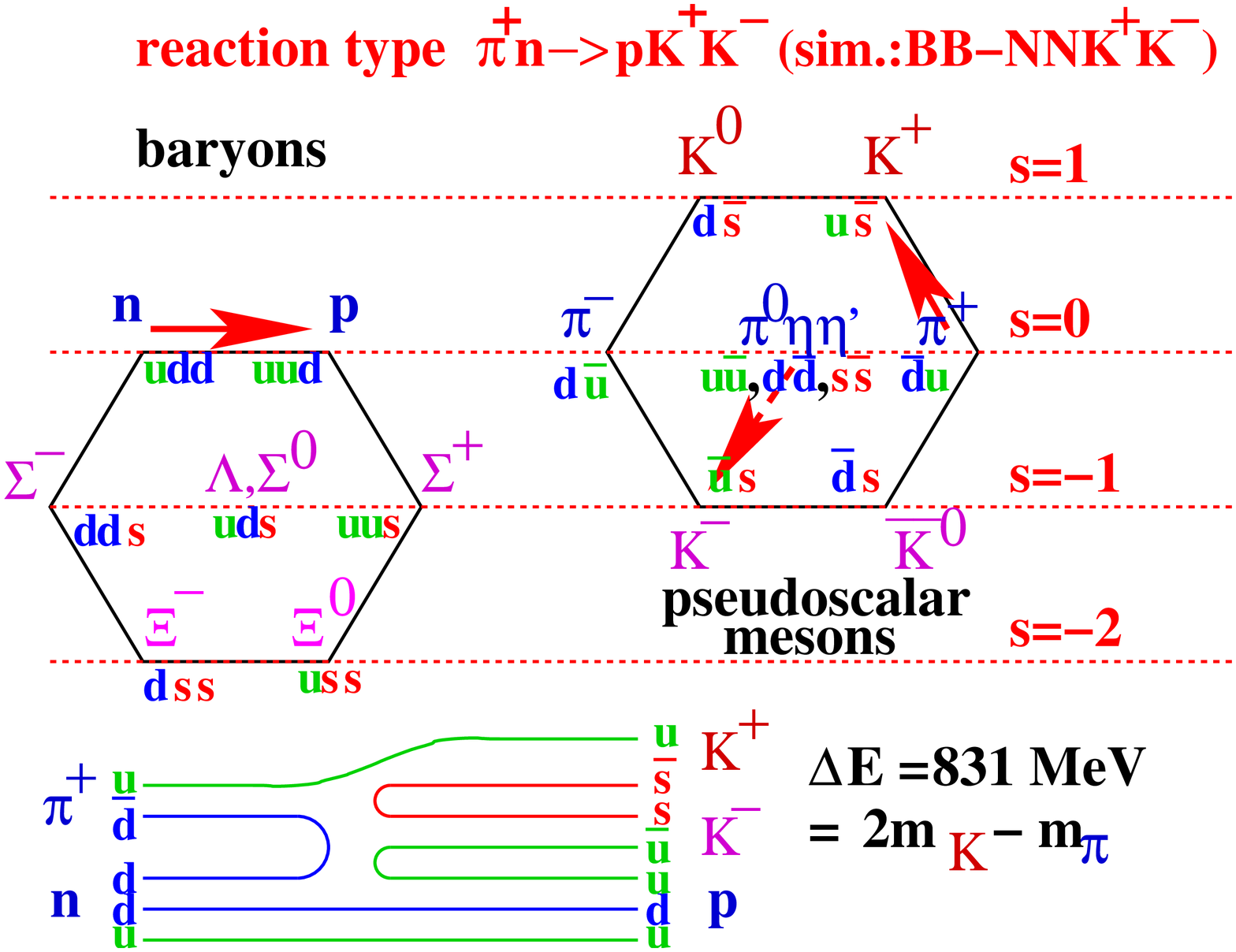,width=0.45\textwidth} \\
\end{tabular}
 \caption{Representation of the
 major elementary production channels of $K+$ mesons in a $\pi^+n$ collision
 in the quark language and the SU(3) scheme.}
\Label{kp-prod-multiplett}
\end{figure}
The elementary production of K$^+$ is governed by the conservation
of strangeness. The initial net strangeness is zero, thus strange
quarks can only be produced together with strange anti-quarks. In
Fig.~\ref{kp-prod-multiplett} we show the main production
processes for K$^+$ mesons schematically together with the baryon and
pseudoscalar meson octets. The arrow indicates the transit
in each octet. In this scheme the sum of the red
arrows have to be zero. The most economic way to do this is to
create an $s$-quark which remains in a baryon (and thus transforms
the nucleon into a hyperon) together with a $\bar{s}$-quark which
becomes part of a kaon. An example of such a process is given on
the l.h.s. of  Fig.~\ref{kp-prod-multiplett}, where the reaction
$\pi {\rm n} \to \Lambda {\rm K}^+$ is shown. In this reaction a
$d \bar{d}$-pair annihilates to form an $s\bar{s}$ pair. This
reaction  requires less energy than
the production via NN$\to$ NN\kp \km  (r.h.s. of
Fig.~\ref{kp-prod-multiplett}) (see also Eq.~(\ref{srthchannel})).
>From the quark diagrams at the bottom of
Fig.~\ref{kp-prod-multiplett} we see that the latter process
creates an additional  $u \bar{u}$ pair and should thus be
suppressed by the OZI rule \cite{OZI}.

\begin{figure}[hbt]
\begin{tabular}{cc}
\epsfig{file=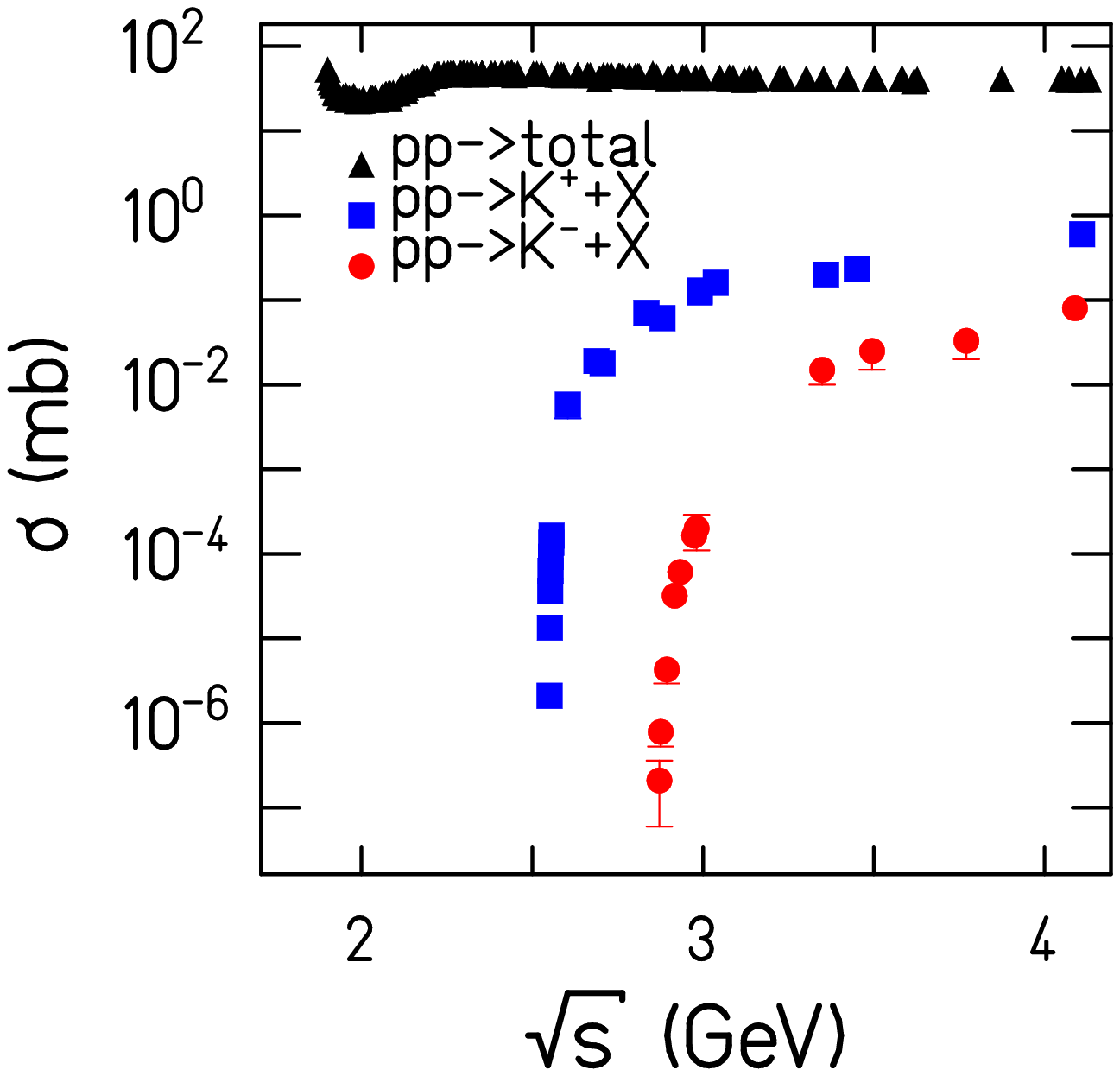,width=0.45\textwidth} &
\epsfig{file=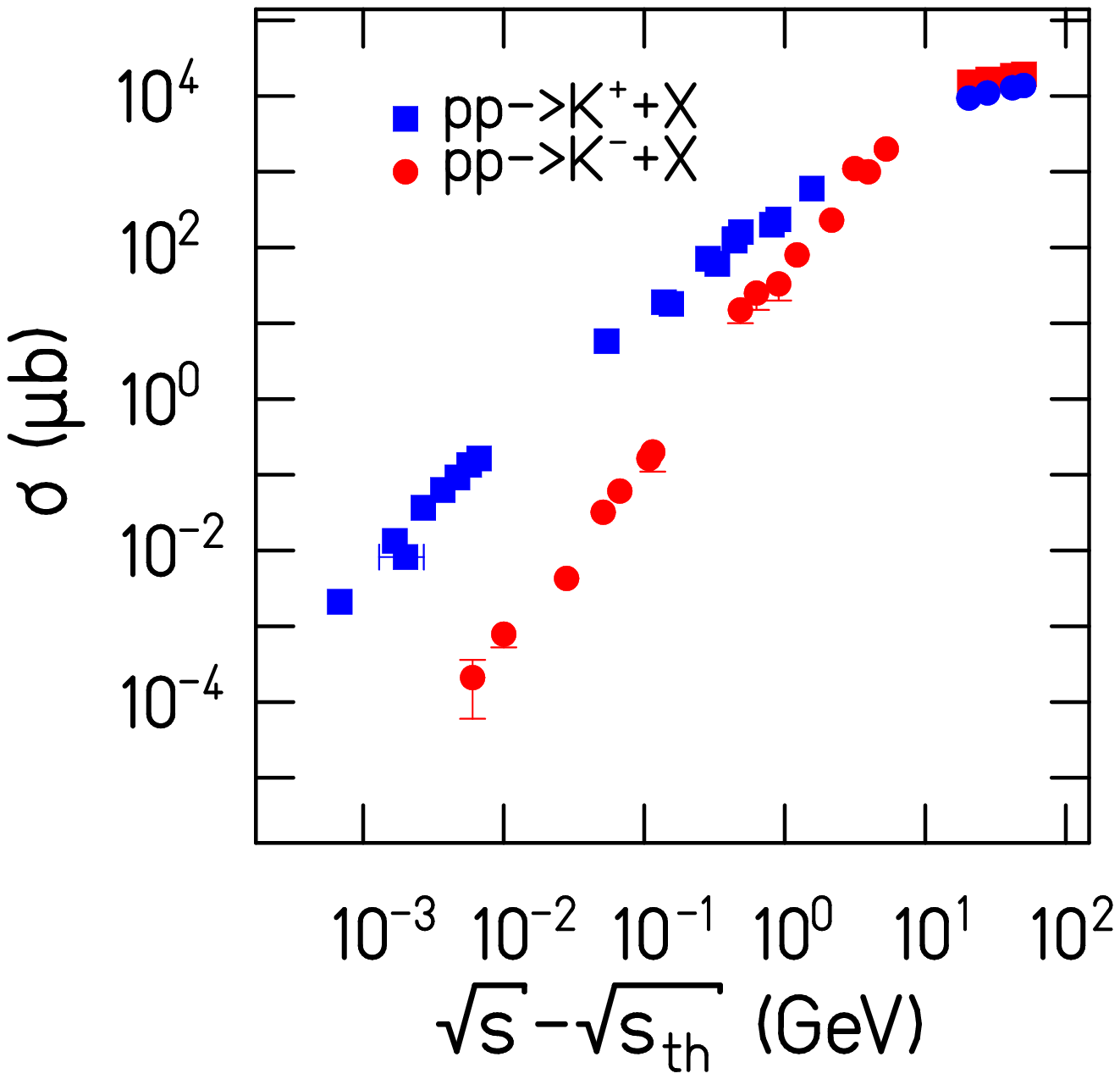,width=0.45\textwidth} \\
\end{tabular}
 \caption{Left: Comparison of the total cross section and of the production cross
 section of K$^+$ and K$^-$ as function of $\sqrt{s}$.
 Right: Near threshold energy dependence of cross sections of the K$^+$ and K$^-$
 production channels in proton-proton collisions.}
\Label{prod-xsection}
\end{figure}

For proton-proton reactions there are experimental data on
strangeness production available. Figure \ref{prod-xsection} shows
on the left hand side a comparison of the cross sections for the
production of a K$^+$ (blue squares) and the production of a K$^-$
(red circles) together with the total reaction cross section
(black triangles) as a function of the c.m.-energy $\sqrt{s}$ in
proton-proton collisions.  We see a steep rise of the cross
sections after passing the kaon production threshold. However, the cross sections
of strangeness production are very small in comparison to the
total cross section, which motivates the description of
strangeness in the framework of virtual particles (see
Fig.~\ref{virtual-particles}). The threshold is lowest in reactions
in which the associate particle is a $\Lambda$. Thus this channel
gives the strongest contribution to the proton-proton induced
K$^+$ production at low energies. Channels with the production of
a $\Sigma$ or of a ${\rm K}^+{\rm K}^-$ pair
contribute only at higher energies due to their higher thresholds.

The right hand side of Fig.~\ref{prod-xsection} demonstrates the
behavior of the production of K$^+$ and K$^-$ near the threshold.
Note that the abscissa gives the excess energy in the c.m.-frame
and both axes are in a logarithmic scale. Both processes strongly
rise with the available excess-energy.  However, the rise of the
production cross section of the K$^-$ is stronger than that
of the K$^+$ when we are in the range of several MeV to
several hundred MeV. This is the dominant range contributing to the
production of kaons and antikaons in the energy domain we are studying.

The cross section for the production of
strangeness in neutron-induced reactions is still a subject
of investigation. In the One-Boson-Exchange model the isospin factor
which has to be employed when calculating the np cross section
from the known pp cross section depends on the nature of the exchanged particle.
It varies between 1 for a pion and 5/2 for a kaon exchange. In IQMD the latter
choice of 5/2 is used which leads to an isospin averaged cross
section of

\begin{equation}
\sigma ({\rm NN} \to {\rm NY} {\rm K}^+)= \frac{1}{4}\bigl (
\sigma _{\rm p \rm p} + 2\sigma_{\rm p \rm n} + \sigma_{\rm n \rm
n} \bigr ) = \frac{1+5+0}{4}\sigma_{\rm p \rm p}=\frac{3}{2}\sigma
(\rm p \rm p \to {\rm NY}{\rm K}^+) .
\end{equation}

\begin{figure}[hbt]
\begin{tabular}{cc}
\epsfig{file=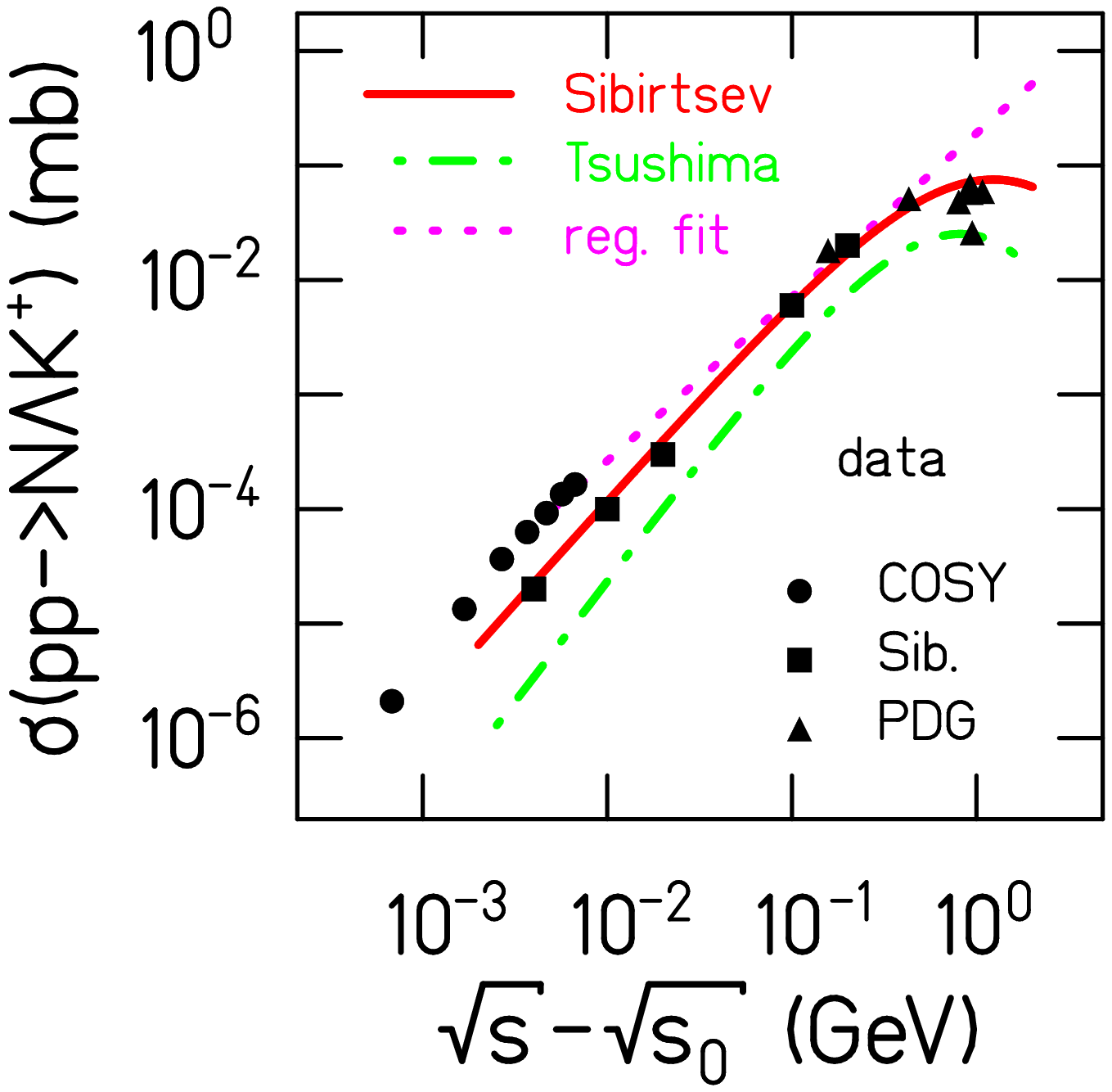,width=0.45\textwidth} &
\epsfig{file=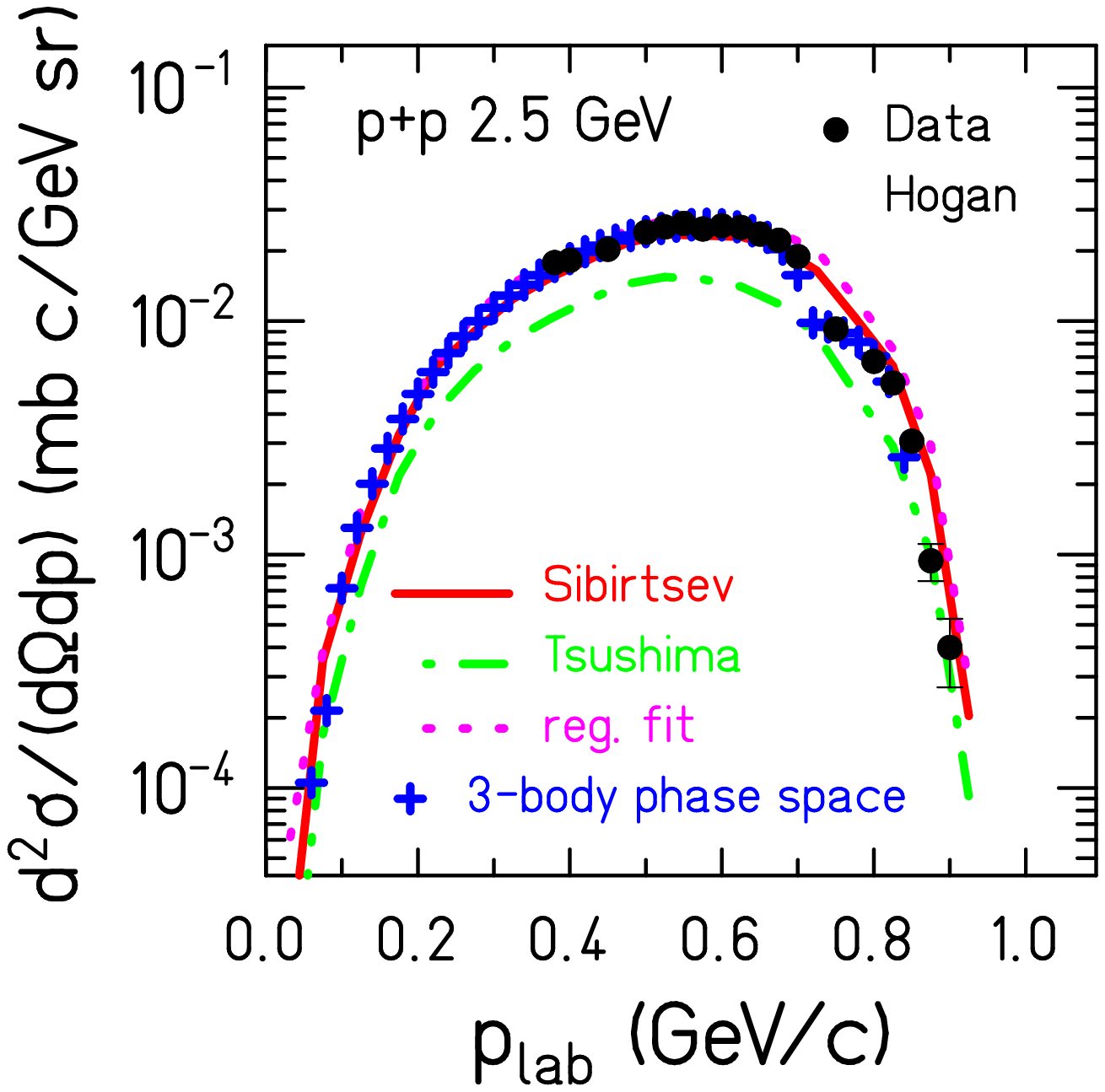,width=0.45\textwidth} \\
\end{tabular}
 \caption{Different parameterizations of the  pp  $\to$ p $\Lambda$ \kp cross section (left)
 and comparison of IQMD results using these parameterizations with experimental
 pp-data at 2.5 GeV measured at $\theta_{\rm lab}$ = 40$^\circ$ from~\cite{Hogan:1968zz}(right).}
\Label{kp-prod-channels}
\end{figure}

There exist different microscopic models for describing the
production of kaons and corresponding parameterizations of these
cross sections. Figure \ref{kp-prod-channels} on the l.h.s. shows
several parameterizations of  the pp  $\to$ p $\Lambda$ \kp
channel used in our simulations. The parametrization of Sibirtsev are obtained
by correcting the experimental data for the final state interaction
\cite{Sibirtsev:1995xb}. We use this set and its parametrization
(\rfl{})
\begin{equation}
\sigma ({\rm N}{\rm N}\to \N\Lambda {\rm K}^+)=1.098 \, {\rm
mb}\cdot \left ( 1- \frac{s_{\rm thres}}{s}  \right )^{1.8} \cdot
\left ( \frac{s_{\rm thres}}{s} \right )^{1.5}
\label{sibirtsev-fit}
\end{equation}
as our standard input.
Tsushima {\rm et al.}\cite{Tsushima:1998jz} have calculated the production cross section
in a resonance model. They obtain
\begin{equation} \sigma ({\rm N}{\rm N}\to {\rm N}\Lambda {\rm K}^+)=1.165 \, \mu{\rm b} \cdot \left (
\frac{s}{s_{\rm thres}} -1 \right )^{2.14} \cdot \left (
\frac{s_{\rm thres}}{s} \right )^{5.024}
\end{equation}
shown as \gml{}. Another simple parametrization is a linear
regression fit in a log-log representation  (magenta dotted line)
 of recent COSY data \cite{Balewski:1998pd} with reads
\begin{equation}
\sigma (\p\p \to \p \Lambda {\rm K}^+ ) =  191 \, \mu{\rm b} \cdot
\left ( \sqrt{s}-\sqrt{s_{\rm thres}}\right )^{1.43} \ .
  \label{refit}
 \end{equation}
This parametrization is valid for energies below
3 GeV, which is above the high energy limit of IQMD because high
mass baryon resonances are not included.

The right hand side of Fig.~\ref{kp-prod-channels} compares
results of experimental pp-data of Hogan {\rm et al.}
\cite{Hogan:1968zz} (black full circles) with IQMD calculations
(full red line) using the cross section \cite{Sibirtsev:1995xb} of
Eq.~(\ref{sibirtsev-fit}). At this energy, the COSY data and the
Sibirtsev parametrization agree and we do not expect a difference.
The agreement shows that in the energy range where data are available
our parametrization of the cross section works well. Changing to the $\sigma
({\rm N}{\rm N}\to {\rm N} \Lambda {\rm K}^+)$ cross section by
Tsushima \cite{Tsushima:1994rj,Tsushima:1998jz}, (green dash-dotted
line) results in a too low yield. The regression-fit
parametrization (magenta dotted line, Eq.~(\ref{refit})) shows
again a spectrum quite comparable to that using the Sibirtsev
parametrization. The blue crosses indicate the results of a fit
using a 3-body phase-space distribution. The agreement shows that
at this energy the production of a kaon proportional to the 3-body
phase-space decay is compatible with the data. We adopt
this 3 body phase space production in our
simulations for all baryon-baryon channels. Other models, like UrQMD
\cite{Bass:1998ca}, generate the kaon via a two-step process
producing first a resonance (e.g. the $N^*(1650)$) which decays
then into a hyperon-kaon pair. As said, results are reported to be
compatible to experimental data as well \cite{bleicher}.

In heavy-ion collisions kaons may be produced in baryon-baryon
collisions in which at least one of the reaction partners is a
baryonic resonance. As already discussed for pp collisions, ${\rm
B}_1{\rm B}_2 \to $B$_3 {\rm Y} {\rm K}^+$ reactions have lower
thresholds than the channel ${\rm B}_1{\rm B}_2 \to $B$_3$B$_4
{\rm K}^+{\rm K}^-$. Here B may be a nucleon N or a $\Delta$, and
Y a $\Lambda$ or a $\Sigma$.

Whereas the cross sections for \kp production in the pp and $\pi$
N channel are known, the production  cross section for the
$\Delta$ N channel is experimentally not directly accessible. One
has to rely on theoretical calculations which have produced quite
different results. As far as the predictive power of transport
calculations is concerned (where these elementary cross section
are input quantities) these uncertainties add to those due to the
unknown cross section in the pn channel.

\begin{figure}[hbt]
\epsfig{file=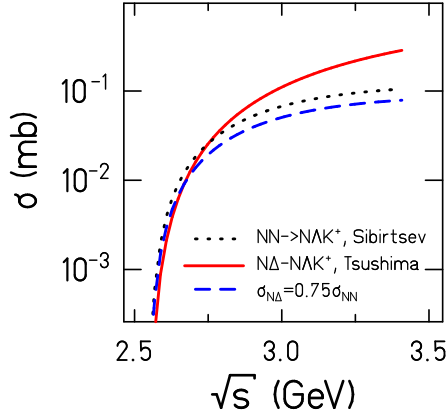,width=0.45\textwidth} 
\caption{ 
Different parameterizations of the
N$\Delta\to $N$\Lambda$ K$^+$ cross section as compared to the NN
$\to$ N$\Lambda$ K$^+$ cross section (see text). } \label{xsections-NDN}
\end{figure}

For the NN $\rightarrow \kp \Lambda$ N channel, both IQMD and HSD
use a parametrization of the experimental pp data from Sibirtsev
\cite{Sibirtsev:1995xb}, discussed above,
applying only the mentioned isospin corrections to the np channels. For the  N$
\Delta\rightarrow \kp \Lambda$ N cross section two approaches have
been advanced which differ substantially, as may be seen in
\figref{xsections-NDN}. Randrup-Ko \cite{Randrup:1980qd} proposed
to scale the NN cross section by the appropriate isospin factor:
 $\sigma({\rm N
\Delta} \rightarrow {\rm K^+ \rm N \Lambda}) = 0.75 \ \sigma({\rm
NN} \rightarrow {\rm K^+ N} \Lambda )$ whereas Tsushima calculates
this cross section in a resonance model
\cite{Tsushima:1994rj,Tsushima:1998jz}. The latter is used in
present-day simulation programs, especially in the actual
version of the HSD approach.  Former HSD calculations used,
however, the cross section of  Randrup and Ko and therefore a
comparison of the present results with the former ones has to be
done with care. As can be seen in Fig.~\ref{xsections-NDN} for
$\sqrt{s}$ values above 2.7 GeV, corresponding to beam energies
above 2 GeV in pp collisions, there is quite a difference between
the two parameterizations.

Furthermore, kaons may also be produced in pion-induced
reactions. Table~\ref{list-k-channels} presents all channels in
IQMD which contribute to the K$^+$ production.

\begin{table}[hbt]
\begin{tabular}{cccc}
${\rm N}{\rm N}\to{\rm N}\Lambda {\rm K}^+$ & ${\rm N}{\rm
N}\to{\rm N}\Sigma {\rm K}^+$ & ${\rm N}{\rm N}\to \Delta \Lambda
{\rm K}^+$ &
${\rm N}{\rm N}\to \Delta \Sigma {\rm K}^+$ \\
${\rm N}\Delta \to{\rm N}\Lambda {\rm K}^+$ & ${\rm N}\Delta
\to{\rm N}\Sigma {\rm K}^+$ & ${\rm N}\Delta \to \Delta \Lambda
{\rm K}^+$ &
${\rm N}\Delta \to \Delta \Sigma {\rm K}^+$ \\
$ \Delta \Delta \to{\rm N}\Lambda {\rm K}^+$ & $ \Delta \Delta \to
{\rm N}\Sigma {\rm K}^+$ & $ \Delta \Delta \to \Delta \Lambda {\rm
K}^+$ &
$ \Delta \Delta \to \Delta \Sigma {\rm K}^+$ \\
$ \pi{\rm N}\to \Lambda {\rm K}^+$ & $ \pi{\rm N}\to \Sigma {\rm
K}^+$ & $ \pi \Delta \to \Lambda {\rm K}^+$ &
$ \pi \Delta \to \Sigma {\rm K}^+$ \\
${\rm N}{\rm N}\to{\rm N}{\rm N}{\rm K}^+ {\rm K}^-$ & ${\rm
N}\Delta \to{\rm N}{\rm N}{\rm K}^+ {\rm K}^-$ & ${\rm N}\Delta
\to{\rm N}\Delta {\rm K}^+ {\rm K}^-$ &
$ \Delta \Delta  \to{\rm N}{\rm N}{\rm K}^+ {\rm K}^-$ \\
$ \Delta \Delta  \to \Delta \Delta {\rm K}^+ {\rm K}^-$ & $
\pi{\rm N}\to{\rm N}{\rm K}^+ {\rm K}^-$ &
$ \pi \Delta \to{\rm N}{\rm K}^+ {\rm K}^-$ & \\

\end{tabular}
\caption{List of the K$^+$ producing reactions parameterized in
IQMD.} \label{list-k-channels}
\end{table}

In these channels different isospin combinations for $N, \Delta,
\pi$ are possible which lead to a further subdivision of these
channels. Note that we only use the $\Delta (1232)$ resonance which is the
dominant resonance channel in the interesting energy domain. For each channel
the production threshold is given by the sum of the outgoing
masses. In free space we find
\begin{eqnarray} \sqrt{s_{\rm thres}}({\rm NN}\to {\rm N\Lambda K})&=&
m_{\rm N}+M_\Lambda+m_{\rm K}=2.55 \,\rm GeV, \nonumber \\
\sqrt{s_{\rm thres}}({\rm NN}\to {\rm N\Sigma K})&=&
m_{\rm N}+M_\Lambda+m_{\rm K}=2.62 \,\rm GeV, \label{srthchannel}\\
\sqrt{s_{\rm thres}}({\rm NN}\to {\rm NNK\overline{K}})&=& 2m_{\rm
N}+2m_{\rm K}=2.87 \,\rm GeV \nonumber
\end{eqnarray}
where we use isospin averaged masses $m_{\rm N}=938$~MeV,
$m_\Lambda=1115$~MeV, $m_\Sigma=1189$~MeV, $m_{\rm K}=495$~MeV,
$m_\pi=138$~MeV, $m_\Delta^{\rm pole}=1232$~MeV. Note that for
channels having a $\Delta$ in the outgoing channel we set its mass
to its pole mass. However, these channels do not contribute
significantly. In the medium the free masses are replaced by the
effective masses at the production density.

For the channels with an incoming $\Delta$ no information from experiment
exists which leads to uncertainties in the interpretation of
heavy ion data. This was already mentioned above for the $N\Delta \to N\Lambda K$ cross section.
We will discuss this point later.

\subsection{Elementary production of  K$^-$ }
As mentioned above in proton-proton collisions near threshold a K$^-$  is always
produced together with a \kp via ${\rm NN}\to {\rm NN}\km\kp$.
Only a few experimental points are available for this reaction as
can be seen in Fig.~\ref{prod-xsection}. We parameterize the
production cross section by \begin{equation} \sigma=63 \mu {\rm b}
(\sqrt{s}-\sqrt{s_{\rm thres}})^{1.6} ,
\end{equation}
with $\sqrt{s}$ given in GeV.

Similar to the K$^+$ also the K$^-$ may be produced in reactions
with resonances or pions in the entrance channel. For the
resonance-induced channels we use the same cross sections as for
the corresponding ${\rm NN}$ induced channel with isospin factors
of 0.75 for $N\Delta$ and 0.5 for $\Delta\Delta$. The fact that some
combinations (like $\Delta^- \Delta^-$ ) have no or only
restricted outgoing channels is taken into account by the
appropriate isospin factors. For the
pion-induced reaction $\pi N\to{\rm N}{\rm K}^+{\rm K}^-$ we use
the cross section parametrization of Sibirtsev
\cite{Sibirtsev:1996rh}.

In heavy-ion collisions, discussed later in detail,  the major
contributions to the K$^-$ production are strangeness exchange
reactions of the type $\pi {\rm Y} \to{\rm N}{\rm K}^-$ and BY
$\to {\rm NN} {\rm K}^-$. The cross section for the pion-induced
strangeness exchange is deduced from the kaon absorption cross section
via detailed balance. For the baryon-induced strangeness exchange
channel we use the parametrization of Ko {\rm et al.}:
$\sigma(N\Lambda)=0.8 {\mbox mb} (E-E_{\rm thres}) $ where the
energies are taken in GeV in the hyperon rest frame. For N$\Sigma$
this cross section is scaled by a factor of 1.5. For $\Delta$ Y
reactions we assume the same energy dependence of the cross
section, but apply an isospin factor of 0.75, taking also into account
that several combinations (like $\Delta^{++}\Lambda$) have
restricted outgoing channels.

\subsection{Nucleon scattering of K$^+$ and   K$^-$ }
The produced kaons may scatter in the nuclear medium. For the
scattering of kaons on protons experimental data exist, as shown
in Fig.~\ref{kaon-elastic}.
\begin{figure}[hbt]\vspace*{1ex} \hfill \\
\begin{tabular}{cc}
\epsfig{file=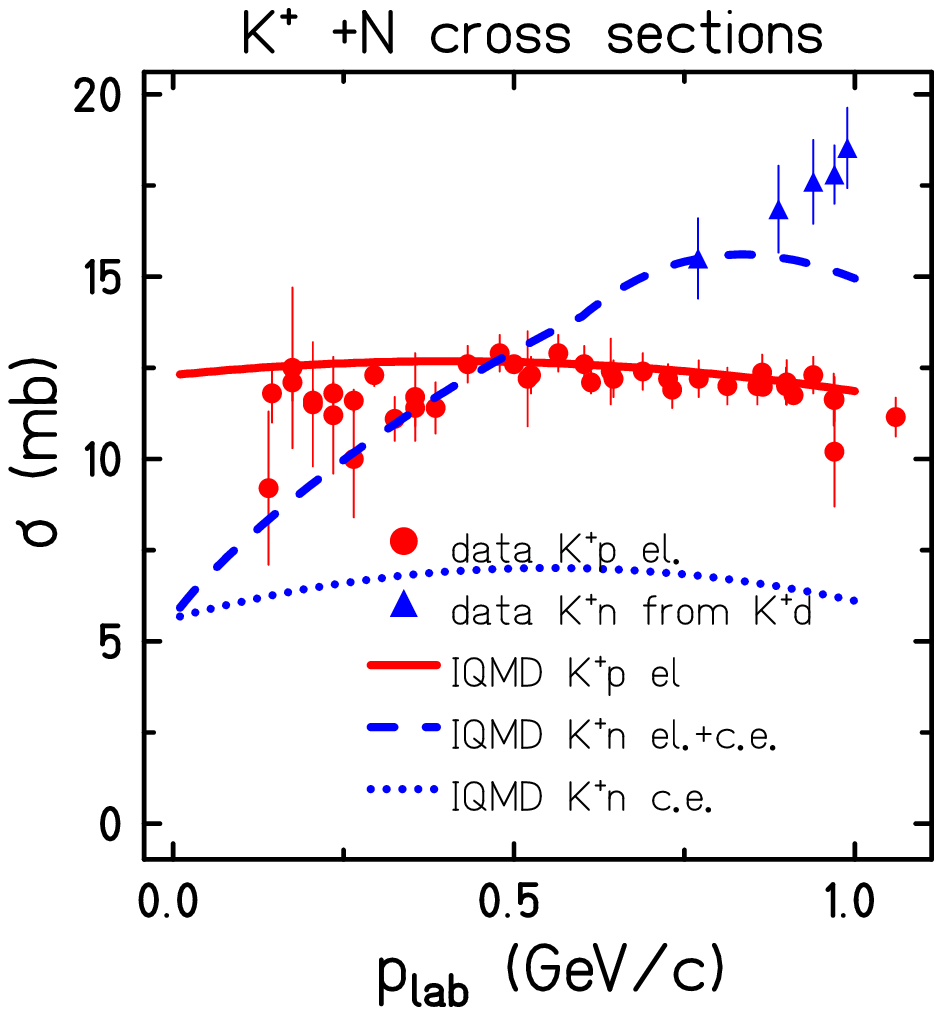,width=0.3\textwidth} &
\epsfig{file=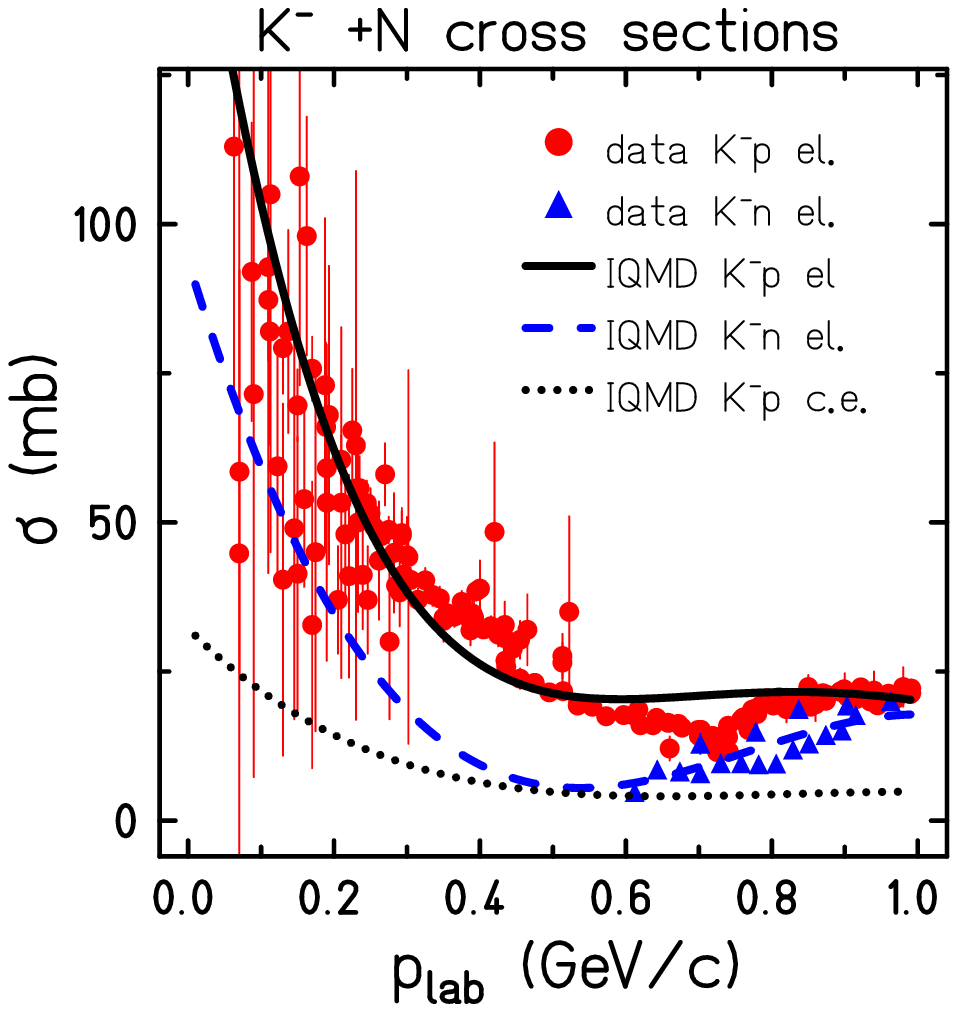,width=0.3\textwidth}
\end{tabular}
 \caption{Total cross section for ${\rm K}^\pm$ p and ${\rm K}^\pm$ n elastic scattering and charge exchange (c.e.) reactions \cite{totkm} and the parameterizations used in IQMD calculations (see text). }
\Label{kaon-elastic}
\end{figure}
The red bullets show experimental cross sections of elastic scattering
with a proton.
The left hand side presents data of elastic scattering of
the K$^+$ with a proton, the right
hand side elastic scattering of the K$^-$ with a proton.

We parameterize these cross sections as $\sigma(\rm K^+\p)/{\rm
mb}$ = $12.3+1.88 p-2.32 p^2$ (red solid line, left hand side) and
$\sigma(\rm K^-\p)/{\rm mb} $=$ 169-809 p+1657 p^2-1585 p^3+706
p^4-118 p^5$ (black solid line, right hand side) where $p$ is the
kaon momentum in the nucleon rest frame in GeV/$c$. For $p>2$ GeV
we set $\sigma(\rm K^-\p)=2.5$ mb.

Similarly there exist cross
sections for the scattering with neutrons (blue triangles).
The left hand side shows the total
scattering cross section of a K$^+$ with a neutron.
The data (blue triangles) are deduced from scattering with deuterons.
The total cross section includes elastic scattering $ {\rm K^+}
{\rm n} \leftrightarrow {\rm K^+} {\rm n}$ and the charge exchange
$ {\rm K^+}{\rm n} \leftrightarrow {\rm K^0} {\rm p}$.
The IQMD parametrizations for the total cross section are given by the blue
dashed lines, those of the charge exchange by the blue dotted lines.
The difference of both curvses corresponds to the elastic cross section.

The right hand side of  Fig.~\ref{kaon-elastic} presents the data of
elastic scattering of  K$^-$ with neutrons (blue triangles).
The IQMD parametrization is shown as blue dashed line on the right hand
side. The black dotted line shows the
parametrization of the charge-exchange reaction
 $ {\rm K^-} \p \leftrightarrow \overline {\rm
K^0} {\rm n}$.

Finally antikaons may be absorbed by a nucleon, producing a pion
and a hyperon.

\subsection{The standard set of our IQMD  and HSD calculations}\label{IQMD-standardset}
If not referenced explicitly the calculations performed with IQMD
are using a soft equation of state with momentum dependent
interactions (SM, see Table~\ref{eostab}). For the HSD a semi-soft
equation of state is used with an incompressibility $K\simeq
250$~MeV. For the \kp nucleus potentials we use in both models the
parametrization discussed in subsection \ref{knpot-desc}, assuming
$\alpha=1$ in Eq.\ref{kpotfactor}. Please note that this potential
differs from that originally used in the HSD calculations
\cite{Cassing:1999es}. For the \km nuclear potential in IQMD we
use as well the parametrization of subsection \ref{knpot-desc},
assuming $\alpha=1$ in Eq.~(\ref{kpotfactor}) whereas in HSD a
G-matrix approach \cite{Cassing:2003vz} is employed. The nucleons
collide with free nucleon-nucleon cross sections without any
medium modification. The kaons are produced using the
parametrization of Sibirtsev (Eq.~\ref{sibirtsev-fit}) for pp-NYK
collisions and those of Tsushima {\rm et al.}
\cite{Tsushima:1994rj,Tsushima:1998jz} for the $\Delta$-induced
collisions. For pn-collisions we apply a factor of 5/2 to the
pp-cross section. Therefore the parameterizations of the dominant
\kp production cross sections (${\rm BB}\to {\rm BYK } $ and $\pi
{\rm B}\to {\rm YK } $ are identical for both models.

When comparing IQMD and HSD calculations the following main features
should be taken into account:
\begin{description}

\item[KN elastic scattering:] ~~~~ \\
The elastic scattering of (anti)kaons with nucleons in IQMD treats the 
collsion partners as free on-shell particles while HSD performs the collision
with effective masses.

\item[Production of $\km$:] ~~~~

\begin{itemize}
\item In IQMD there is no propagation of the finite width of the
spectral function of the $\km$, as it is done in HSD. IQMD assumes
the $\km$ to be on shell. \item In HSD the treatment of off-shell
effects also influences the production processes, e.g. $\pi \rm Y
\to N \km$. Resonances below the threshold of the free production
(like the $\Lambda(1405)$) may change  the production cross
section in the medium. This is incorporated in HSD. In IQMD the
production process is assumed to be the free production cross
section taken at $\sqrt{s}$ modified by the in-medium mass of the
$\km$.

\end{itemize}

\end{description}

\subsection{Dynamics of heavy-ion collisions}

As an introduction to the dynamics of kaon production we briefly
sketch the evolution of a heavy-ion collision in this energy
range, in which the production and propagation of strange
particles is embedded.

We describe the reaction of Au+Au at 1.5 \AGeV incident beam energy
at an impact parameter of $b=0$ fm.  The time of the whole reaction is less
than $10^{-22}$ seconds and we describe the time evolution in
units of fm/$c$:
$
1~ \mbox{fm/$c$} = \frac{10^{-15} \mbox{m} }{3 \cdot 10^{8}
\mbox{m/s} } \approx 3.3 \cdot 10^{-24} \mbox{s}
$

The time $t=0$ fm/$c$ is chosen such that projectile and target have first contact. The
first nucleon-nucleon collisions take place and first resonances
are produced. The nuclear matter
starts to decelerate by collisions between projectile and target nucleons.

\begin{figure}[hbt]
\begin{tabular}{cc}
\epsfig{file=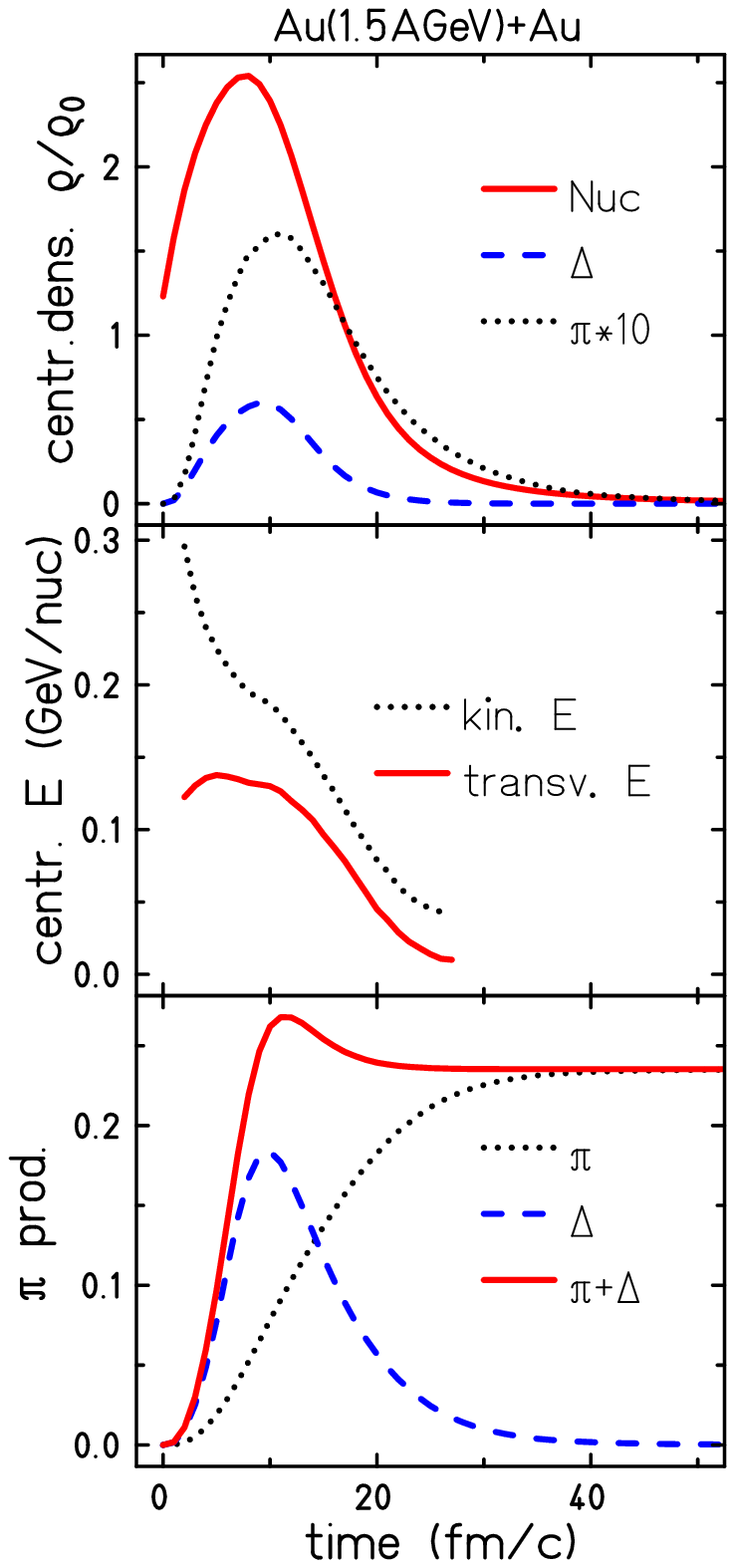,width=0.3\textwidth} &
\epsfig{file=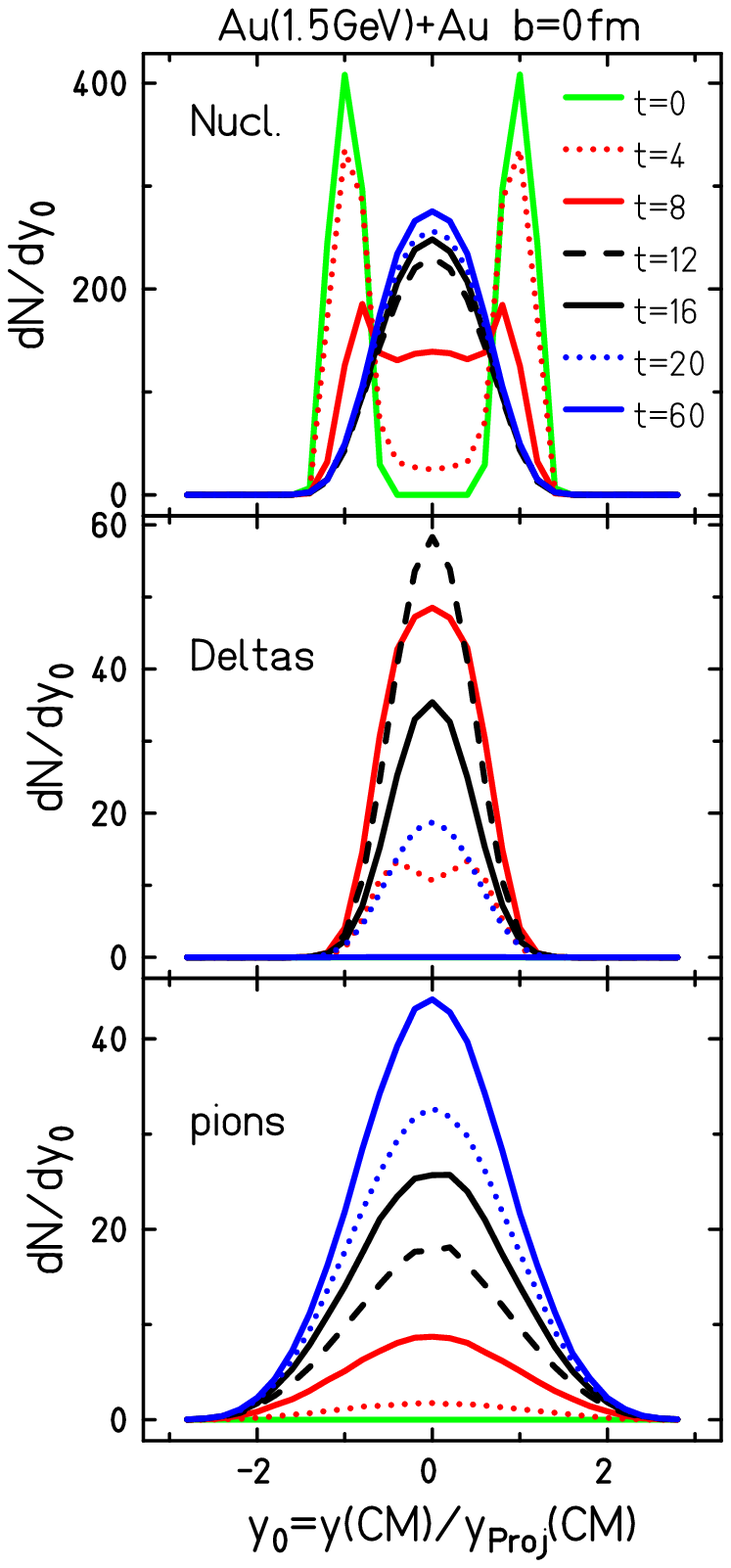,width=0.32\textwidth} \\
\end{tabular}
 \caption[Time evolution of density, energy, pion number and
 rapidity distributions]{Time evolution of a Au+Au collision at 1.5 \AGeV and  at $b$=0 fm.
 Left: Evolution of central density, central energy, pion and Delta multiplicities
 as a function of time. Right: Rapidity
 distributions for nucleons, Deltas, and pions for different times of the
 evolution (in fm/$c$).}
\Label{time-evol-dns}
\end{figure}

Figure \ref{time-evol-dns} shows, on the left hand side, the time
evolution of the densities of N, $\Delta$, and $\pi$ in the
central region (top), the mean kinetic and transverse energies in
the central region (mid) and the number of deltas and pions
(bottom). The nucleon density (top, red full line) grows  rapidly,
reaches a maximum of about 2.5 times saturation density at about 8
fm/$c$ and falls afterwards. The density of $\Delta$'s (black
dotted line) begins to rise later, reaching a maximum  of nearly
half saturation density at about 9 fm/$c$ and falls rapidly
afterwards. The pion density (multiplied by 10, blue dashed line)
starts to rise even later and reaches  a maximum  of around one
sixth of the saturation density at about 10 fm/$c$.

In the central region the kinetic energy (\lhs, middle, black
dotted line) drops rapidly. In the center-of-mass frame the
incident energy leads to a large initial longitudinal momentum.
This energy is partly consumed by the creation of resonances
($\Delta$'s), but another part of it is redirected by nuclear
collisions into the transverse direction. Therefore, the
transverse energy (red full line) is built up rapidly in the
central region. Equilibrium is not obtained, as will be discussed in section
VI. In the final expansion phase the fast (high
energy) particles are leaving the central region very rapidly,
leaving the slower particles behind. Thus, the energies are
falling. Finally, the particles are leaving the central region.

The fast decrease of the energy is accompanied by a fast increase
of the number of deltas (bottom, blue dashed line). As already
noted, the production of resonances consumes a large share of this
energy, which will be released later in the form of pions (black
dotted line). Their number is continuously increasing to reach the
final value at about 30-40 fm/$c$. Pions are strongly interacting
with nucleons. Therefore, pions propagating in the dense medium
have a high chance to be reabsorbed and to feed the number of
deltas. This effect explains the slow increase of the pion yield
when the density is high. The deltas themselves are also
reabsorbed by the nuclear matter in N$\Delta \to$ NN collisions.
This effect can be seen looking at the total number of deltas and
pions (full red line). It shows a maximum at about 10-12 fm/$c$
and decreases slightly afterwards. From 20 fm/$c$ on it stays
roughly constant. Now only the relative contribution of deltas and
pions to the total yield is changing.

The right hand side of Fig.~\ref{time-evol-dns} shows the time
evolution of the rapidity distribution of nucleons (top), deltas
(middle) and pions (bottom) at different times. The rapidity has
been scaled to the projectile rapidity in the center-of-mass.
Thus, a value of 1 corresponds to projectile rapidity, -1 to
target rapidity.

At $t=0$ fm/$c$ (\gfl) projectile and target show their incident
momentum distribution as peaks at projectile and target rapidity.
The width of the peaks is due to the momentum distribution
of the nucleons in the nuclei. Deltas and pions do not exist at this time. At
$t=4$ fm/$c$ (red dotted line) first nucleons have been stopped to
mid-rapidity, and first deltas have been produced. The stopped
nucleons collide with incoming nucleons of the projectile and
target causing a slight double peak in the delta rapidity
distribution about halfway between the cm-rapidity (0) and the
projectile (1) resp. target rapidity (-1). There are almost no
pions.

At $t=8$ fm/$c$ (\rfl) when maximum density is reached, the nuclei
still have not been completely stopped. There are remnants of the
peaks at projectile and target rapidities. The number of deltas
has strongly increased, their distribution is now peaked around
mid-rapidity. The rather narrow width is due to the high mass of
the delta which takes up a big part of the energy available in the
    first collisions. In the decay of deltas a large amount of momentum is
given to the light pions. Therefore, the rapidity distribution of the pions becomes
rather broad.

At $t=12$ fm/$c$ (black dashed line) the rapidity distribution of nucleons is
peaked at mid-rapidity. The delta distribution shows its maximum
values while the pion distribution rises continuously. At $t=16$
fm/$c$ (full black line)  and $t=20$ fm/$c$ (blue dotted line) the
nucleon rapidity distribution is slightly growing in the center.
This is due to the feeding by the decay of the deltas whose
yield is continuously falling. For kinematic
reasons in the $\Delta$-decay the nucleon receives only little
energy while most of the energy is given to the light pion whose
broad rapidity distribution is still rising. At $t=60$ fm/$c$
(full blue line) the reaction is in the final state. There are no
more deltas. The system is expanding and the particles are
directed outwards and will finally hit the detectors.

\subsection{Pion production in heavy-ion collisions}
As we have seen, the production of deltas and their decay into
pions is  very important for the understanding of the dynamics of
heavy-ion collisions. Thus we briefly report on comparisons of
experimental pion observables with IQMD calculations. For more
details of the description of pions in IQMD we refer to
Ref.~\cite{Bass:1995pj}. Details on experimental observations of
pions at these energies can be found in
\cite{Wagner:2000ak,Stoicea:2004kp,Averbeck:2000sn}. For more
detailed comparisons of experiment and theory concerning the
production of pions we refer to \cite{Reisdorf:2006ie}.

\begin{figure}[hbt]
\begin{tabular}{cc}
\epsfig{file=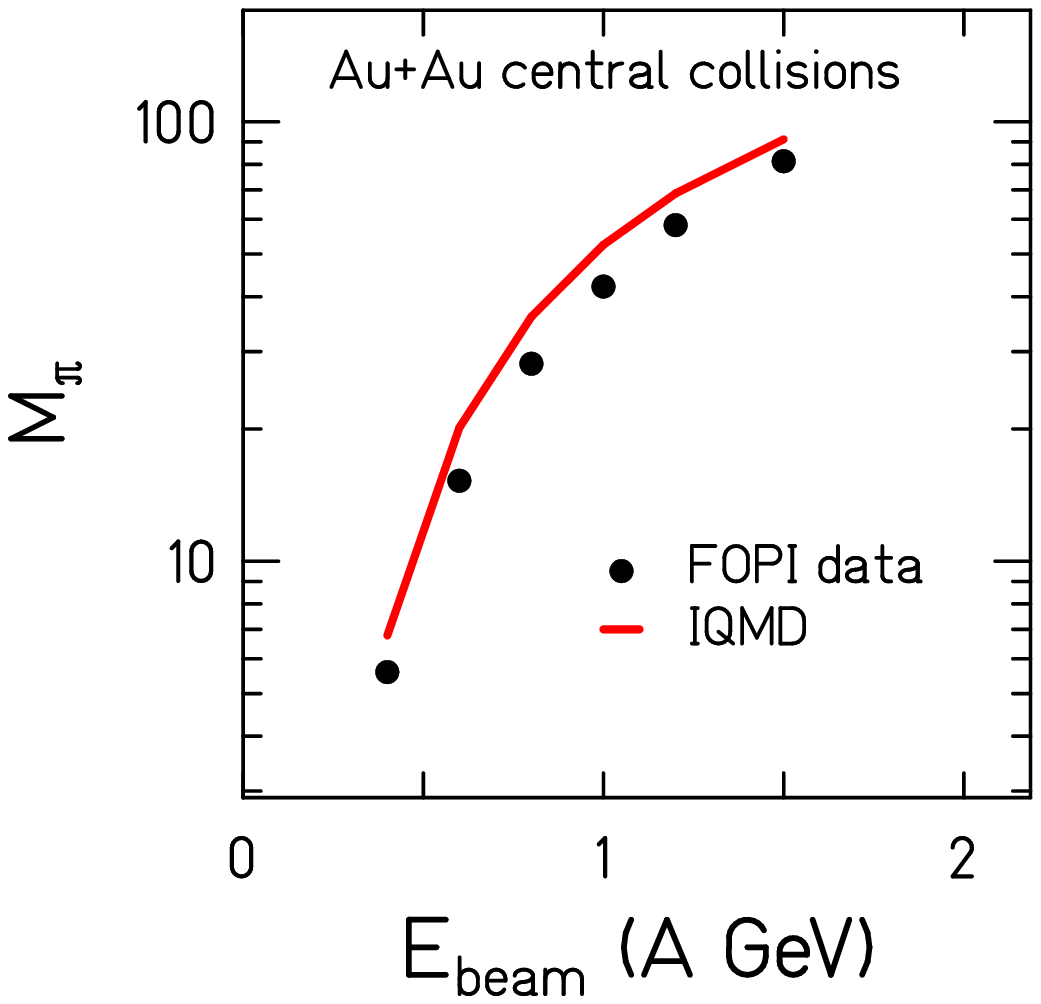,width=0.45\textwidth} &
\epsfig{file=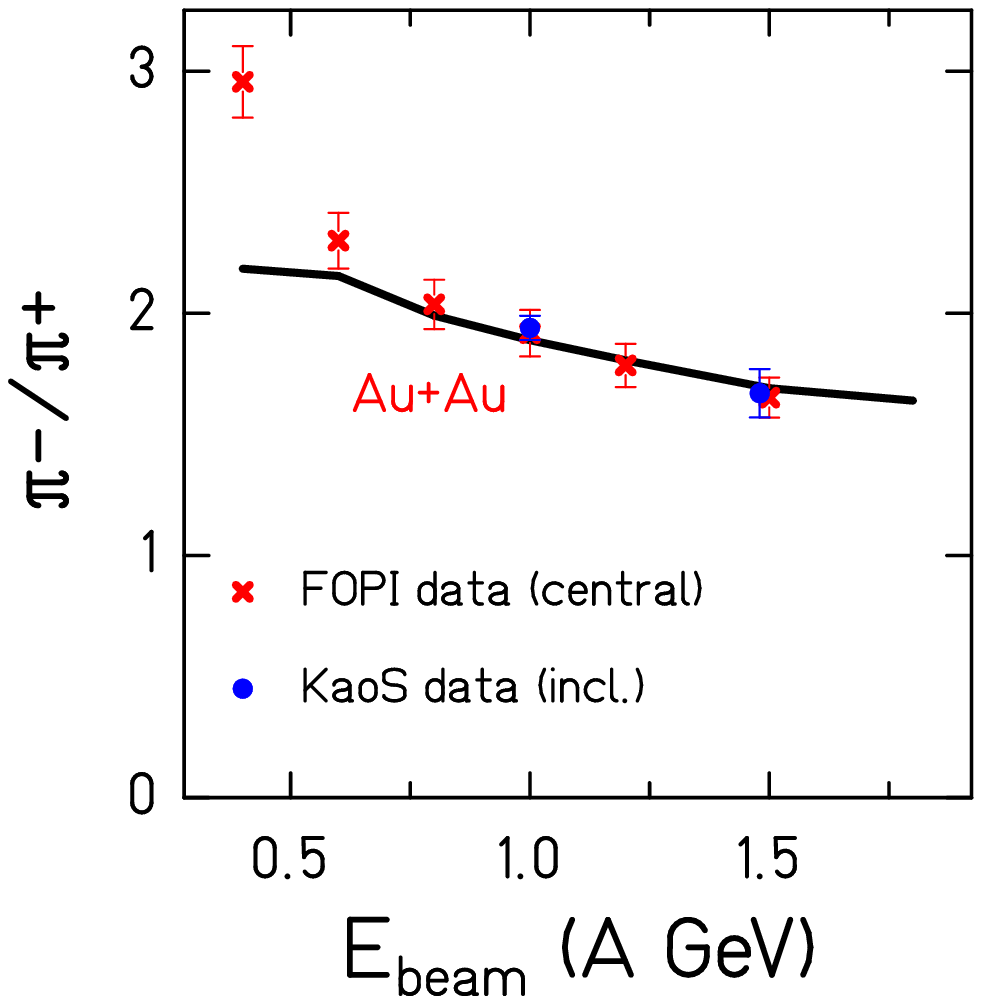,width=0.45\textwidth} \\
\end{tabular}
 \caption{Comparison of the experimental \cite{Reisdorf:2006ie} excitation function of the pion yield
 (left) and of the $\pi^-/\pi^+$ ratio (right) \cite{Tschuck,Wagner:2000ak} with IQMD calculations (full line)
  for central Au+Au collisions.}
\Label{pioncompa}
\end{figure}

Figure \ref{pioncompa} shows on the left side the comparison of
the excitation function of the pion yield in central Au+Au
collisions, measured by the FOPI collaboration
\cite{Reisdorf:2006ie}, with calculations performed by IQMD (red
full line). The calculated pion yield exceeds the experimental one
which is found to be mainly due to description of the reabsorption of
the delta in the dense medium.  The right side compares to the
ratio of the multiplicities of negatively and positively charged
pions (full line) to experimental data of Au+Au collisions at
different energies from the FOPI Collaboration (crosses, central
collisions) and from the KaoS Collaboration (circles, inclusive
data from~\cite{Tschuck,Wagner:2000ak}) . For energies above 500
$A$ MeV a good agreement between data and calculations is
observed.
\begin{figure}[hbt]
\begin{tabular}{cc}
\epsfig{file=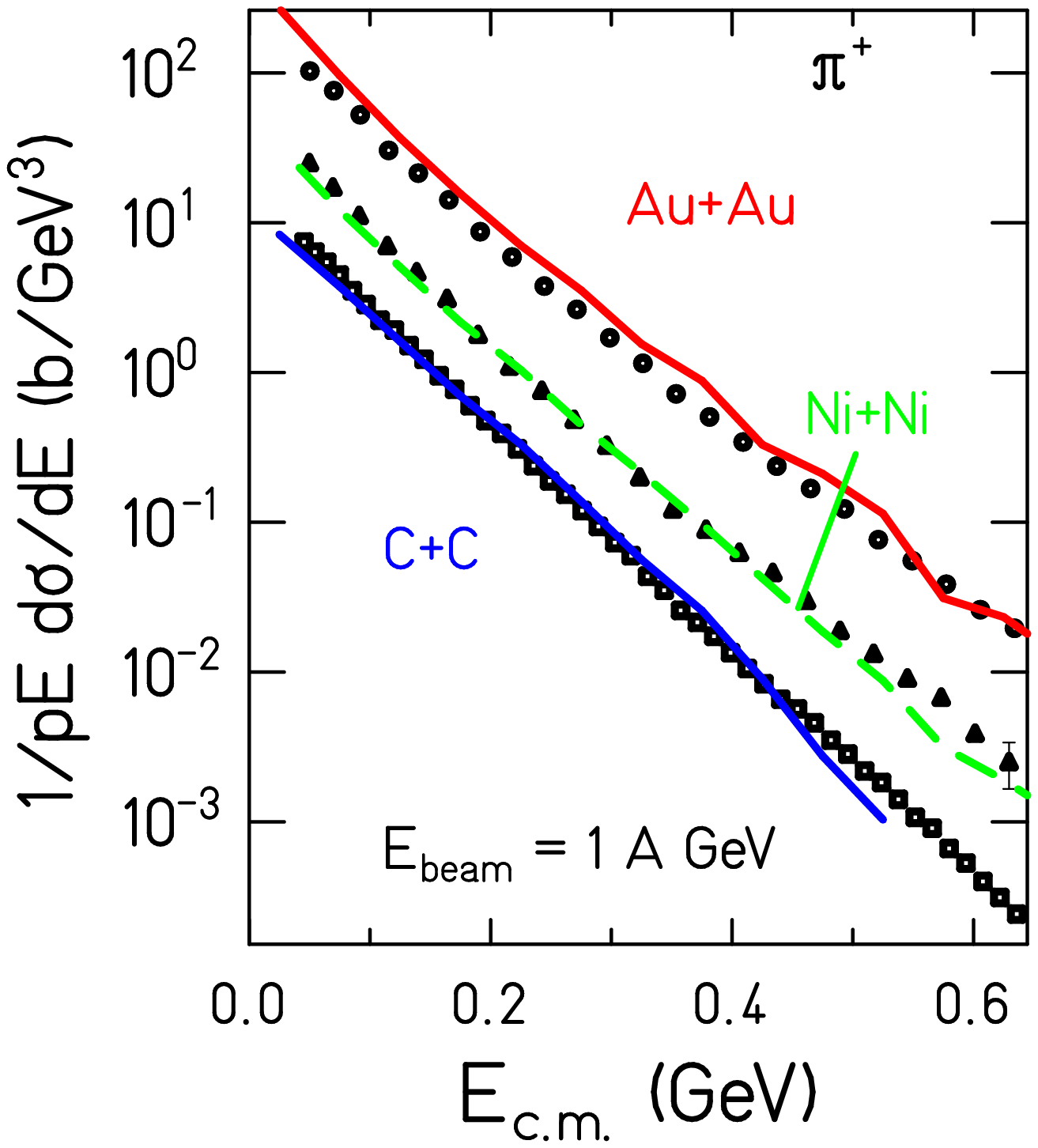,width=0.45\textwidth} &
\epsfig{file=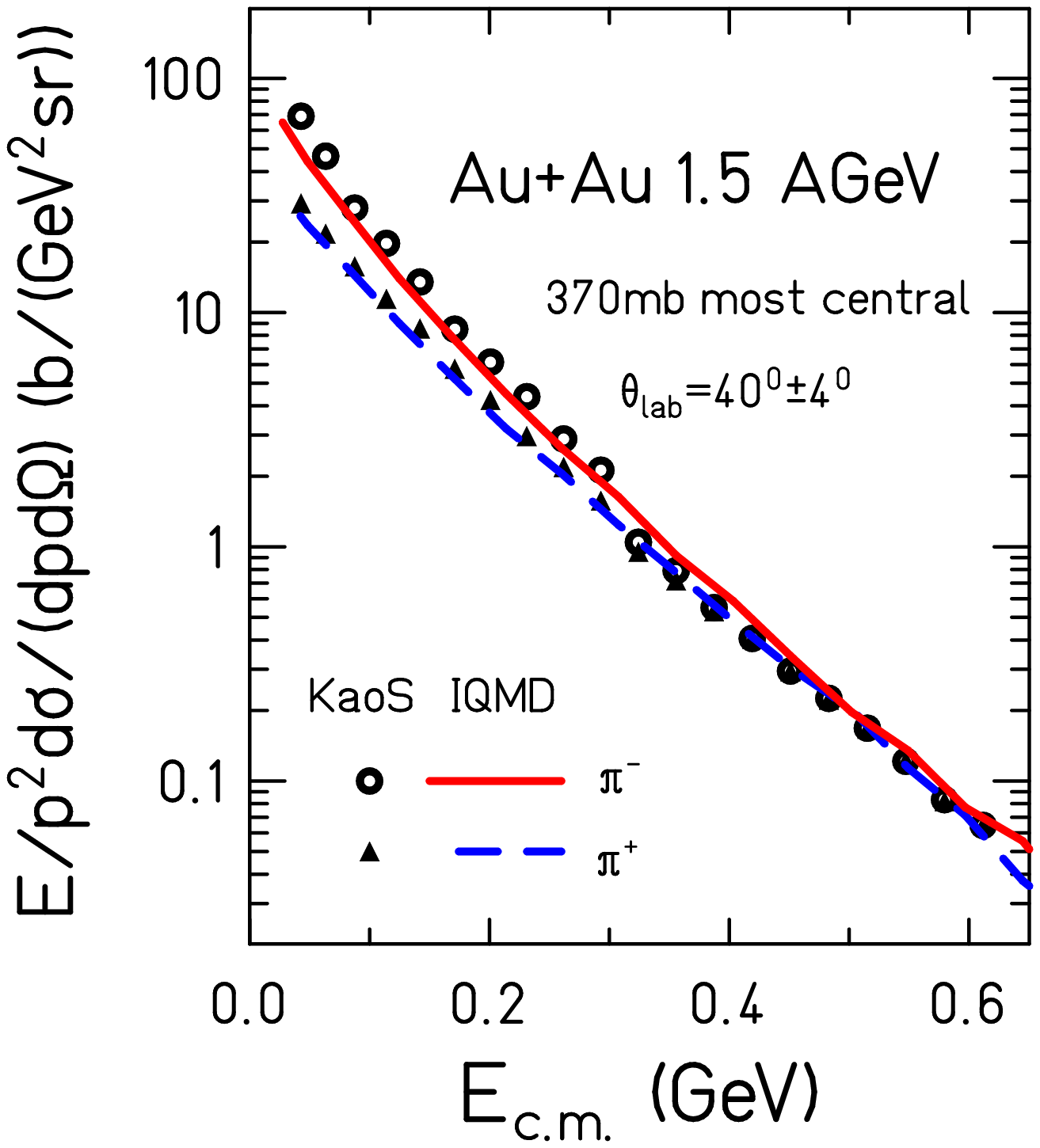,width=0.45\textwidth}  \\
\end{tabular}
 \caption{$\pi^+$ spectra of KaoS  at 1.0 \AGeV \cite{Oeschler:2002ch,Sturm:2000dm}
 compared with IQMD calculations
 (left). Comparison of the $\pi^\pm$ spectra at
 1.5 \AGeV beam energy  \cite{Tschuck} with IQMD calculations (right).}
\Label{pionKaoS}
\end{figure}

Figure \ref{pionKaoS} compares the spectra of $\pi^+$ from
inclusive collisions of C+C, Ni+Ni and Au+Au at 1.0 \AGeV incident
energy (left side) and of  $\pi^+$  and  $\pi^-$ in central
collisions (370 mb most central) of  Au+Au at 1.5 \AGeV incident
energy (right side) of IQMD calculations (lines) with KaoS data
(symbols)~\cite{Tschuck}. The absolute yields at kinetic center of
mass energies above 0.1 GeV as well as the slopes are well
described. It should be noted that subthreshold kaon production is
influenced by the behavior of high-energy pions.
Furthermore, it should be kept in mind that the pions are strongly
interacting with the nucleons. Thus, the pion spectra are
influenced by rescattering and their shape is determined mainly at
times later than the time of the production of kaons. The rapidity
distribution of pions is also reasonably well reproduced as can be
seen in \figref{pionFOPIdndy}. All transport theories have
problems to reproduce low energy $\pi$'s at mid-rapidity (see
\figref{pionKaoS}). These do not play, however, an important role,
neither for the strangeness multiplicity nor for the strangeness
dynamics.

\begin{figure}[hbt]
\epsfig{file=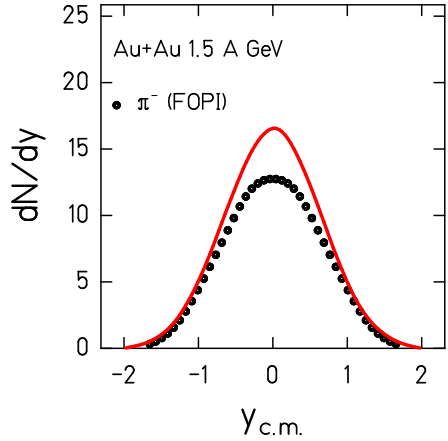,width=0.45\textwidth}
\caption{Comparison of the d$N$/d$y$ distribution of $\pi^-$ of
central Au+Au collisions ($b <$ 2 fm) at 1.5 \AGeV from the FOPI
Collaboration~\cite{Reisdorf:2006ie} with IQMD calculations.}
\Label{pionFOPIdndy}
\end{figure}

Last but not least we present in \figref{protonFOPIdndy} the
rapidity distribution of free participant protons. It is obtained
by subtracting from the total yield of participants those which
are bound in fragments. It should be stressed that for this
analysis the results of IQMD have undergone the same treatment as
the experimental data concerning filtering, triggering, event
selection, etc The width of the distribution is well reproduced.
Around mid-rapidity one observes slightly more protons in the IQMD
simulations as compared to experiment. This is due to the
underprediction of the yield of light composites like t, $^3\rm
He$, $\alpha$, which  are difficult to model in semiclassical
approaches. The difference is less than 10\% and therefore does
not essentially perturb our conclusions.

\begin{figure}[hbt]
\epsfig{file=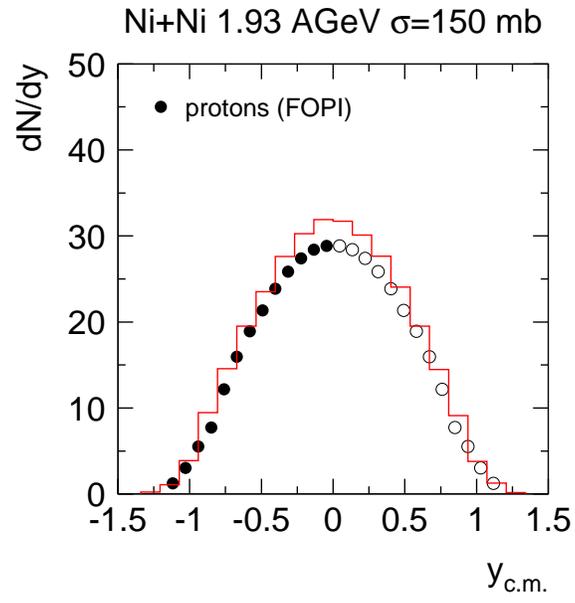,width=0.45\textwidth}
\caption{Comparison of the d$N$/d$y$ distribution of free participant protons
of Ni+Ni collisions at 1.93 \AGeV measured by the FOPI
collaboration \cite{Hong:1997mr} with IQMD calculations (red
histogram). } \Label{protonFOPIdndy}
\end{figure}

\clearpage
\section{Production of \kp mesons}
\label{Kplus}
\setcounter{figure}{0}
%

\subsection{\kp production in p+A collisions}

At first glance, p+A collisions seem to be a nice intermediate
step from p+p to A+A studies. Indeed, they offer the opportunity
to investigate the kaon properties in a well-defined environment
as the charge distribution of the nucleons in nuclei is known
from electron scattering. However, in going from p+p to p+A
collisions, several changes happen simultaneously:
\begin{enumerate}
\item Strange particles can be created at sub-threshold energy,
i.e.~at an incident energy at which a production in a free p+p
collision is not possible due to energy-momentum conservation. The
K$^+$-production threshold is constrained by the conservation of
strangeness. The lowest threshold in p+p reactions is given by the
channel $\p+\p \to \p+\Lambda+\K^+$ with the threshold
\begin{equation}
\sqrt{s}= \sqrt{\sum P_\mu P^\mu} \ge m_{\rm p} + m_\Lambda +
m_{\rm K^+}=2.55 \, {\rm GeV};\qquad P=p_1+p_2
\end{equation}
where $p_1$ and $p_2$ are the four-momenta of the colliding
particles. This implies a minimum incident kinetic energy of 1.58
GeV for the proton projectile. In p+A collisions a K$^+$  can in
principle be produced  starting from a beam energy of \be E_{\rm
p} = \frac{s_{\rm thres} - m_{\rm p}^2 -m_A^2}{2m_A}; \qquad
\sqrt{s_{\rm thres}}=m_A+m_\Lambda+m_{\rm K^+} \ee where $m_A$ is
the mass of the target nucleus.  At the absolute threshold, \kp
creation is very improbable because the available energy and
momentum has to be transferred in one single nucleon nucleon
collision producing a \kp.
Therefore, the cross section decreases rapidly at energies below
the nucleon-nucleon threshold. \item Due to the Fermi momentum of
the nucleons in the target nucleus, the effective c.m.
energy in a p+A collision may be higher than that in a p+p
collision at the same beam energy. This reduces the threshold
incident energy in single NN collisions down to a beam energy of
around 1 GeV. \item Resonances may serve as an energy storage
since they transform kinetic energy into mass. The projectile
proton which collides with a target nucleon may create a $\Delta$.
Subsequently, the $\Delta$ may collide with another target nucleon
producing a K$^+$.
\item There is a potential interaction between the \kp and the surrounding
nucleons which modifies the \kp properties in matter.
\end{enumerate}
\begin{figure}[hbt]
\begin{tabular}{cc}
\epsfig{file=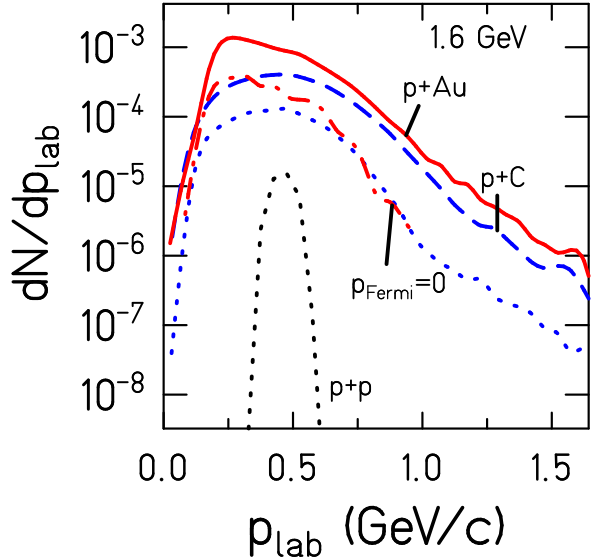,width=0.6\textwidth} &
\end{tabular}
 \caption{Inclusive momentum spectra of \kp mesons produced in p+p, p+C and p+Au
collisions at 1.6 GeV. We compare calculations for p+C and p+Au,
including (solid and dashed lines) and excluding (dotted and
dash-dotted lines) the Fermi motion, with p+p calculations.}
\Label{pa-nopot}
\end{figure}
Figure~\ref{pa-nopot} shows the \kp momentum distribution in the
laboratory system in p+p and p+A collisions at 1.6 GeV
incident energy, i.e.~slightly above the threshold. In the p+p
case (\bpl) the emitted \kp mesons are limited to a very small
momentum range in the c.m. system and therefore the
kinematically allowed region in the laboratory is also a quite
narrow and centered around $p_{\rm lab} = 0.5$ GeV. Already  in
the p+C collision (\bdl) there is a much wider distribution which
is also peaked around $p_{\rm lab} = 0.5$ GeV. Surprisingly, this
widening is only in part due to the Fermi momentum of the nucleons
in the carbon target where the projectile may collide with a
nucleon with a Fermi momentum in the opposite direction of its
momentum and thus enhance the available energy of the \kp. This
can be seen when comparing the p+p curve (\bpl) with a calculation
for p+C without Fermi momentum ( $p_{\rm Fermi}=0$, blue dotted
curve). The Fermi motion enhances the cross section but does not
change the form of the momentum distribution obtained in
calculations which include the Fermi motion (\bdl ).

The major part of the difference between p+p and p+C stems from
the opening of additional channels due to multi-step processes:
$\N_1\N_2 \to \N\Delta, \quad \N_3 \Delta \to \N\Y\K$ ($\N\Delta $
channel) and $\N_1\N_2 \to \N \Delta, \quad \Delta \to \N \pi
\quad \pi \N_3 \to \Y\K$ ($\N\pi$ channel) which in the case of
p+C contribute 30\% or 15\% to the total kaon multiplicity at
$E_{\rm lab}$ = 1.6 GeV  with zero or with full Fermi momentum,
respectively. The relative contributions of these channels are
strongly beam energy dependent: At energies below $E_{\rm lab} =
1.3$ GeV, the $\N\pi $ channel dominates as illustrated by
Fig.~\ref{pc-e-nc1}.

\begin{figure}[hbt]
\epsfig{file=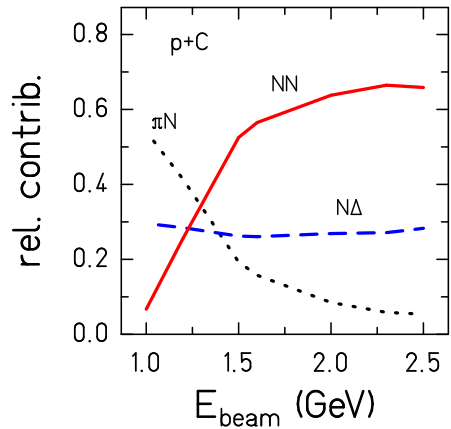,width=0.45\textwidth}
 \caption{Contribution of different production channels to the \kp production in p+C collisions
 as a function of the incident beam energy. }
\Label{pc-e-nc1}
\end{figure}

Because the cross section increases strongly close to the
threshold (see Fig.~\ref{pc-e-nc1}) the Fermi motion
allows for a strong enhancement of the \kp production due to
collisions in which the Fermi motion of the target nucleon points
toward the projectile. Finally, if the first collision is elastic (the cross
section at these energies is strongly forward-backward peaked)
there is a chance for producing a \kp in a second or even third
collision of the projectile  with a target nucleon. All these
effects enhance the \kp production cross section with
respect to p+p collisions and explain the huge increase of the \kp
multiplicity when going from p+p to p+C collisions. When going
from p+C (\bdl) to p+Au (\rfl) \figref{pa-nopot} shows an
additional increase at low laboratory momenta. This is due to \kp
production in second or third collisions of the projectile after
it had already lost a part of its energy. It may still produce a
\kp if the Fermi momentum of the hit target nucleon supports this.
In the Au case there is a much higher chance to find such a
particle. The \kp is then not anymore produced in the p+p
c.m. system but in the c.m. of the slowed-down
projectile with a target nucleon. This explains the shift toward
lower momenta which can be seen in calculations with (\rfl) and
without (\rml) Fermi momentum.

In Fig.~\ref{pA_ANKE} we compare the simulations in the standard
version
with experimental p+C~$\rightarrow$~\kp data measured at COSY by
the ANKE Collaboration \cite{Buescher:2001qu,Buescher:2002np} at very small laboratory angles
($0^\circ \le \theta \le 6^\circ$), and at energies around
threshold ($E_{\rm beam} = 1.2,\ 1.5,\ 2.0 $ and $ 2.3$ GeV).
There is an uncertainty in the absolute values of about 20\% due
to the ANKE cross section determination~\cite{Anke_priv}.
Simulations of spectra measured under almost zero degree in the
laboratory are very sensitive to the rescattering
of the kaons.  Isotropic KN collision will deplete the yield
in forward direction and rescattering into this region is rare.
This depletion is directly proportional to the cross section.
Changes of the density distribution of the target nucleons, the \kp nucleus
potential or differences of the momentum distribution of the
nucleons which all have influence on the c.m. energy of the
scattering partners show visible changes of the \kp spectra
because the cross section increases very strongly as a function of
$\sqrt{s}$. The influence of the Coulomb potential is negligible.
The calculations have been filtered to respect the experimental
acceptance.  The simulated spectra show the same form as the
experimental ones but over-predict slightly the experimental
values for the standard (Sibirtsev \cite{Sibirtsev:1995xb}) NN$\to \kp$ cross
section. As seen for the $E_{\rm beam} =  1.5 $ GeV data the other
available NN$\to$\kp cross section parameterization (Tsushima
\cite{Tsushima:1994rj,Tsushima:1998jz}) brings the simulation closer to the data. Besides for
high $p_{\rm lab}$ values, the  \kp production in p+C collisions
is a superposition of elementary p+p $\to$ K collisions (dotted
magenta line, $E_{\rm beam} =  2.0 $ GeV).  Fermi motion of the
nucleons and multiple NN collisions tend to increase the \kp
yield, the \kp nucleus potential has to opposite effect. At high beam
energies the influence of the Fermi motion is visible at large
momenta only. At the two lower energies the influence of the \kp nucleus
potential is almost compensated by the Fermi motion. Consequently,
at low momentum the \kp spectrum in p+C collision is very close to
a superposition of p+p collisions with a scaled isospin averaged
p+p cross section ($\sigma({\rm NN} \rightarrow {\rm K^+ \Lambda
N}) = 1.5 \ \sigma({\rm p+p} \rightarrow {\rm K^+ \Lambda p}$).
Due to poorly known pn$\rightarrow$ \kp Y nn cross section and due
to the uncertainty of the in-matter modification of the pp $\to
$\kp Y pn cross section which are different in the
parameterizations of Sibirtsev and Tsushima
it is presently difficult to interpret the p+A data. If we switch
off the Fermi momentum (green dashed-dotted line) the cross
section drops by a factor of 2 (10) for the reaction at 1.5 (1.2)
GeV. This demonstrates another time the importance of the Fermi
motion for the total yield of the sub-threshold particle
production.

\begin{figure}[hbt]
\epsfig{file=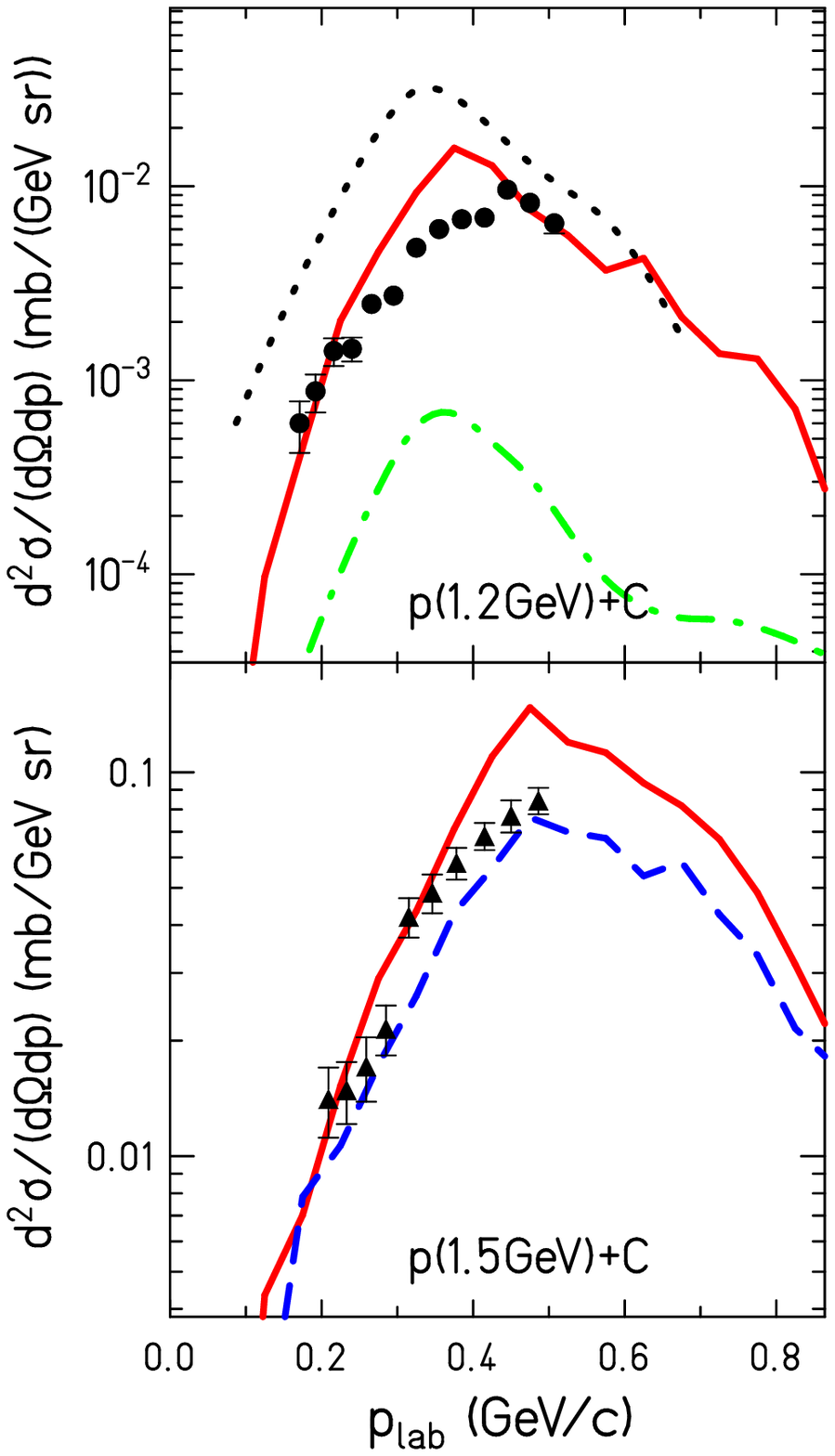,width=0.49\textwidth}
\epsfig{file=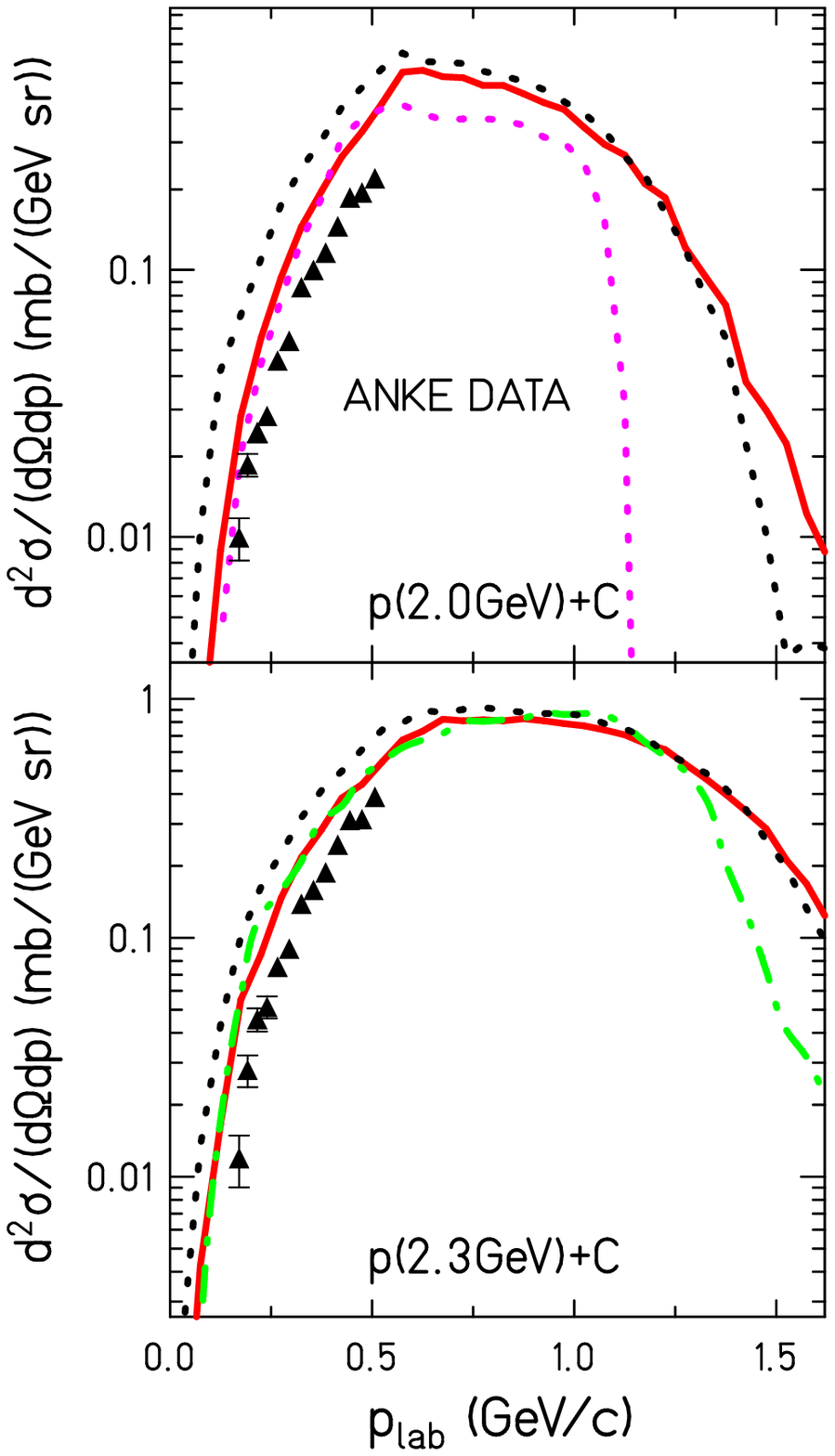,width=0.49\textwidth}
\caption{
Calculated p+C spectra
 at energies below and above the threshold as compared with the
 experimental results of the ANKE Collaboration \cite{Buescher:2001qu,Buescher:2002np}. Full red line: Standard
 version, dotted black line: Standard without \kp nucleus potential,
 green dashed-dotted: Standard without Fermi motion,
 blue dashed: Standard but Tsushima parametrization of the NN$\to $\kp cross
 section, dotted magenta: Superposition of p+p cross section.}
\Label{pA_ANKE}
\end{figure}

Spectra at larger laboratory angles are much less sensitive to
these details. Here rescattering plays already an important role.
In Fig.~\ref{pA_data} we compare IQMD calculations with the only
other data set which is available for \kp production in p+A
collisions. The KaoS Collaboration has measured \kp at laboratory
angles of $\theta_{\rm lab} = 32^\circ$ and
$40^\circ$~\cite{Scheinast:2005xs}. These values correspond to an
emission close to $90^\circ$ in the center of mass of the p+p
system. We present results using two different NN $\rightarrow$
\kp cross-section parameterization (Tsushima \cite{Tsushima:1994rj,Tsushima:1998jz},
Sibirtsev\cite{Sibirtsev:1995xb}) to demonstrate the uncertainty of the result
due to the limited knowledge of this input quantity.

\begin{figure}[hbt]
\begin{tabular}{cc}
\epsfig{file=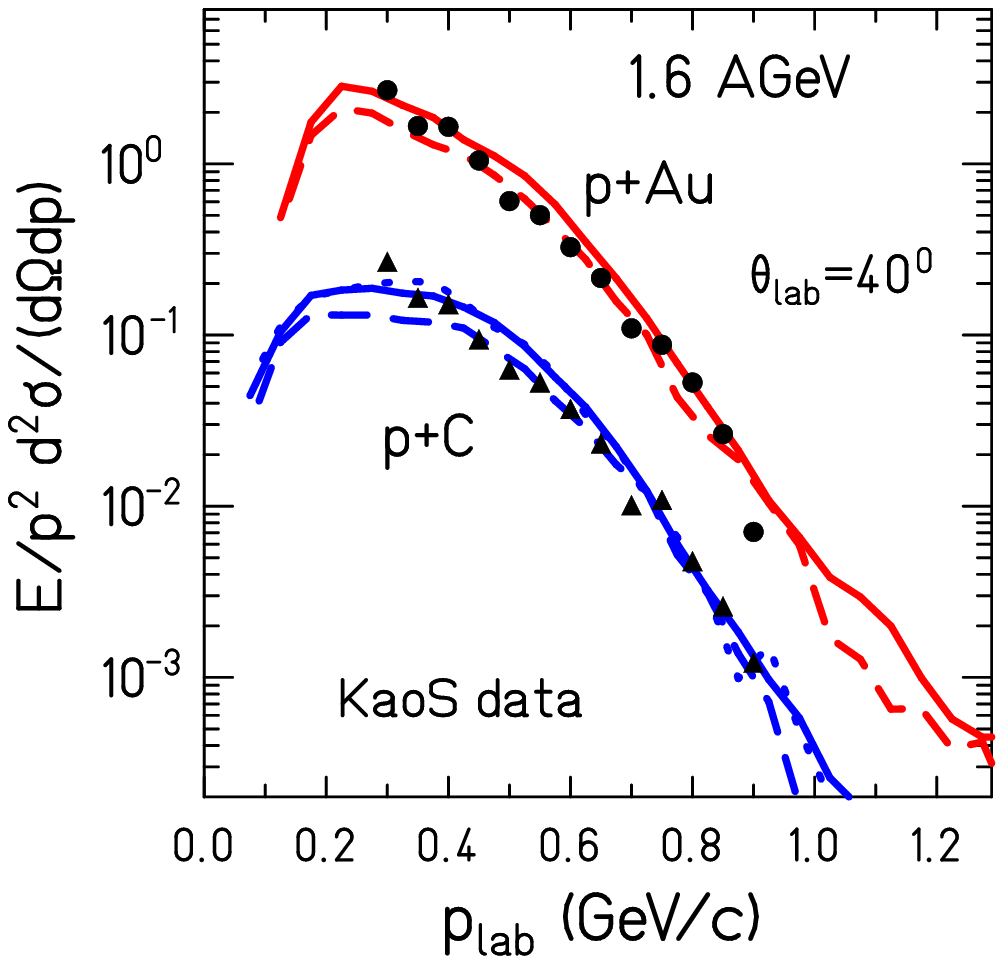,width=0.45\textwidth} &
\epsfig{file=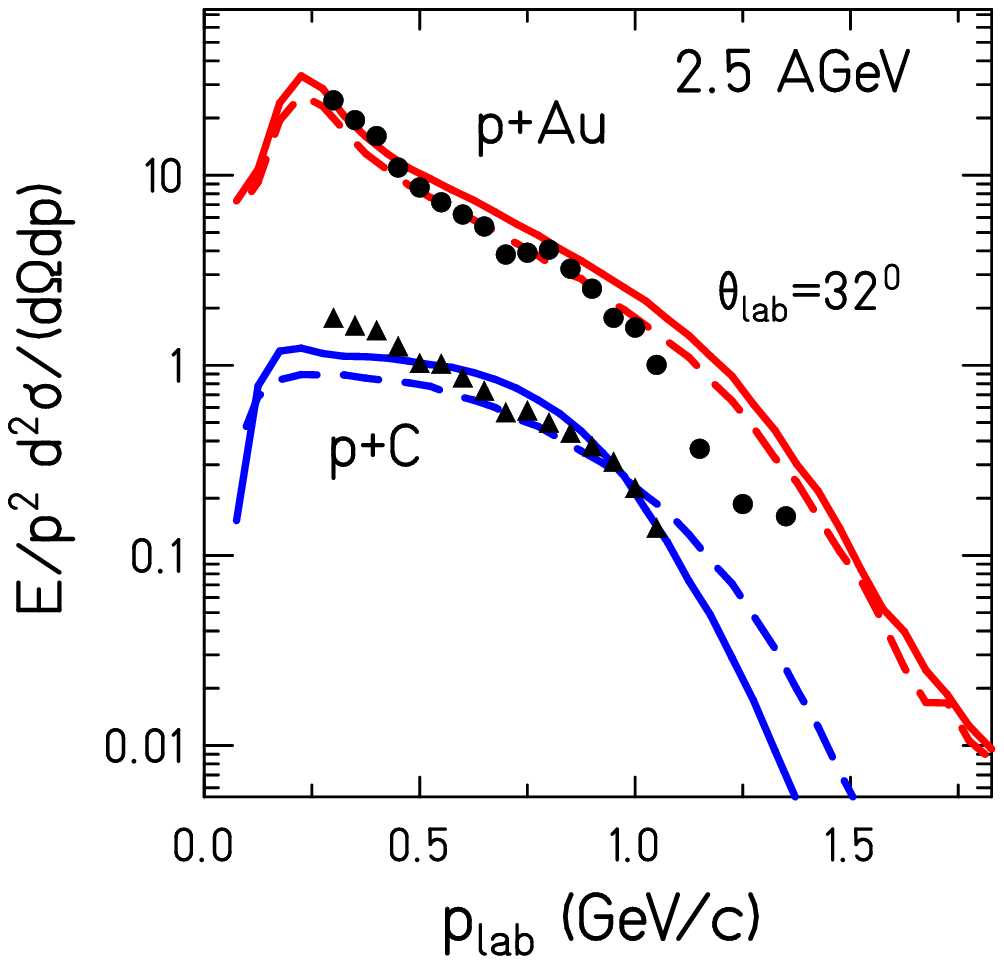,width=0.45\textwidth}
\end{tabular}
 \caption{Comparison of p+C and p+Au
 momentum spectra of the KaoS Collaboration~\cite{Scheinast:2005xs} with calculations.
The solid lines refer to the cross section parametrization of
Sibirtsev {\rm et al.}~\cite{Sibirtsev:1995xb} and the dashed ones
to that of Tsushima {\rm et
al.}~\cite{Tsushima:1994rj,Tsushima:1998jz}.} \Label{pA_data}
\end{figure}
The standard IQMD simulation reproduces these data quite well,
both the shape as well as the absolute value, a pre-requisite to
apply the model to A+A collisions.

\subsection{The major production channels in A+A collisions}
By going from p+A to A+A collisions the importance of the
different production channels changes. Whereas in p+A collisions
NN interactions are the most important ones for the \kp production
(Fig.~\ref{pc-e-nc1}), in A+A collisions the channels which
involve a $\Delta$ dominate, as seen in
\figref{AuAu_CC_excitation_function}, left. In heavy-ion reactions
the chance increases that a $\Delta$, produced in a collision
between a projectile and a target nucleon, hits a second
projectile or target nucleon. In such a collision the available
c.m. energy, $\sqrt{s}$, is quite high due to the large momentum
of the nucleon on the one side and due to the high mass of the
$\Delta$ on the other side. Because all cross sections increase
with $\sqrt{s}$, we expect a high production yield in this
channel. Figure \ref{AuAu_CC_excitation_function} shows on the
left hand side  the excitation function of the contributions of
the different channels to the \kp yield in central Au+Au
collisions. The right hand side displays the excitation function
of \kp production via all channels and via NN $\rightarrow \kp
\Lambda$ N alone for central Au+Au and C+C collisions. We observe
that for both systems the NN channel represents only a small
fraction.

\begin{figure}[hbt]
\epsfig{file=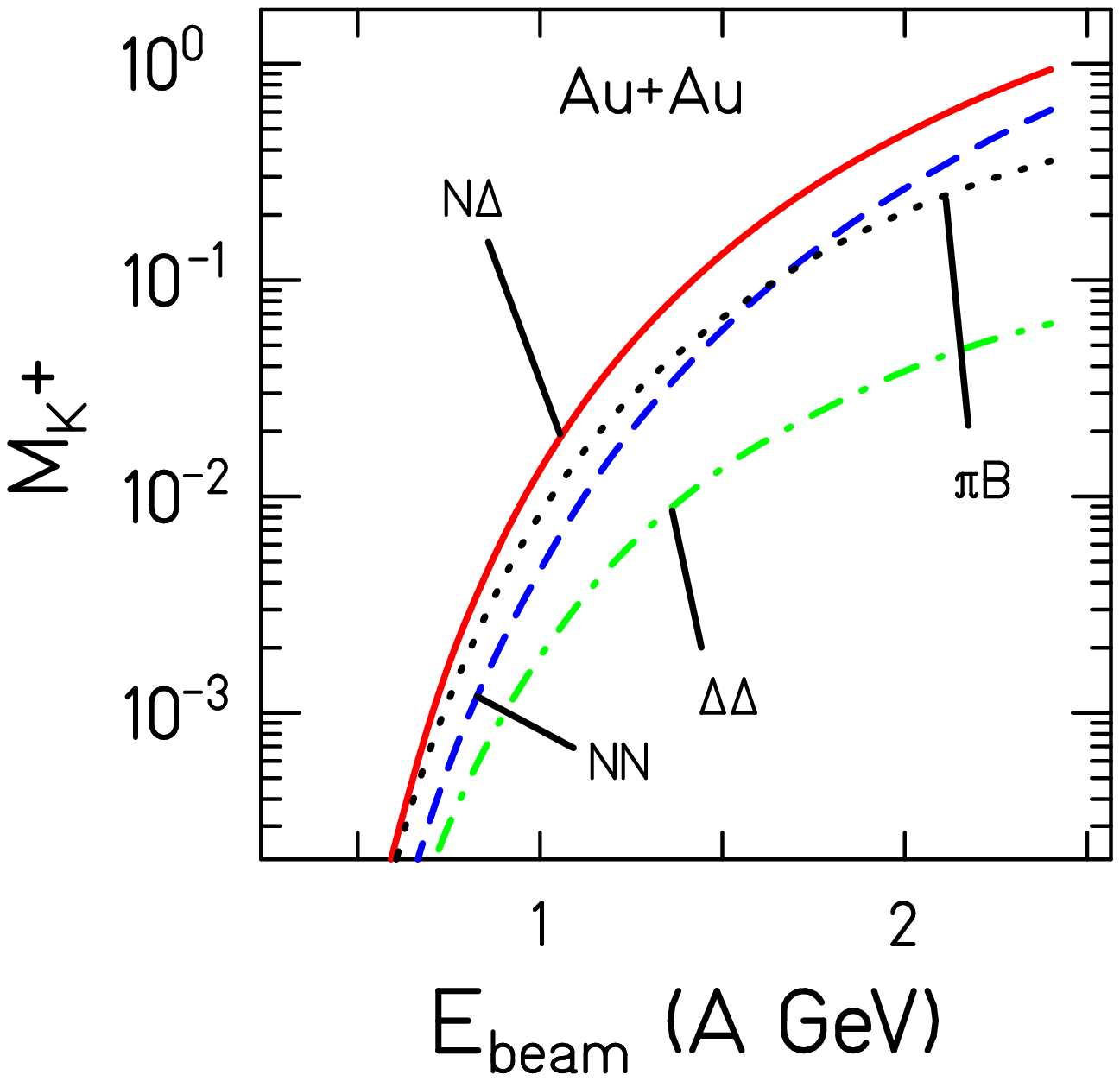,width=0.45\textwidth}
\epsfig{file=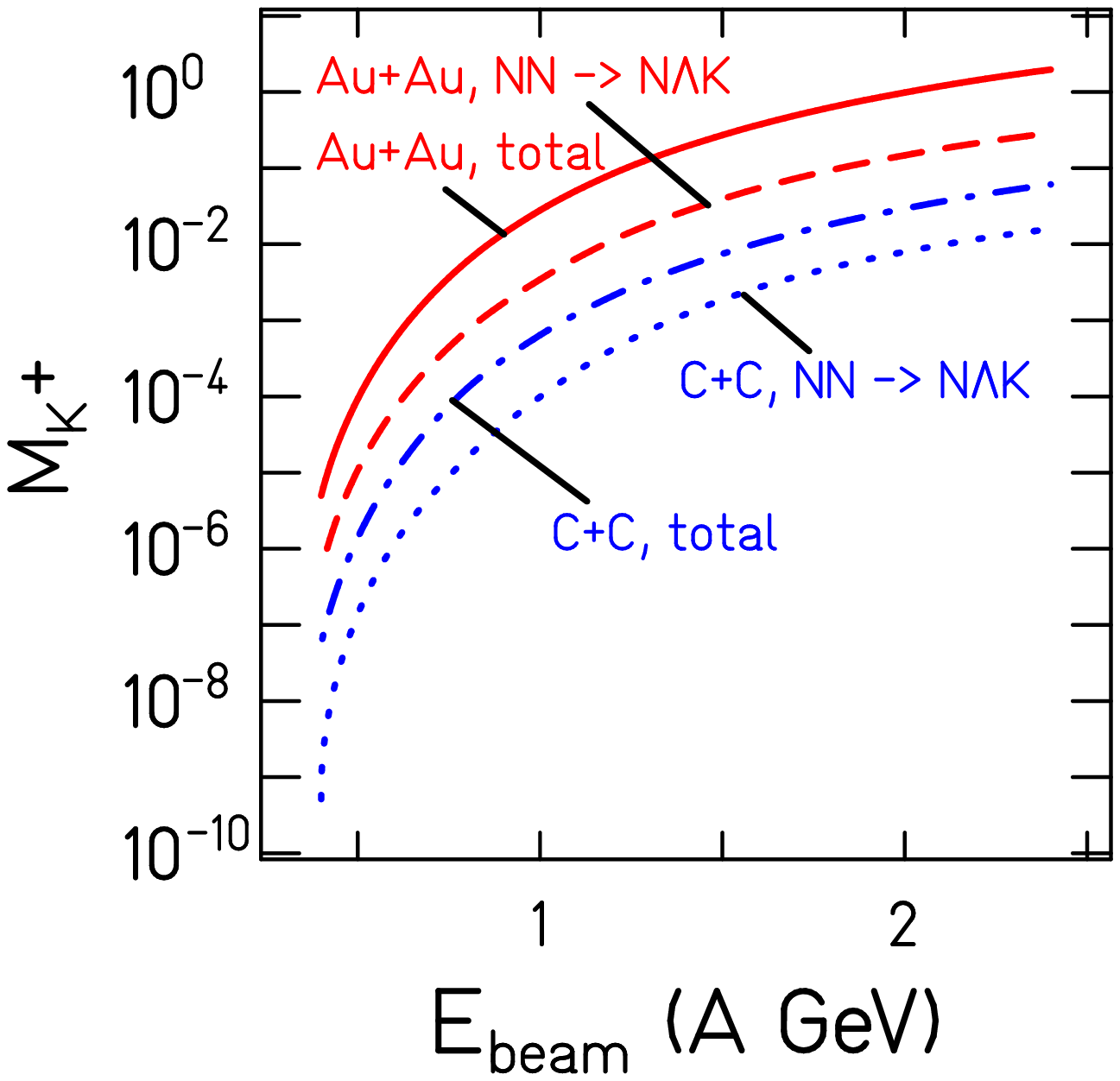,width=0.45\textwidth}
 \caption{Excitation function of \kp meson production in central (b=0) collisions.
 Left: Excitation function of the
 different channels for Au+Au collisions. Right: Comparison of the total yield
 of Au+Au and of C+C collisions with the contribution of
 the elementary channel NN $ \to$ N$\Lambda $K.
 }
\Label{AuAu_CC_excitation_function}
\end{figure}

The N$\Delta$ channel is dominating even well above the NN
threshold at 1.58 GeV incident energy. Contributions from the
$\Delta \Delta$ channel cannot be neglected at the rather low beam
energies. The relative channel decomposition of the \kp production
as a function of the beam energy (for $b$=0 events in Au+Au
collisions), as a function of the impact parameter (for Au+Au
collisions at 1.5 $A$ GeV) and as a function of the system size
(for $b$=0 at 1.5 $A$ GeV beam energy) are shown in \Figref{nk-b}.
The excitation functions, shown on the left, demonstrate that at
different energies the importance of the different channels varies
substantially as already visible in
Fig.~\ref{AuAu_CC_excitation_function}. At low beam energies a \kp
can hardly be produced in NN collision, the Fermi momentum does
not provide sufficient energy. Even in $\Delta$N collisions the
c.m. energy is usually not sufficient. In $\Delta\Delta$
collisions the available c.m. energy is larger and - although
these collisions are rare - their contribution (as well as the
$\pi$B cross section) gets significant at the lowest beam energy.
This means as well that the prediction of the \kp yield at this
energy depends on the quality of the theoretical prediction of the
$\Delta \Delta$ cross section which is experimentally not
accessible. At higher energies, the N$\Delta$ cross section
dominates more and more until the direct production in a NN
collision takes over, yet only very much above the threshold.

\begin{figure}[hbt]
\epsfig{file=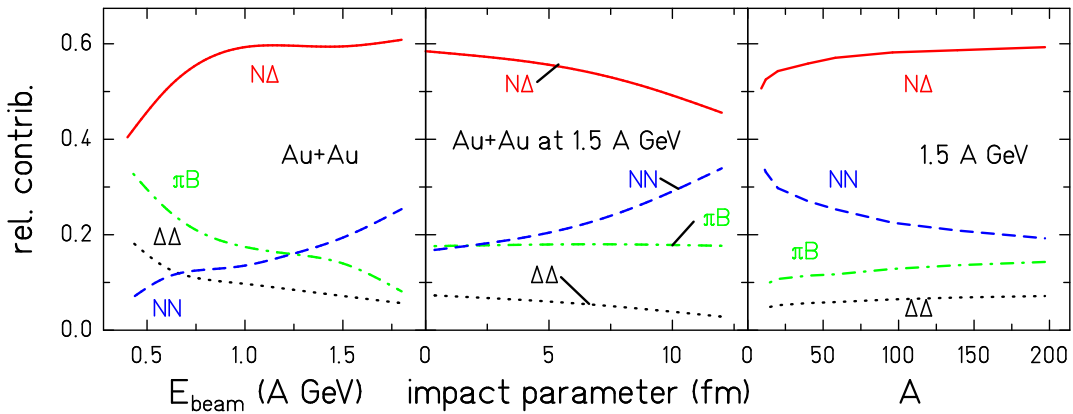,width=0.95\textwidth}
 \caption{The contribution of different production channels.
 Left: As a function of the beam energy in Au+Au reactions at $b$=0.
 Middle: As a function of the impact parameter in Au+Au reactions at 1.5 $A$ GeV.
 Right: As a function of the system size  for $b$=0 at 1.5 $A$ GeV.}
 \Label{nk-b}
\end{figure}

\subsection{\kp production via the N$\Delta$ channel}
As shown in Fig.~\ref{xsections-NDN}, for $\sqrt{s}$ values above
2.7 GeV, corresponding to beam energies above 2 GeV in p+p
collisions, there is quite a difference between the two
parameterizations for the N$\Delta \to \kp $channel. In heavy-ion
collisions well below 2 \AGeV this difference is even amplified
because the most energetic collisions contribute most to the
yield. In non-peripheral collisions the $\Delta$ channels are
dominant and hence this difference to the elementary cross section
transforms directly into a difference of the \kp yield in A+A
collisions which increases with increasing system size
(Fig.~\ref{xsections-ND}, right) as expected from the relative
importance of the $\Delta$ channels. As the difference between the
various parameterizations increases with incident energy
(Fig.~\ref{xsections-ND}, left), the excitation functions of the
\kp production in Au+Au collisions exhibit an increasing
difference with beam energy.

\begin{figure}[hbt]
\epsfig{file=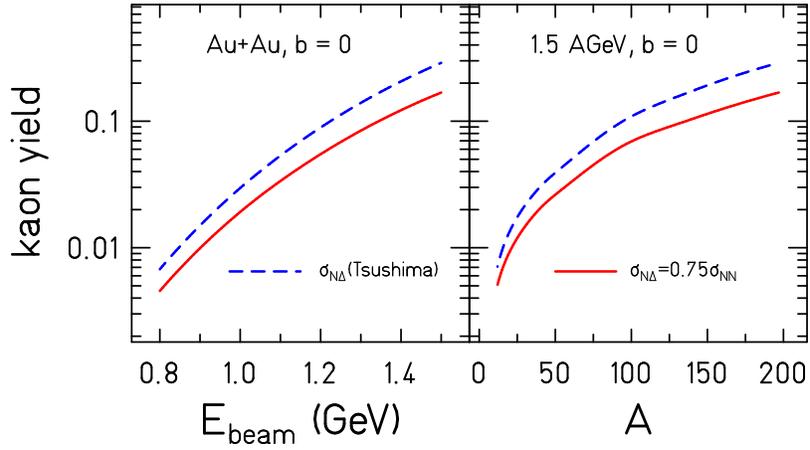,width=0.75\textwidth}
\caption{ The influence of the various
parameterizations of the elementary N$\Delta \rightarrow
$N$\Lambda \kp$ cross section (see Fig.~\ref{xsections-NDN}) on the kaon yield in A+A collisions.
Left we display the excitation function, right the system-size dependence.
} \Label{xsections-ND}
\end{figure}

\subsection{The \kp nucleus optical potential}

As discussed in section \ref{theory}, \kp mesons interact with the
nuclear environment. The resulting IQMD distribution of \kp rest energy,
$\omega({\bf{k}}=0)$, at the moment of their production in
a central Au+Au reaction at 1.5 $A$ GeV  is shown in
Fig.~\ref{opt-pot-3}. An average mass shift of about 57 MeV is
seen, corresponding to an average baryonic density at creation of
around $2\rho_0$ (see \figref{opt-pot-33}).
\begin{figure}[hbt]
\begin{tabular}{cc}
\epsfig{file=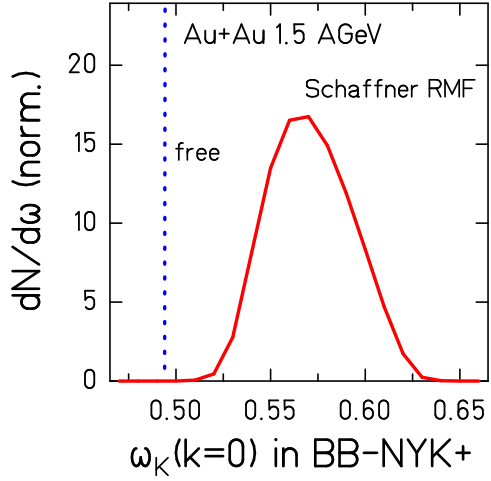,width=0.5\textwidth} \\
\end{tabular}
 \caption{Mass ($\omega({\bf{k}}=0)$) distribution of the \kp at production in central ($b$ = 0 fm) Au+Au
 collisions at 1.5 $A$ GeV  as compared to the free \kp mass.}
\Label{opt-pot-3}
\end{figure}
A larger mass means a larger threshold and therefore the \kp nucleus
potential lowers the \kp yield. This is demonstrated in
Fig.~\ref{pot-yields}, left, where in central Au+Au reactions the
yield at the lowest beam energy is suppressed by a factor of three
whereas at the highest energy the suppression is only 40\%. The
system-size dependence is rather weak as shown in
Fig.~\ref{pot-yields}, right. Light systems become less compressed
during the reaction and consequently the influence of the \kp nucleus
potential decreases.

Please note that the modification of the yield due to the
different parameterizations of the $\rm N\Delta \rightarrow \kp N
\Lambda$ cross section (Fig.~\ref{xsections-ND}) is of the same order of magnitude as that
of the \kp nucleus potential (Fig.~\ref{pot-yields}). Hence, a smaller \kp production cross section has the
same influence on the \kp multiplicity as a larger \kp nucleus potential.
Also the Fermi motion of the nucleons modifies the production
yield considerably, as seen in Fig.~\ref{pot-yields}. Both graphs show as well the importance
of the Fermi motion for the \kp yield.

These compensatory effects make it highly questionable if not
impossible to use the absolute \kp yield to separate cross-section
parametrization  and potential effects. Initially this problem has
not been realized. Different groups advanced calculations using
different cross sections (\cite{Randrup:1980qd},
\cite{Tsushima:1994rj,Tsushima:1998jz}) and came to different
conclusions, especially about the importance of the \kp nucleus
potential for the explanation of the experimentally measured \kp
yield in A+A collisions. In the meantime this has been solved
\cite{Kolomeitsev:2004np}.

\begin{figure}[htb]
\epsfig{file=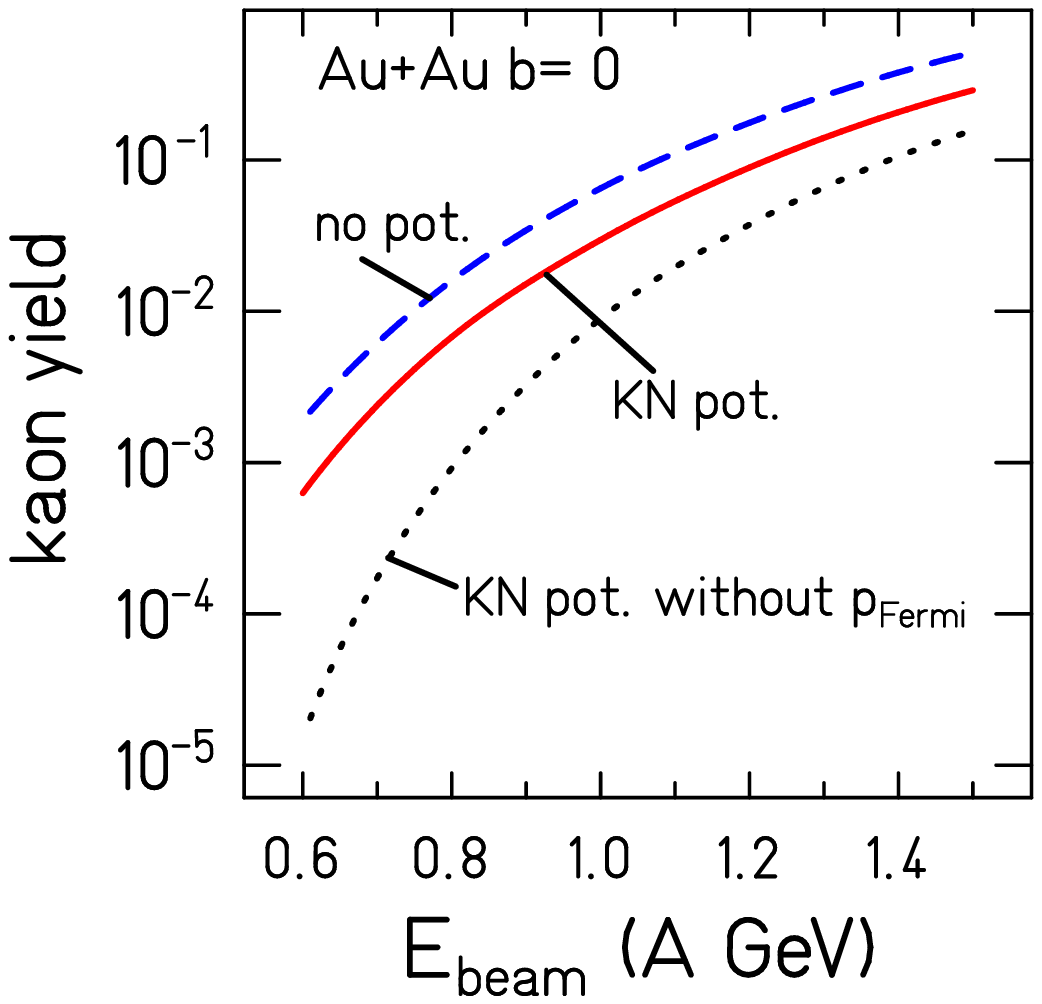,width=0.45\textwidth}
\epsfig{file=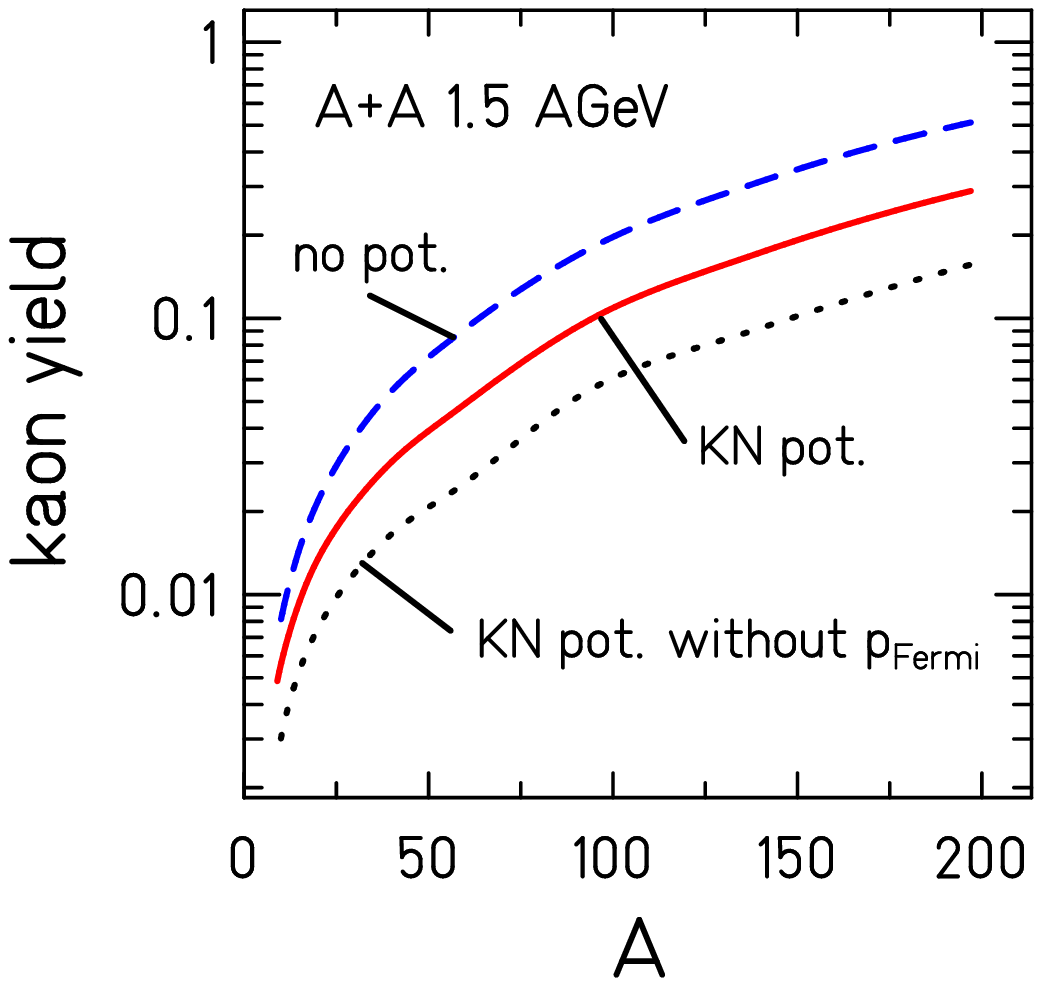,width=0.45\textwidth}
\caption{
Excitation function
(left) and system-size dependence (right) of the \kp yield with
and without \kp nucleus potential and with \kp nucleus potential but no Fermi
momentum.} \Label{pot-yields}
\end{figure}

Despite of their relatively small cross section with nucleons of
about 13 mb (the sum of elastic and charge-exchange collisions)
most of the \kp rescatter with nucleons. We add in passing  that
we have not considered the details of the charge-exchange
reactions \kp n $\to $K$^0$ p in this calculation. We have
parameterized this cross section, as it is shown in
Fig.\ref{kaon-elastic}, but we do not apply the charge exchange
itself. We simply treat these collisions like elastic ones without
changing the charge of the kaon. This may cause an uncertainty on
the \kp yield of at most 10\% in the Au case. These collisions do
not change the total yield because all other particles which
contain an $\bar s$ quark are beyond reach.

Figure~\ref{dns-tim-resc}, left, shows the density at the time
when the \kp mesons are created, and the mean free path of the \kp
as a function of the distance of the production point from the
center of the reaction.
On the right hand side we display the distribution of the radial
position at which \kp mesons are produced (upper part) and at
which the \kp have their last interaction (lower part), selected
according to the number of rescatterings $N_C$. (Note that $N_{\rm
coll}$  refers to the number of NN collisions prior to \kp
production, while $N_C$ gives the number of KN collisions.)
Those with many rescattering interactions originate more from the
interior; those without collision ($N_C = 0$, corresponding to
18\% of the produced \kp) more from the outer region but not from
the surface. This can easily be understood from the mean free
path, shown left. Due to the high density in the center, the mean
free path is there about 2 fm and \kp, which are created there,
scatter quite frequently.

Dividing the time by the system size, one can see that the scaled
time evolution of the \kp production in a C+C collision is similar
to that of a Au+Au collision as shown in \Figref{AuAu_CC_kp},
left. The corresponding distributions of densities
(\Figref{AuAu_CC_kp}, right) at which \kp mesons are produced and
at which the last collision between a \kp and a nucleon takes
place, differ strongly. This reflects the much larger stopping
(and consequently the higher baryon density) in Au+Au as compared
to C+C collisions.

\begin{figure}
\epsfig{file=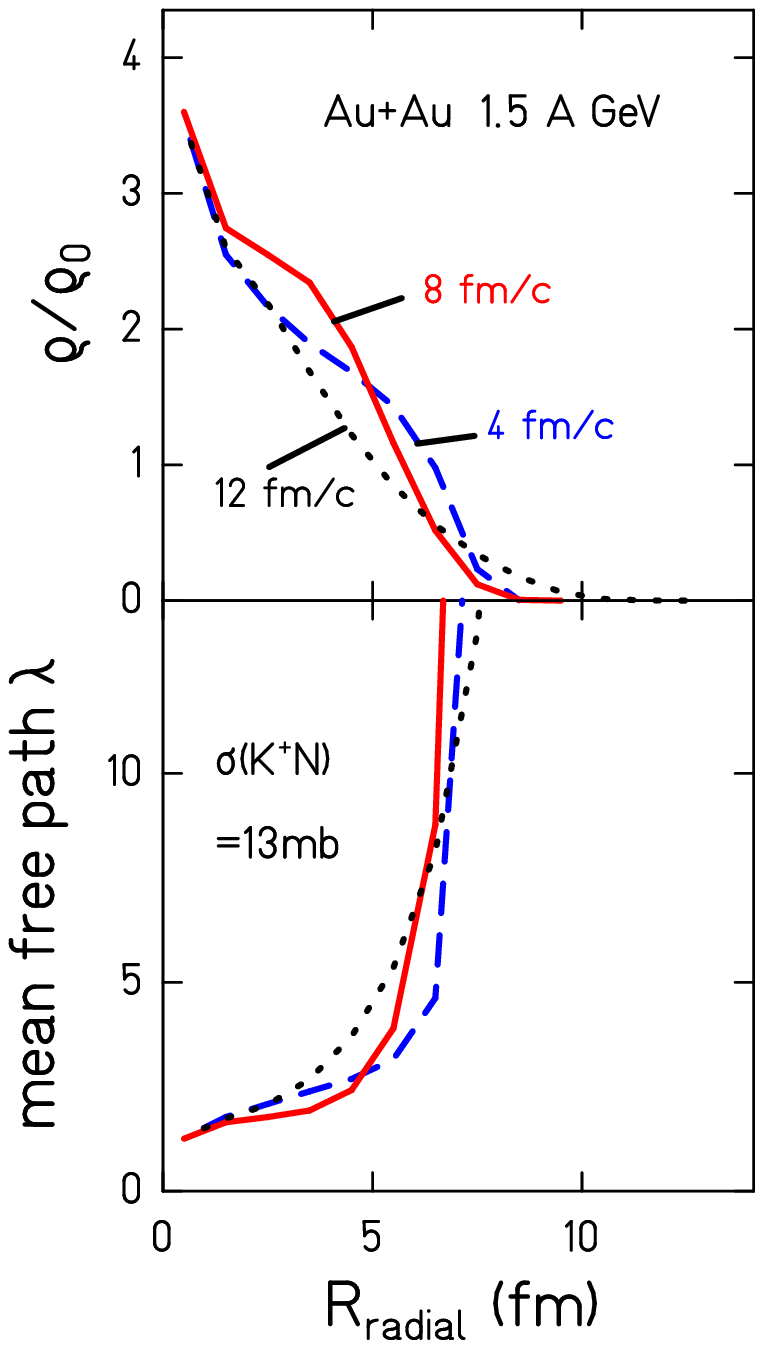,width=0.45\textwidth}
\epsfig{file=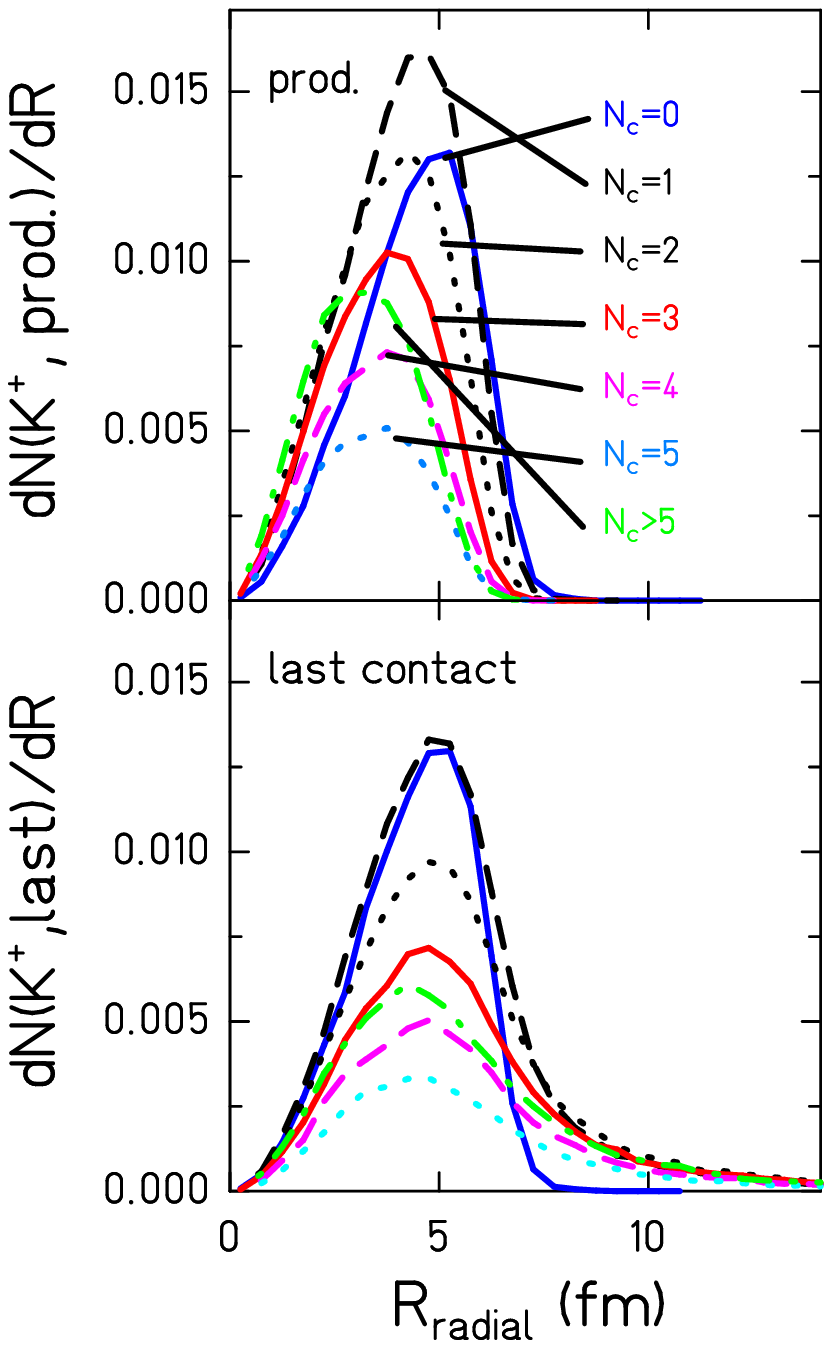,width=0.45\textwidth} \caption{Left:
Density of nucleons and the mean free path of the \kp  as a
function of the distance of the production point from the center
of the reaction for various time steps. Right: Density profiles
for production and last contact selected according to the number
of rescattering collisions $N_C$.} \Label{dns-tim-resc}
\end{figure}
\begin{figure}[hbt]
\epsfig{file=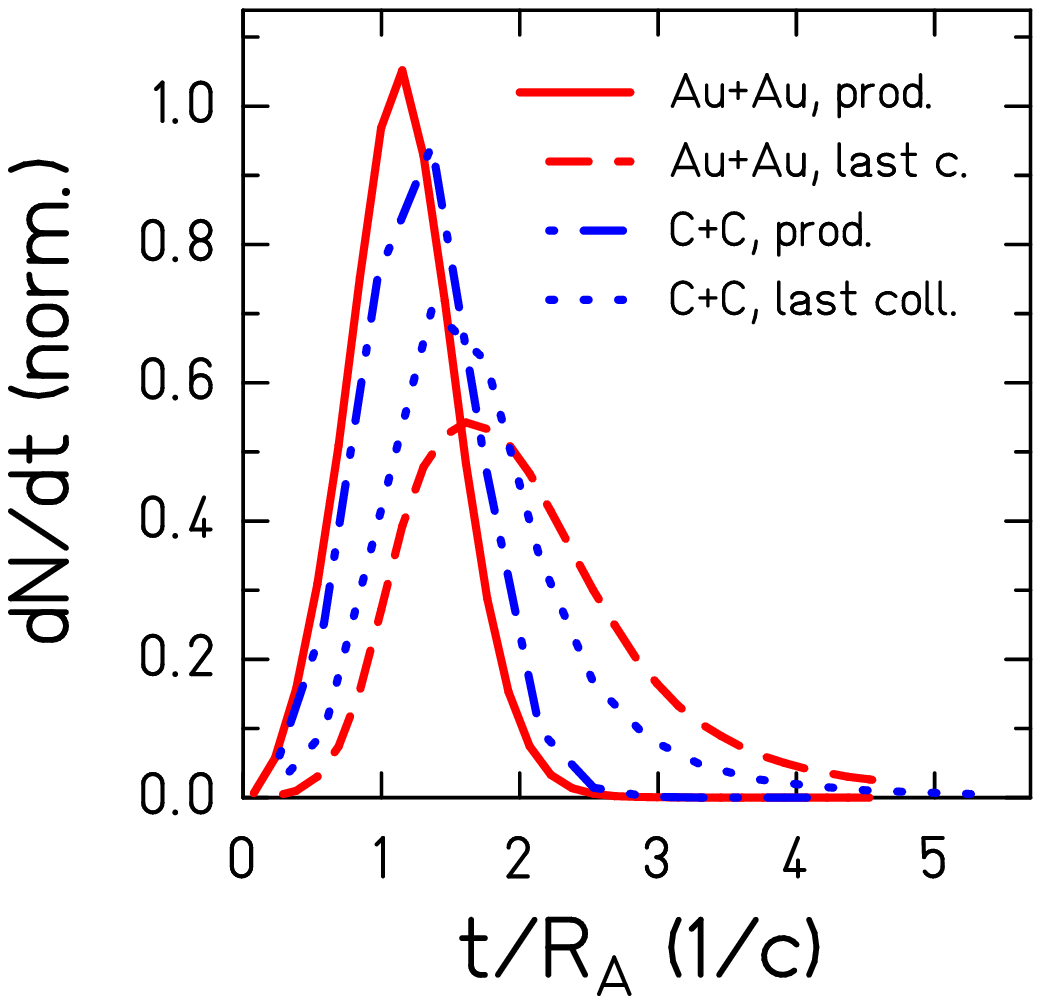,width=0.45\textwidth}
\epsfig{file=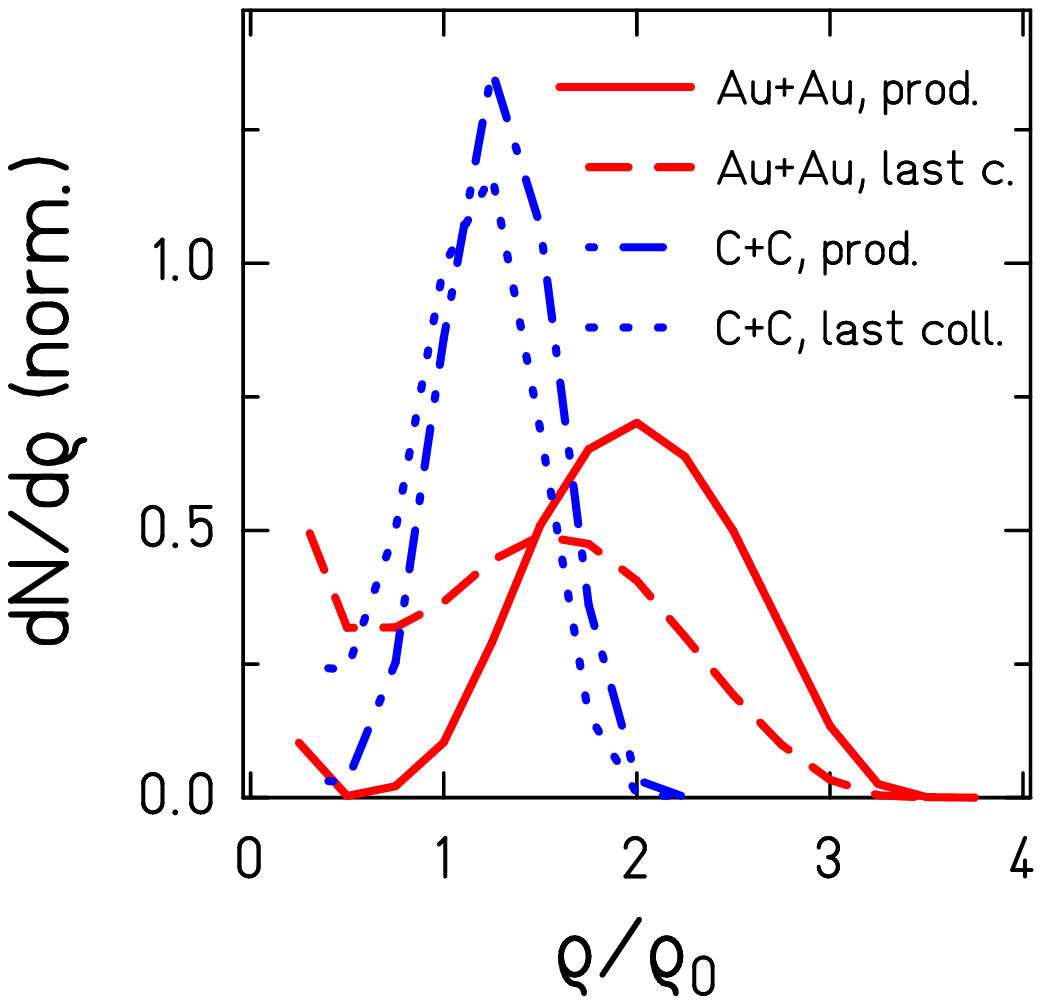,width=0.45\textwidth}
\caption{Comparison of central ($b$ = 0) collisions of C+C and
Au+Au at 1.5 \AGeV. On the \lhs ~the yields are shown as a
function of the production/emission time divided by the  radius
$R_A$ of the respective nuclei. The \rhs ~exhibits the respective
densities at production and at the last collision.}
\label{AuAu_CC_kp}
\end{figure}

\subsection{Reaction Dynamics}
Close to threshold \kp are rarely produced in a first collision
between a projectile and a target nucleon.  Figure~\ref{nuc-col},
left, shows the probability distribution of the number of
collisions prior to the \kp production of those nucleons which
produce finally a \kp meson. Its production in first-chance
collisions is a negligible process even at a beam energy of 1.5
$A$ GeV. In order to gain sufficient energy the nucleons have to
collide before, usually quite often. These collisions start to
thermalize the momentum distribution, as is seen from the average
$Q_{zz} = 3 p^2_z - p^2 = 2 p^2_z - p^2_t$ value for protons,
shown on the mid part of Fig.~\ref{nuc-col} which approaches the
value of zero, expected for a system with an isotropic momentum
distribution.

\begin{figure}[htb]
\begin{tabular}{ccc}
\epsfig{file=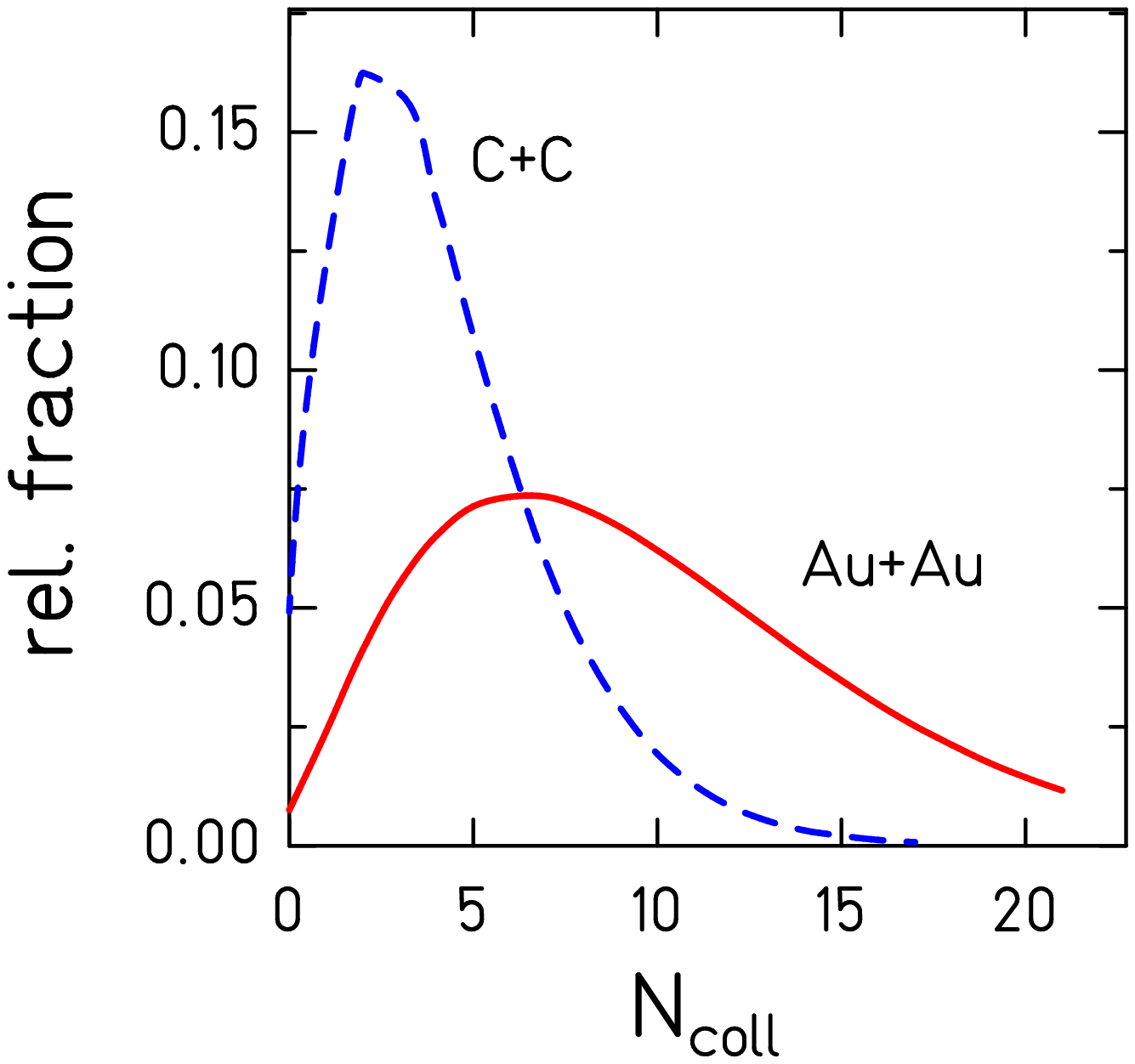,width=0.33\textwidth} &
\epsfig{file=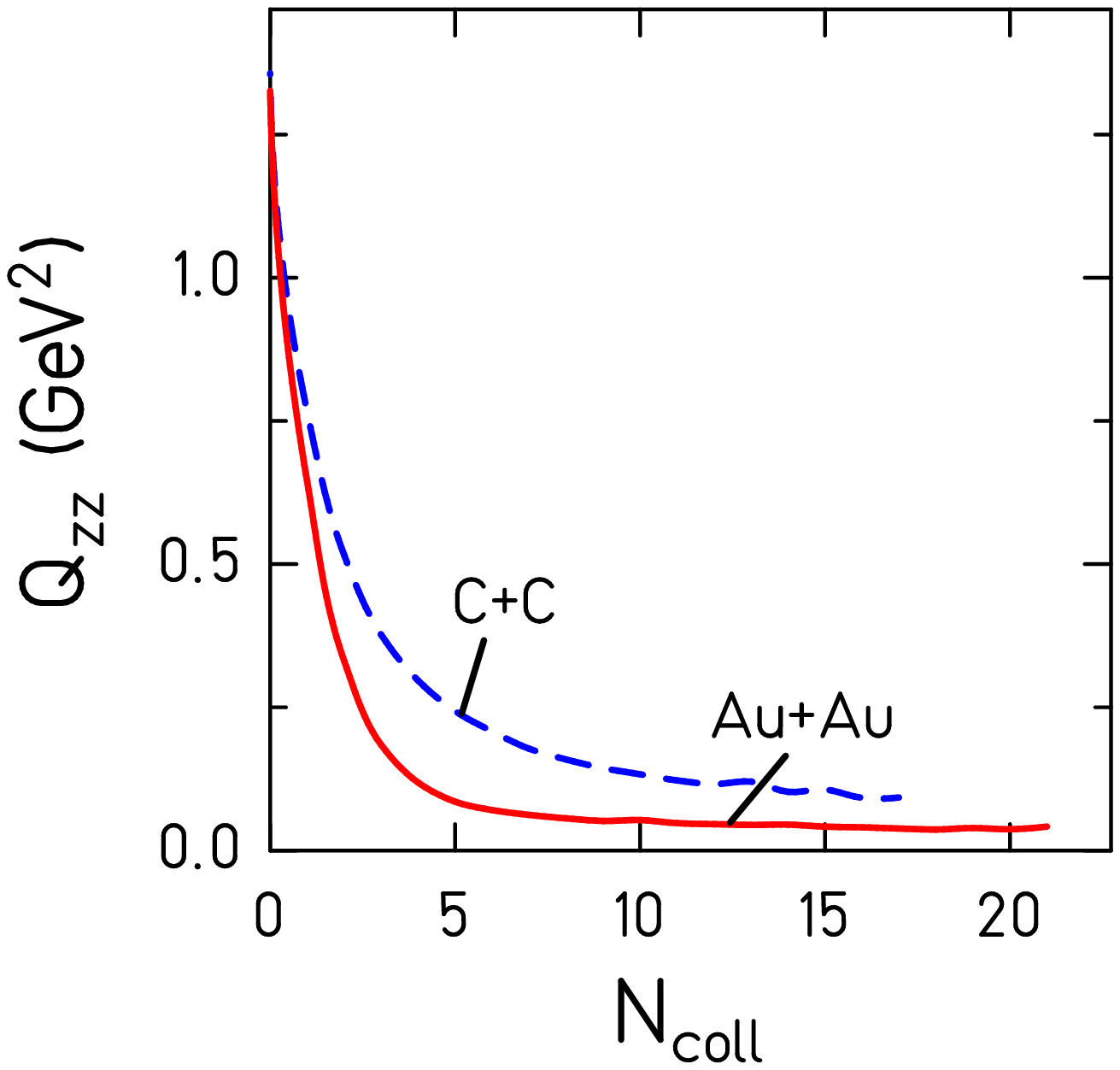,width=0.33\textwidth} &
\epsfig{file=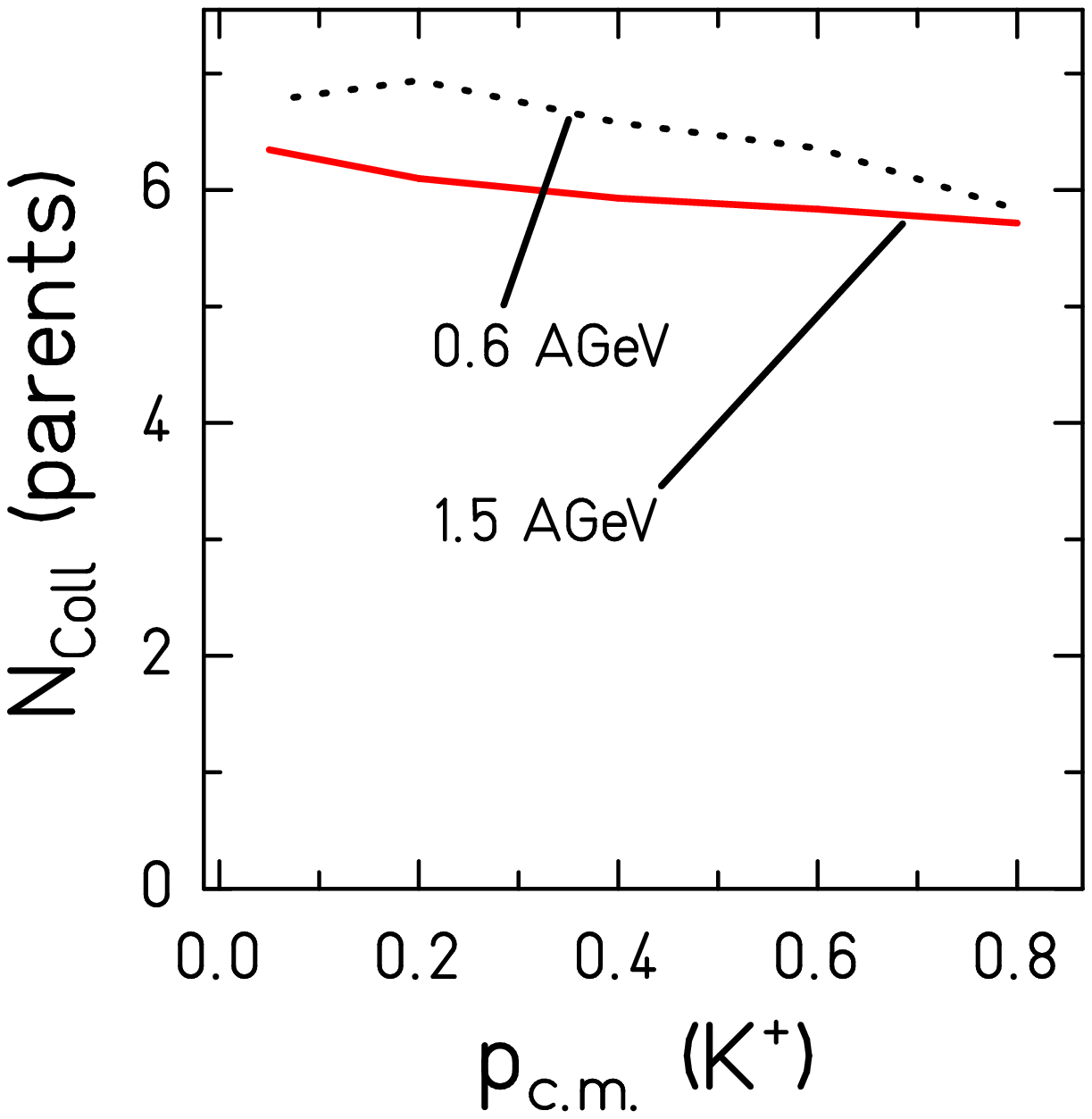,width=0.31\textwidth} \\
\end{tabular}
\caption{Properties of the parent baryons which create a \kp for
Au+Au and C+C at $b$=0 fm and for 1.5 $A$ GeV. Left: Distribution
of the number of collisions prior to the \kp production. Middle:
Sphericity $Q_{zz}$ as a function of the collision number. Right:
Average number of collisions prior to the creation
of a \kp with a given momentum $p_{\rm {c.m.}}$ in central AuAu collisions.} \Label{nuc-col}
\end{figure}

The average number of collisions prior to \kp creation depends on
the reaction channel and it increases with the number of $\Delta$
in the entrance channel because these particles have to be
produced first. These collisions modify the momentum distribution
of the nucleons or - in other words - these nucleons carry
information on their environment which they can communicate to the
\kp. However, different \kp momenta are not selective to a
different number of prior collisions of the parents as
demonstrated in Fig.~\ref{nuc-col}, right. The inverse slope parameters of \kp
mesons from the various channels are very similar. We note,
however, that first-chance NN collisions show a steeper slope than
the average over all collisions as will be discussed later.

\begin{figure}[htb]
\begin{tabular}{cc}
\epsfig{file=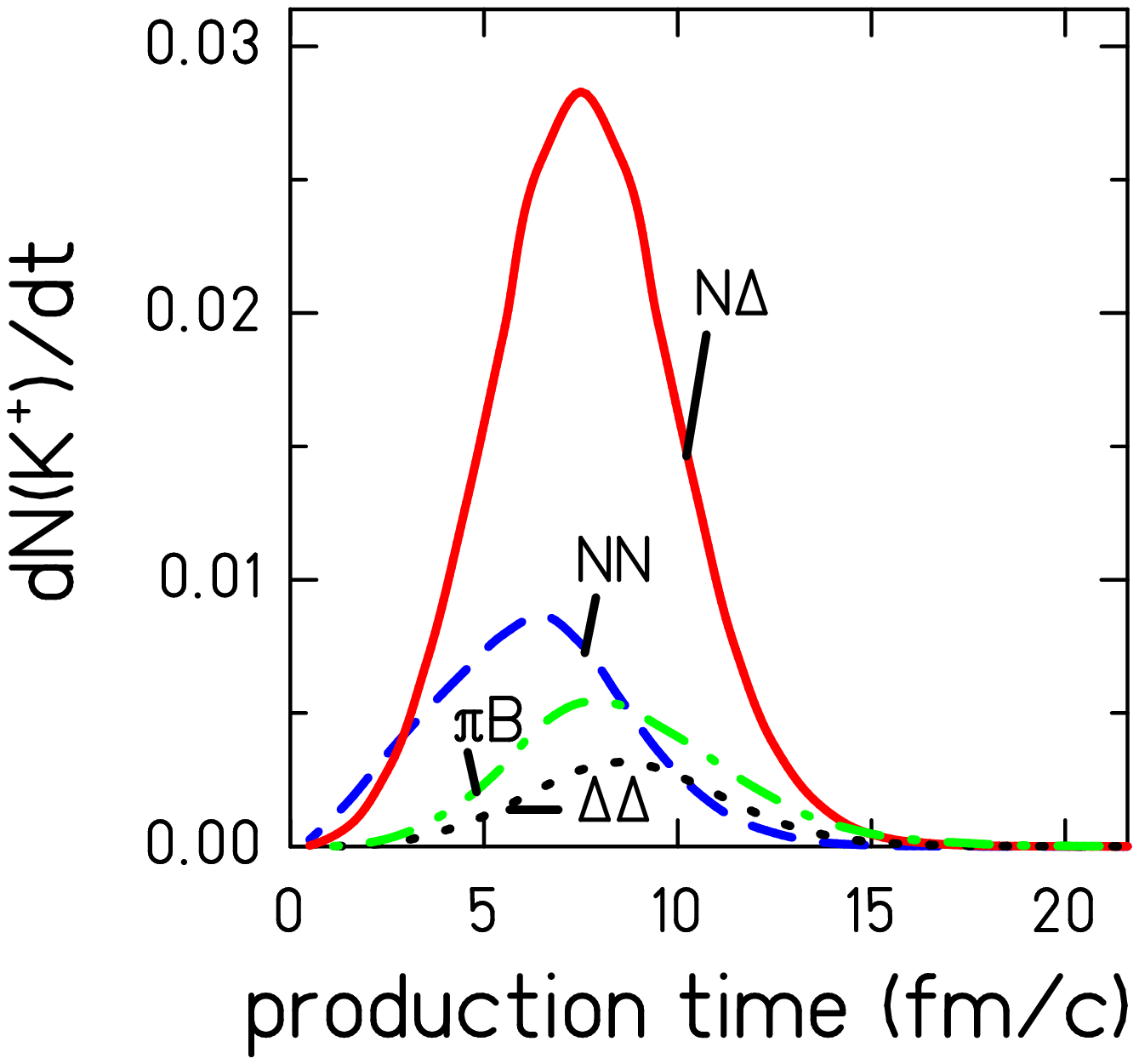,width=0.4\textwidth} &
\epsfig{file=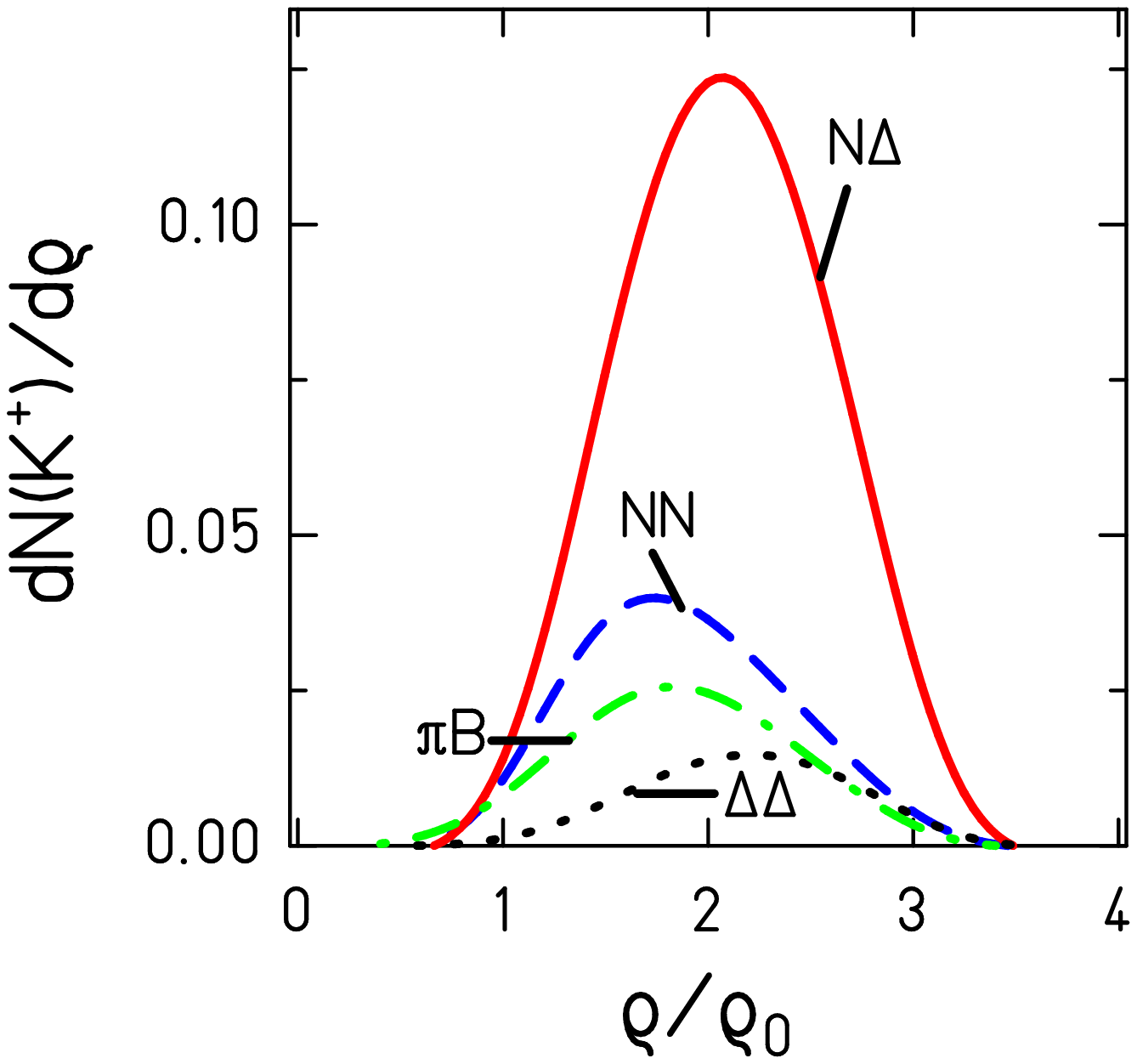,width=0.4\textwidth} \\
\end{tabular}
 \caption{Evolution of the \kp yield for the various production
channels as a function of the time (left) and in respect to the
density at which they are produced (right) for central Au+Au
collisions at 1.5 $A$ GeV.} \Label{time-rho-prod}
\end{figure}

The \kp mesons are produced quite early ($\approx$ 7 fm/$c$) in
the Au+Au reaction (\Figref{time-rho-prod}, left). The difference
between the different channels reflects how many collisions the
parents have to have before a \kp meson is created giving a
sequence starting with NN, then $\rm \Delta N$, $\pi \rm B$ and
$\Delta\Delta$ collision. The average baryon density at the
production of the \kp is around $\rho = 2 \rho_0$
(\Figref{time-rho-prod}, right) as observed already in
\Figref{opt-pot-3}. As expected, \kp with a $\Delta$ in the
entrance channel a) are produced later because the $\Delta$ has to
be created first and b) are produced at a higher density because
the lower the density the higher is the chance that the $\Delta$
decays before producing a \kp{}.
This density dependence is of great importance for our later discussion
on the nuclear equation of state, section~ \ref{differences}.

In \figref{nk-b}, middle, we have shown that with increasing
impact parameter, this means with decreasing number of
participants, the importance of the two-step channels implying a
$\Delta$ decreases. Whereas in central collisions the two-step
reactions dominate, in peripheral collisions the NN channel
becomes almost as important.
Studying the contributions as a function of the system size (right
hand side) a very similar behavior is seen: The smaller the system
the less important becomes the two-step N$\Delta$ or three-step
$\Delta\Delta$ channel. In conclusion, \kp production in A+A
collisions is very different from that in p+A collisions.
The in-medium modification of the \kp properties change, due to different
densities. Furthermore the dominant
production channels are different and rescattering is an important
issue if we want to understand the spectra. The influence
of these effects depends on the size of the system.
Consequently, the \kp observables allow to study many features of
the heavy-ion reaction which are not accessible by non-strange
particle observables. We continue with a detailed study of these
observables and --- by comparing the experimental results with IQMD
calculation --- with the interpretation of these observations.

\subsection{Excitation functions}
\begin{figure}[hbt]
\epsfig{file=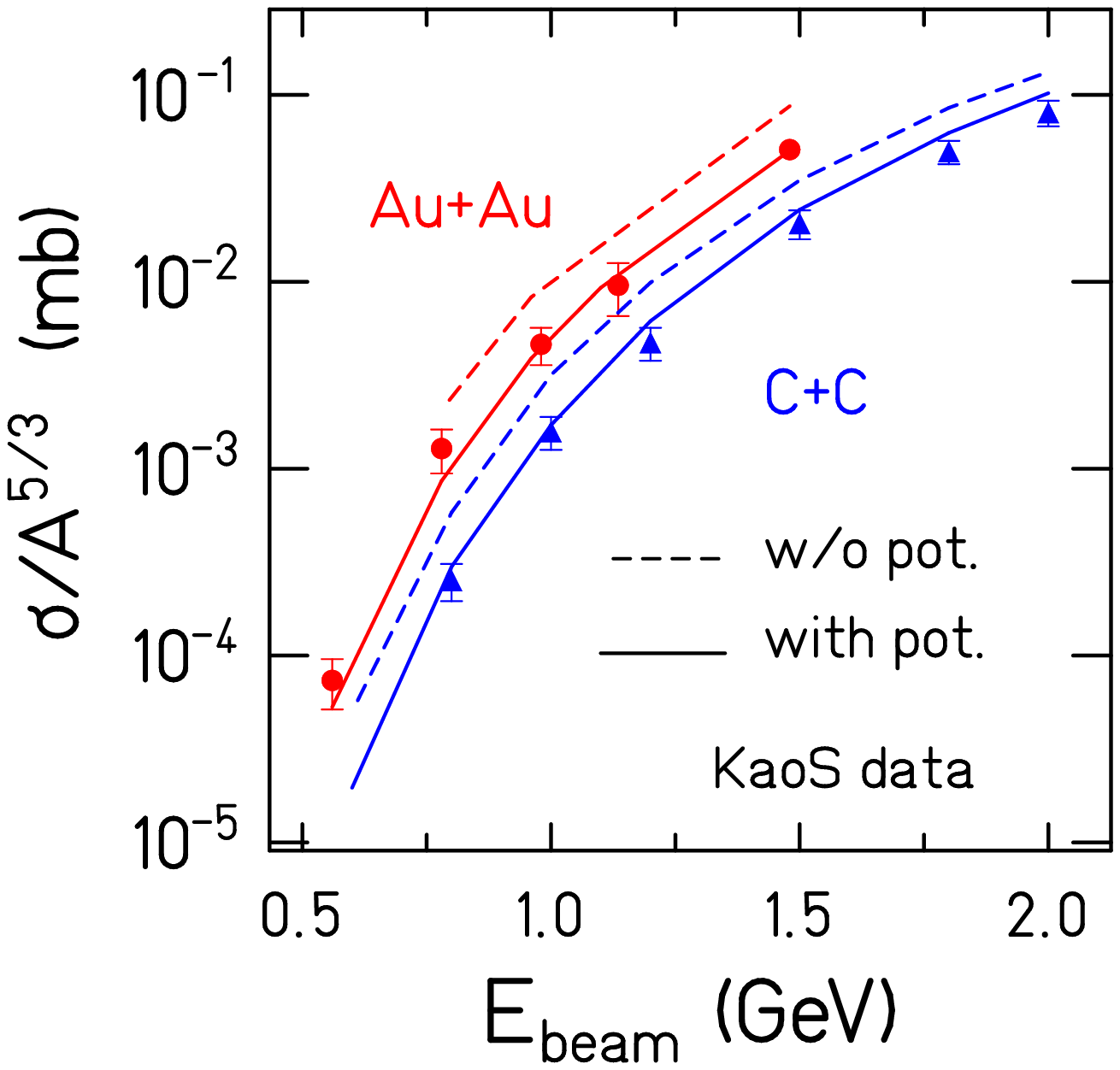,width=0.48\textwidth}
\epsfig{file=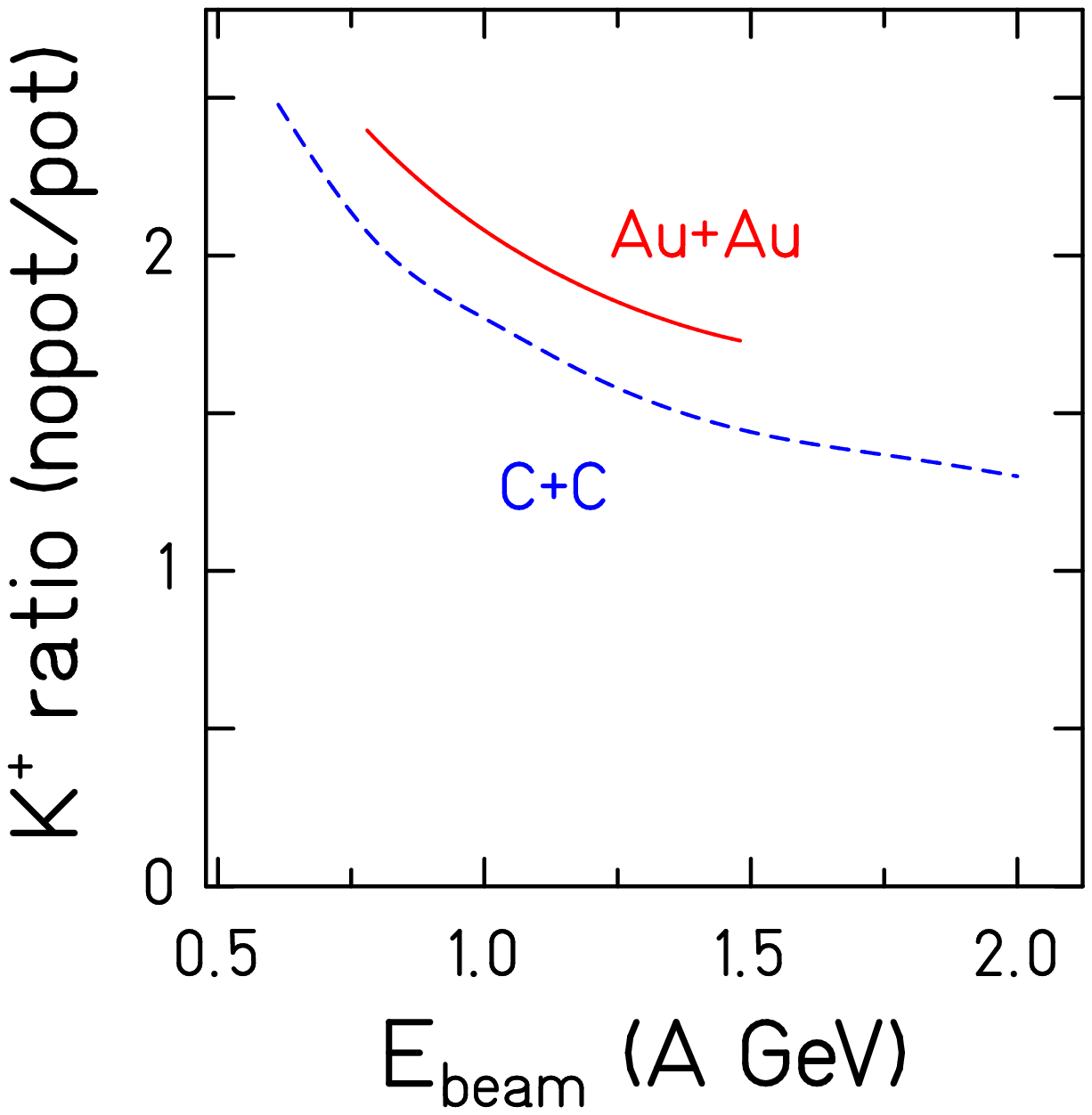,width=.48\textwidth}
\epsfig{file=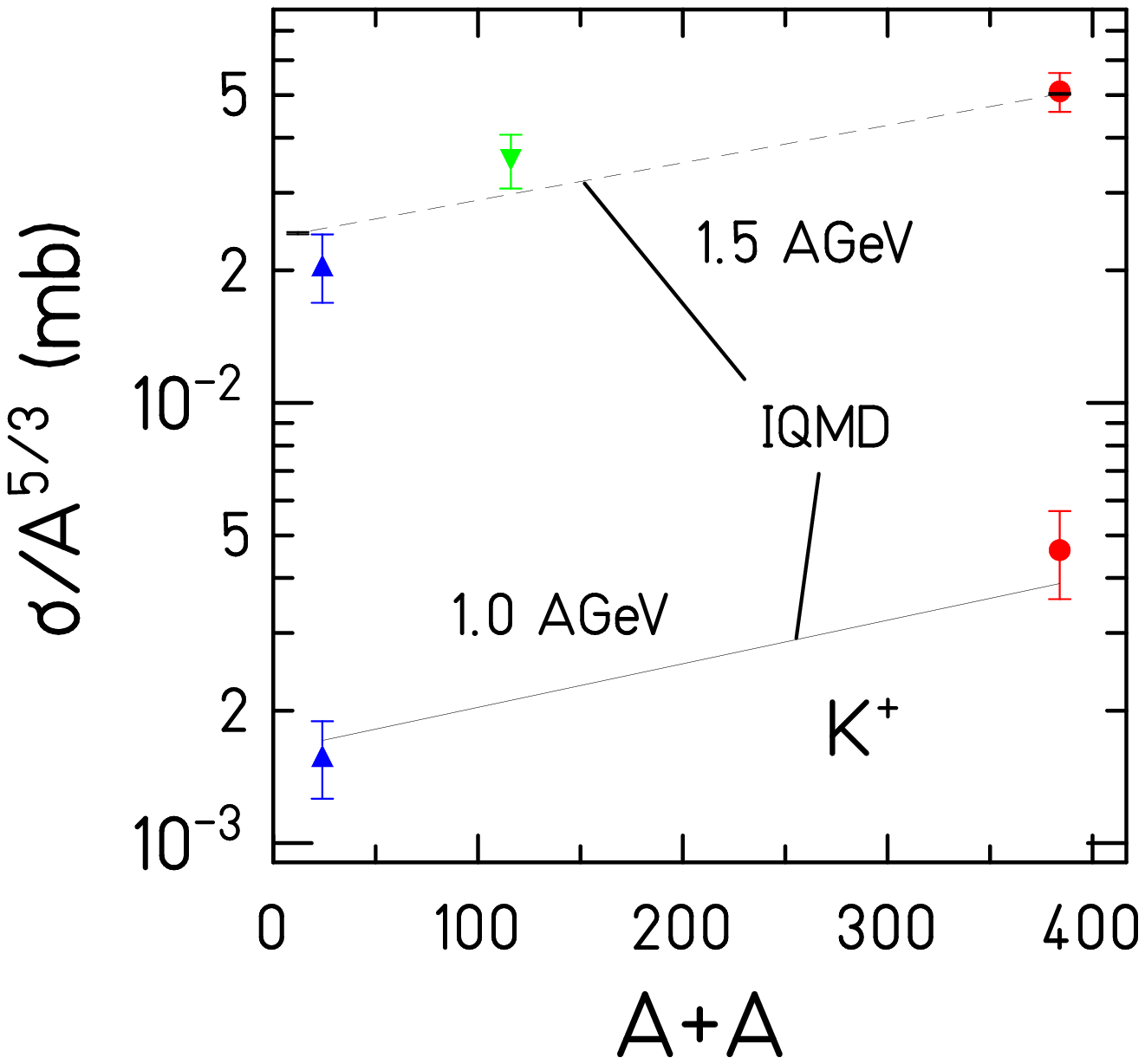,width=0.48\textwidth}
\epsfig{file=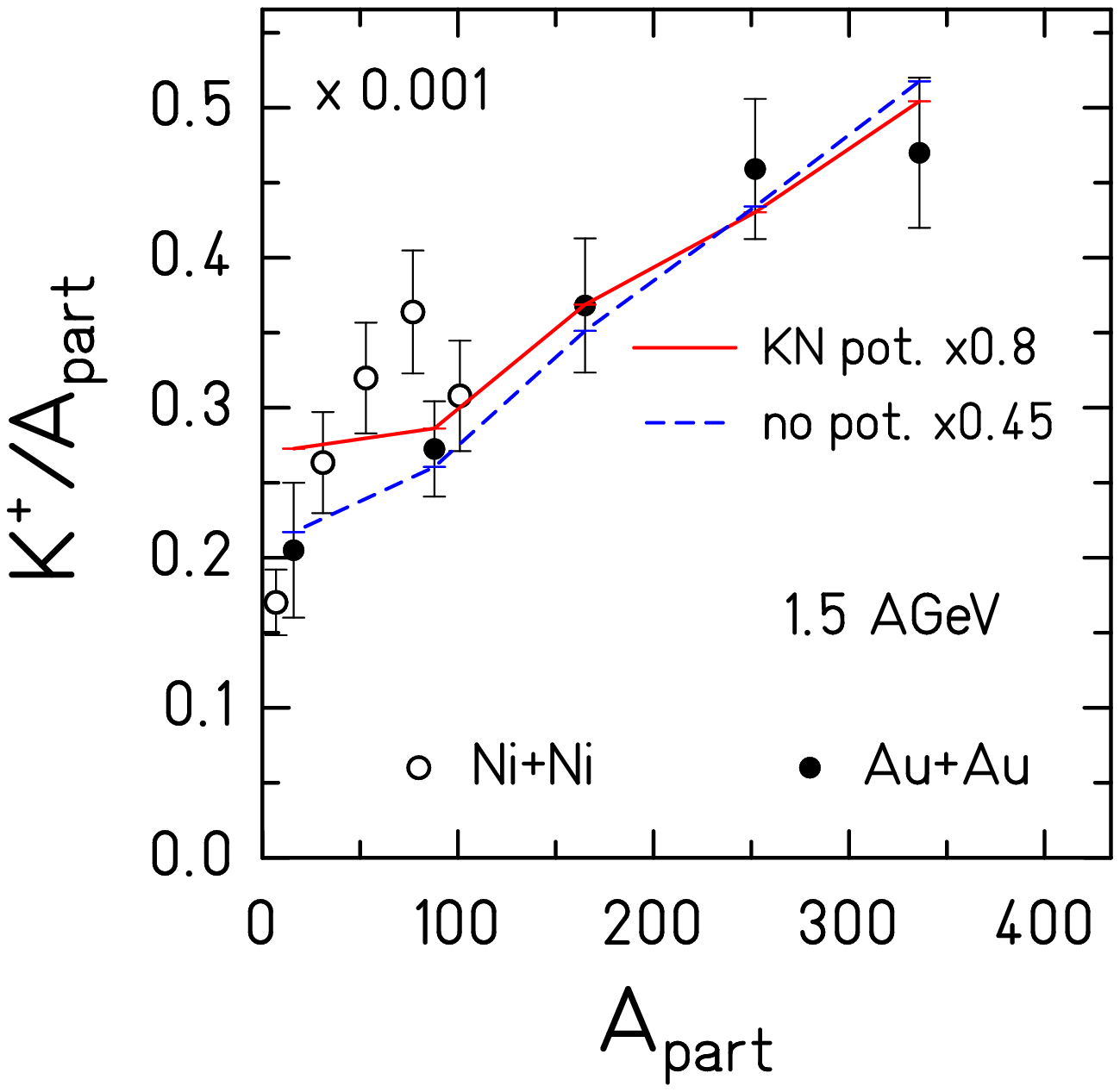,width=0.48\textwidth} \caption{Upper
left: Excitation function of the inclusive \kp cross section
in C+C and Au+Au collisions divided by $A^{5/3}$ with and without the \kp nucleus potential. The
symbols represent experimental results of the KaoS
Collaboration~\cite{Forster:2007qk}. Upper right: Influence of the
\kp nucleus potential demonstrated by the ratio of the \kp yields without
potential divided by those with potential for two collision
systems.
Lower left: Production cross section divided by $A^{5/3}$ as a
function of the system size $A+A$. The symbols represent data from
the KaoS Collaboration~\cite{Forster:2007qk}, the lines connect
the calculations for C+C and Au+Au. Lower right: Multiplicity per
\Apart of \kp mesons as a function of \Apart for Au+Au and Ni+Ni
collisions  at 1.5 \AGeV ~\cite{Forster:2007qk} as compared to
IQMD simulations. The calculations are scaled by factors of 0.8
and 0.45, respectively.} \label{exfct}
\end{figure}
The excitation function of the \kp yield in
experiment~\cite{Forster:2007qk} and in theory for inclusive Au+Au  and
C+C reactions is shown in Fig.~\ref{exfct}, top left. The dashed lines
present the theoretical results assuming a vanishing \kp nucleus potential,
the full ones include the \kp nucleus potential. In order to demonstrate
the high collectivity of the \kp production, the cross sections in
this figure are divided by $A^{5/3}$.   If the \kp multiplicities
are proportional to the system size, i.e.~to the number of
participating nucleons, $\sigma/A^{5/3}$ is constant and the
results of both systems coincide. The top and bottom left figure shows
that for the heavier system $\sigma/A^{5/3}$ is up to 5 times
higher than for the lighter one. The origin of this enhancement is
the higher compression which leads to a smaller mean free path
and, as a consequence, to more many-step processes as will be
discussed in the next section. The influence of the \kp nucleus potential
depends on the system size and on the beam energy,
Fig.~\ref{exfct}, top right. While a considerable change is observed as
a function of the beam energy, the influence of the system size is
rather small.

Including the \kp nucleus potential brings the calculation close to the
experimental data (top left). One may ask whether one can turn this argument
around and claims that the experimental data are a ``smoking
gun'' for the mass change of the \kp mesons in the medium. As
discussed in the last section, at high beam energies a smaller
$\sigma_{\Delta {\rm N} \rightarrow {\rm K^+}}$ cross section
leads to a similar increase of the yield as a smaller \kp nucleus
potential. This is, however, not true any more for the excitation
function. The influence of the \kp nucleus potential on the \kp yield
increases with decreasing energy (top right) whereas the
contribution of the $\N\Delta$ channel stays rather constant down
to a beam energy of 750 $A$ MeV, see \Figref{nk-b}. On the basis
of the present calculation one can identify
the \kp nucleus potential as the reason for the agreement of the
experimental and theoretical excitation function of the cross
sections of C+C and Au+Au systems, although for a confirmation
data at lower incident energies and for different system sizes are
necessary.

The collectivity of the \kp production is quantified in
\figref{exfct}, bottom. The left side shows the scaling with
system size. If the cross section would be proportional to the
number of participating nucleons, i.e.~the nucleons which are
located in the overlap zone between projectile and target, we
would expect a cross section $\propto A^{5/3}$. For an easy
comparison in this presentation we have divided the inclusive
cross section by
 $A^{5/3}$. Both, the data
and the calculation (dashed and full lines), show an increase of
this quantity with the system size $A+A$. A similar comparison is
presented in the bottom right figure where for reactions at 1.5
\AGeV the multiplicity per \Apart for the different centrality
bins are displayed and compared with theory. IQMD calculations
reproduce quite well the trend but over-predict the data by 20\%.
Without \kp nucleus potential the over-prediction is more than a factor of
2.

\subsection{Kaon spectra}
The \kp momenta at production are modified by two
processes before they are registered in the detectors: Elastic
rescattering off the nucleons and potential interactions. Both effects are demonstrated in
Fig.~\ref{spectra-effects}. The spectra are displayed as $E/p^2\,
{\rm d} N/{\rm d} p$ and can be described by
an exponential function $\exp{(-E_{\rm c.m.}/T)}$. The fitted inverse slope
parameters $T$ are given in this figure.
\begin{figure}[hbt]
\epsfig{file=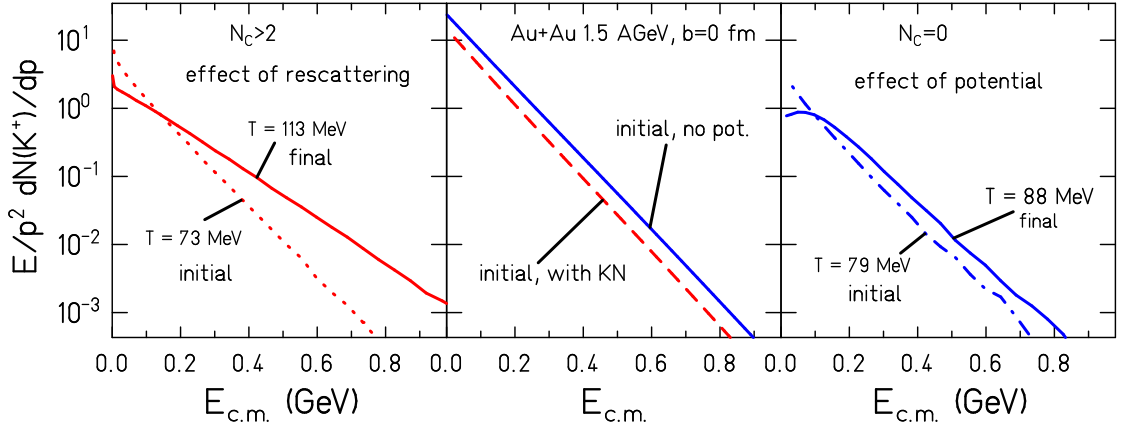,width=0.99\textwidth}
\caption{Influence of rescattering and of the \kp nucleus potential
demonstrated for central Au+Au collisions at 1.5 $A$ GeV. Left:
Influence of the rescattering of \kp mesons by selecting kaons
which have scattered twice or more, showing their initial and
final distribution. Middle: Influence of the \kp nucleus potential on the
spectra at production. Right: Influence of the \kp nucleus potential on the
spectral shape demonstrated by selecting kaons which never
scattered ($N_C$ = 0) and comparing the initial and final
spectra.} \Label{spectra-effects}
\end{figure}
At production the shape of the spectra is determined by the
three-body phase space, which is not exponential. Averaging over
the distribution of $\sqrt{s}$ and of the center of mass
velocities of all collisions which produce a \kp the distribution
becomes almost exponential. Due to the (in the average) lower
c.m.-energy the slope of the \kp is slightly steeper if two
nucleons are in the entrance channel ($T$ = 70 MeV) as compared to
a $\Delta$ N combination ($T$ = 78 MeV). As the $\Delta$N channels
dominates, the inverse slope parameter at production is $T$ = 75
MeV. This value is lower than that of the final baryons and
subsequent KN rescattering causes the slope of the \kp mesons to
be shallower as seen in Fig.~\ref{spectra-effects}, left. The
fitted inverse slope parameter for central Au+Au reactions at 1.5 \AGeV is
initially $T$ = 73 MeV for \kp with more than two rescattering
collisions and increases to 113 MeV (Fig.~\ref{spectra-effects},
left). In more detail, we find that the slope of the \kp spectra
increases with the number of rescattering collisions ($T$ = 88,
99, 121, 113 MeV for $N_C = 0, 1, 2, >2$).

The \kp nucleus potential lowers the \kp yield, as we have discussed. It
has, however, little influence on the shape of the \kp spectrum as
shown in Fig.~\ref{spectra-effects}, middle. Between production
and registration in the detector the \kp mesons have to loose
their excess mass and this leads --- due to the inherent momentum
and energy conservation --- to an acceleration of the \kp. This
acceleration shifts the spectrum to higher momenta and distorts it
only for small momenta as can be seen in
Fig.~\ref{spectra-effects}, right, where events with $N_C=0$ are
selected. At higher momenta we observe a marginal increase of the
slope.

Figure~\ref{au148-pep}, left, displays the distribution of the number of
rescatterings $(N_C)$ the \kp suffers before leaving the collision zone. The distribution
is quite independent on the \kp nucleus potential. Only about 17.5\% do not
suffer a rescattering collision. Thus, it is evident that the
scattering of the \kp with the environment determines the slope of
its spectra. This is demonstrated on the right hand side where we
display the influence of the \kp nucleus potential and of the \kp
rescattering on the shape of the \kp energy spectra. The inverse
slope parameters of the \kp at production (dotted line), at last
contact (last \kp N collision or production if $N_C =0$, solid line)
and after the \kp is out of range of the \kp nucleus
potential (dashed line) are shown with respect to the number of KN
scatterings  for Au+Au collisions at 1.5 \AGeV. The values
at production  vary with $N_C$ as different densities are
selected by this way.
 This
figure demonstrates that in this heavy system the influence of
scattering on the slope is huge as compared to the acceleration by
the potential and that after two collisions the inverse slope
parameter of the \kp mesons reaches a saturation value. Even an
artificial increase of the cross section for rescattering does not
cause higher values. Thus two collisions bring the \kp into a
kinetic equilibrium with its environment.
\begin{figure}[htb]
\epsfig{file=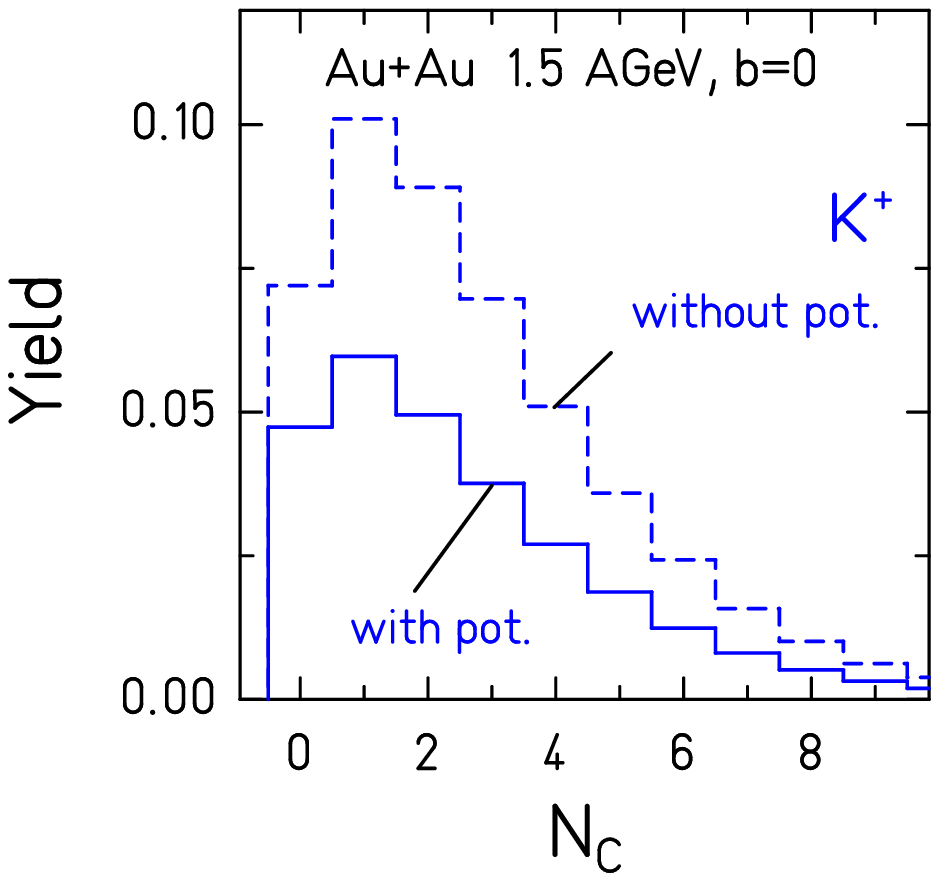,width=0.45\textwidth}
\epsfig{file=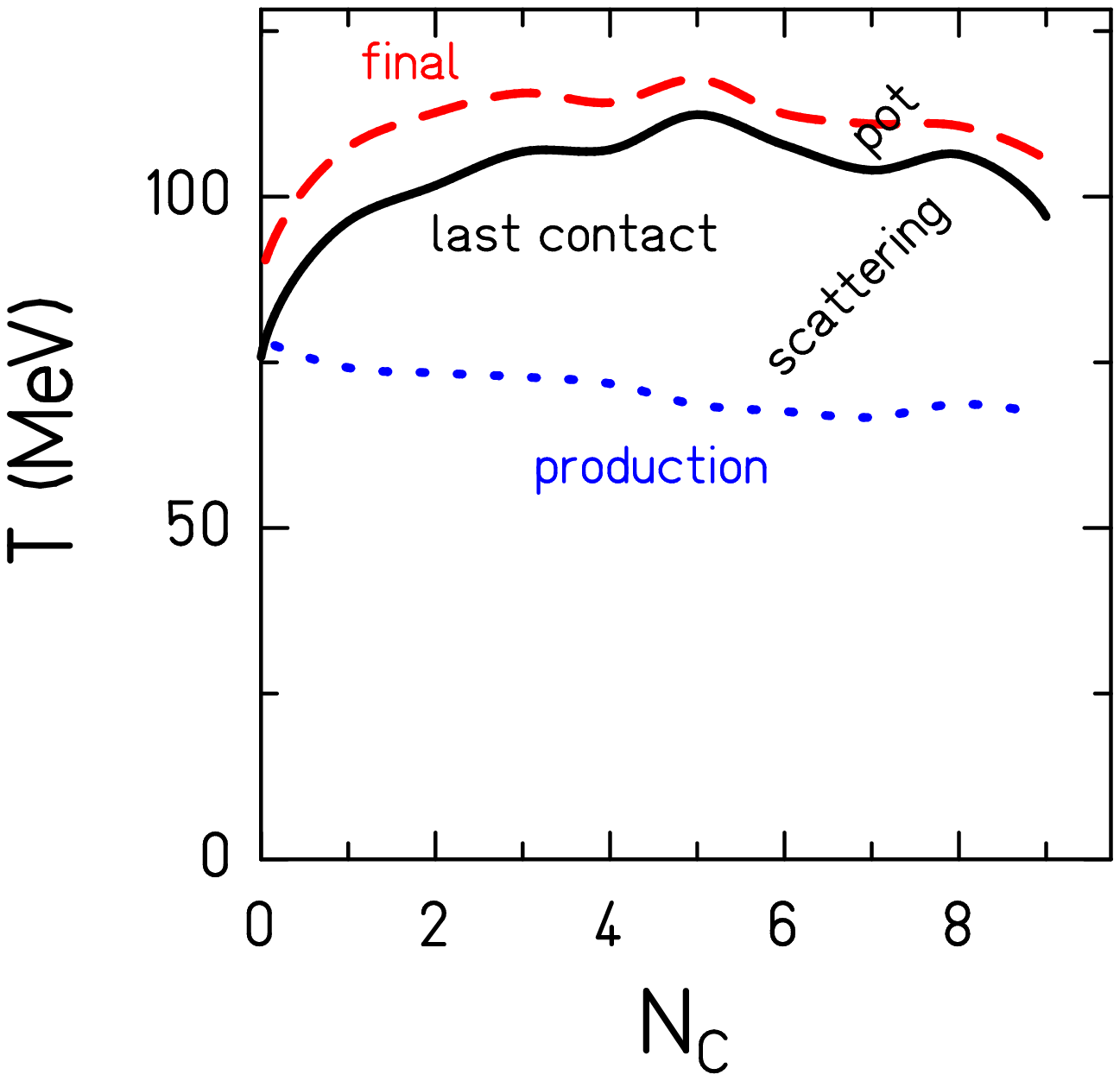,width=0.45\textwidth}
\caption{Left: Number of K rescattering
Right: Change of the
inverse slope parameters due to the different processes discussed
in the text as a function of the number of rescatterings $N_C$.}
\Label{au148-pep}
\end{figure}
\begin{figure}[hbt]
\epsfig{figure=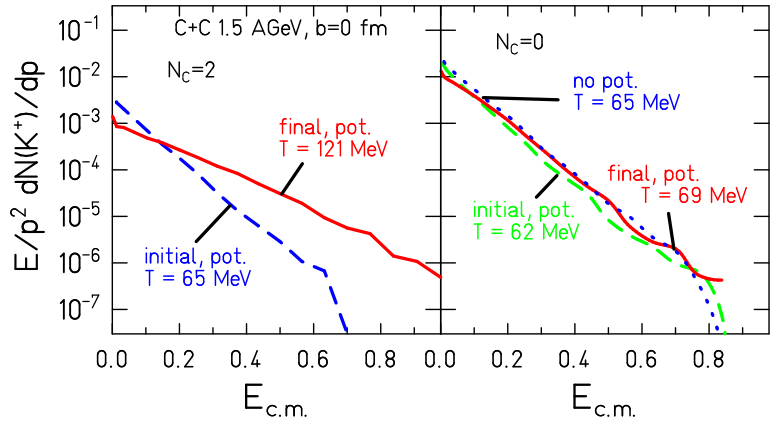,width=0.7\textwidth}
\caption{Influence of rescattering and of the \kp nucleus potential
demonstrated for central C+C collisions at 1.5 $A$ GeV.
Left: Influence of the rescattering of \kp mesons by
selecting kaons which have scattered twice, showing their initial
and final distribution. Right:
Influence of the \kp nucleus potential on the \kp spectrum for \kp which do
not rescatter. Without \kp nucleus potential initial and final distribution
is identical.  } \label{CC_slopes}
\end{figure}

For the lighter C+C system we expect that rescattering is less
important. Consequently, in C+C collisions the slope of the energy
spectra should be steeper. Indeed, as shown in
Fig.~\ref{CC_slopes}, the slope at production of the \kp is around
65 MeV and thus about 10 MeV lower than in the Au+Au reaction
(Fig.~\ref{spectra-effects}) because in a gold nucleus the fraction of NN
collisions as compared to those with a $\Delta$ in the entrance
channel is lower (see Fig.~\ref{nk-b}), 20\% as compared to 35\% in
the C+C system.
In the right panel of Fig.~\ref{CC_slopes}
the effect of the potential
is demonstrated by selecting $N_C=0$. That effect is marginal.
Rescattering increases the slope to a value of 121 MeV for $N_C=2$
as it can be seen from the left part of the figure. In the
discussion of Fig.~\ref{spectra-effects} we found in the Au case a
similar slope value for the final momenta for the same $N_C=2$
selection.

Due to the lower density the \kp mass shift is smaller and hence
the influence of the \kp nucleus potential on the spectra is, especially at
low energies, weaker as compared to Au + Au reactions.
Consequently, for the same number of rescattering collisions the
spectra in Au+Au and C+C collisions are rather similar. Thus a
different slope of the spectra in C+C as compared to Au+Au is
predominantly due to the much smaller number of rescattering
collisions in the former case. We have also to take into account,
that - as we will see in Fig.~\ref{au15-a2-c-au} - kaons show a
strong polar anisotropy, preferring an emission towards the beam
axis. This may indicate that kaons may not only have rescattered
with "thermalized matter" but also with spectators. This might
explain the high values of the slope especially for the C+C case.

As already indicated in Eqs.~(\ref{uoptk}) and (\ref{schaf}) the
\kp nucleus potential contains a part depending on the relative velocity
${\mathbf{\beta}}_{rm KN}$ of the kaon with respect to the nuclear
medium. One might thus question whether the influence of that
momentum dependence might show a significant effect on the
spectra. Indeed the deactivation of those momentum-dependent
interactions would yield a slight increase at high energies for
the Au+Au case and thus enhance the temperature by about 1-2 MeV.
For the C+C no significant effect is visible. The reason for that
moderate influence is related to several aspects: the momentum
dependence shows only strong effects for large
${\mathbf{\beta}}_{\rm KN}$. These kaons should be dominantly seen
at the high-energy part of the spectra. However, these kaons are
dominated by a (late) rescattering of the kaons with the nucleons
which takes place at quite moderate densities. Furthermore,
rescattering normally reduces the relative momenta with respect to
the nuclear medium. Therefore, the effect of the momentum
dependence becomes quite small.

\begin{figure}[htb]
\epsfig{file=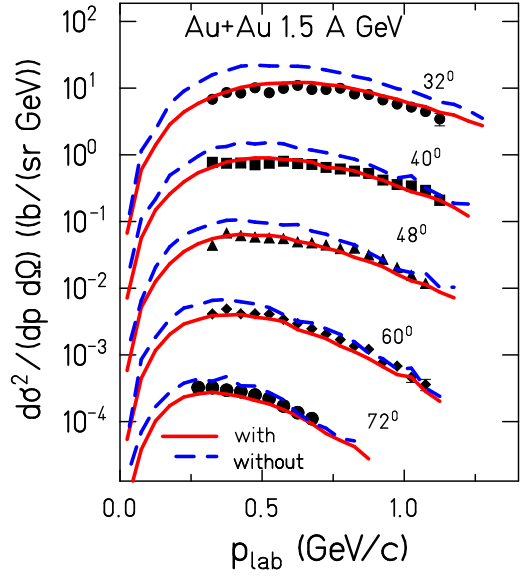,width=0.5\textwidth}
\caption{Comparison of experimental laboratory momentum spectra
with IQMD calculations including and excluding the \kp nucleus potential. The
data are from the KaoS Collaboration~\cite{Forster:2007qk}.}
\label{au15-pep-compa}
\end{figure}

After these theoretical considerations we now compare our results
with data. Figure~\ref{au15-pep-compa} shows the experimental
inclusive momentum distributions for 1.5 $A$ GeV Au+Au by the
KaoS Collaboration~\cite{Forster:2007qk} at various angles in the
laboratory system in comparison with calculations, including and
excluding the \kp nucleus potential. The influence of the potential is
visible and gives a suppression by about a factor of 1.5 as
compared to the calculation without potential. As expected the
influence of the potential is largest at low momenta. The theory
reproduces the data quantitatively at all emission angles but the
measured angular distributions exhibit a somewhat stronger
asymmetry than the calculated ones.
\begin{figure}[hbt]
\epsfig{file=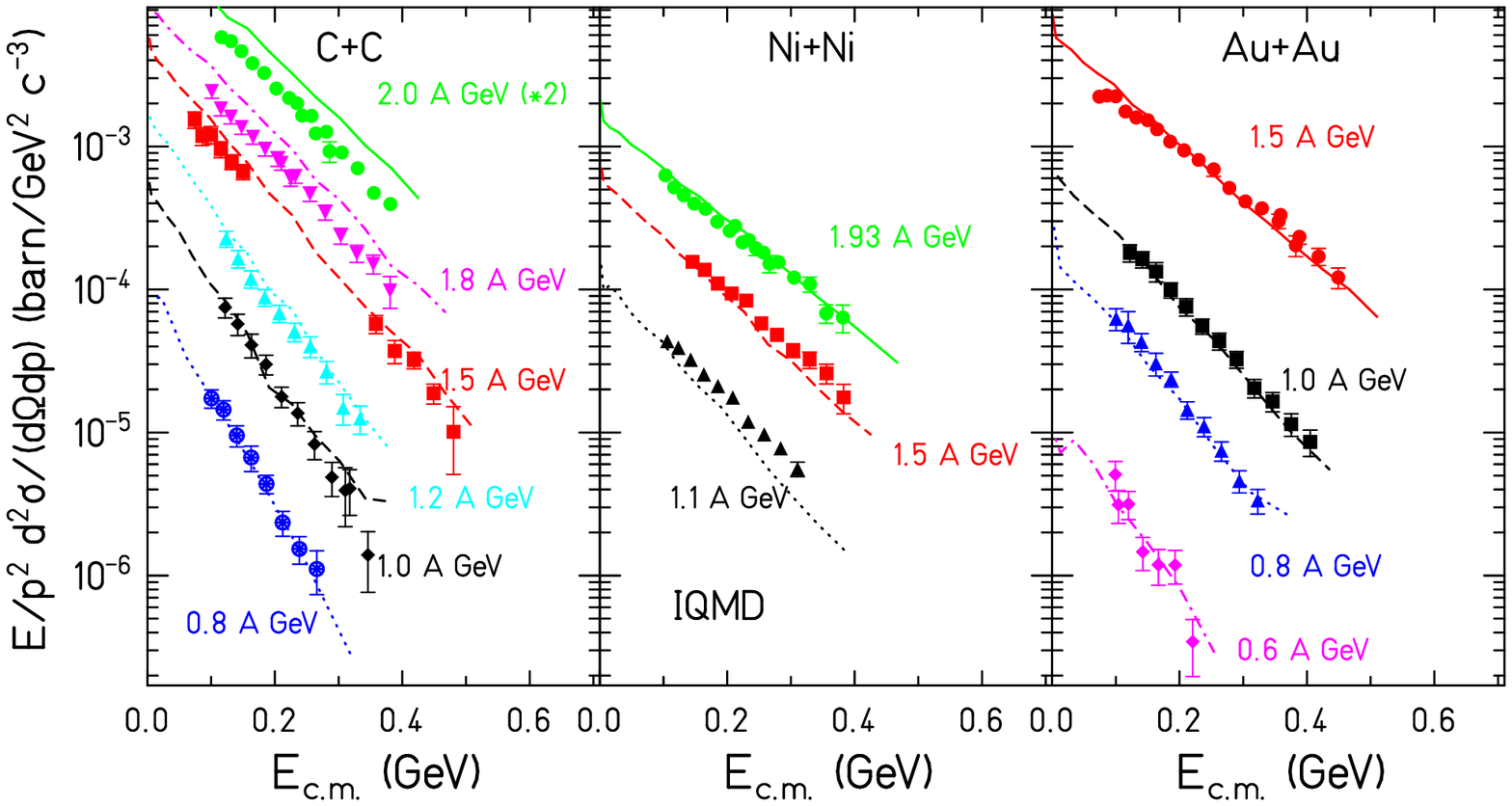,width=.9\textwidth}\\
\epsfig{file=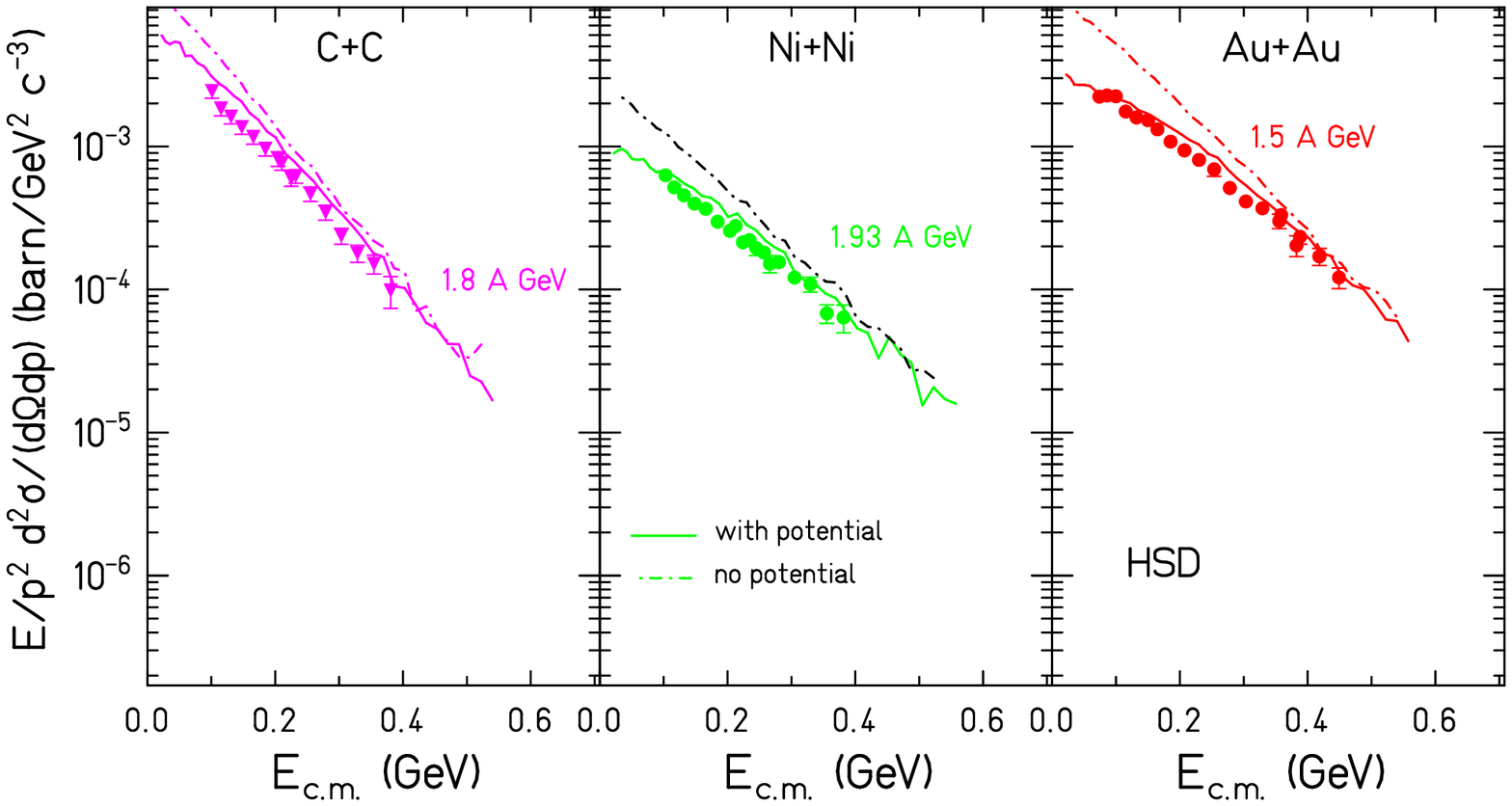,width=.9\textwidth}\\
 \caption{Top: Measured (symbols) and calculated (IQMD, lines) inclusive
 invariant cross sections of \kp at mid-rapidity
 as a function of the kinetic energy $E_{\rm c.m.}$
 for C+C (left), Ni+Ni (middle) and Au+Au (right)  reactions at various beam energies.
 The mid-rapidity condition is a selection
 of $\theta_{\rm c.m.} = 90^\circ \pm 10^\circ$ both for the data and the calculations.
 The data are from the KaoS Collaboration~\cite{Forster:2007qk}, the lines represent calculations with
 \kp nucleus potential included. Bottom: The same data compared to HSD calculations \cite{Mishra:2004te}.
 Here the dashed lines refer to the option without \kp nucleus potential, the solid ones to that with \kp nucleus potential.}
\label{midrap}
\end{figure}

The inclusive spectra at mid-rapidity for different symmetric
systems and at various incident energies as measured by the KaoS
Collaboration~\cite{Forster:2007qk} are compared in
\Figref{midrap} with IQMD calculations (top) and with HSD
calculations \cite{Mishra:2004te} (bottom), both in their standard version.
All these spectra are well reproduced by an exponential function
with one inverse slope parameter $T$.

\begin{figure}[hbt]
\epsfig{file=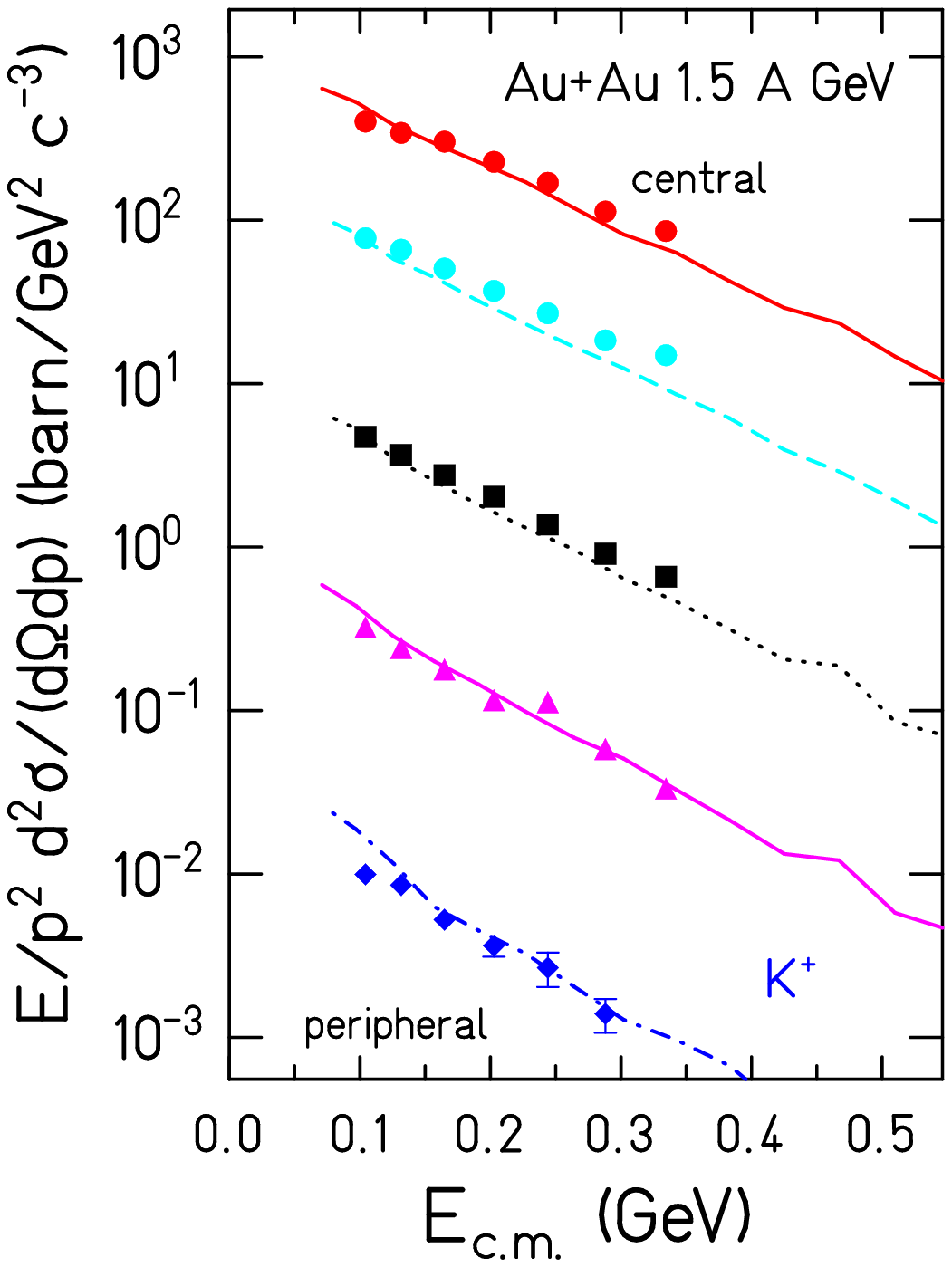,width=0.45\textwidth}
\epsfig{file=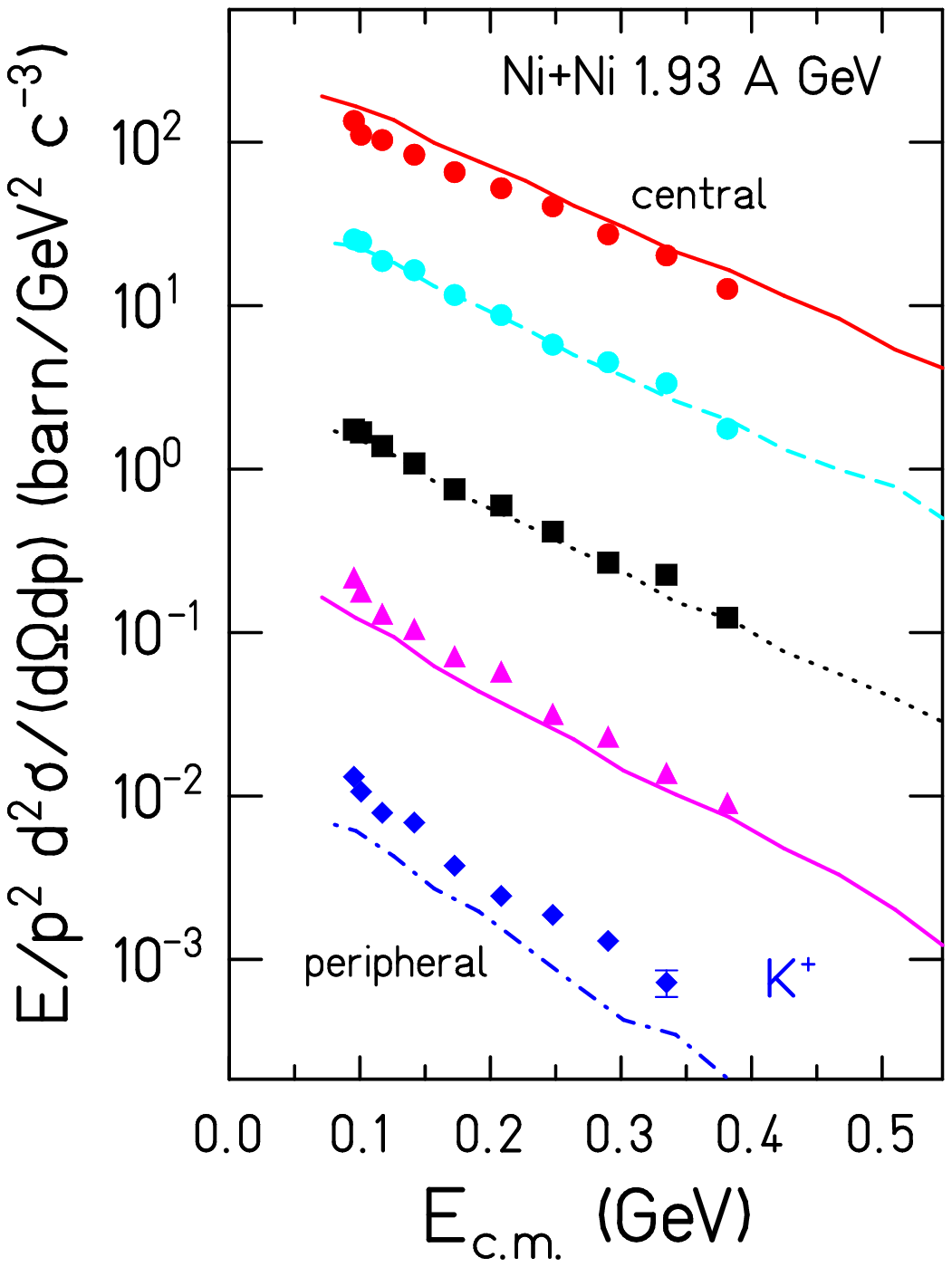,width=0.45\textwidth}
\caption{
Centrality selected energy spectra for
  \mbox{Au+Au} at $1.5$~$A$~GeV  and for
  \mbox{Ni+Ni} at $1.93$~$A$~GeV both at $\theta_{\rm lab}=40^\circ$(right) as compared with IQMD calculations.}
\label{spec_cm}
\end{figure}

The KaoS Collaboration has also
analyzed the \kp spectra at $\theta_{\rm lab}=40^\circ$ for
different centrality classes and in \Figref{spec_cm} we compare
these data with IQMD calculations. The slopes are
increasing towards central collisions both in theory as well as in
experiment. This is expected due to the increasing number of \kp which have
scattered. A very good agreement between theory and experiment for
all centrality classes can be seen.

These spectra can be well described by an exponential function
with an inverse slope parameter $T$.  These values, both from
experiment and from IQMD calculations (lines), are displayed in
\Figref{tmidrap}, on the upper left hand side as excitation
function for three different systems. The inverse slope parameters
increase both with system size and with increasing incident
energy. This trend can be explained by the increasing fraction of
\kp mesons which rescatter. Increasing energy means more available
energy in the elementary production but also increasing density
and therefore more rescattering. It is interesting to compare the
results for \kp with the corresponding values for high-energy
pions~\cite{Sturm:2000dm}(upper right panel). Both exhibit a remarkable resemblance.

\begin{figure}[hbt]
\epsfig{file=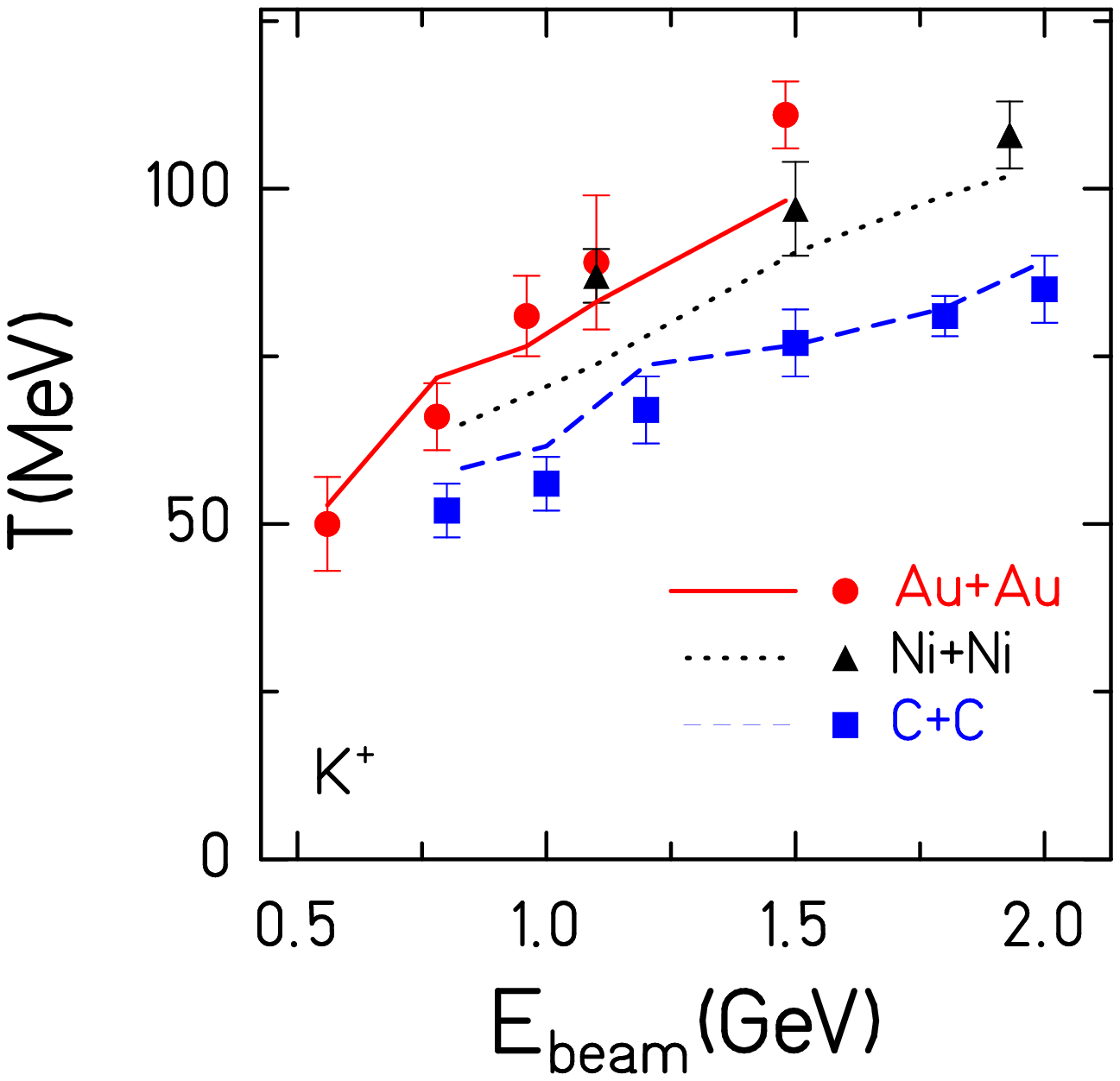,width=0.48\textwidth}
\epsfig{file=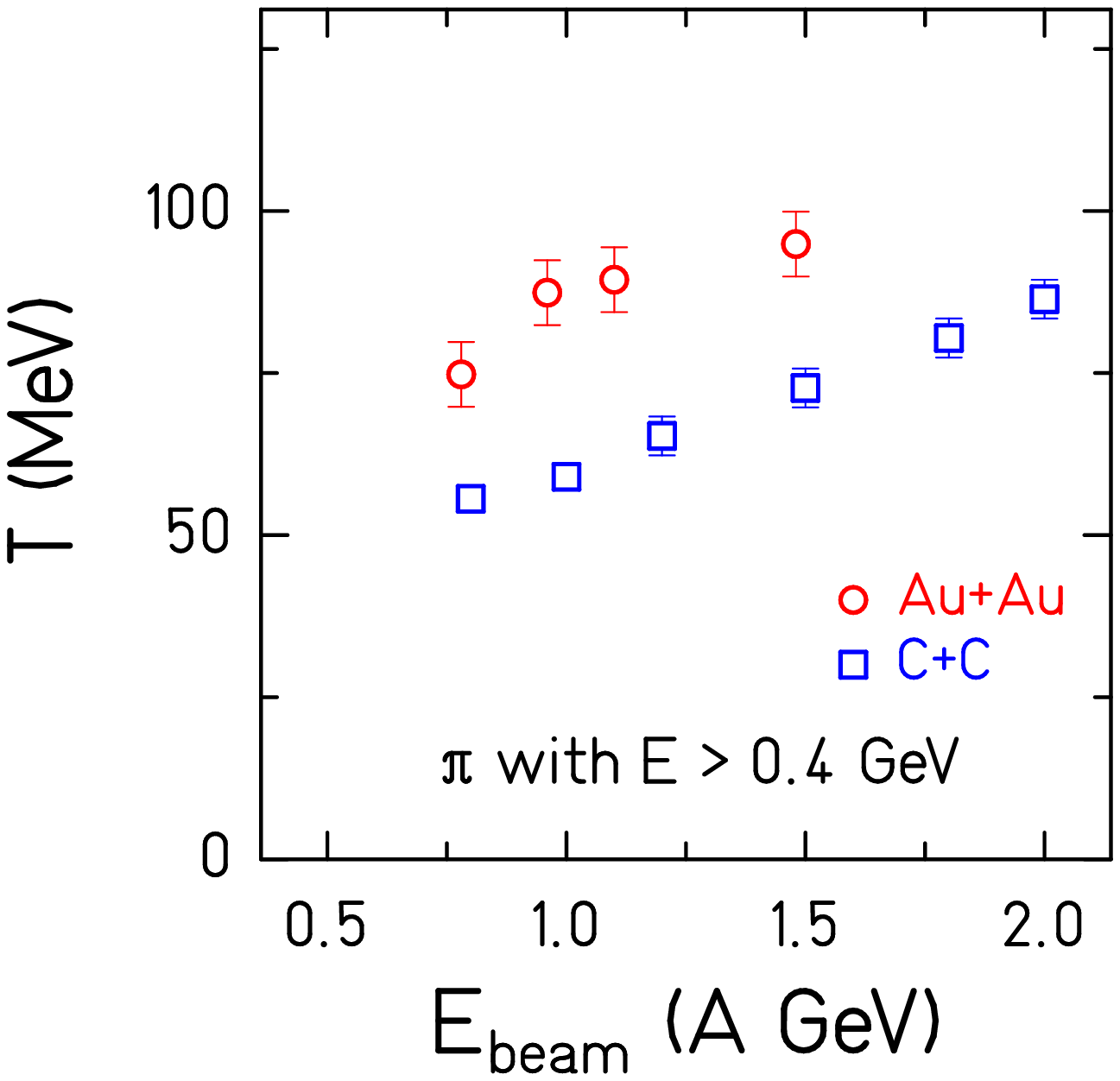,width=0.48\textwidth}
\epsfig{file=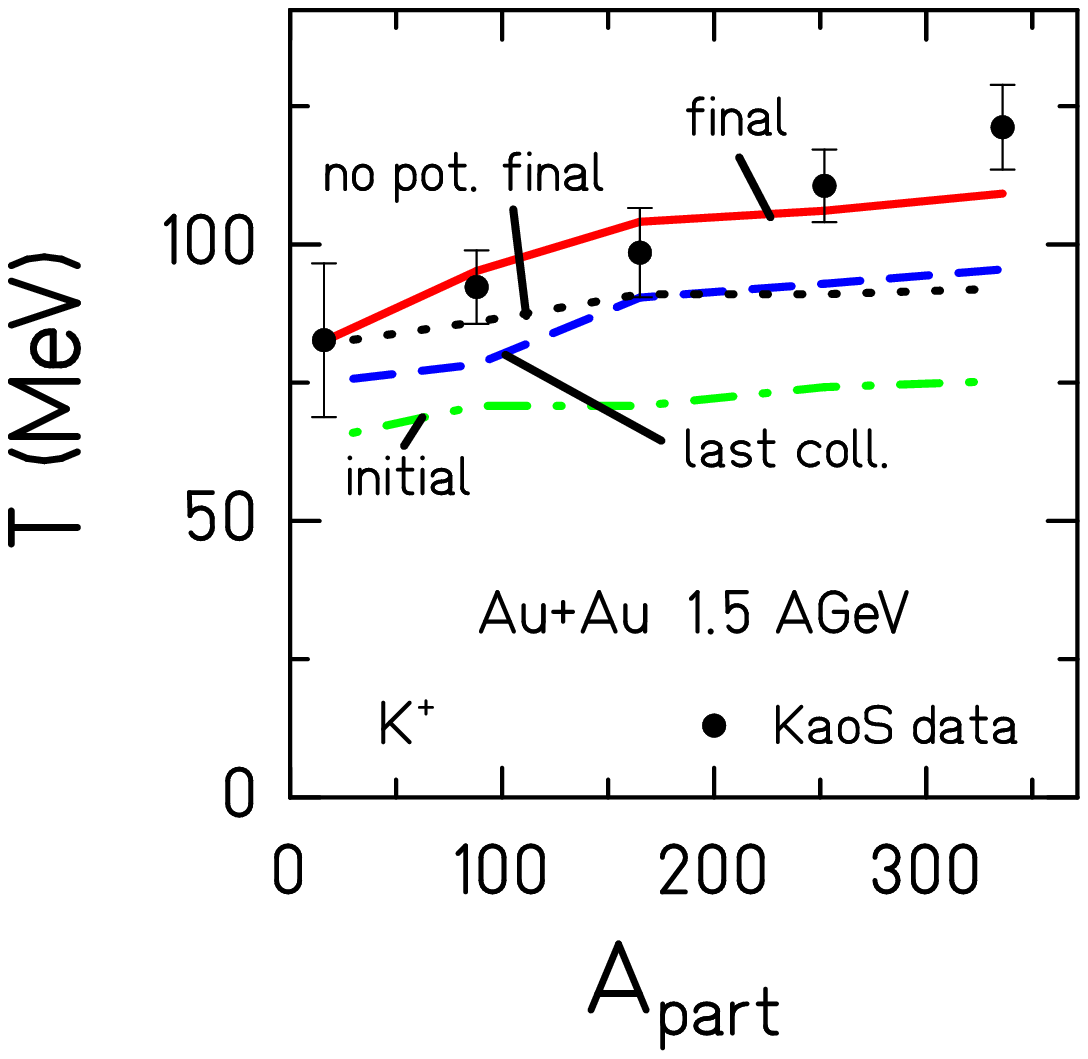,width=0.48\textwidth}
\caption{Dependence of the inverse slope parameter on incident
energy (top) and on centrality (bottom). All results are for
inclusive collisions and at mid-rapidity ($\theta_{\rm c.m.} =
90^\circ \pm 10^\circ$). The upper left panel displays the \kp
inverse slope parameter which can be compared to those of the
high-energetic pions (upper right). The IQMD results for \kp are
given as lines. The data (symbols) are from the KaoS
Collaboration~\cite{Forster:2007qk,Sturm:2000dm}. at mid-rapidity.
Bottom: Dependence of the slope parameter on the centrality for
\mbox{Au+Au} at $1.5$~$A$~GeV.} \Label{tmidrap}
\end{figure}

On the bottom part of \Figref{tmidrap}, the centrality dependence
of the inverse slope parameters is given for Au+Au collisions at
1.5 \AGeV{}. This figure also elucidates the influence of both,
the \kp nucleus potential and rescattering, on the final slopes.
The effect of rescattering is important, yet only if both are
present the slope of the experiment can be reproduced. For the
Ni+Ni case (not shown) the theoretical (experimental) slope
parameters (going from the most central to the most peripheral
reactions) are 106.1 (109.1), 101.8(96.4), 97.8(91.7), 91.4(81.0),
79.2(69.3) MeV, reflecting as well the decreasing number of
rescattering collisions. The uncertainty of the fitted inverse
slope parameters is of the order of 10\%. In passing we would like
to mention that also the magnitude of the nucleon-nucleon cross
section has a mild influence on the \kp slope. If we assume that
the in-medium nucleon-nucleon cross section is only half of the
free cross section the \kp yield decreases by 5\%.

\begin{figure}[bt]
\epsfig{figure=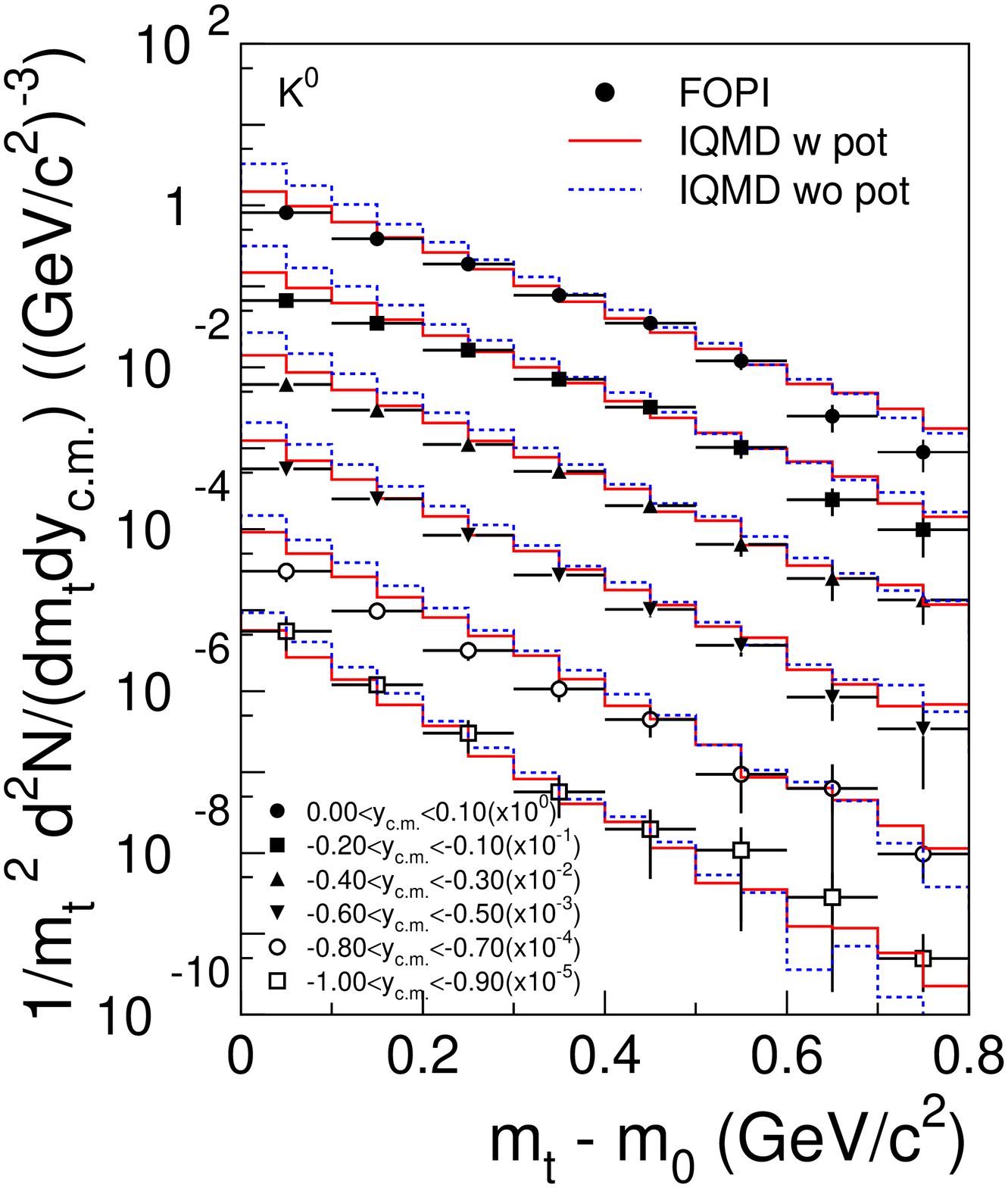,width=0.4\textwidth}
\hspace{2cm}
\epsfig{figure=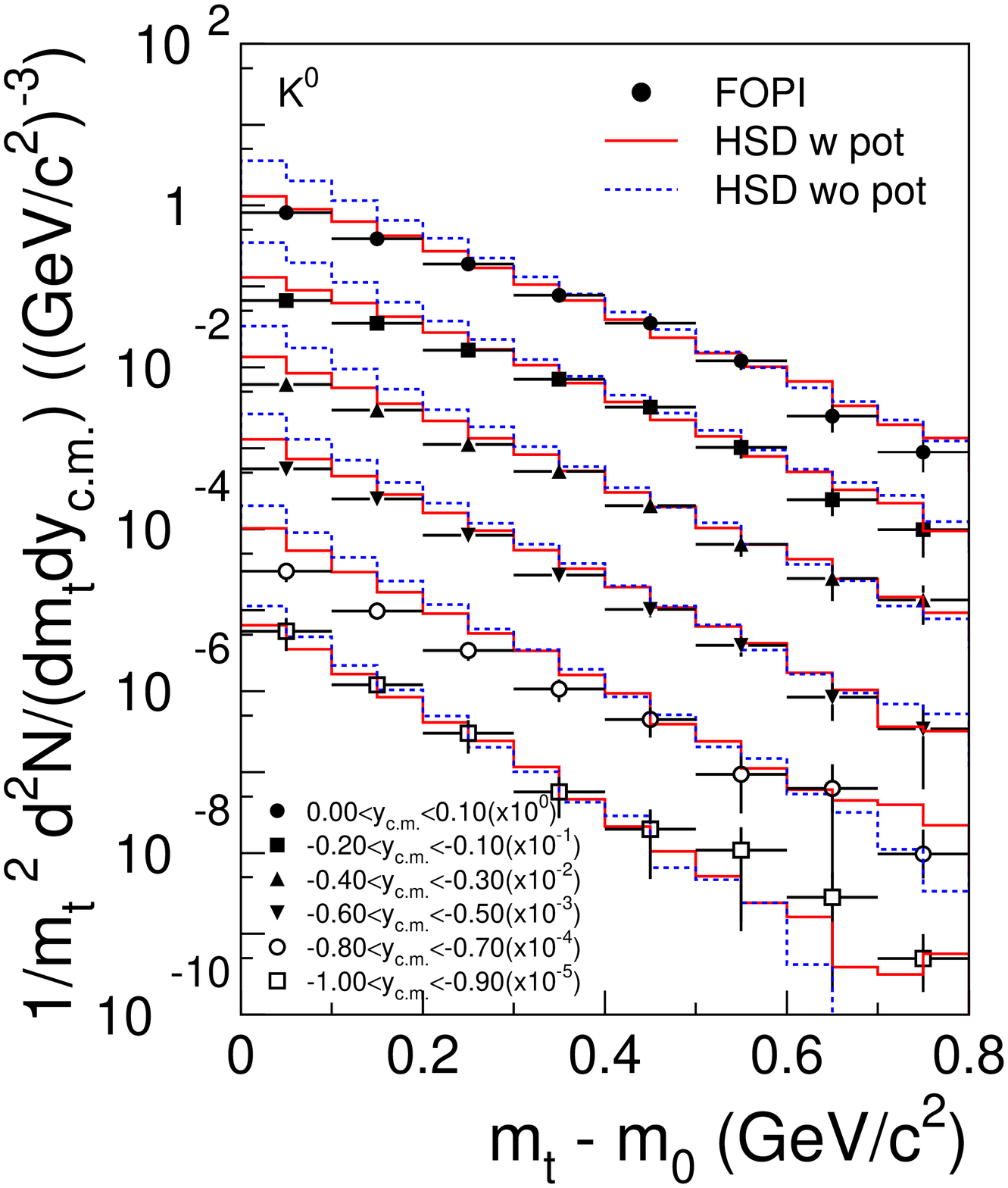,width=0.4\textwidth}
\caption{$m_t-m_0$ spectra of K$^0$ measured
by the FOPI Collaboration for central Ni+Ni at 1.93 \AGeV
\cite{Merschmeyer:2007zz} in comparison with IQMD (left) and HSD (right) calculations
(histogram) for different values of the rapidity $y_{\rm c.m.}$.
We present calculations with and without \kp nucleus potential.} \label{spec_k0}
\end{figure}

The FOPI Collaboration has measured the $m_t-m_0$ spectra of K$^0$
mesons for central Ni+Ni collisions ($\sigma \approx 350$ mb, this
value equates to an impact parameter in a sharp cut-off model of
$b$=3.2\,fm) at 1.93 \AGeV{}. The spectra  for different rapidity
bins are shown in \Figref{spec_k0} and compared with IQMD and HSD
calculations. In the experiment central events are selected by
requiring high charged-particle multiplicity in the forward
hemisphere of the detector setup. The IQMD output was used as an
event generator for a detector simulation based on the GEANT
package to obtain a realistic impact-parameter distribution for
the multiplicity selection cut \cite{Merschmeyer:2007zz} which was
then used for the comparisons between data and model predictions.

\begin{figure}[tb]
\begin{center}
\epsfig{file=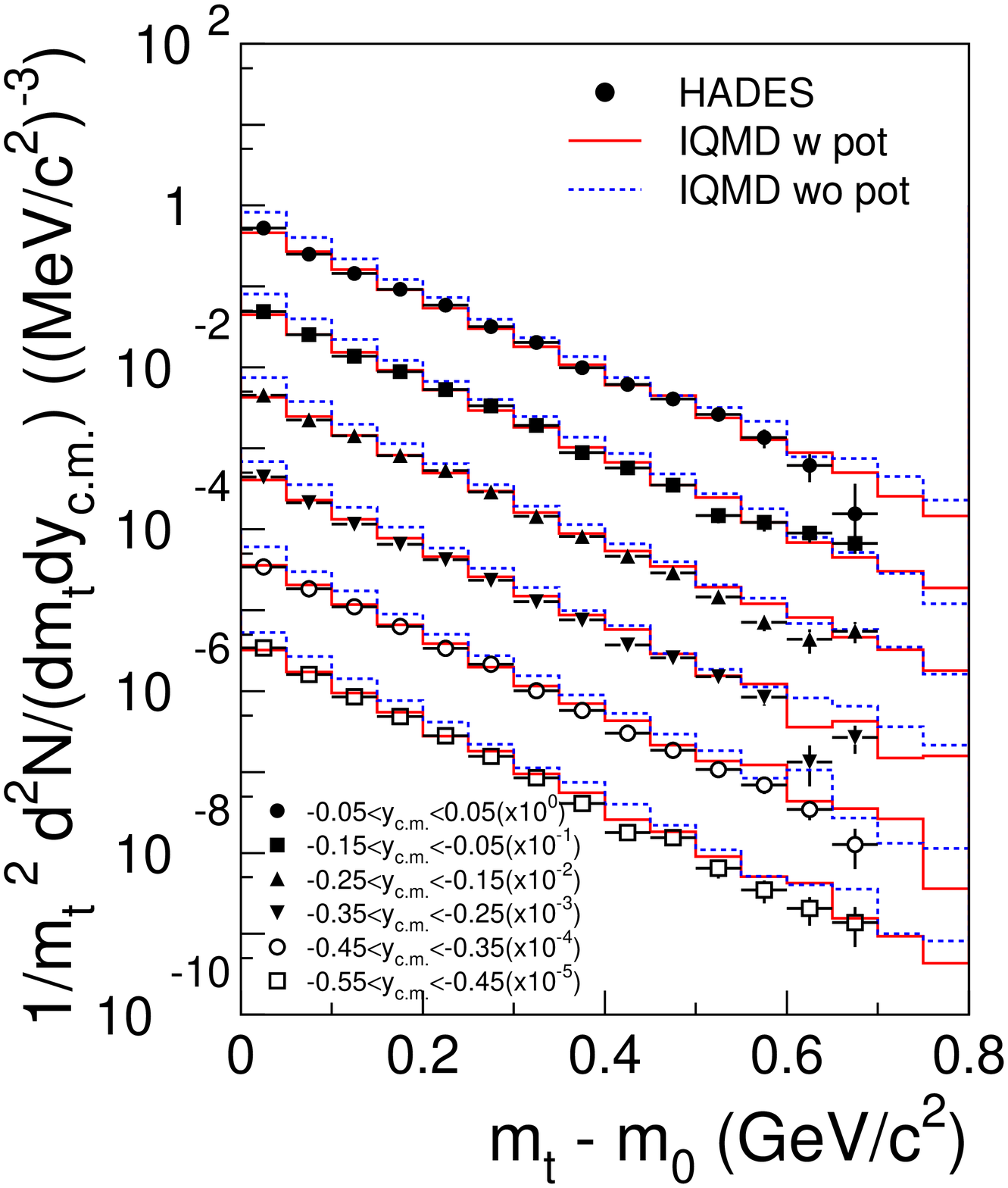,width=0.4\textwidth}
\epsfig{file=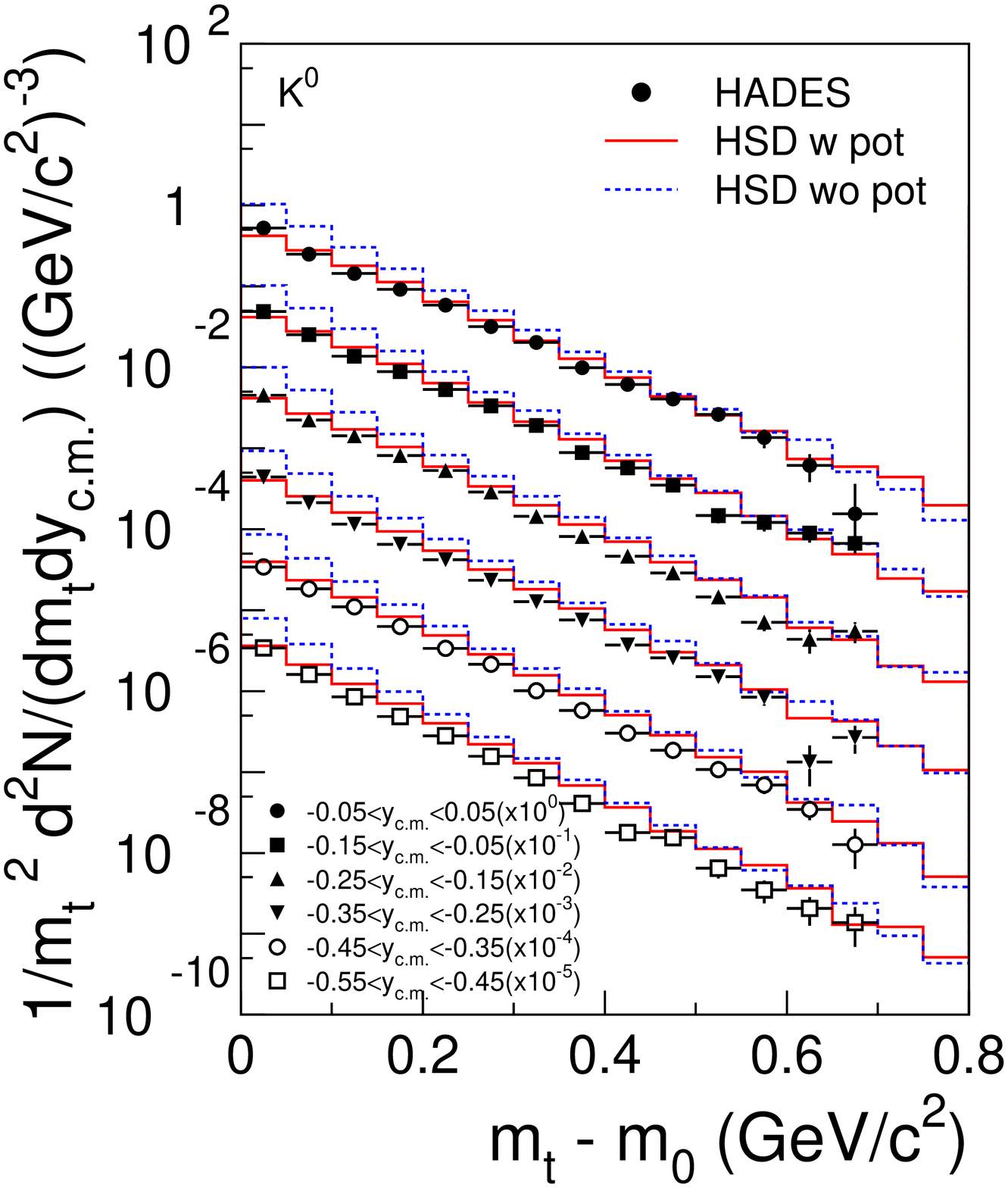,width=0.4\textwidth}
\epsfig{file=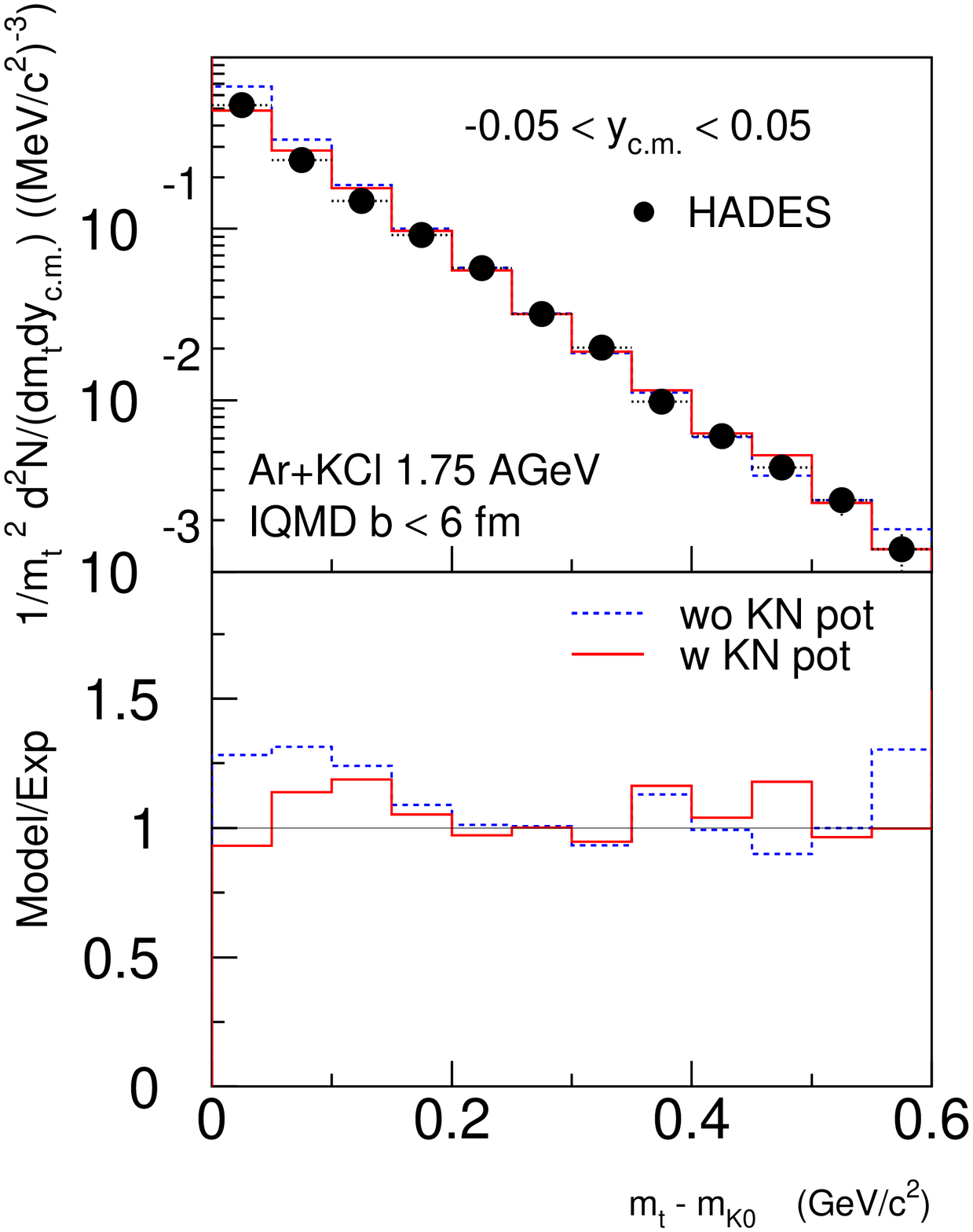,width=0.4\textwidth}
\epsfig{file=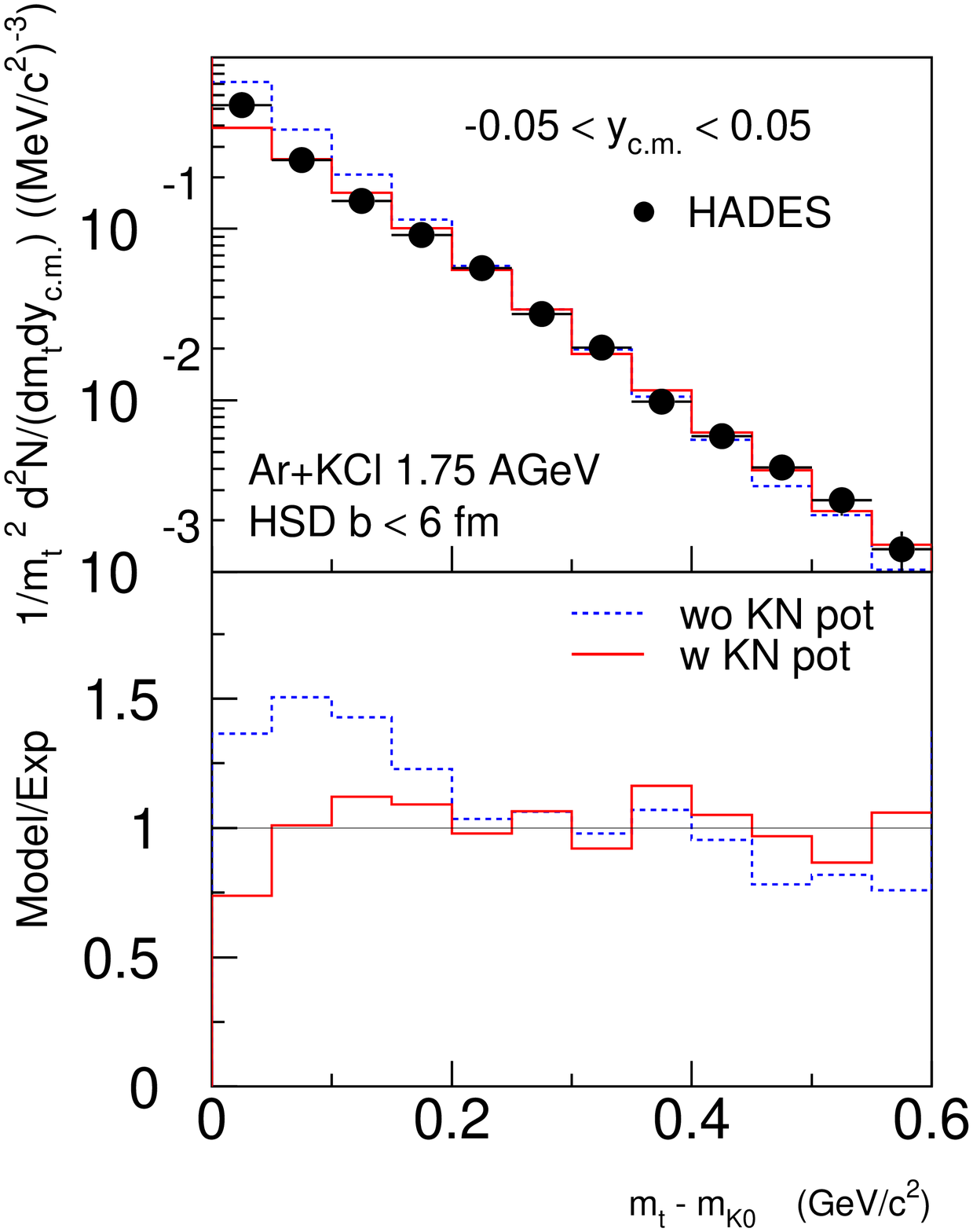,width=0.4\textwidth}
\caption{Upper panel: $m_t-m_0$ spectra of K$^0$ measured by the HADES
Collaboration for Ar+KCl at $1.75$ \AGeV \cite{Agakishiev:2010zw}
in comparison with IQMD and HSD calculations for different values of the
rapidity. Lower panel: Top: Comparison of measured K$^0$ spectrum at
mid-rapidity with IQMD and HSD calculations with and without \kp nucleus
potential. Bottom: Ratio of the calculations divided by
the measured spectrum, enhancing the sensitivity. The model calculations
are normalized to the experimental data beyond
$m_t - m_0 > 0.3$ GeV/c$^2$.} \label{hadk0}
\end{center}
\end{figure}

More recently the HADES Collaboration has obtained $m_t-m_0$
spectra of K$^0$ for the lighter Ar+KCl system at a slightly lower
beam energy of $1.75$ \AGeV\ . These results are compared with IQMD
and HSD calculations in \figref{hadk0}, upper panel.
In the calculation we have chosen a sharp
cut-off of $b$ = 6 fm. This gives $<b> = 4$ fm, a value which is
compatible with $<b> = 3.5 $ fm, the value reported by the HADES
collaboration using a rather soft central trigger.

In the top panel the measured $m_t-m_0$ spectra are depicted in
  comparison to the predictions of the IQMD and of the HSD model, with and without a
\kp nucleus potential. In the
  lower panel the ratio of IQMD model predictions and the experimental data is
  shown. Calculations incorporating an \kp nucleus potential are better describing the
  experimental data.

\begin{figure}[htb]
\epsfig{file=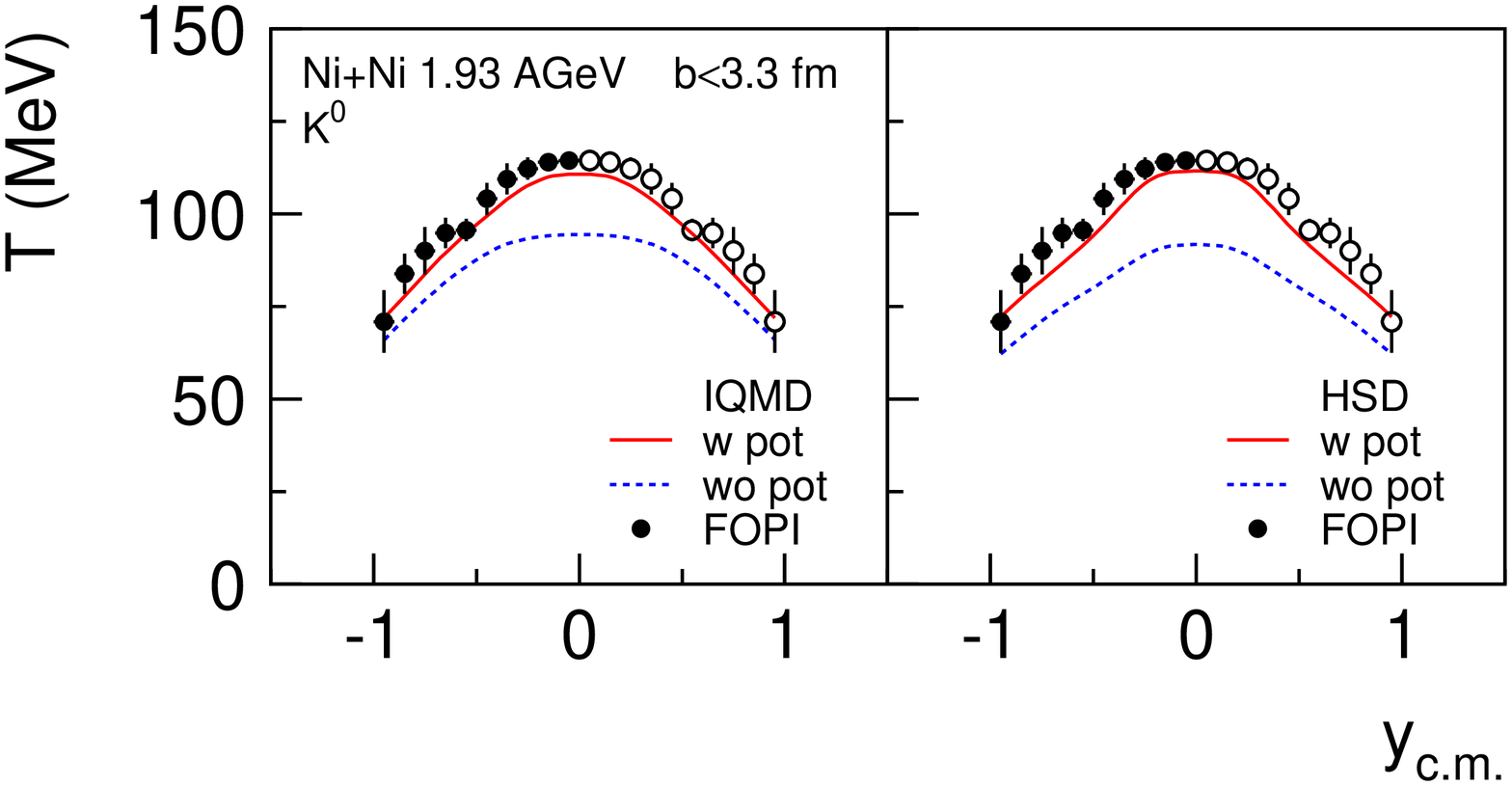,width=0.65\textwidth}
\epsfig{file=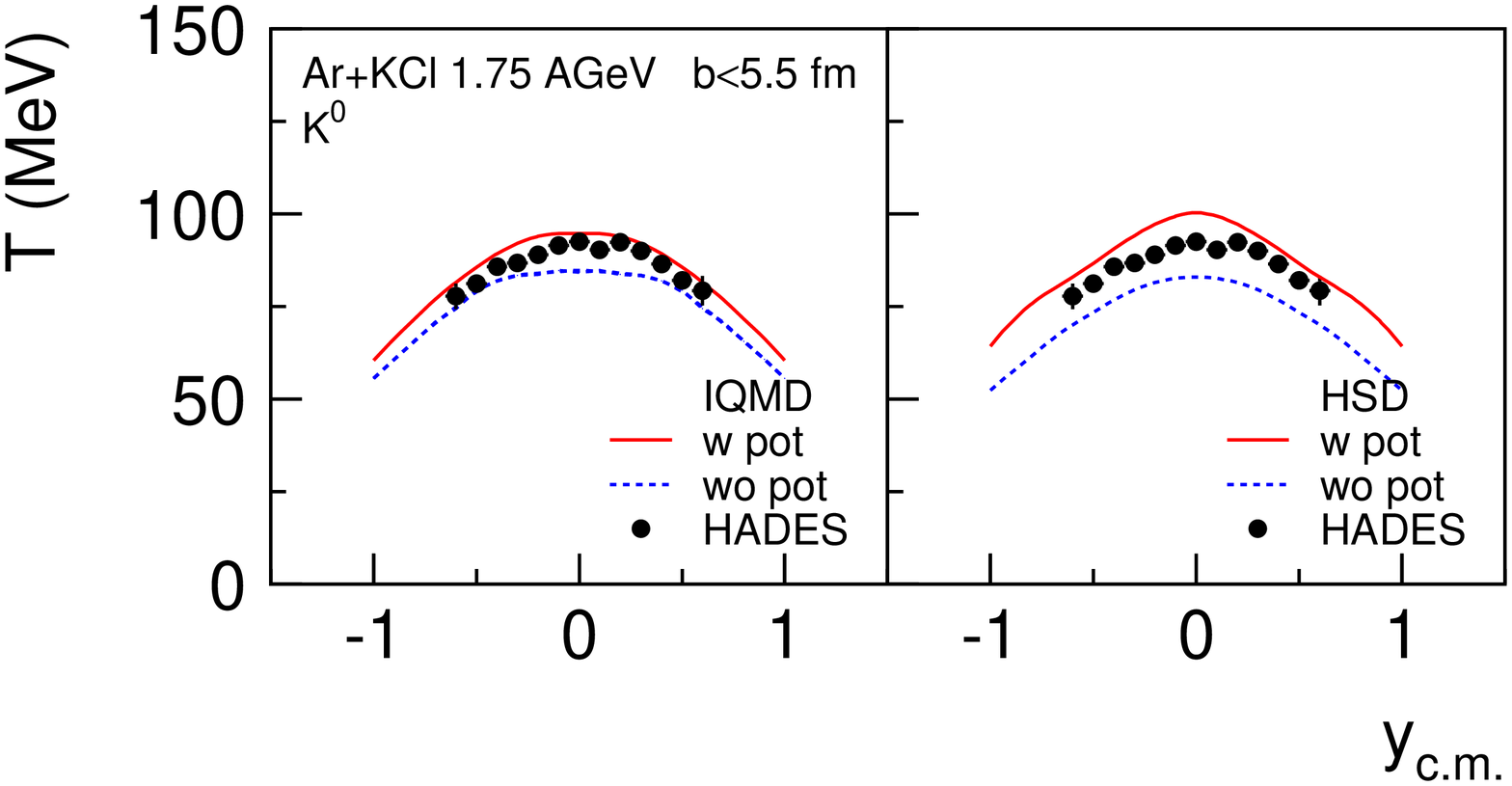,width=0.65\textwidth}
\caption{Inverse slope parameter T of the spectra figs. \ref{spec_k0}  and \ref{hadk0}. Upper
part: For central ($b <$ 3.3 fm) Ni+Ni collisions at 1.93 $A$ GeV
as measured (and extrapolated to 4$\pi$) by the FOPI collaboration
in comparison with those obtained in IQMD (left) and HSD (right)
calculations. The error bars contain only the statistical error.
Lower part: Same for central Ar+KCl collisions at 1.75 \AGeV as measured by the HADES
collaboration \cite{Agakishiev:2010zw}.
}\label{T_dndy}
\end{figure}

Fitting the measured and the calculated 
$m_t-m_0$ spectra of K$^0$ of \Figref{spec_k0} and \Figref{hadk0}
with an exponential function and extrapolating towards vanishing
$m_t-m_0$ values allows for the determination of the rapidity
dependence of the slopes and of the total yield d$N$/d$y$. As seen
in \Figref{spec_k0} the spectra of K$^0$ are not really
exponential and therefore the extrapolation depends on the range
of $m_t$ which is used for this fit. It was taken care that the
fit ranges for the experimental data and the theoretical
distributions are the same. Therefore the theoretical and
experimental inverse slope parameters can be compared but neither
the slope nor the d$N$/d$y$ are free from the uncertainty of the
fit. For the HADES data \Figref{hadk0} the situation is different.
Here theory as well as experiment show an exponential form of the
spectrum. Because the HADES acceptance is close to mid-rapidity we
expect more thermalization.

Figure~\ref{T_dndy} displays the experimental inverse slope
parameters measured by FOPI (top) and by HADES (bottom) as a function of
the rapidity in comparison with the two model predictions.
The Ni+Ni data at 1.93 $A$ GeV as well as the Ar+KCl
data at 1.75 \AGeV are well described by both models approaches.

\subsection{Rapidity distribution}

\begin{figure}[hbt]
\epsfig{figure=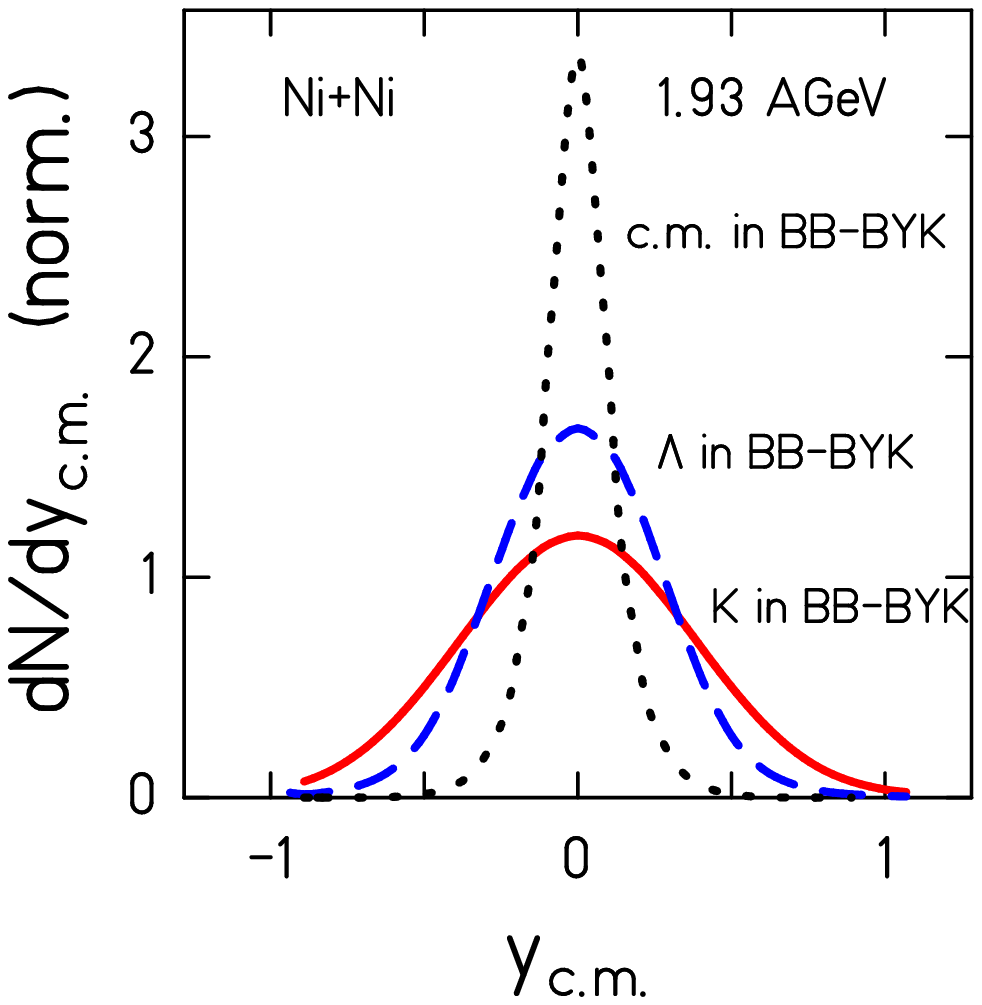,width=0.4\textwidth}
\epsfig{figure=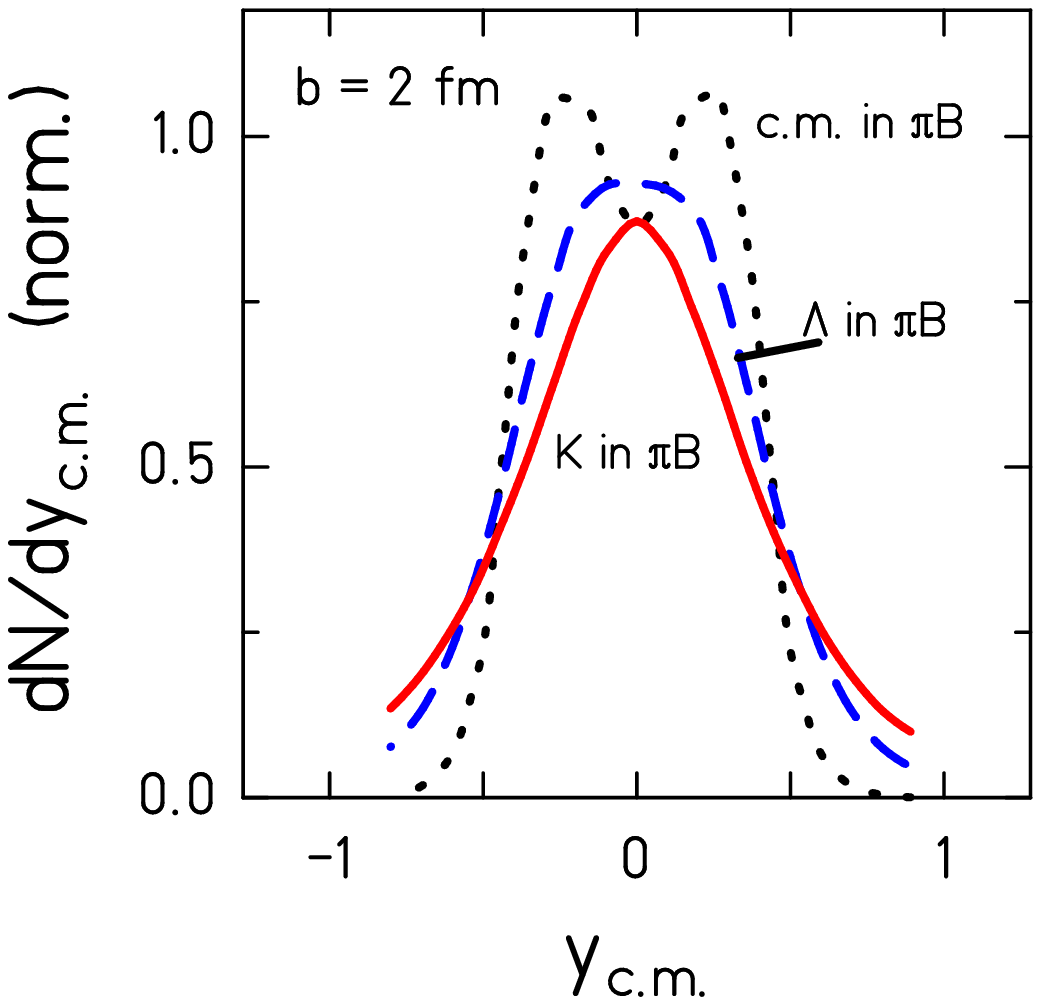,width=0.4\textwidth}
\caption{Initial rapidity distribution d$N$/d$y$ of \kp and
$\Lambda$ separated for the two principal \kp production processes for
Ni+Ni at 1.93 \AGeV.} \label{dndyt}
\end{figure}

The rapidity of the center of mass of the two baryons, which
create a \kp, is centered around $y_{\rm c.m.}=0$, as shown in
\Figref{dndyt}, left. When \kp and $\Lambda$ are created according
to the three-body phase space the rapidity distribution of the
$\Lambda$ is broader than that of the sources and that of the \kp
becomes quite wide. This will have consequences for the in-plane
flow discussed in Section J. \kp mesons produced in $\pi$-induced
reactions exhibit a similar width as if produced in BB collisions,
\Figref{dndyt}, right. The center of mass rapidity of these
collisions is not centered around mid-rapidity because the $\pi$
are decay products of $\Delta$ resonances.
\begin{figure}[htb]
\epsfig{file=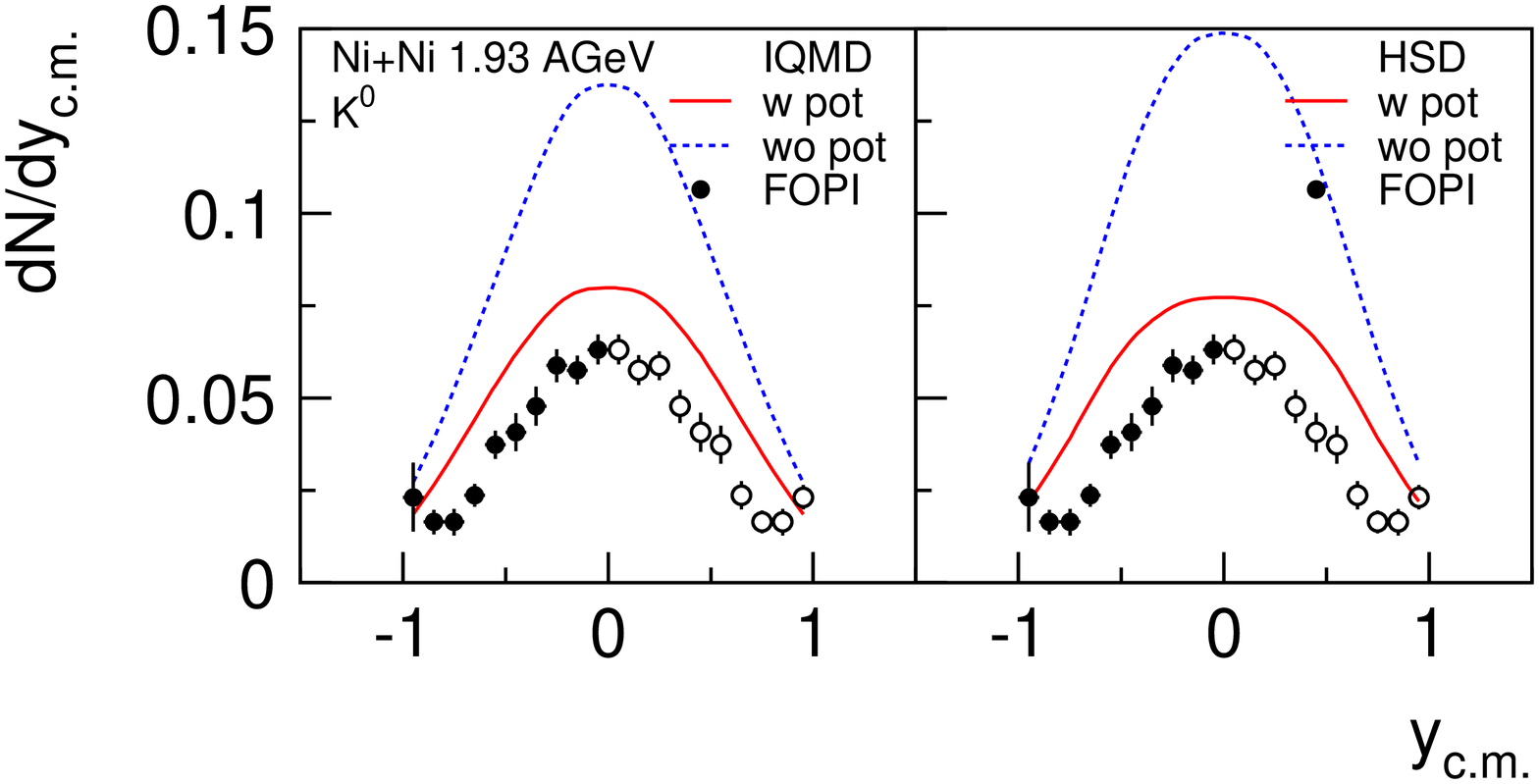,width=0.65\textwidth}
\epsfig{file=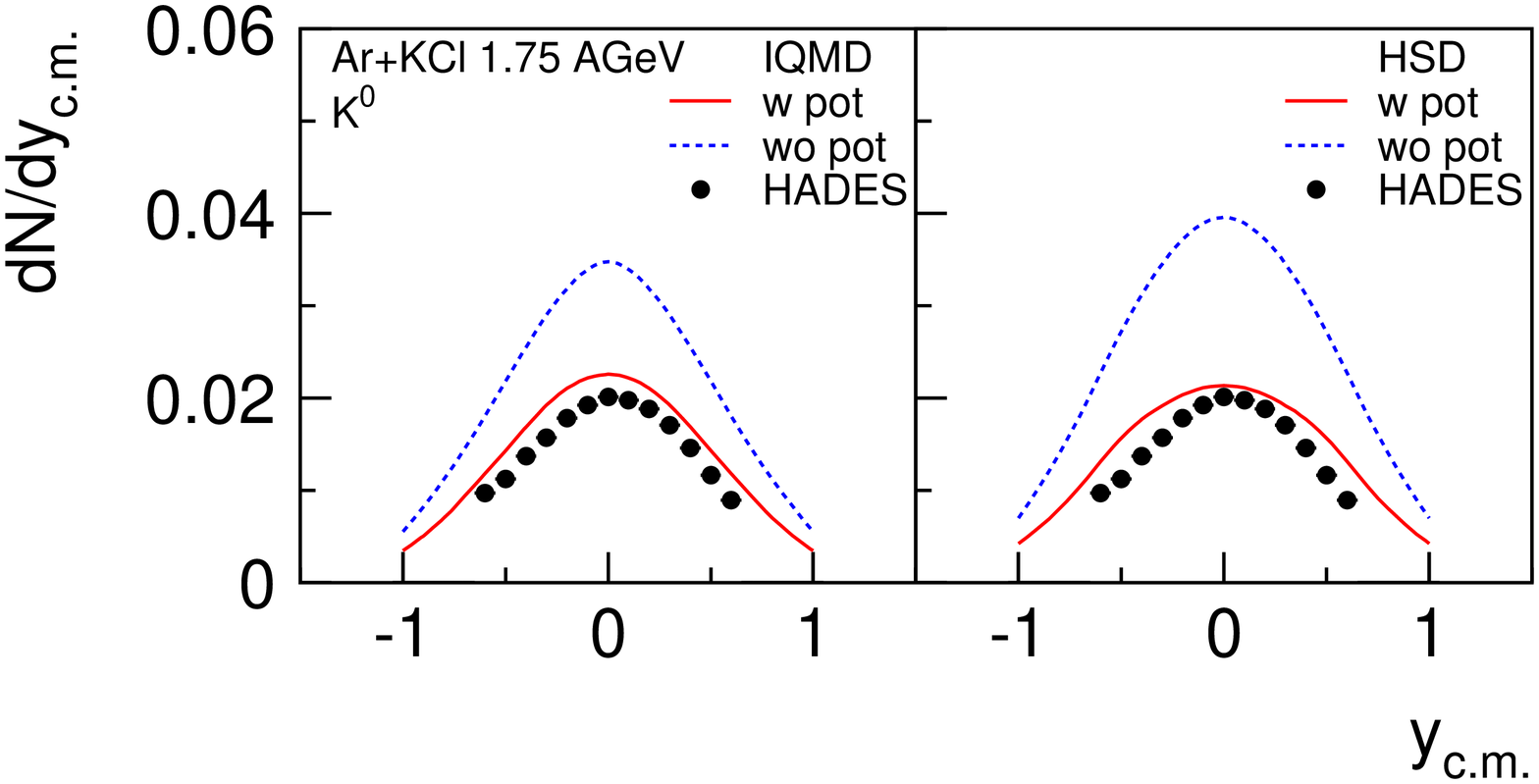,width=0.65\textwidth}
 \caption{Rapidity distributions  d$N$/d$y$ of K$^0$. Upper
part: For central ($b <$ 3.3 fm) Ni+Ni collisions at 1.93 $A$ GeV
as measured (and extrapolated to 4$\pi$) by the FOPI collaboration
in comparison with those obtained in IQMD (left) and HSD (right)
calculations. The error bars contain only the statistical error.
Lower part: Same for central Ar+KCl collisions at 1.75 \AGeV as measured by the HADES
collaboration \cite{Agakishiev:2010zw}.
} \label{y_dndy}
\end{figure}
Figure \ref{y_dndy} displays the experimental d$N$/d$y$
distributions of K$^0$ (which should be - besides isospin effects
- identical to that of the \kp), in comparison with IQMD as well
as with HSD calculations. The free d$N$/d$y$ is larger in IQMD due
to different isospin assumptions. The agreement between the
experiments and IQMD calculations is reasonable if the \kp nucleus
potential is employed. Without the \kp nucleus potential IQMD
over-predicts the yield by a factor of two. As discussed, in HSD
calculations the density dependence of the \kp nucleus potential
is weaker. Therefore the difference between calculations with and
without potential is smaller. HSD calculations also reproduce
quite well the data when employing an
 \kp nucleus potential for the kaons.
\begin{figure}[htb]
\epsfig{file=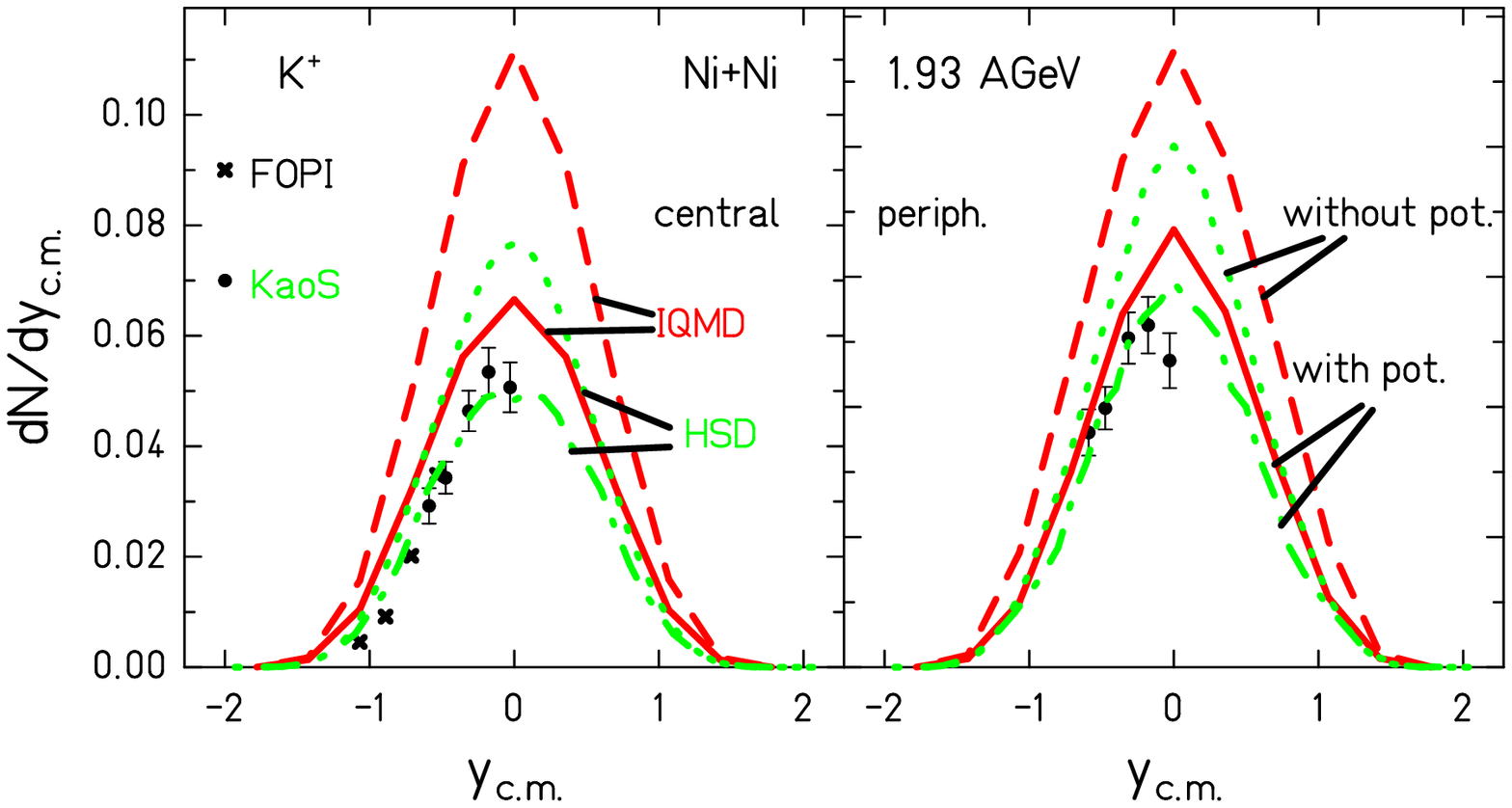,width=0.8\textwidth}
\caption{$4\pi$ extrapolated rapidity distribution of the \kp in Ni+Ni reactions at 1.93 $A$ GeV
as measured by the FOPI and KaoS Collaborations for central ($b <$
4.5 fm) and peripheral (4.5 fm $< b <$ 7.5 fm) collisions. Only the statistical errors are shown.}
\label{dndykpk0}
\end{figure}

Figure~\ref{dndykpk0} compares the experimental results of the
FOPI \cite{Best:1997ua} and the KaoS Collaboration for the \kp
rapidity distribution. The results from the FOPI Collaboration,
measured at $b <$ 3.3 fm, have been scaled with a factor of 1.22
to match with the central bin of the KaoS Collaboration measured
at $b <$ 4.5 fm. This factor has been obtained by comparing IQMD
calculations with the two impact parameter ranges. As for the
K$^0$ a good agreement between the experiments and
IQMD calculations is found if the \kp nucleus potential is employed.

\subsection{Polar distribution}

IQMD predicts that at creation the polar distribution of  \kp
mesons in central collisions is rather flat. This is a consequence
of the dominance of the BB channel and of the creation of the \kp
according to phase space.

\begin{figure}[htb]
\epsfig{file=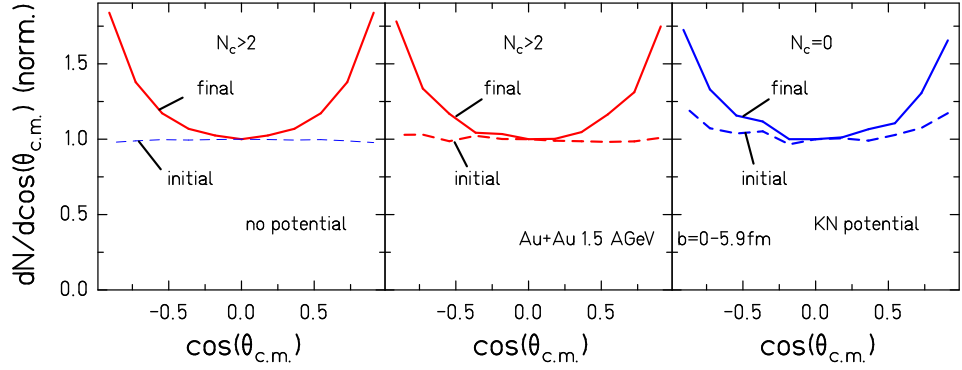,width=.85\textwidth}
\caption{Influence of rescattering and of the \kp nucleus potential on the
polar distribution for central ($b < 5.9$ fm) Au+Au reactions at 1.5 $A$ GeV in IQMD calculations.}
\label{au15-a2-c-au}
\end{figure}

Figure \ref{au15-a2-c-au} addresses the question of how this
distribution is modified later by KN collisions and by the \kp nucleus potential.
In central collisions Au+Au collisions
($b<5.9$ fm) the polar distribution is almost flat at creation.
The \kp nucleus potential causes a forward-backward enhancement
demonstrated by selecting \kp that did not rescatter ($N_C = 0$,
right panel of Fig.~\ref{au15-a2-c-au}). It is important to note
that rescattering ($N_C > 2$) alone creates the same or an even
stronger forward-backward enhancement as can be seen from the left
hand side of Fig.~\ref{au15-a2-c-au}. If we combine both effects
by switching on the \kp nucleus potential for $N_C> 2$ the enhancement does
not change as shown in the center panel.

\begin{figure}[hbt]
\epsfig{file=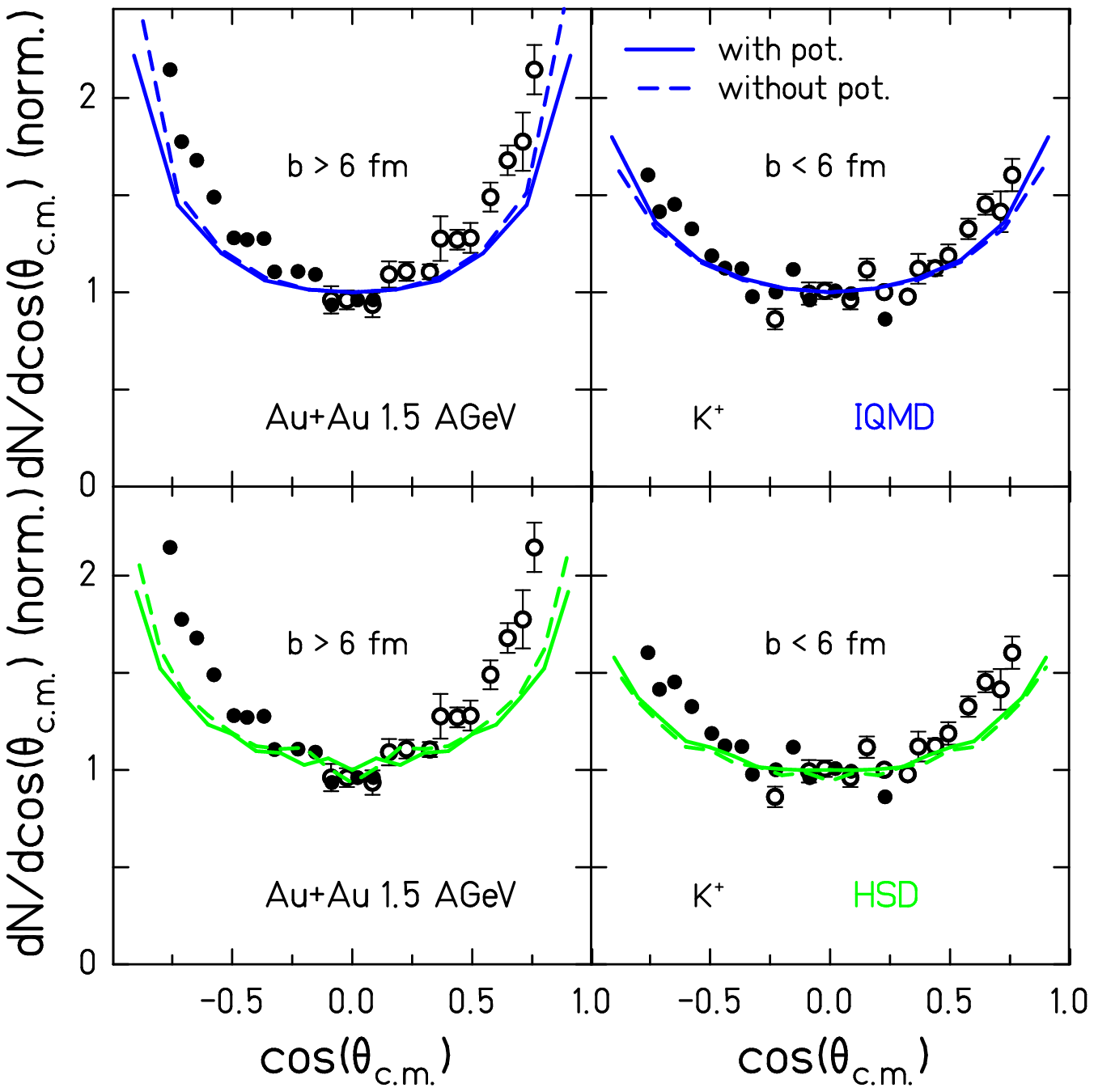,width=0.65\textwidth}
\epsfig{file=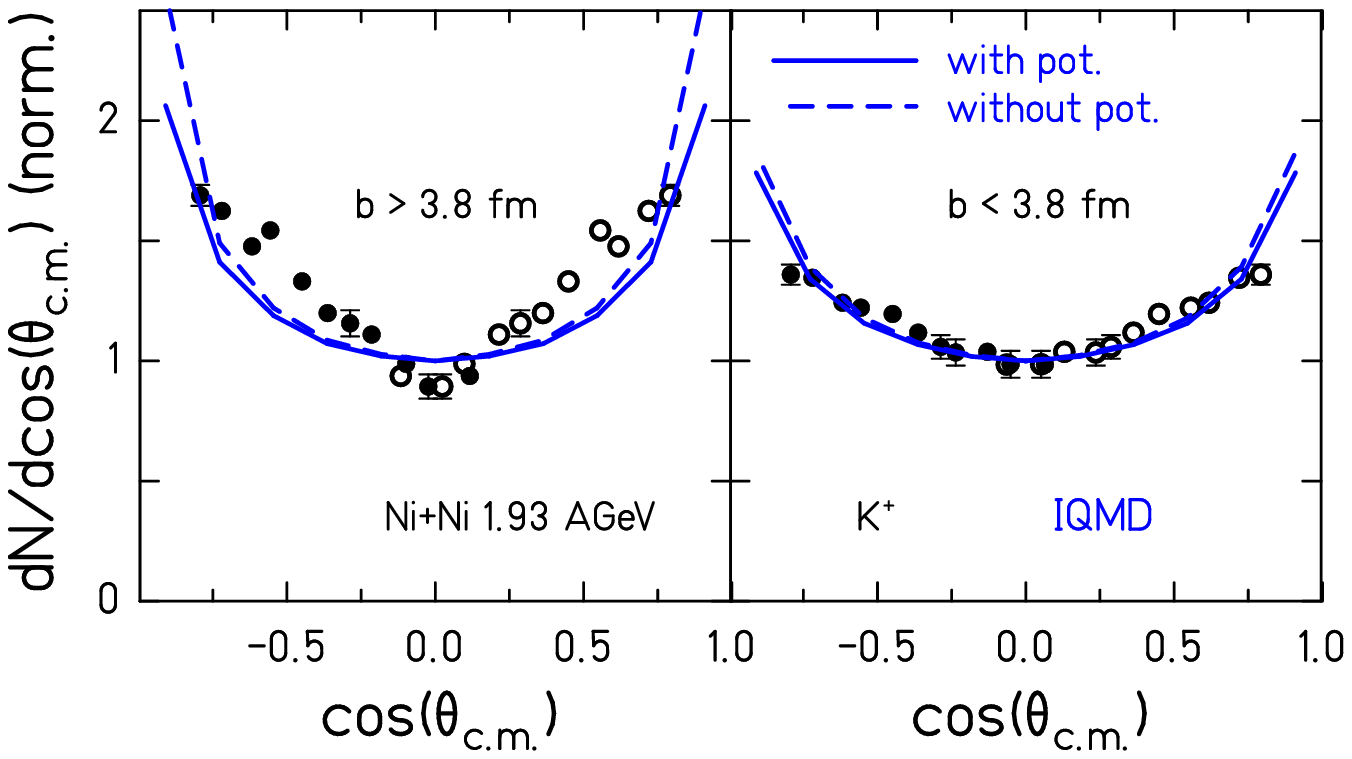,width=0.65\textwidth} \caption{The
measured polar angle distribution for Au+Au reaction at 1.5 $A$
GeV (upper part) and for Ni+Ni reactions at 1.93 $A$ GeV (lower
part) for two impact parameter ranges are compared to calculation
using IQMD and HSD \cite{Mishra:2004te} (only for Au+Au). The data are
from the KaoS \cite{Forster:2007qk}. The solid lines refer to
calculations with and the dashed ones to those without \kp nucleus
potential. The distributions are normalized to 1 for $\cos
\theta_{\rm c.m.}=0$.} \label{au15-a2-comp}
\end{figure}

Figure~\ref{au15-a2-comp} compares the model predictions with data
for central and peripheral impact parameters for Au+Au at 1.5
\AGeV and for Ni+Ni at 1.93 $A$ GeV. In theory as well as in
experiment \cite{Forster:2003vc} peripheral collisions yield a
stronger asymmetry due to collision of the \kp with less stopped
(spectator) target or projectile baryons. For central collisions
theory and experiment agree. For the peripheral collisions the
experimental distribution is slightly more forward/backward peaked
than theory predicts.

In conclusion, the anisotropy of the polar distribution is a
consequence of the interaction of the \kp with nucleons. At
production the distribution is rather flat. If one switches off
the \kp nucleus potential the anisotropy does not change considerably but
for heavier systems the \kp nucleus potential alone creates a similar
asymmetry. Because the effects of rescattering and \kp nucleus potential do not add
linearly polar distributions are of little help to disentangle
both.

\subsection{Azimuthal distribution}

The azimuthal distribution of the emitted particles is often
parameterized in a Fourier series \be \frac{{\rm d}N(y)}{{\rm
d}\phi} \, = \, C \, [ 1 + 2 v_1(y)\cos(\phi) + 2
v_2(y)\cos(2\phi)+ ...], \label{fourier} \ee with $y$ being the
rapidity and $v_1$ and $v_2$ the parameters to be determined.
\subsubsection{In-plane flow $p_x(y)$ and $v_1$}
\begin{figure}[htb]
\epsfig{file=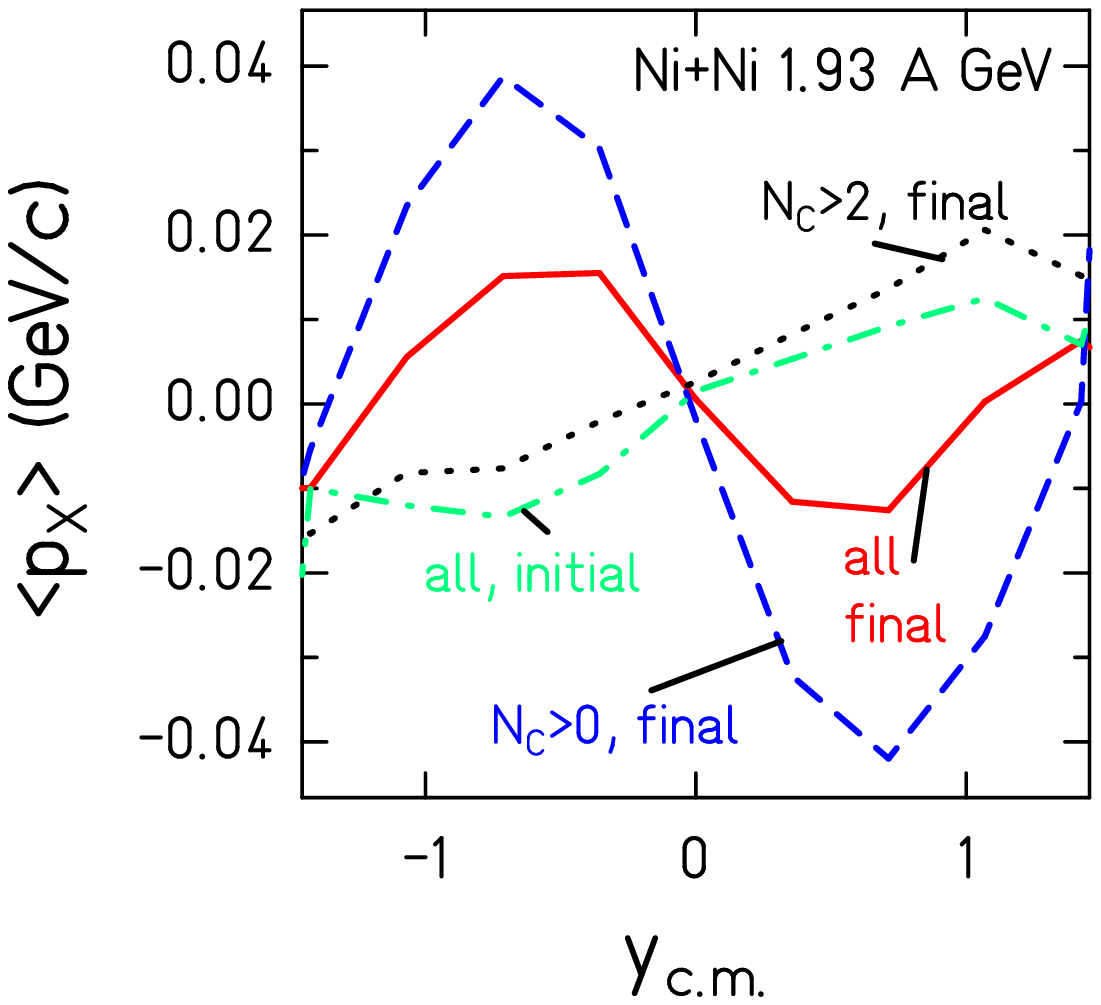,width=0.4\textwidth}
\caption{$<p_x>$ of \kp  as a function of the rapidity $y_{\rm
c.m.}$. We compare the distribution at creation with the final
distribution, separated according to the number of collisions. All
calculations are performed with \kp nucleus potential included.}
\label{kp-v1-ini-fin}
\end{figure}

The \kp are created according to the three-body phase space. We
therefore expect that their in-plane $v_1(y)=<p_x/p_t>$ or $<p_x(y)>$
distribution is that of the sources, i.e.~that of
the producing baryons. In the calculations this is not the case.
Directly after creation we see a much smaller $<p_x(y)>$
of the \kp as compared to the sources. This is shown in
\figref{kp-v1-ini-fin} and is a consequence of the large rapidity
shift of the \kp with respect to the source rapidity (see
\figref{dndyt}). Therefore, at a given \kp rapidity bin, the
positive and negative $v_1(y)$ values from the different source
rapidity bins cancel each other almost. The repulsive \kp nucleus potential
accelerates the \kp to the side opposite to that of
the projectile/target remnant and $<p_x(y)>$ changes its sign.
Rescattering collisions have the opposite effect by aligning the
\kp to the flow of the nucleons. The final $<p_x(y)>$ distribution
is hence a combination of both effects and is sensitive to the
number of rescattering collisions as well as to the \kp nucleus potential.
\begin{figure}[hbt]
\epsfig{file=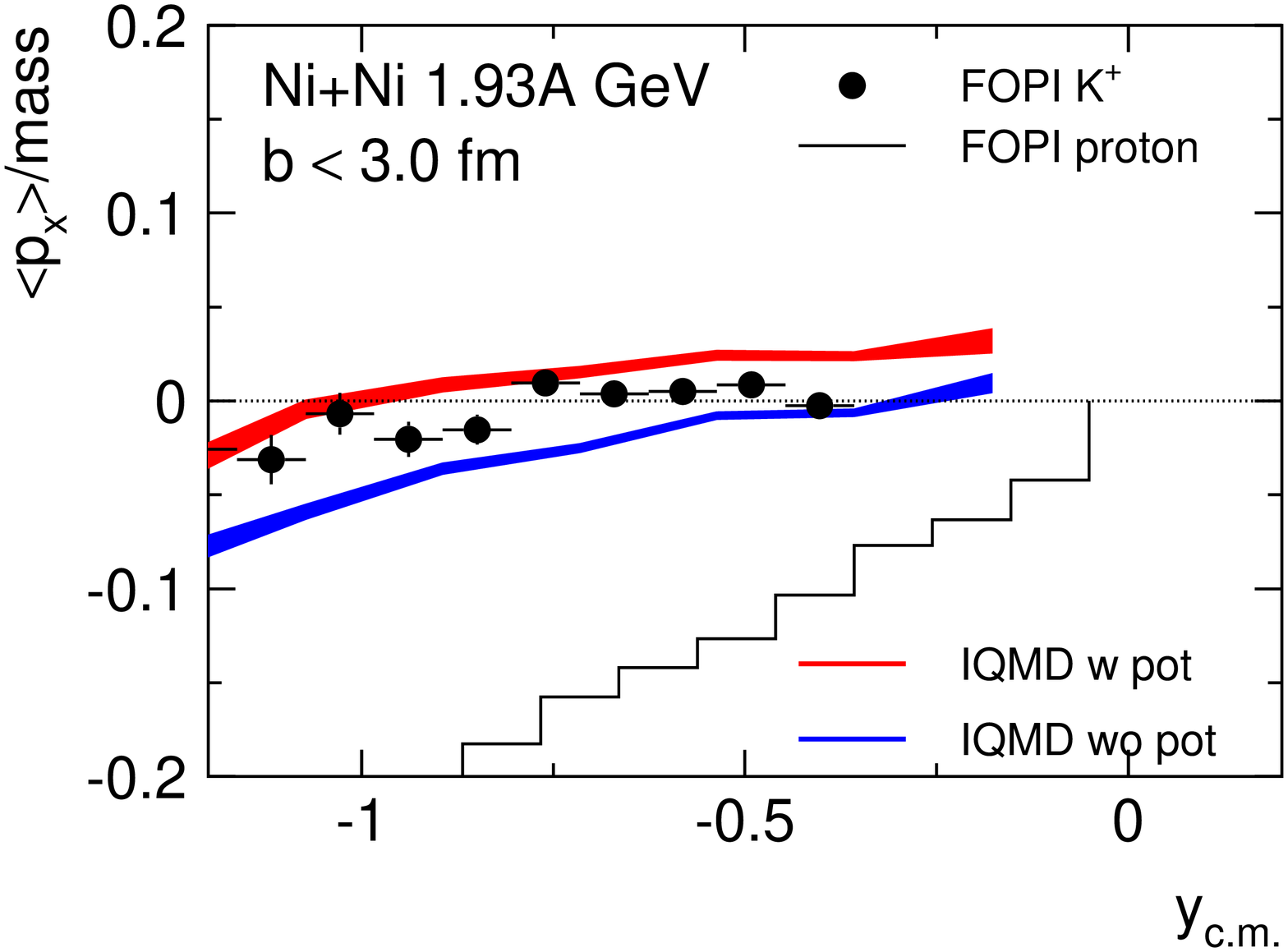,width=0.4\textwidth}
\epsfig{file=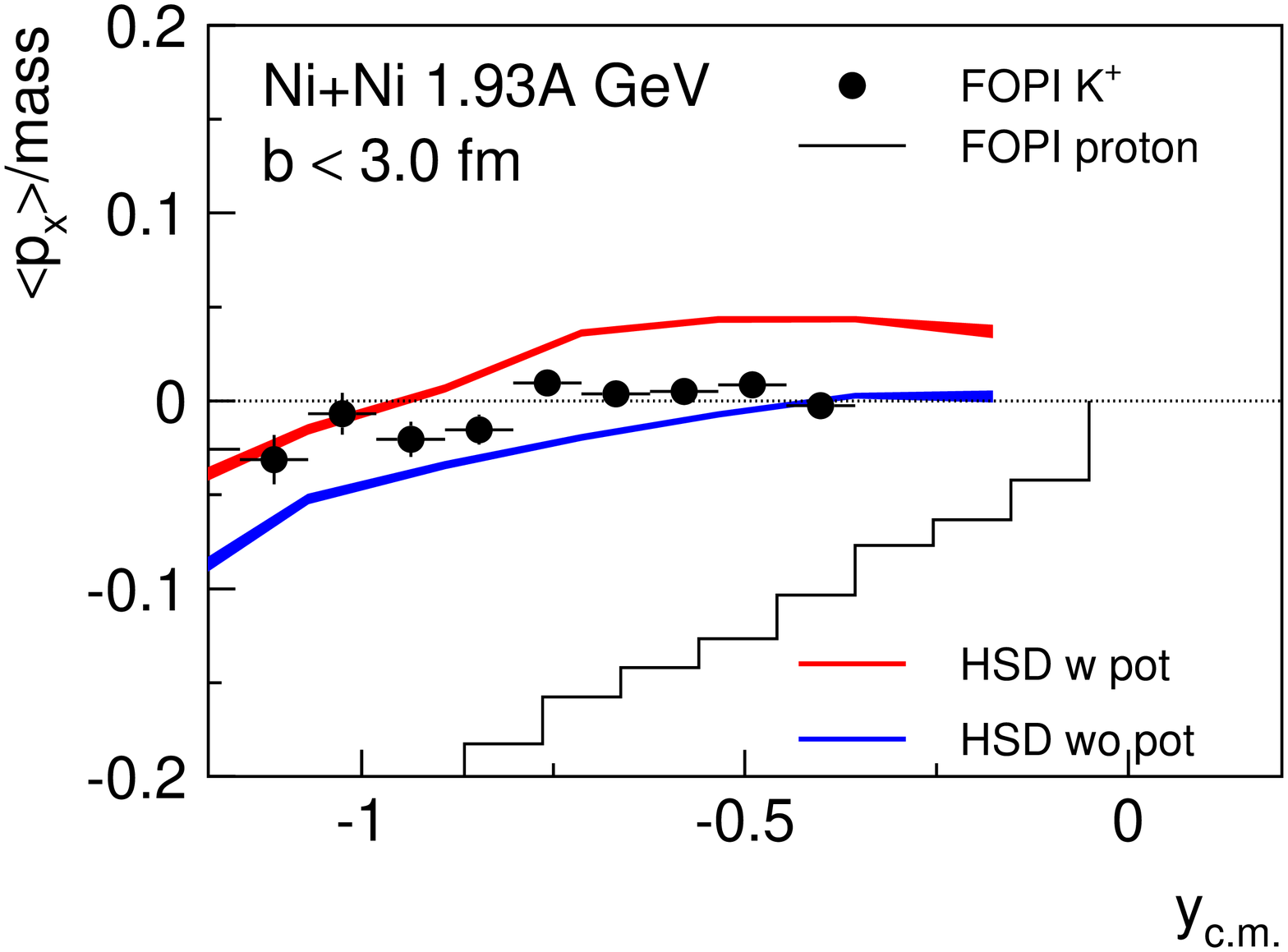,width=0.4\textwidth}
\caption{Experimental data for $<p_x>$ as a function of rapidity in central
Ni+Ni collisions in comparison to IQMD (left) and HSD (right) calculations. Data points are for \kp,
full black histogram for protons. Both distributions were generated with a
transverse momentum cut of 0.5 $p_t/m$. The data are from the
FOPI collaboration \cite{Workshop2000}.} \label{v11-fopi}
\end{figure}

The experimental distributions of the directed transverse
momentum $<p_x>$ as a function of the center of mass rapidity
$y_{\rm c.m.}$ from the FOPI collaboration for central Ni+Ni
collisions are shown in Fig.~\ref{v11-fopi} for \kp mesons in
comparison to protons \cite{Workshop2000}. The data points tend to
an almost vanishing flow which is
reproduced by the IQMD as well as by the HSD model when employing a
\kp nucleus potential. The sensitivity of this particular
observable to the \kp nucleus potential is, however, weak. The situation is different
if studying the $p_t$ differential distribution.

\begin{figure}[htb]
\epsfig{file=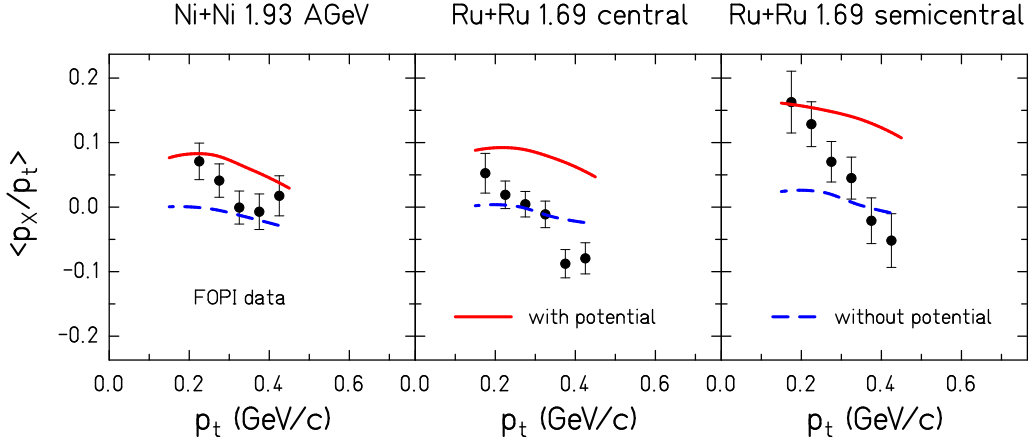,width=0.9\textwidth}
\caption{$v_1= <\frac{p_x}{p_t}>$ of the \kp as a function of the transverse
momentum, $p_t$, for Ni+Ni (left) and Ru+Ru (middle and right).
The experimental data of the FOPI
Collaboration~\cite{Crochet:2000fz} are compared with IQMD
calculations.} \label{kp-v1pt}
\end{figure}

\begin{figure}[htb]
\epsfig{file=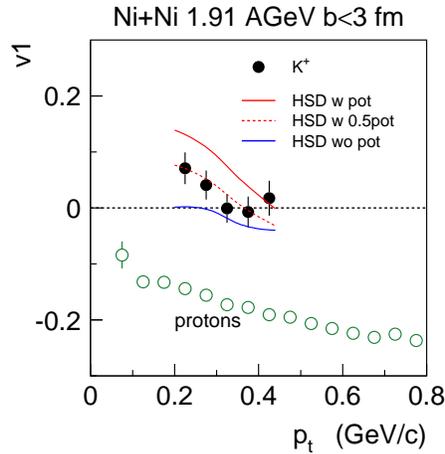,width=0.35\textwidth}
\caption{$v_1$ of the \kp as a function of the transverse
momentum, $p_t$, for Ni+Ni. The experimental data of the FOPI
Collaboration~\cite{Crochet:2000fz} are compared with HSD
calculations.} \label{kp-v1pthsd}
\end{figure}

The FOPI Collaboration has measured $v_1$ as a function of the
transverse momentum $p_t$ of the \kp mesons at target rapidities
for two systems ~\cite{Crochet:2000fz}. In addition, the
impact-parameter dependence was studied in  the intermediate heavy
system Ru+Ru. The overall magnitude of the experimental data is
quite well described by the IQMD model (Fig.~\ref{kp-v1pt}). The
strong variation of the $v_1$ as a function of transverse momentum
$p_{\rm t}$ is not reproduced by the calculations. The strength
\kp nucleus potential changes the absolute value but not the form
of the distribution, neither do different choices of the
rescattering cross section. Within the IQMD model the difference
between theory and experiment cannot be related to either
potential or rescattering and is not understood yet. Employing the
HSD model and using a similar approach for treating the in-medium
properties of \kp mesons, the strong $p_t$ dependence of $v_1$ is
described using a repulsive \kp nucleus potential of
$\approx$~20~MeV (half of the standard value of 40 MeV) see
Fig.~\ref{kp-v1pthsd}. The data are from
Ref.~\cite{Crochet:2000fz}, note that in that paper BUU
calculation have been shown.

\subsubsection{Out-of-plane flow $v_2$}

The out-of-plane distribution is best studied at mid-rapidity, when
in symmetric systems the contribution of the in-plane flow $v_1$
is zero by definition and therefore the problem to separate $v_1$
from $v_2$ is not given.

Geometry is an important factor for the \kp
out-of-plane distribution, as demonstrated in \figref{au15-phi-det}.
The \kp are produced in the participant
zone. If they have to pass spectator matter (lying by definition
in-plane)  they suffer in the average from more collisions than those emitted
under $90^\circ$ with respect to the reaction plane.
Therefore, by selecting \kp with $N_C$ = 0 preference is given to
those emitted perpendicular to the reaction plane (dotted line in
Fig.~\ref{au15-phi-det}, left). The \kp nucleus potential (dashed line)
enhances the out-of-plane flow.  This class represents only 17.5\%
of all \kp. For $N_C > 2$ the initial distribution is slightly
in-plane enhanced. By rescattering the positive $v_2$ value of
the baryons is transferred to the K$^+$. By this mechanism
rescattering contributes to an increase of the out-of-plane
enhancement.

In \Figref{au15-phi-c}, left, we display the influence of
rescattering and of the \kp nucleus potential on the azimuthal distribution
${\rm d}N/{\rm d}\phi$ at mid-rapidity for Au+Au collisions at 1.5
\AGeV (left) and for Ni+Ni collisions at 1.93 \AGeV (right) (data
from ~\cite{Uhlig:2004ue}). Without \kp nucleus potential  and
rescattering (sources, purple dash-dotted line) the distribution
is quite flat. Rescattering without the \kp nucleus potential (no pot.,
resc., green dotted line) causes an enhancement of the out-of-plane emission ($v_2=
-0.092$). If we switch on the \kp nucleus potential but ignore the
rescattering (pot., no resc., blue dashed line) the enhancement is
of a similar strength ($v_2= -0.117$). Combining the two
contributions (pot., resc., \rfl) results in an enhancement of
$v_2$ by about only 30\% ($v_2 = -0.1408$). Both, rescattering and
\kp nucleus potential cause an asymmetry, yet combining both sources
results only in an hardly visible stronger asymmetry, similar to what has been observed
for the polar distribution.
\begin{figure}[hbt]
\epsfig{file=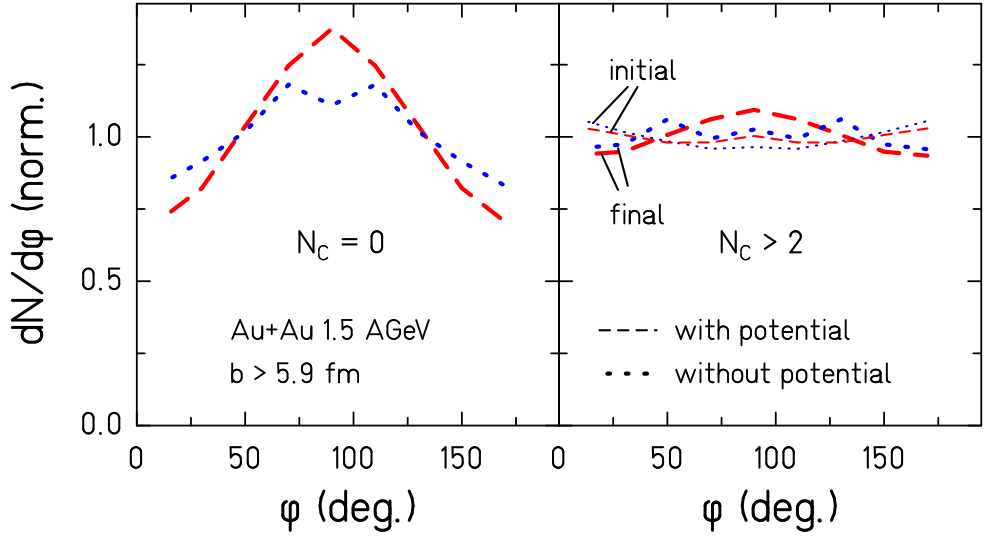,width=0.9\textwidth}
\caption{Initial and final out-of-plane distribution of the \kp selected according to
 the number of rescattering collisions, $N_C$ after production. For $N_C = 0$ and no
 potential the initial and final distributions are identical and
 similar to the initial with potential.}
\Label{au15-phi-det}
\end{figure}
\begin{figure}[hbt]
\epsfig{file=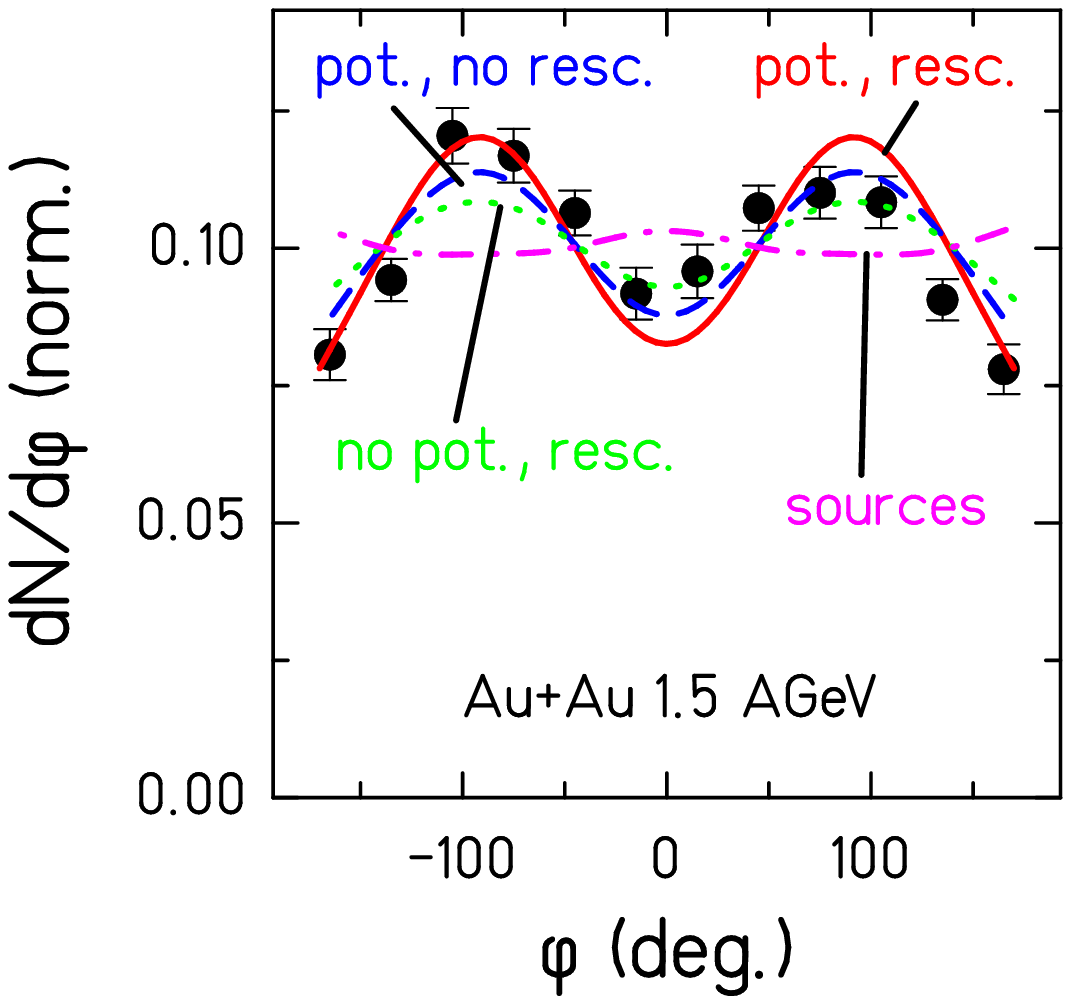,width=0.48\textwidth}
\epsfig{file=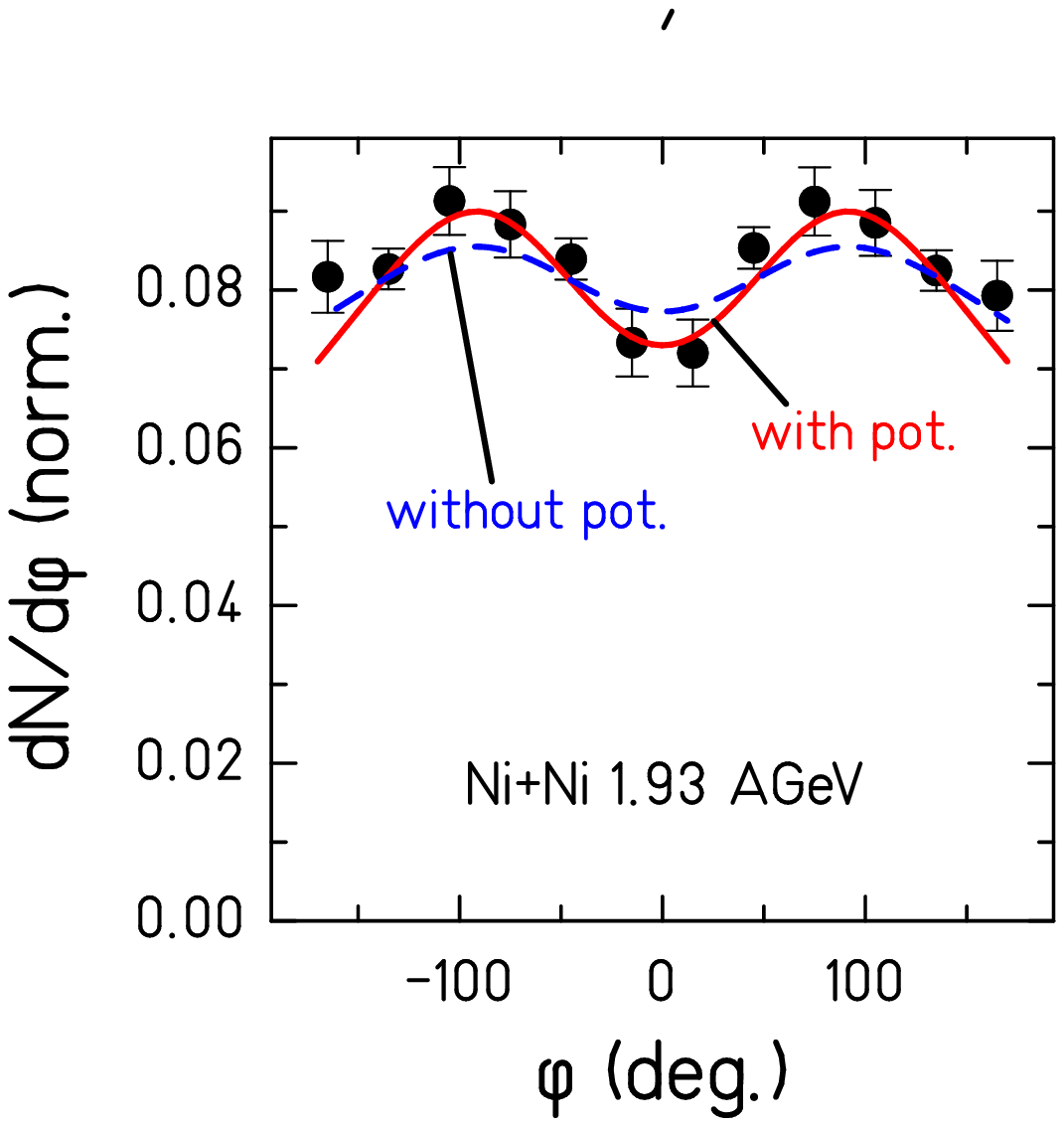,width=0.48\textwidth} \caption{The
influence of the \kp nucleus potential and the KN rescattering on the
azimuthal distribution of the \kp in the experimental acceptance region for
Au+Au collisions at 1.5 \AGeV with impact parameter $b >$ 7.8 fm
(left) and for Ni+Ni at 1.93 \AGeV with impact parameter $b >$ 3.8
fm (right). Data from \cite{Uhlig:2004ue}.} \Label{au15-phi-c}
\end{figure}
\begin{figure}[hbt]
\epsfig{file=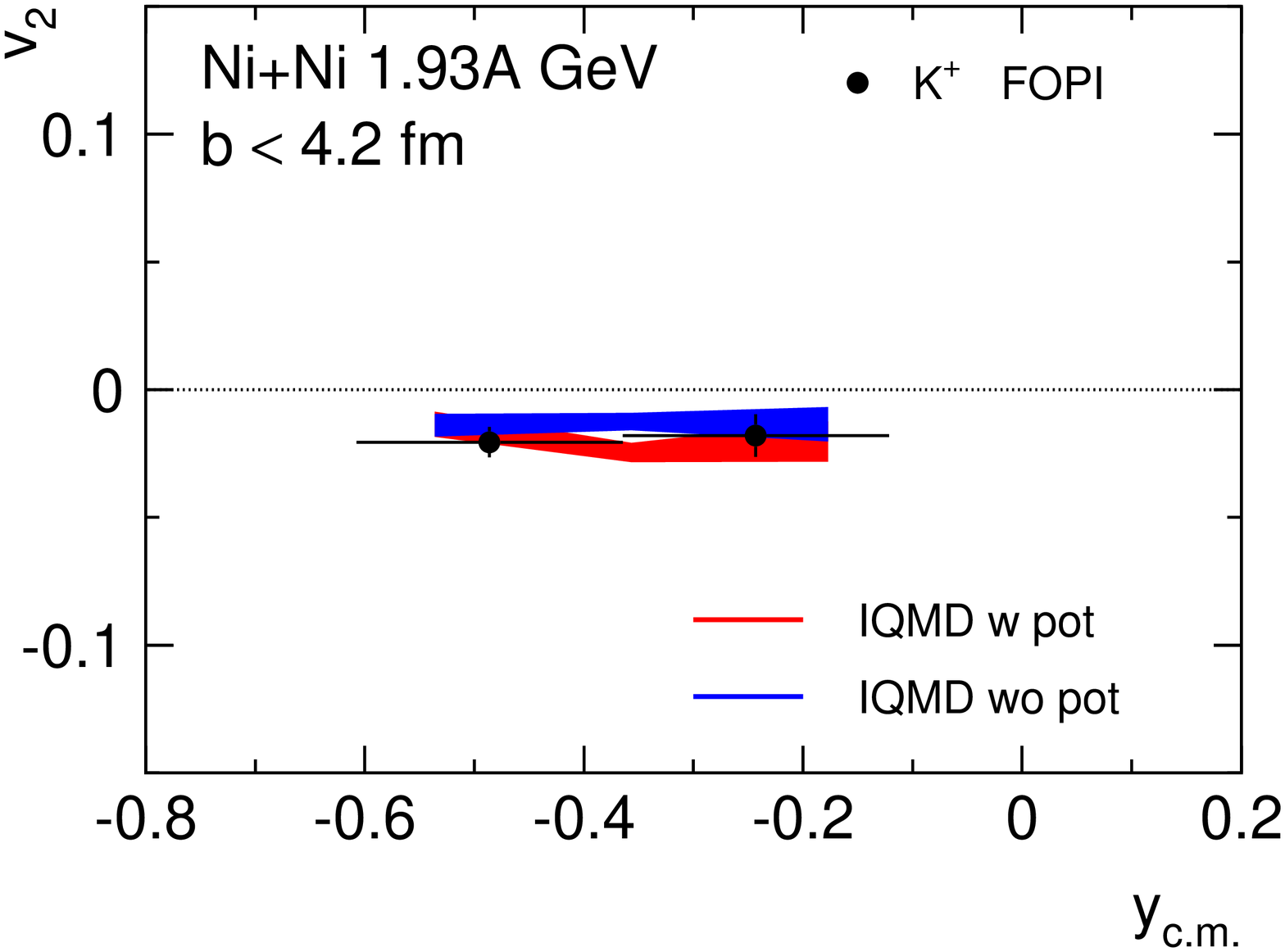,width=0.4\textwidth}
\epsfig{file=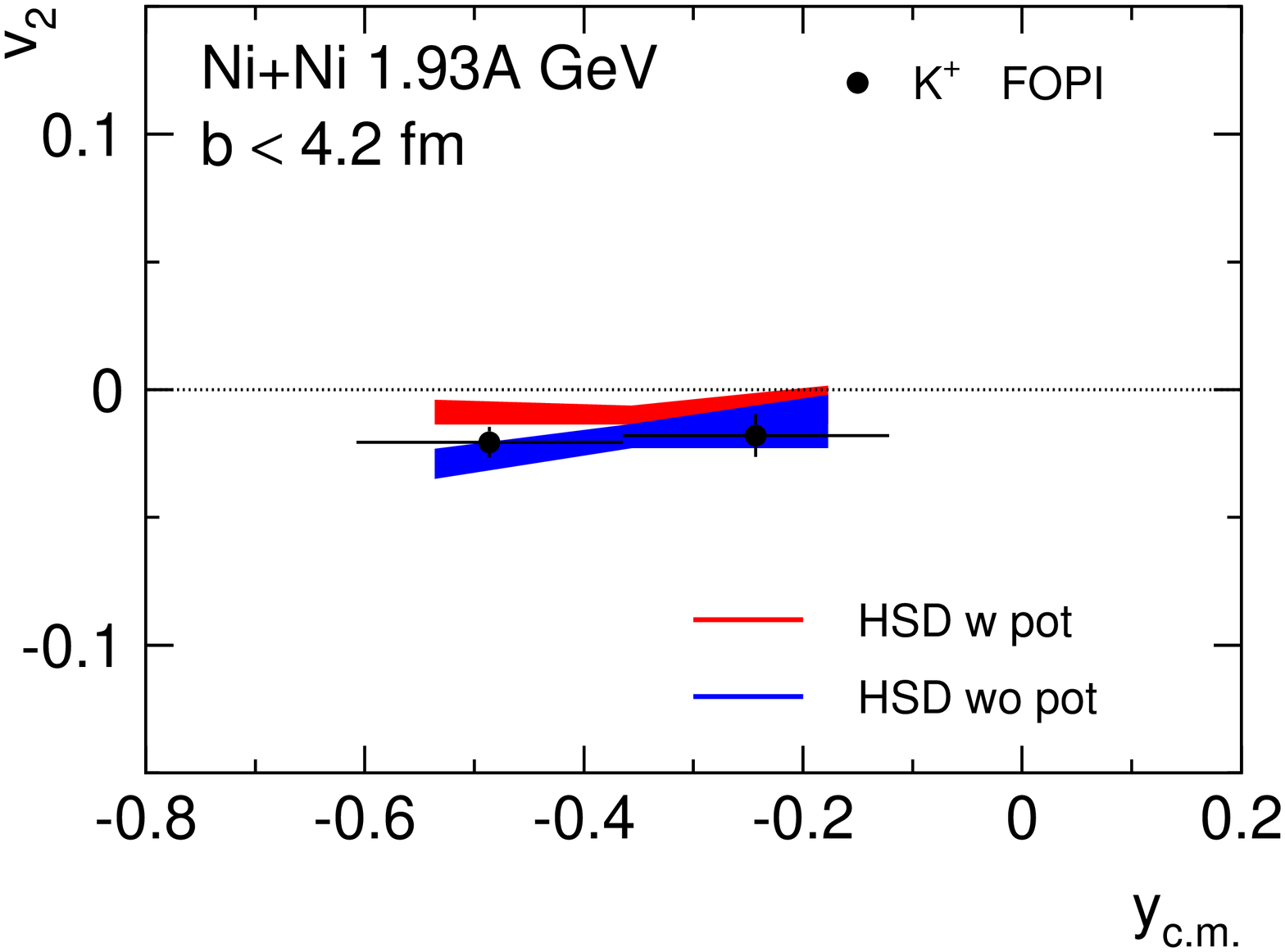,width=0.4\textwidth}
\caption{Comparison of $v_2$ of \kp as a function of the rapidity between
preliminary FOPI data and IQMD (left)  as well as  HSD  (right) calculations.} \label{v2-fopi}
\end{figure}

For central Ni+Ni collisions at 1.93 \AGeV the FOPI Collaboration
has measured $v_2$ as a function of the rapidity and compared with
IQMD calculations (\figref{v2-fopi}). This distribution is rather
flat and the sensitivity on the \kp nucleus potential which we have seen
for the large Au system (Fig.~\ref{au15-phi-c}), has disappeared. IQMD as well as HSD calculations
reproduce  the experimental findings.

The excitation function of  $v_2$ around mid-rapidity for different
options and for the Au+Au system is shown in \Figref{au15-phi-a}.
Without rescattering and without \kp nucleus potential $v_2$ is around zero.
At all energies the azimuthal anisotropy of the \kp is not created
in their production process but by rescattering and by
the \kp nucleus potential. A significant out-of-plane enhancement is
already caused by rescattering. The \kp nucleus potential
alone creates as well a significant out-of-plane enhancement. The
combined effect of rescattering and of the \kp nucleus potential
interaction enhances, however, only slightly that created by the
\kp nucleus potential only. The $b$ = 7 fm calculations are compared with
experimental data
\cite{Shin:1998hm,Uhlig:2004ue,Ploskon:2005qr}. The two results at
1.5 \AGeV agree within error bars.
\begin{figure}[htb]
\epsfig{file=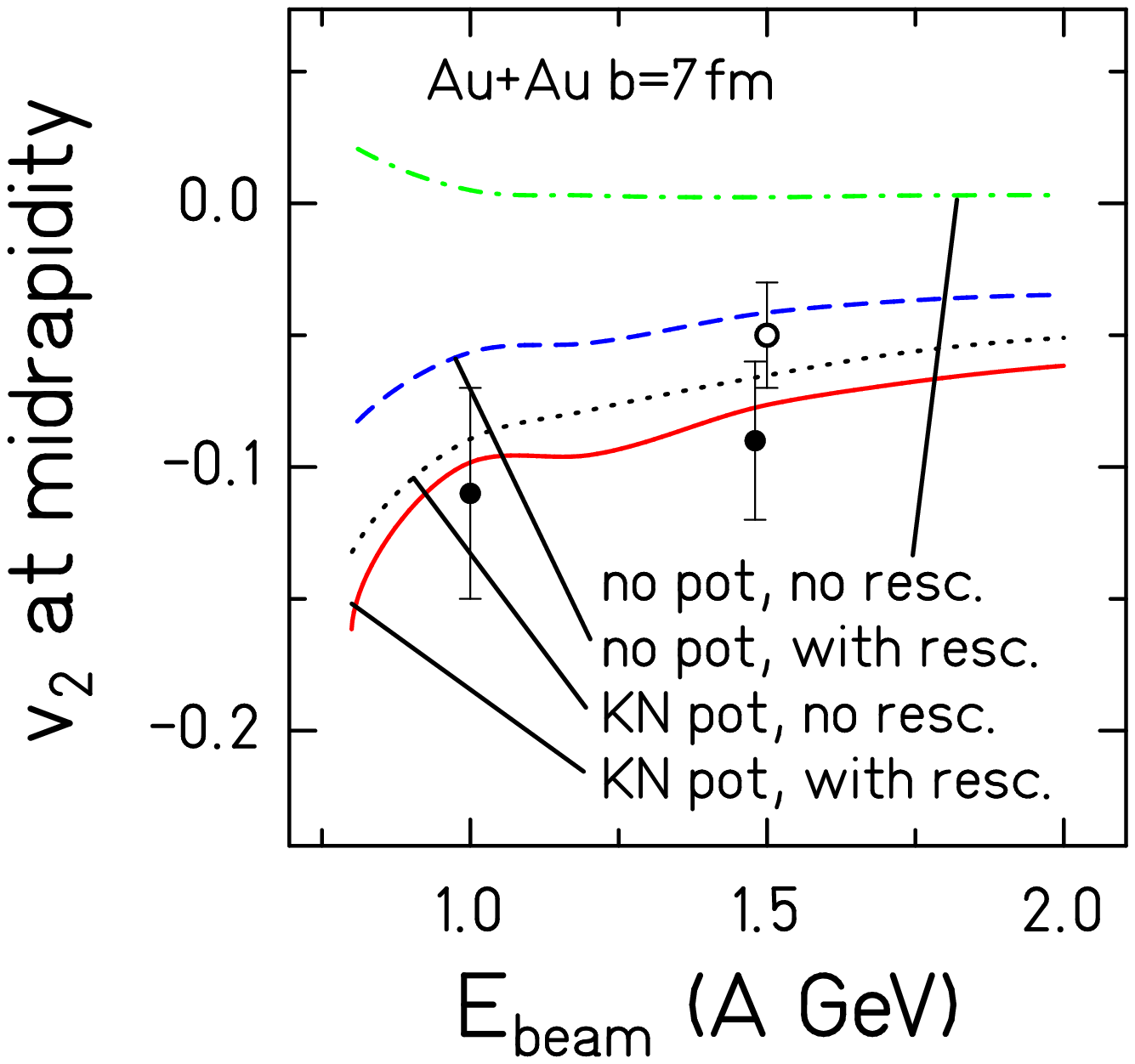,width=0.6\textwidth} \caption{Excitation
function of the $v_2$ coefficient at mid-rapidity for Au+Au reactions and with different options for the simulation. We
compare here IQMD calculations at $b$ = 7 fm with peripheral KaoS data
(points \cite{Shin:1998hm,Uhlig:2004ue}, open circles
\cite{Ploskon:2005qr})} \Label{au15-phi-a}
\end{figure}

The origin of the $v_2$, created by the \kp nucleus potential, is
visible in Fig.~\ref{au15-v2-pt} where we study $v_2$ as a
function of the last contact time (equal to the production time if
the \kp do not rescatter, otherwise equal to the time of the last
collision). The early produced \kp which do not rescatter ($N_c =
0$) have initially a moderate negative $v_2$ (dashed dotted line)
but finally a very strong $v_2$ (dashed line). The difference
originates from the potential interaction. Thus early created \kp
show a squeeze out perpendicular to the reaction plane created by
the presence of the (in-plane) spectator matter. Rescattering
weakens this squeeze by randomizing the \kp direction (full line).
Late emitted \kp show an opposite trend because they cannot
interact with the fast moving spectator matter any more. They
follow the $v_2$ of the sources. In contradistinction to $v_1$
(Fig.~\ref{kp-v1pt}) the value of $v_2$ depends (in theory as well
as in experiment) only slightly on the momentum of the \kp as seen
in \Figref{au15-v2-pt}, right~\cite{Ploskon:2005qr}. As already
seen for the integrated value (\figref{au15-phi-a}) IQMD
calculations with \kp nucleus potential give a larger negative
$v_2$ value for this reaction. The present error bars are too
large to allow for conclusions on the \kp nucleus potential from
the excitation function of $v_2$ at mid-rapidity and hence also
from the momentum dependence.

\begin{figure}[htb]
\epsfig{file=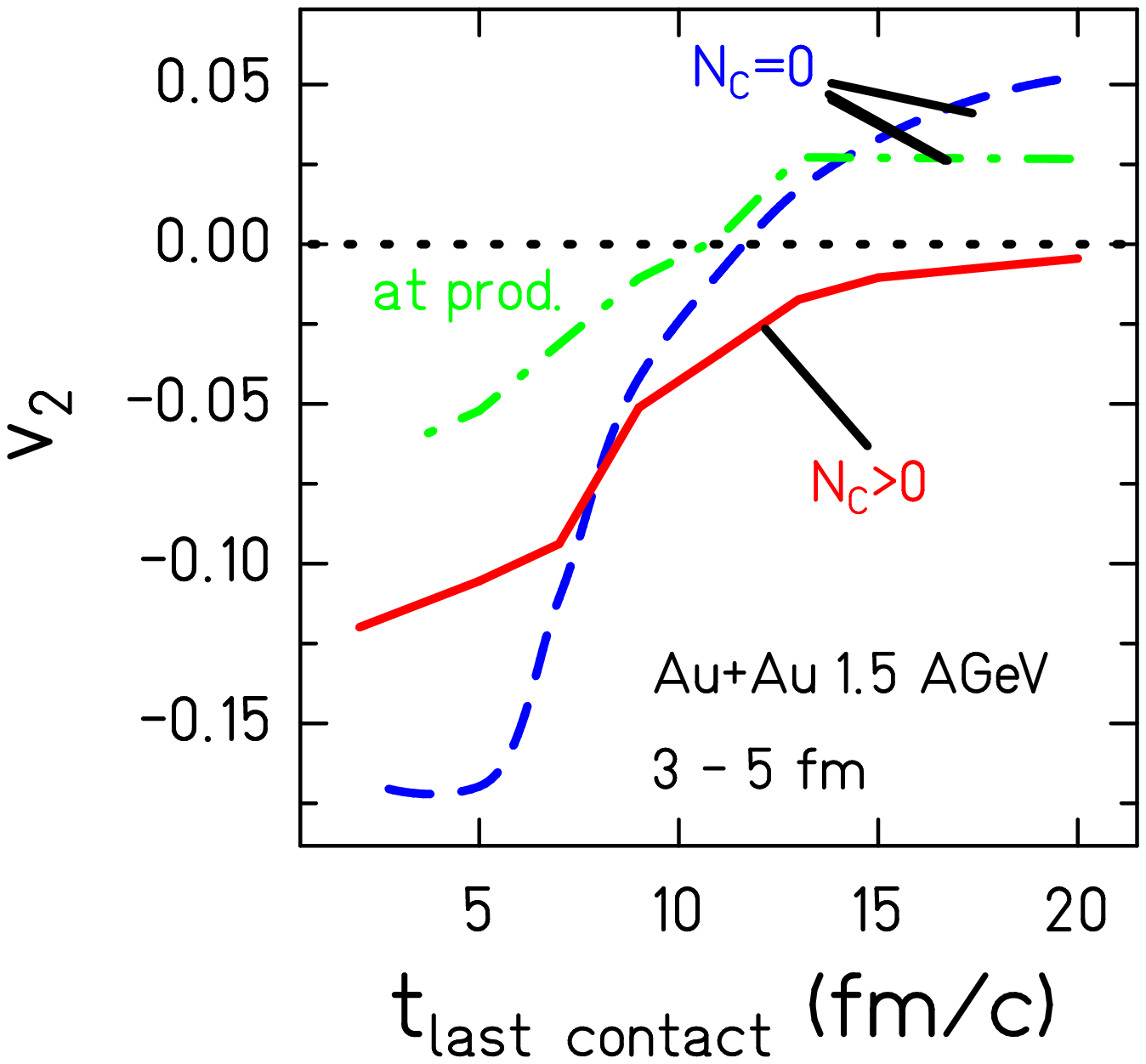,width=0.45\textwidth}
\epsfig{file=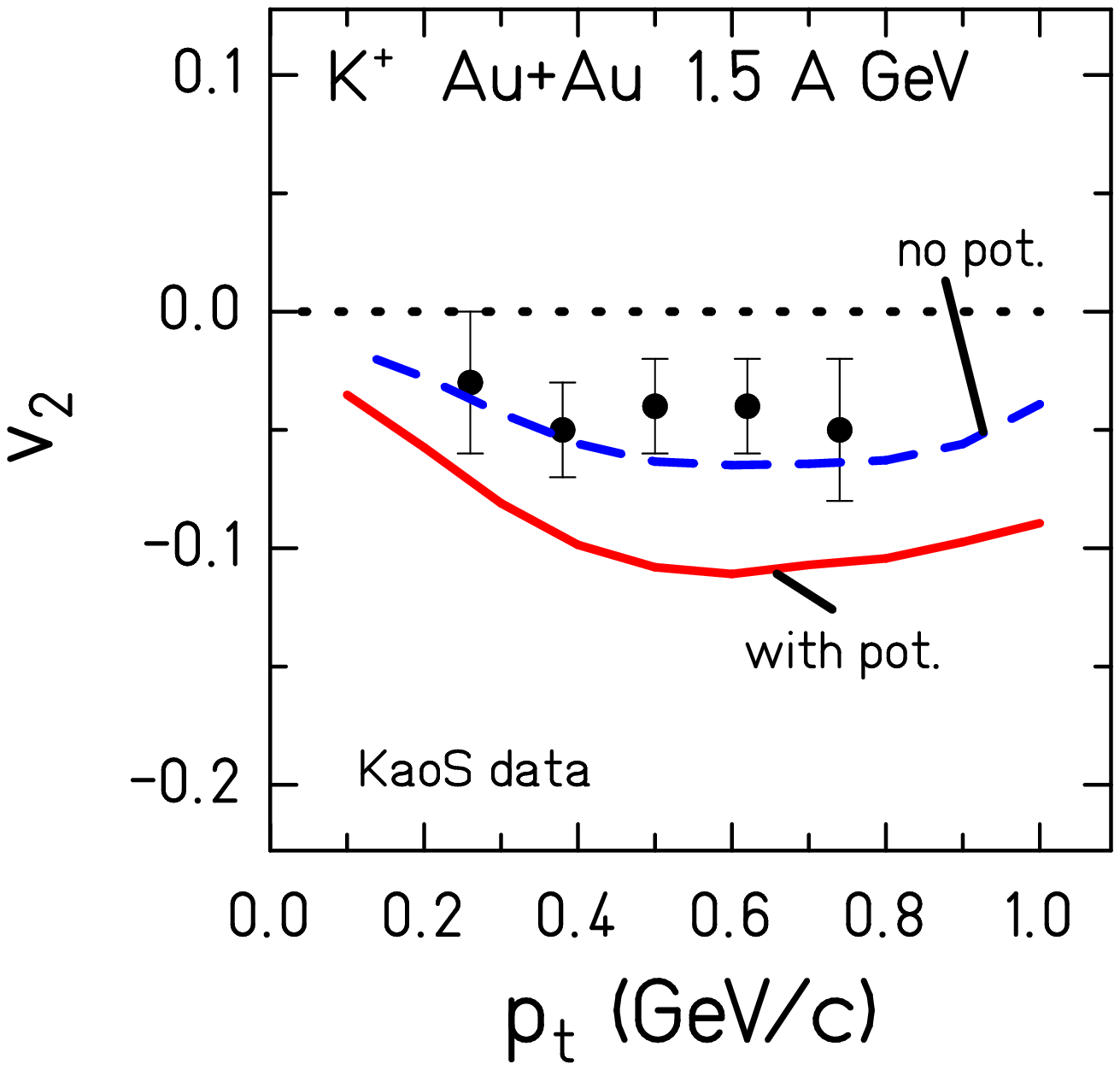,width=0.45\textwidth}
\caption{Left: $v_2$ as a function of the time when the \kp had
its last contact with the system for Au+Au at 1.5 \AGeV. Dashed
dotted line: $v_2$ of \kp which do not rescatter ($N_c=0$) at
production, dashed line $v_2$ of the same \kp at the end of the
reaction, full line: $v_2$ of \kp with $N_c>0$ at the time point
of their last collision. Right: Comparison of the theoretical and
experimental \cite{Ploskon:2005qr} $v_2$ values as a function of
the \kp transverse momentum, $p_t$, for Au+Au reactions at 1.5
\AGeV.} \Label{au15-v2-pt}
\end{figure}

Thus we can conclude that the \kp mesons at creation have a flat
in-plane and out-of-plane distribution, i.e.~$v_1 = v_2 = 0$. The
experimentally observed finite values of these coefficients are
due to rescattering and due to the repulsive \kp nucleus potential which
add in a non linear way. This shows that \kp interact with the
medium after production but this implies as well that these
observables reflect the complex reaction dynamics. Therefore, to
use these data to gain additional information on either of them is
difficult and not possible without including other observables.
Besides the dependence of $v_1$ on $p_t$ the measured coefficients
are quantitatively reproduced by theory.

\subsection{$\Lambda$ production}

Being unstable, strange baryons are more difficult to detect.
Therefore there are much less data on $\Lambda$ and $\Sigma$
production than on strange meson production. Their multiplicity
corresponds to that of the \kp and K$^0$, the associated
particles, and is therefore of little importance. The
understanding of the dynamics of the $\Lambda$ in matter is
important for the understanding of the \km mesons which at these
energies are almost exclusively created by a strangeness-exchange
reaction between a strange baryon and a pion. The $\Lambda$N cross
section is larger than the \kp N cross section and therefore the
$\Lambda$ adopt rapidly the properties of their environment. It is
expected that their final distribution differs substantially from
that at production.

\begin{figure}[htb]
\epsfig{figure=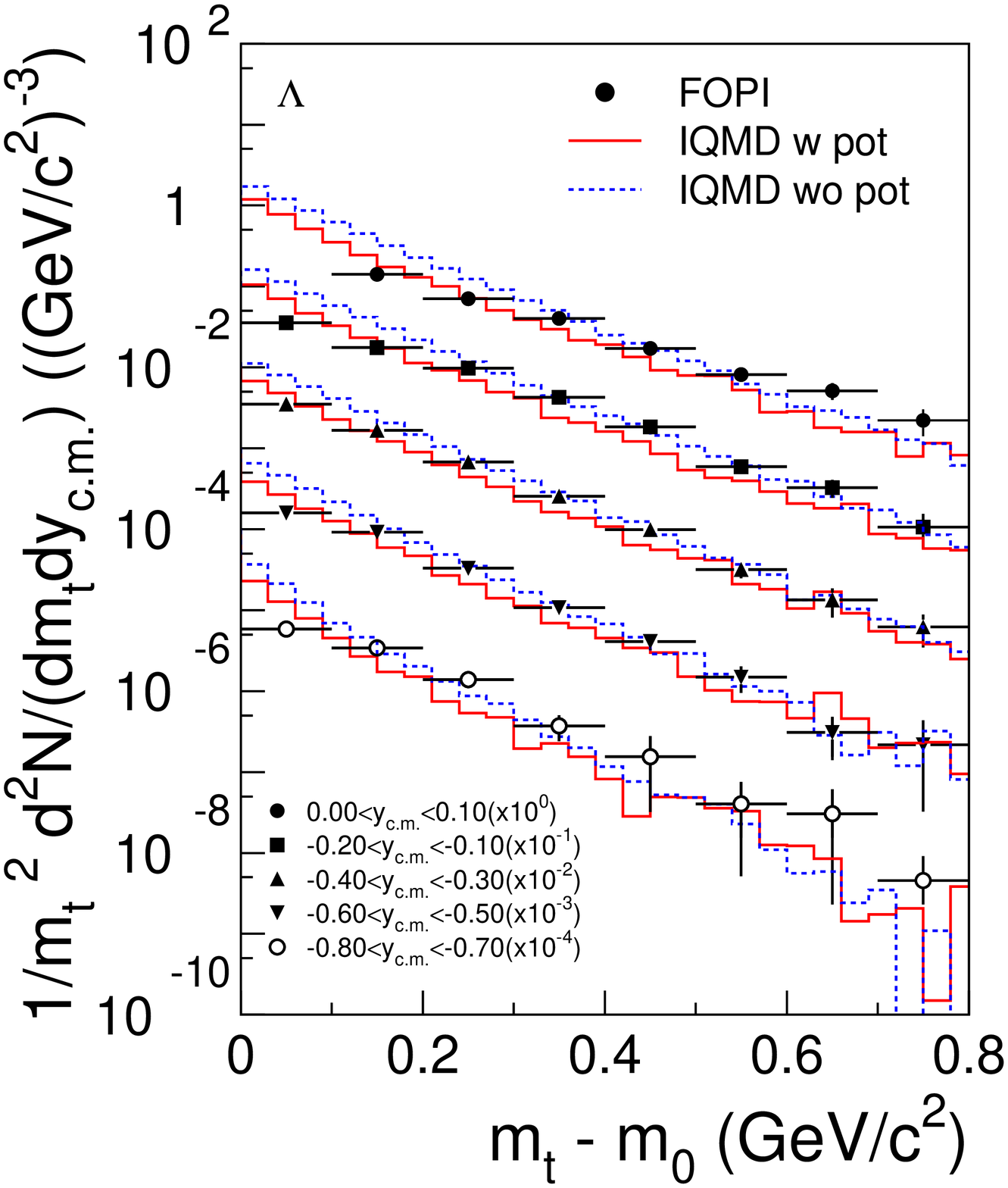,width=0.4\textwidth}
\epsfig{figure=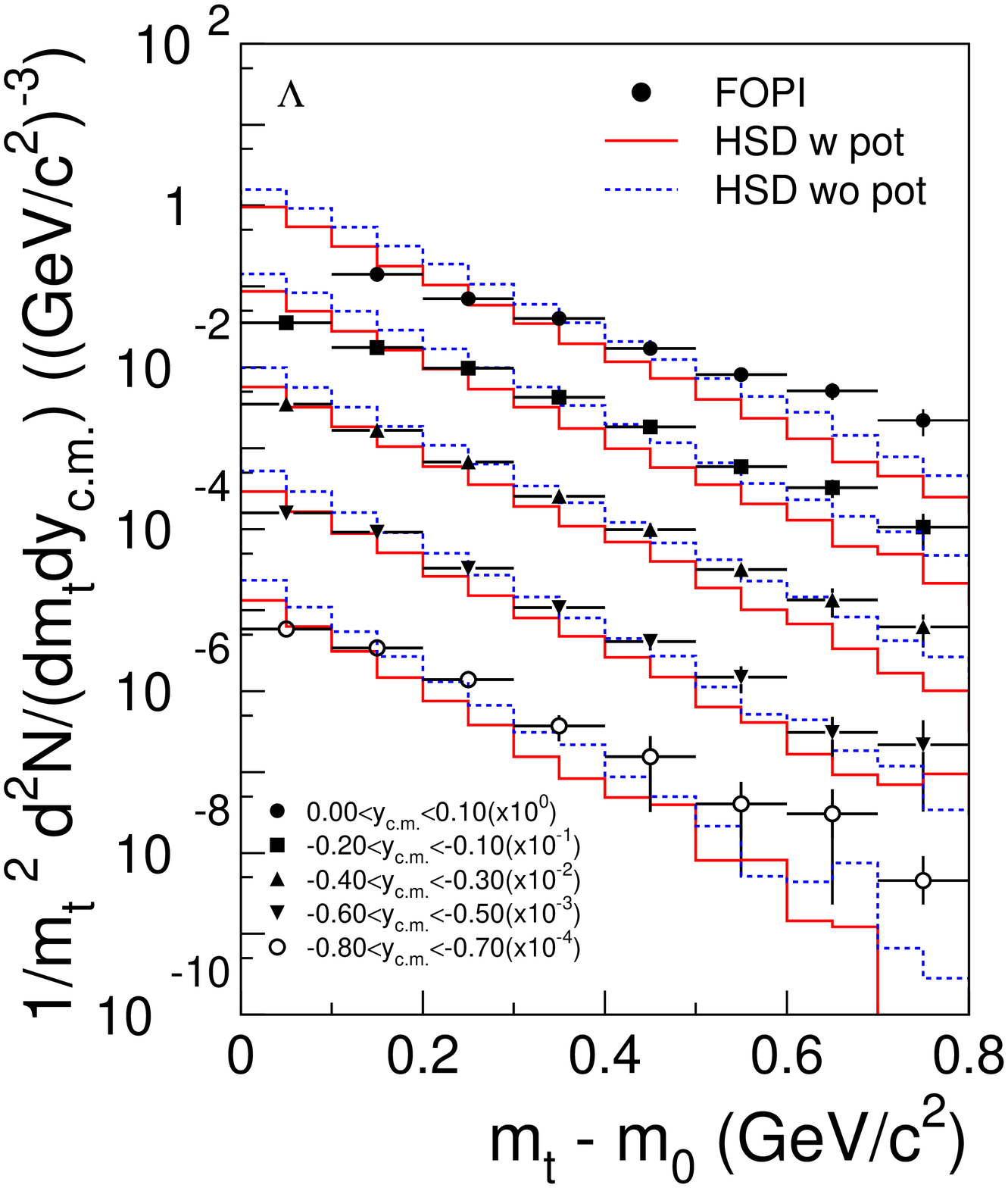,width=0.4\textwidth}
\caption{Central Ni+Ni collisions at 1.93 \AGeV: $m_t$ spectra of $\Lambda$ for
different rapidity bins. The experimental results are compared
with IQMD calculations (left) and with HSD calculations (right).}
\label{spectraLambda}
\end{figure}

\begin{figure}[htb]
\epsfig{file=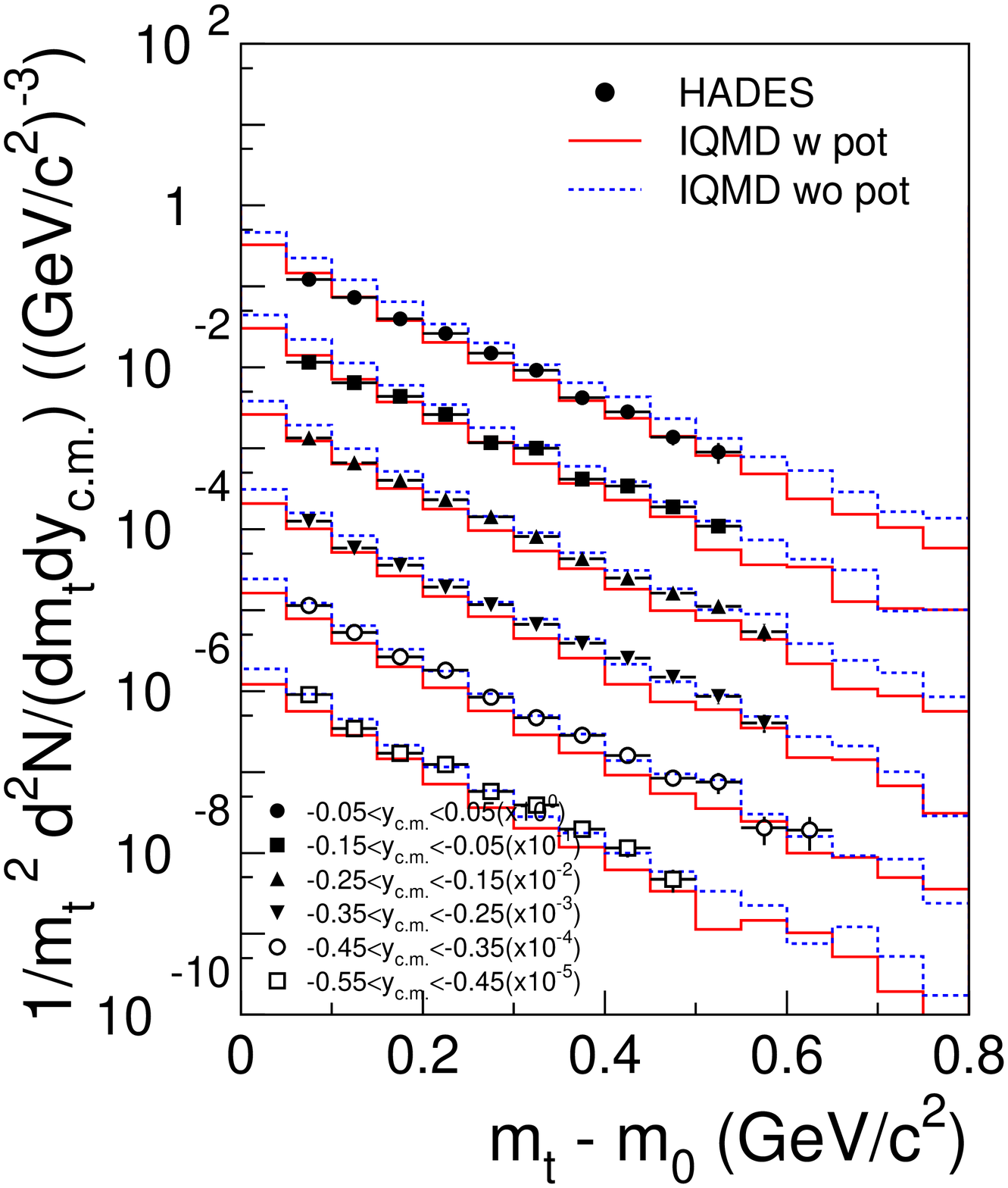,width=.45\textwidth}
\epsfig{file=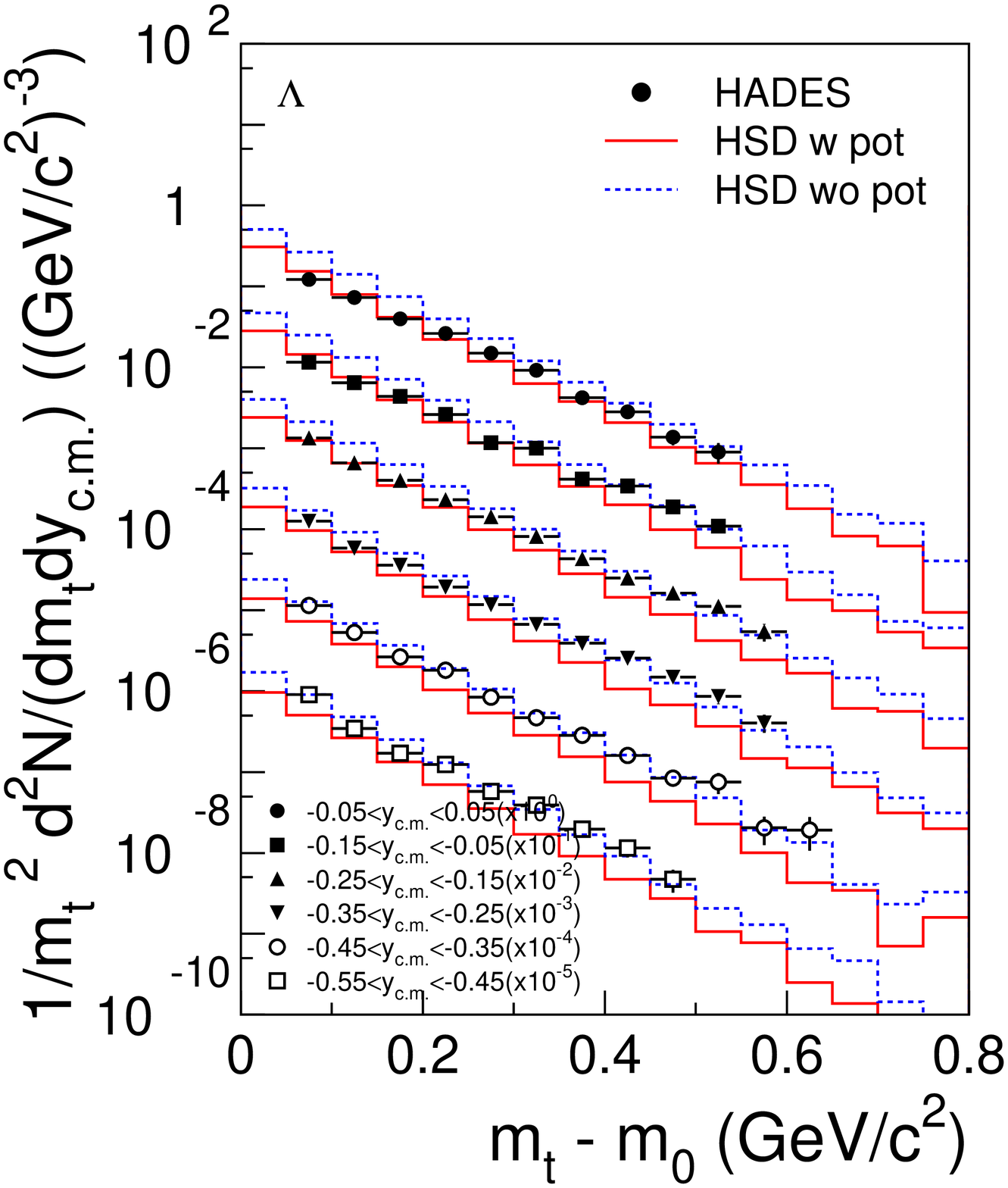,width=.45\textwidth}
\caption{$m_t$ spectra of $\Lambda$ at different rapidity bins
for the reaction Ar+KCl at 1.75 \AGeV. The experimental results of
the HADES Collaboration~\cite{Agakishiev:2010rs} are compared with
predictions of IQMD and HSD calculations with (red full line) and without
(blue dashed line) \kp nucleus potential.}
\label{Lahades}
\end{figure}

Figure ~\ref{spectraLambda} shows the experimental $m_t$ spectra
for different rapidity  bins for Ni+Ni at 1.93 \AGeV
\cite{Merschmeyer:2007zz} and Fig.~\ref{Lahades} the same
distribution for Ar+KCl at 1.75 \AGeV\cite{Agakishiev:2010rs}.
Both sets of spectra have an exponential form. The
experimental data are compared to IQMD (left) and
HSD (right) calculations, presented by the histograms.
The FOPI data are filtered using the FOPI filter, the HADES data have the same normalization
as the K$^0$ shown in Fig.~\ref{hadk0}. For the 1.93 \AGeV ($b \le
3.3 $ fm) data of the FOPI Collaboration we see at low-$m_t$
values a quantitative agreement between calculations and the data for
almost all rapidity bins. At high values of $m_t$ differences
between simulations and data appear, especially close to
mid-rapidity. These differences are more pronounced in HSD calculations
than in IQMD calculations. For the  Ar+KCl at 1.75 \AGeV the agreement
between theory and experiment is good even for large $m_t.$ values.

The influence of the \kp nucleus potential is visible. It
increases the threshold and reduces therefore the number of
$\Lambda$ which are created together with the K$^{+/0}$ in BB$\to$
YNK$^{+/0}$ reactions.

\begin{figure}[htb]
\epsfig{figure=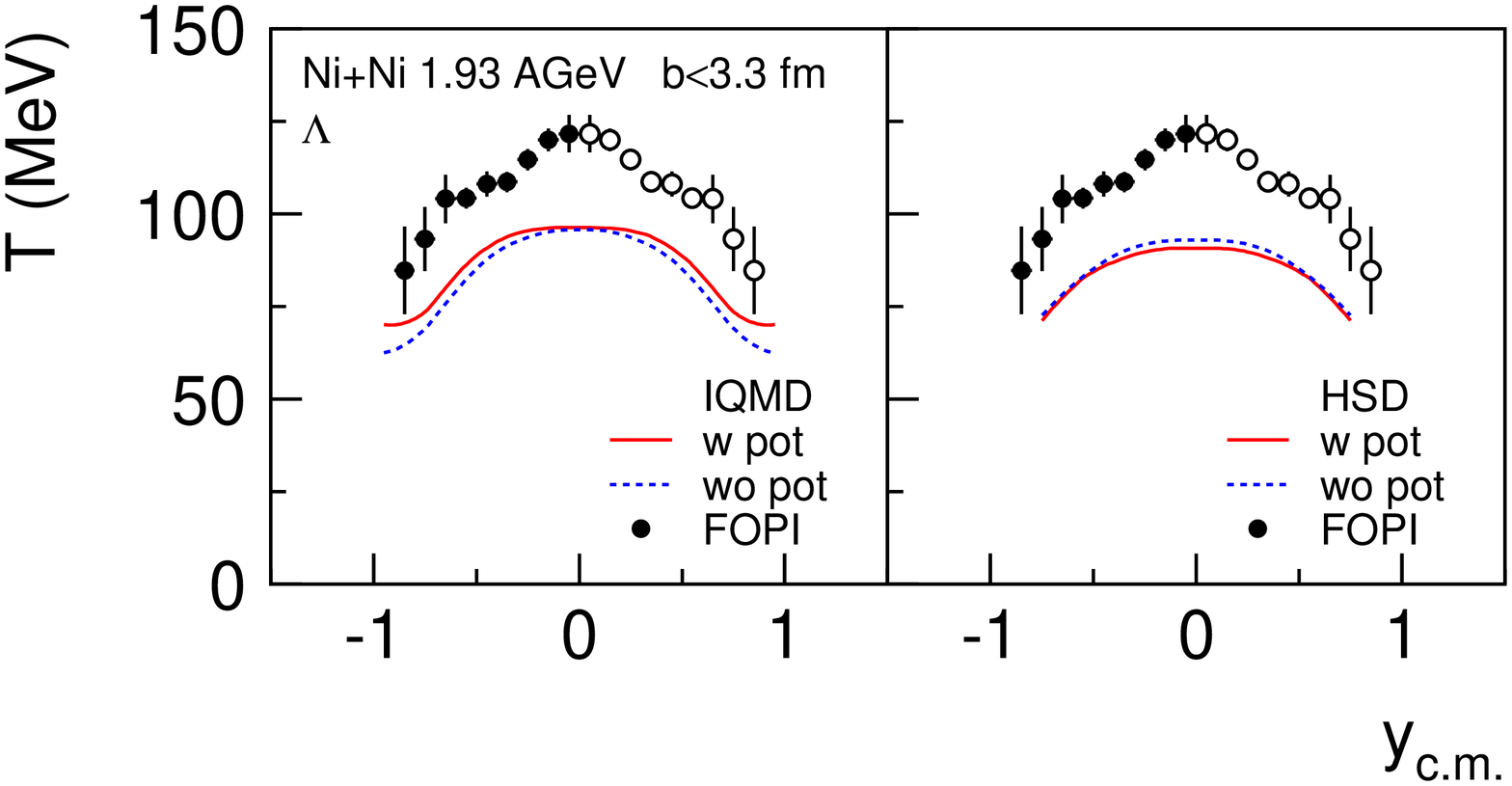,width=0.65\textwidth}
\vspace*{-1cm}
\epsfig{figure=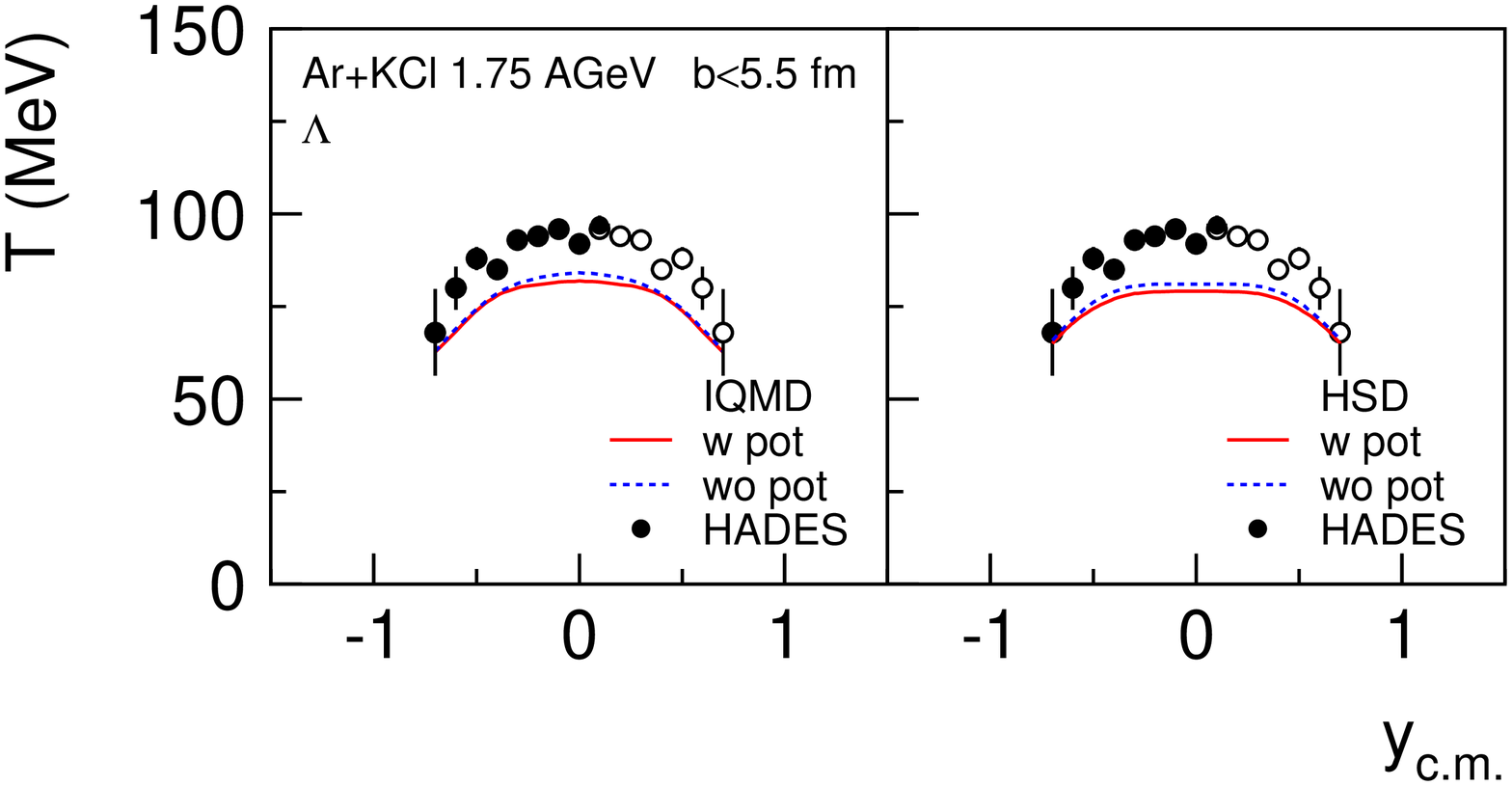,width=0.65\textwidth}
\caption{ Inverse slope parameter $T$ of the $\Lambda$ spectra as
a function of rapidity. We compare the FOPI data for central Ni+Ni
collisions \cite{Merschmeyer:2007zz} at 1.93 \AGeV  and
preliminary results for Ar+KCl at 1.75 \AGeV
\cite{Agakishiev:2010rs} from the HADES collaboration with results
of the IQMD and HSD transport models. The theoretical predictions
were selected according the impact-parameter filter given by the
experiment: $b < 3.3$~fm and $b < 6.0$~fm for Ni+Ni and Ar+KCl,
respectively.} \label{dyTlambda}
\end{figure}
At production the slope is determined (as that of the \kp which is
well described) by the three body phase space. Due to the large
N$\Lambda$ elastic scattering cross section
\cite{Alexander:1969cx,Anderson:1975rh} the final slope differs
quite substantially from that at production and is therefore
sensitive to the $\Lambda$N dynamics during the reaction. Assuming
an exponential form, the fitted inverse slope parameters $T$ as a
function of the rapidity for 1.93 \AGeV Ni+Ni and 1.75 \AGeV
Ar+KCl are displayed in \figref{dyTlambda} and compared with
theory. The \kp nucleus potential does not change the slope of the
distribution.  We see quite a discrepancy between both models and
the data. The FOPI collaboration measured inverse slope parameters
of $\Lambda$ are typically higher than those predicted by the
models as can be inferred already from fig. \ref{spectraLambda}.
Hyperon rescattering is treated in a similar way in both models.

For the Ar+KCl data at 1.75
\AGeV the experimental slopes are quite well reproduced by theory,
although at mid-rapidity the fit to the experimental spectra yields
slightly higher inverse slope parameter than predicted by the
simulations.

\begin{figure}[htb]
\epsfig{figure=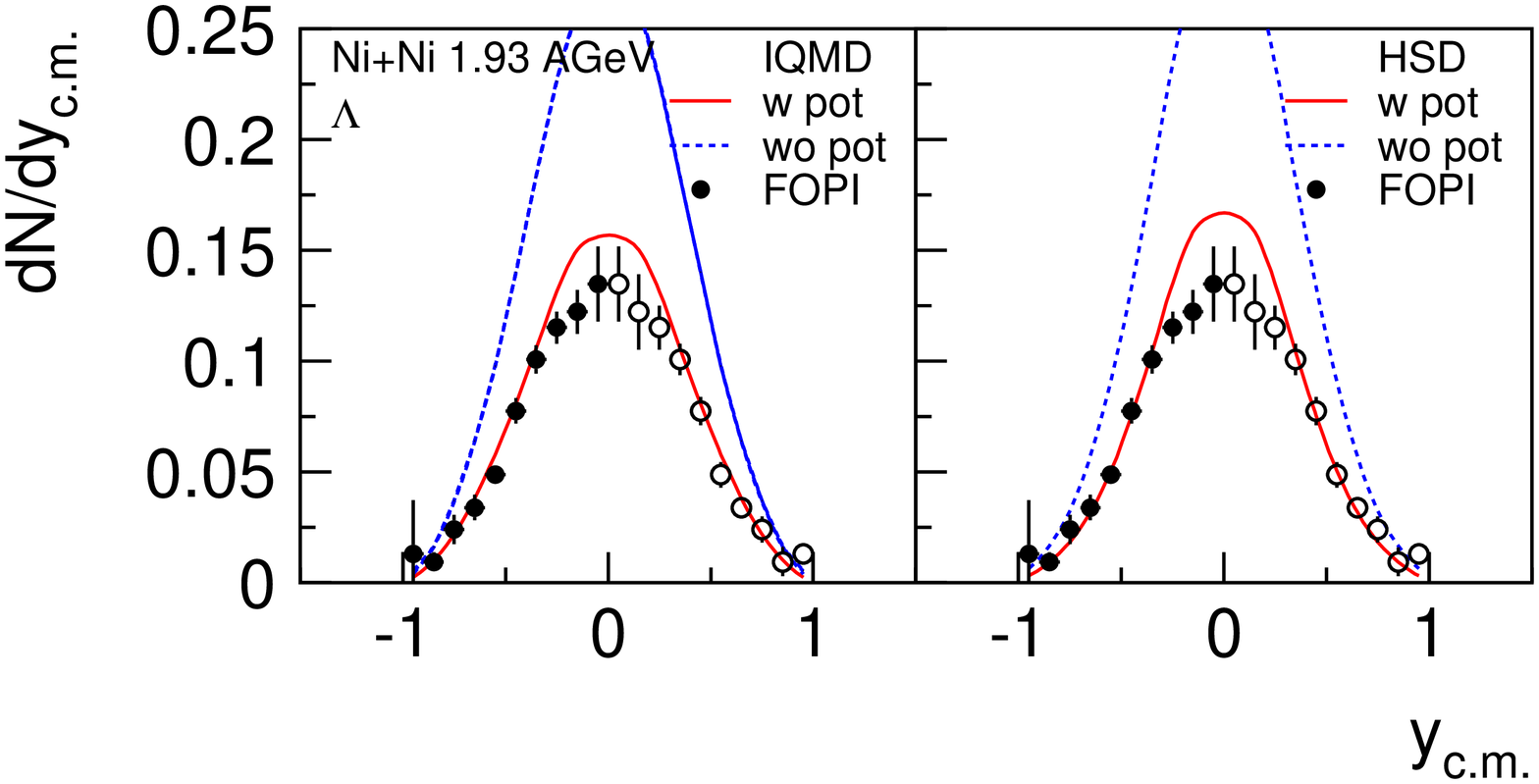,width=0.65\textwidth}
\vspace*{-1cm}
\epsfig{figure=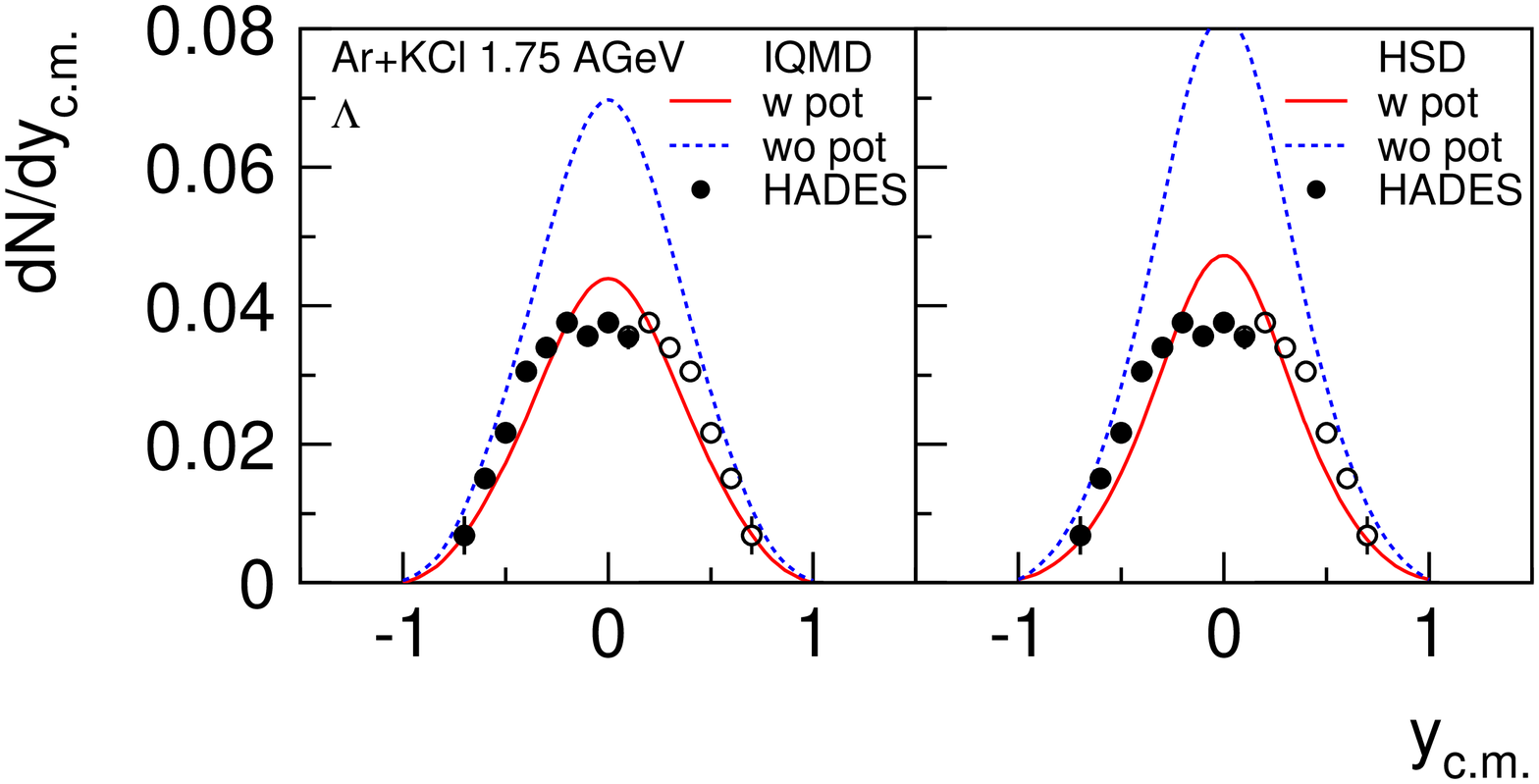,width=0.65\textwidth}
\caption{ $\Lambda$ rapidity distribution, ${\rm
d}N_{\Lambda}/{\rm d}y_{\rm c.m.}$, for data Ni+Ni at 1.93 \AGeV
and Ar+KCl at 1.75 \AGeV
compared with theoretical results of IQMD and HSD.} \label{dndylambda}
\end{figure}

Extrapolating the $m_t$ spectra to $m_t-m_0 = 0$ one obtains the
$\Lambda$ rapidity distribution ${\rm d} N/{\rm d}y$ which is
displayed for the same data in \figref{dndylambda}. We compare the
experimental result,
Ni+Ni at 1.93 \AGeV (measured by FOPI) with IQMD and HSD calculations,
as well as and  Ar+KCl at 1.75 \AGeV (measured by HADES),
including and excluding the \kp nucleus potential.
Due to a stronger \kp nucleus potential the IQMD model
comes only close to the experimental data when this potential is included,
whereas for the HSD
calculations  are close to experiment with and without this potential.

It is not understood why both theories under-predict the
$\Lambda$ slope of the Ni+Ni data set of the FOPI Collaboration by
at least 20\% but come close to that observed in the Ar+KCl
experiment from the HADES collaboration.
The relatively little known elastic rescattering cross
section can  not be the origin of this difference because it is
sufficiently large to bring the $\Lambda$ to equilibrium with
their local environment, as has been tested with IQMD calculations.

\subsection{In-plane and azimuthal distribution of the $\Lambda$}

\begin{figure}[htb]
\epsfig{figure=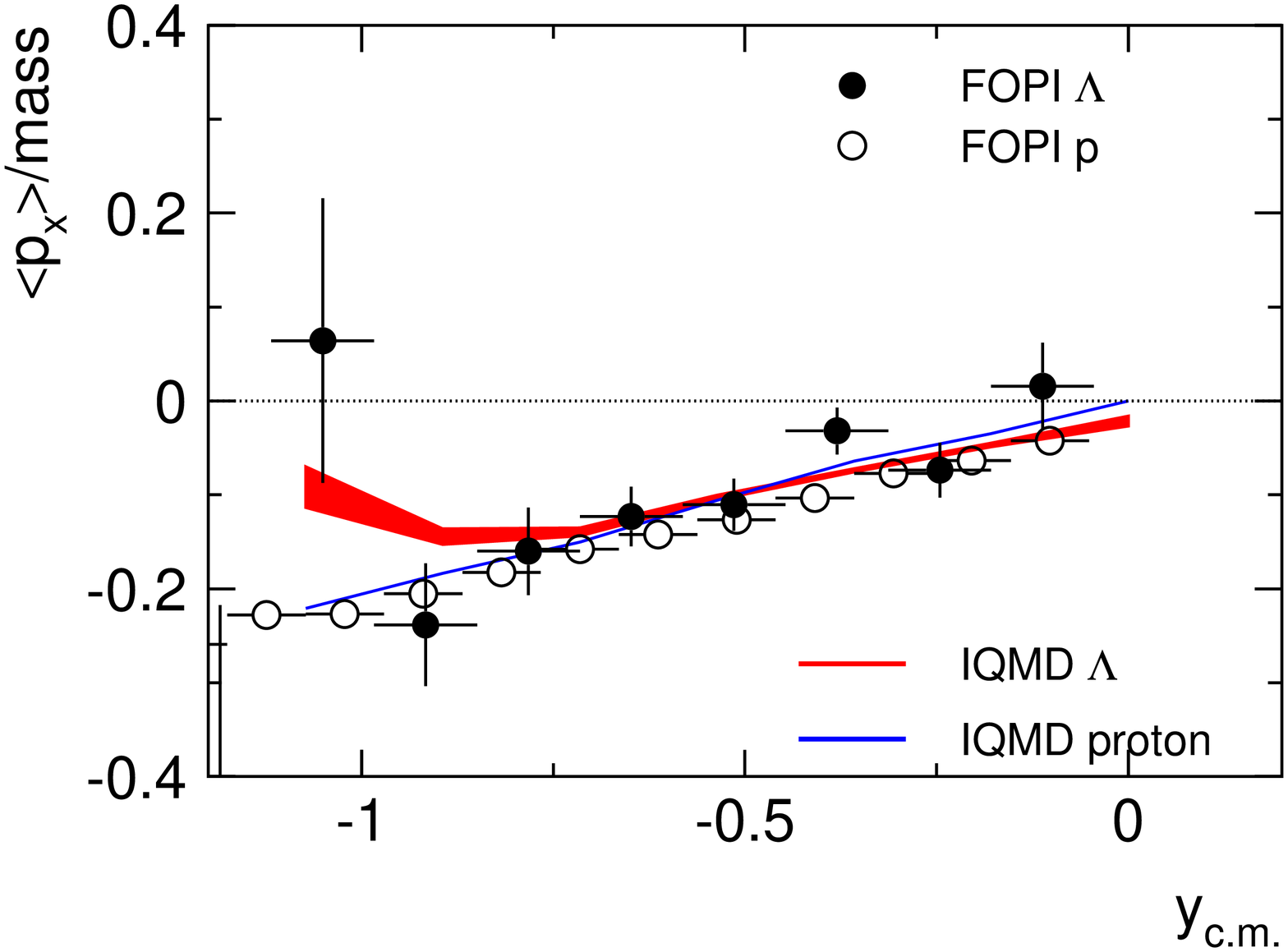,width=0.4\textwidth}
\caption{In-plane flow of $\Lambda$'s and protons for $p_t/m_0 >
0.5$ as measured by the FOPI Collaboration \cite{Ritman:1995tn} in
comparison with IQMD calculations.} \label{lflow1}
\end{figure}

The $v_1$ and $v_2$ values of the $\Lambda$ for $p_t/m_0 > 0.5$
have been measured and have been compared with theory by the FOPI
Collaboration. In Fig.~\ref{lflow1} we display $v_1$ as a function
of the $\Lambda$ rapidity as compared to IQMD calculations. In this acceptance
region the model predicts that $v_1(y)$ of proton and of $\Lambda$
are identical for $y_{\rm c.m.} > -0.4$ and deviate slightly for
$y$-values, close to target rapidity. This corresponds
quantitatively to the experimental findings. At production
$\Lambda$ do not show flow, they acquire it by rescattering with
protons. Figure~\ref{lflow1}  show that $\Lambda$ with a high
$p_t$ have sufficient rescattering collisions to have the same
flow as the scattering partners. A large rescattering cross
section would not increase the $v_1$ value further.

\begin{figure}[htb]
\epsfig{figure=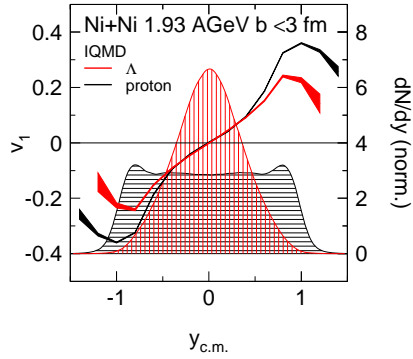,width=0.4\textwidth}
\caption{ Rapidity distribution and in-plane flow, $v_1$, of $\Lambda$ and p as a function
of the rapidity.} \label{lflow22}
\end{figure}

\begin{figure}[htb]
\epsfig{figure=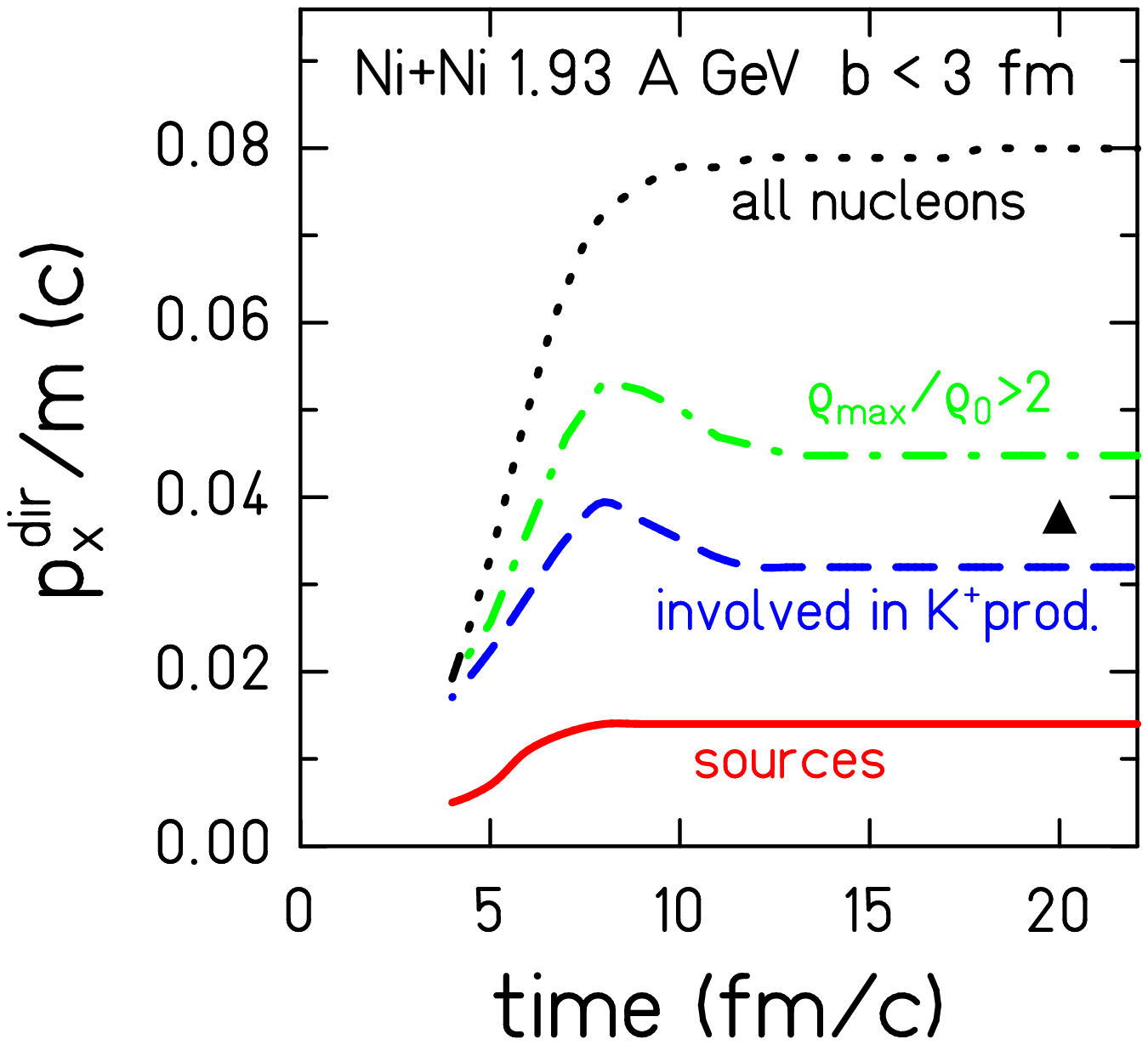,width=0.4\textwidth} \caption{The
final average hyperon in-plane flow as compared to that of the
sources and of different classes of nucleons. The full triangle
corresponds to the value of the $\Lambda$.} \label{lflow2}
\end{figure}
It is interesting to study the in-plane flow of the $\Lambda$
without the above mentioned cut in $p_t$ . This is presented in
\figref{lflow22}.  There we display the rapidity distribution and
the in-plane flow of p and $\Lambda$ as a function of rapidity. We
see that at mid-rapidity the in-plane flow of both baryons agree
whereas at beam and target rapidity a substantial difference can
be observed.  The $\Lambda$ are located at mid-rapidity whereas
the protons show a rather flat distribution.  If one weights the
in-plain flow by the distribution of the particles it is evident
that the rapidity integrated in-plain flow of protons
($p_x^{dir}=(\sum_{\rm all\ particles}{\rm
sign}(y_i)p_x(y_i))/N_{\rm all\ particles}$) is rather large as
compared to that of the $\Lambda$.

Fig. \ref{lflow2} shows this from another point of view. Here we present \cite{David:1998qu}
the in-plain flow averaged over all transverse momenta and rapidities
as a function of time. We see quite a different behaviour than in \figref{lflow1}: The average flow
of the protons is twice as large as that of the $\Lambda$, presented as a triangle.
This result has been experimentally observed at AGS energies \cite{Chung:2001je}.

In order to understand this result one has to realize that the
$\Lambda$ are produced in the high density zone. The average
density for the production of a \kp$\Lambda$ pair is around
$2\rho_0$ - in the case of central Ni+Ni at 1.93 \AGeV as well as
for central Au+Au at 1.5 \AGeV (see for example
Fig.~\ref{density-time}). If one calculates $p_x^{dir}$ only for
those nucleons which in the course of the reaction have passed a
density larger than twice the ground state density, $p_x^{dir}$ is
reduced by a factor of almost two (green dashed dotted line) and
comes close to the value for the $\Lambda$. The $p_x^{dir}$ of
those nucleons which have participated in the creation of \kp is
finally very close to that of the $\Lambda$.

This allows for several conclusions:\\
a)  The observed large in-plane flow of protons  (which is in first order $\propto
(-{\rm d}V/{\rm d}x )t $ where $V$ is the potential (which is a function of the density) and $t$ is the passing time of the nuclei)  is predominantly caused by nucleons close to the surface of the interaction zone
where the density gradient is large but the maximal density,  reached during the collision,  is well below $2\rho_0$.  \\
b) In the interior of the interaction zone the density gradients are moderate and lead to a moderate  $p_x^{dir}$\\
c) The $\Lambda$ carry about the same in-plane flow as the
surrounding nucleons. Since most of the hyperons are produced as
heaviest particle in a three-body phase space decay, the energy of
a hyperon directly after its production is rather low. High $p_t$
$\Lambda$ obtain dominantly their energy from rescattering
with high-energetic nucleons and thus adapt to their kinematical properties.  \\
d) Nucleons with a large $p_t$  come, as the $\Lambda$, from the high density zone. \\
This explains why a cut in $p_t$ yields similar in-plane flow of
the $\Lambda$ and protons, \figref{lflow1} whereas the in-plane
flow is different by a factor of two when we average over all
$p_t$, as shown in \figref{lflow2}.

\newpage
\section{Production of \km mesons}
\setcounter{figure}{0}
\subsection{Subthreshold \km production}

Already in the first experiments, it has been observed that close
to threshold the \kp and \km yields behave completely different as
compared to the extrapolation from the elementary production cross
sections $\rm NN \to NK^+\Lambda$ and $\rm NN \to NN\kp \km$. This
observation can be seen in \Figref{KMelem}, left, which shows the
multiplicity of \kp and \km mesons per participating nucleon
\Apart in nucleon-nucleon and in heavy-ion collisions as a
function of the energy above threshold in the NN system, $\sqrt{s}
- \sqrt{s_{\rm thres}}$. Whereas in heavy-ion reactions the \km
and the \kp yields are very similar around their respective
thresholds, the elementary cross sections show order
of magnitude differences.

The origin of this observation is a new \km production mechanism  in
heavy-ion reactions which is absent in nucleon-nucleon collisions:
\begin{figure}[htb]
\epsfig{file=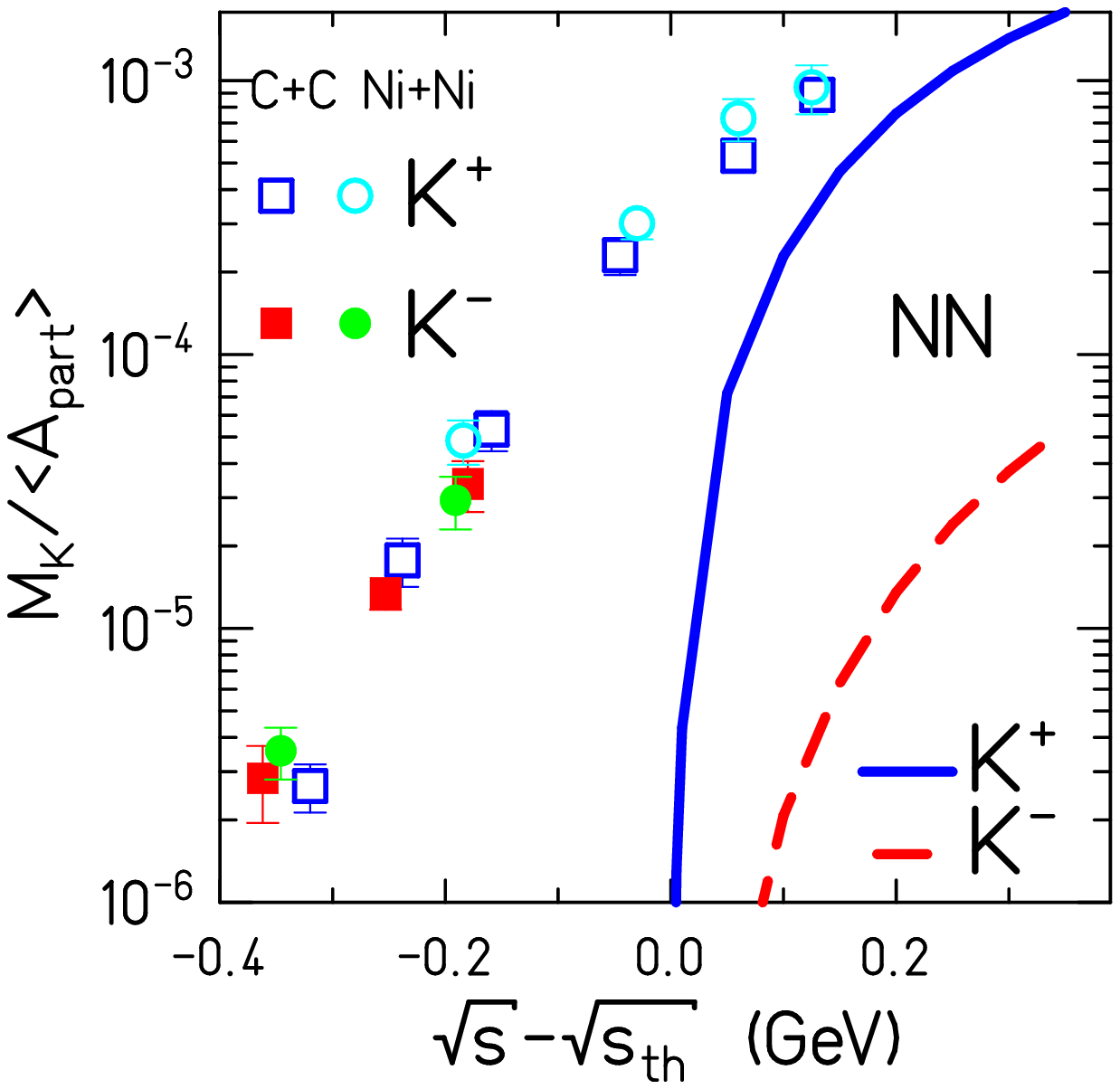,width=.49\textwidth}
\epsfig{file=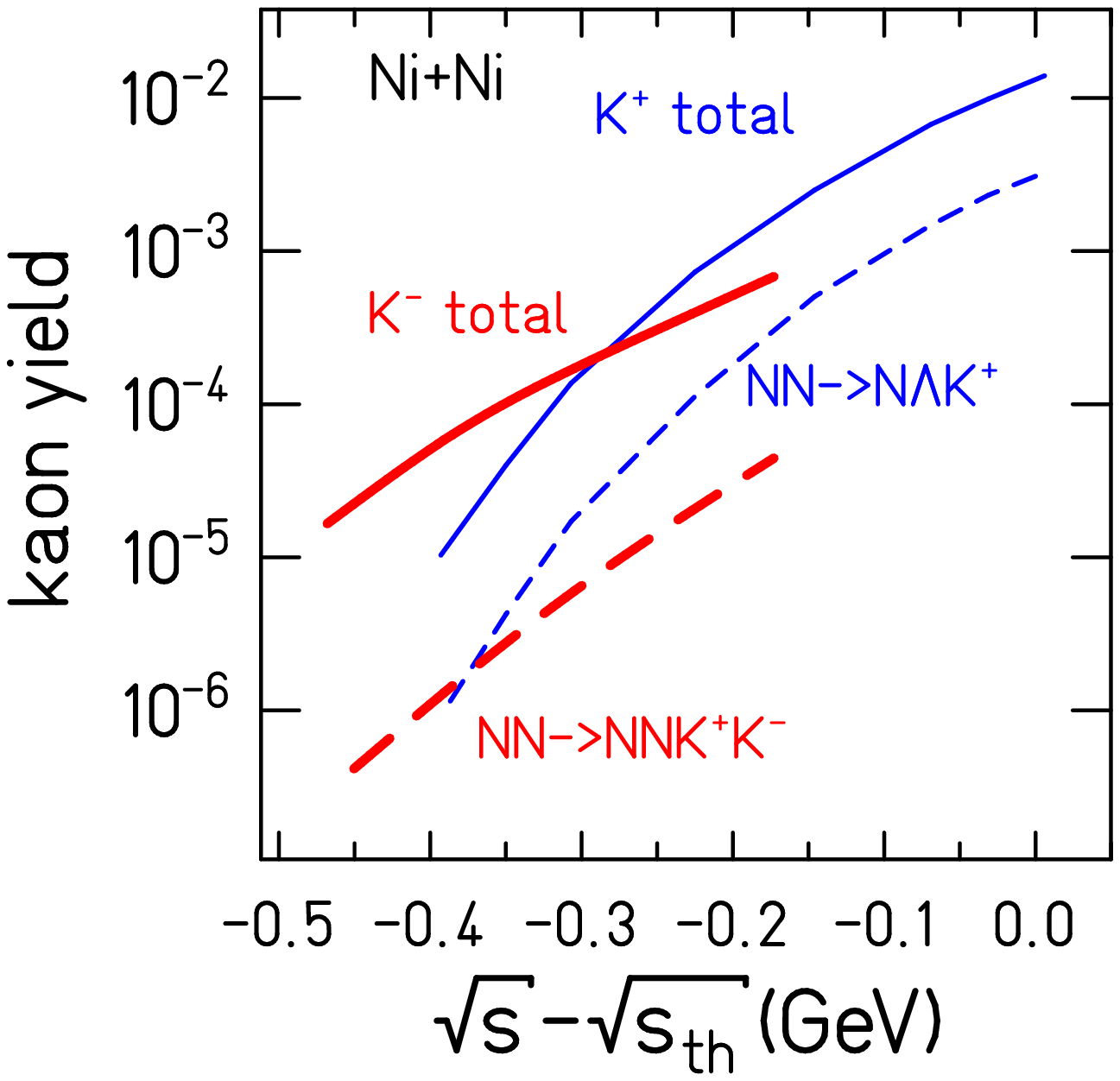,width=.49\textwidth}
\caption{Right:  Comparison of the production of \kp and \km in
heavy-ion reactions in comparison with the production in
elementary collisions. Left: Excitation function of the \kp yield
in $\rm NN \to NK^+\Lambda$ collisions and of the \km yield in
$\rm NN \to NNK^+K^-$ collisions as compared to the total
production of \kp and \km mesons in heavy-ion collisions.}
\label{KMelem}
\end{figure}
The \km meson contains a $s$ quark which can be exchanged with a
baryon. This is in contrast to the \kp whose $\bar s$ does not
find a baryonic exchange partner. The only possible inelastic reaction is
the charge exchange \kp $\leftrightarrow$ K$^0$. Close to
threshold the production of $s$ and $\bar s$ quarks is dominated
by the reaction with the lowest threshold. This is the baryonic
reaction $\rm BB(\Delta) \to N\kp Y$ (with Y being $\Lambda$ or
$\Sigma$) as discussed in Section ~\ref{Kplus}. In heavy-ion
collisions the strange baryon can transfer its strange quark to a
\km  and at threshold energies, indeed, most of the observed \km
mesons are not directly produced, but produced by secondary interactions
$\rm B Y \to NN\km $ or $\rm \Lambda (\Sigma) \pi \to \km N$. This
new production mechanism, only occurring in heavy-ion collisions,
links the \km to the \kp production. The contribution of $\rm NN
\to NN\kp \km$ to the \km yield in central Au+Au reactions at 1.5
\AGeV is even more marginal than that of $\rm NN \to NK^+\Lambda$
for the \kp yield, as shown in \Figref{KMelem}, right.

Hence, there are two dominant mechanisms for the production of
strange mesons in heavy-ion collisions: The strangeness production
via e.g.~BB $\to$ N \kp +$\Lambda(\Sigma)$, which has been
discussed in Section~\ref{Kplus} and the strangeness exchange
reactions $\Lambda (\Sigma) \pi \to \rm \km N $ and B$\Lambda
(\Sigma) \to$ \km NN  discussed here.

\subsection{Dynamics of \km production and emission}

For the exchange of strangeness two channels are important: $\pi$Y
$\to$\km N and BY $\to$ NN\km.  The latter process becomes
important if B is a $\Delta$ because there the less kinetic energy
is needed to overcome the threshold, yet this cross section is
unknown. Following Randrup and Ko~\cite{Randrup:1980qd}, we apply
an isospin factor to the corresponding NN channel
($\sigma(\N\Delta)=0.75 \sigma(\N\N)= 0.6 \,{\rm mb} \, (E-E_{{\rm
thres}}))$, with energies measured in GeV in the hyperon rest
frame). The $\Lambda (\Sigma) \pi \to \rm \km N $ cross section
can be obtained by detailed balance from the measured $\rm \km N
\to \Lambda (\Sigma) \pi $ cross section. Being influenced by the
$\Lambda$(1405) resonance, the exothermic reaction $\rm \km N \to
\Lambda (\Sigma) \pi $ is very strong at low relative momenta
whereas the inverse cross section goes to zero there. From the
magnitude of the cross sections alone the following reaction
scenario emerges: Some of the strange baryons from the primary
reaction BB $\to$ N$\Lambda$ \kp  create in a second reaction a
\km meson which has a high chance to be reabsorbed quickly by the
inverse process N\km $\to \Lambda \pi$. This situation causes a
completely different dynamics as compared to \kp mesons.

\begin{figure}
\epsfig{figure=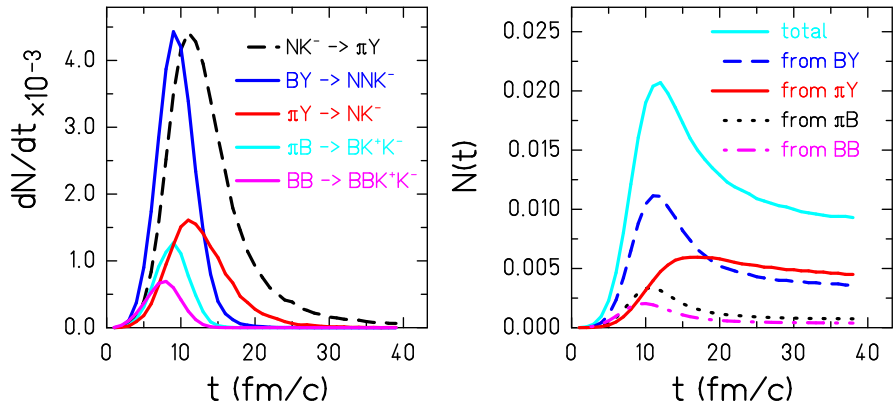,width=0.8\textwidth}
\caption{Contribution of the different channels to the \km
yield as a function of creation time of the \km (full lines) for
central ($b$ = 0) Au+Au collisions at 1.5 \AGeV. On the left side we display the
rate  ${\rm d}N/{\rm d}t$, on the right side the yield $N(t)$.
}
\label{KM_time_channels}
\end{figure}

Using these cross sections we calculate the time evolution of the
production and absorption of \km mesons in the different channels
as displayed in \Figref{KM_time_channels}, left. At early times, a
dominance of the Y $(\Lambda , \Sigma)$ B channel is observed, but
most of these early produced \km mesons are reabsorbed. Later,
when the $\Delta$ are disintegrated the $\Lambda (\Sigma) \pi$
channel becomes dominant. By this time the system is less dense
and the survival chance of the \km mesons has therefore increased.
Counting the net number of \km  mesons which are present in the
system as a function of time, one observes that the $\rm BY \to
NK\Lambda(\Sigma)$ reaction contributes finally less than 50\% to
the surviving \km mesons as can be seen in
\Figref{KM_time_channels}, right.

\begin{figure}
\epsfig{figure=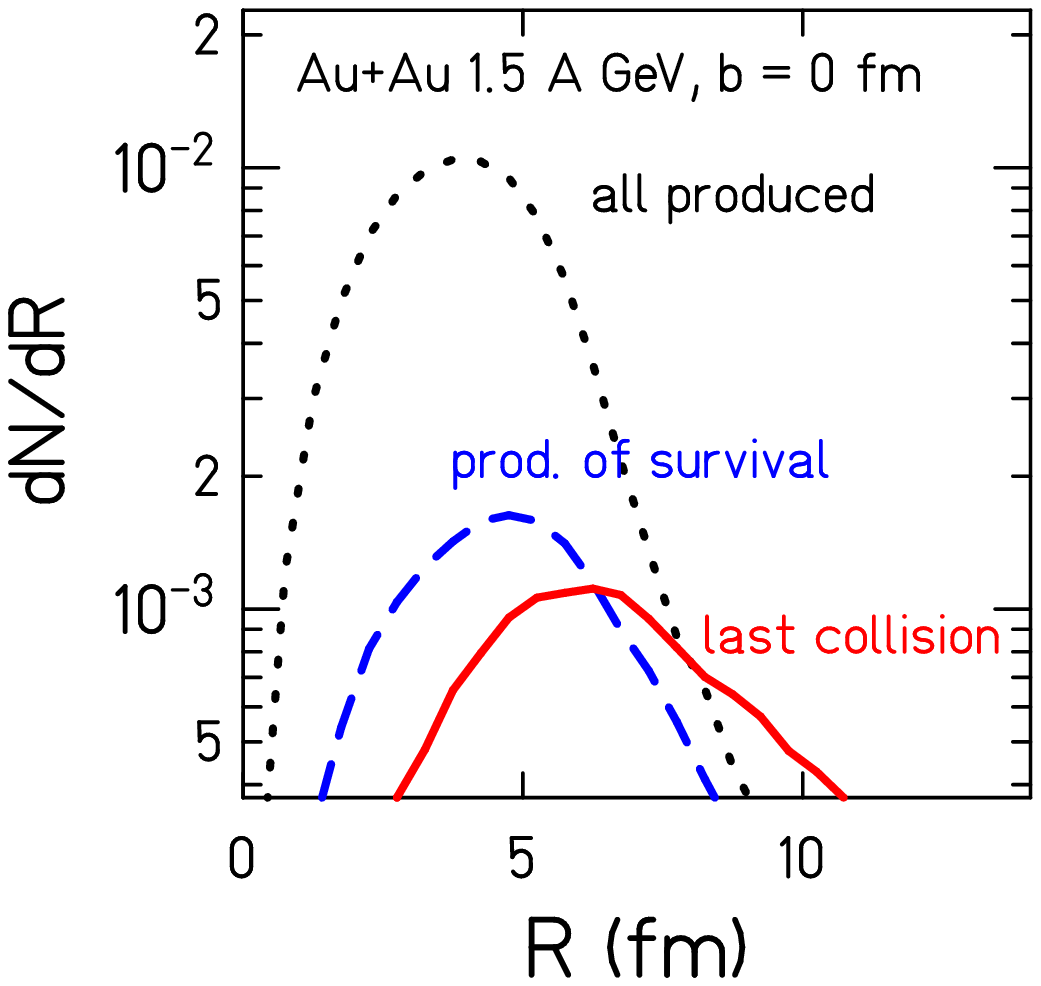,width=0.49\textwidth}
\epsfig{file=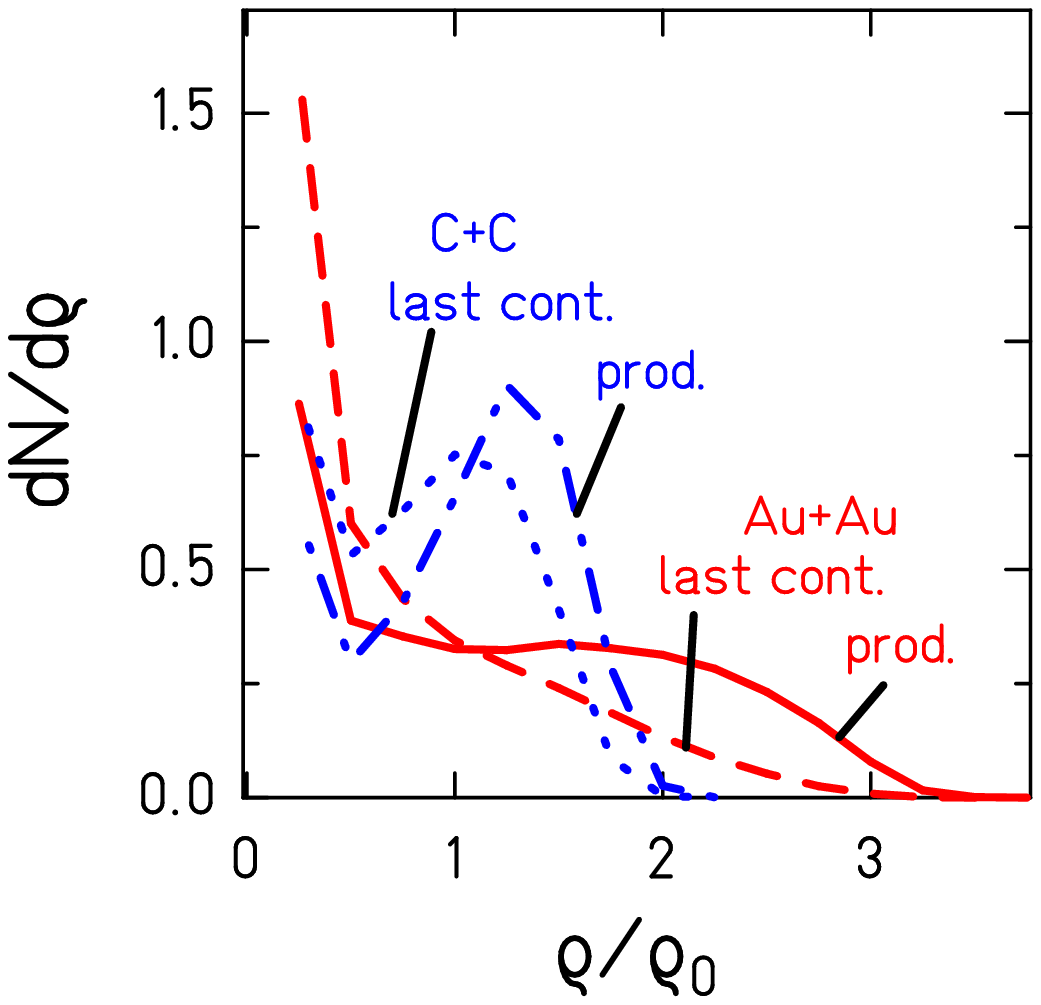,width=0.49\textwidth}
\caption{
Left: Radial profile of
the production points for all produced \km mesons (dotted line)
and of those which are not reabsorbed (dashed line). For those
which survive the radial profile of their last interaction point
(full line) is displayed as well. Right: Distribution of the
density at creation and at the point where the surviving \km have their last
interaction for Au+Au and C+C collisions at 1.5 \AGeV.}
 \label{KM_survival}
\end{figure}

The radial profile of the production points of all \km mesons and
of the surviving ones is seen in \Figref{KM_survival}, left, which
displays also the radial profile of the last collision points
of the \km. These correspond to the production points if the surviving \km
does not rescatter or to the point of the last elastic collision.
The survival probability is obviously depending on the production
point. Only about 20\% of all produced \km survive in Au+Au
collisions at 1.5 \AGeV. In C+C collisions at the same incident
energy this fraction reaches 47\%.

The surviving \km mesons are produced late or/and close to the
surface of the reaction zone. The consequence of both is that the
surviving \km mesons are produced at low densities.  This is
detailed on the right hand side of \Figref{KM_survival} showing
the density at the points at which the surviving \km have been
produced and where they had their last collision. The surviving
\km mesons produced in the interior have a high probability to
suffer from elastic \km\N~collisions. Therefore, for the majority
of the \km mesons the last interaction takes place at less than
0.5 $\rho_0$. Surprisingly, the \km in C+C reactions come from
higher densities than those in Au+Au collisions where the \km
interact still with the expanding fireball, as can be seen in
Fig.~\ref{KM_survival}, right. This observation is not influenced
by the \km potential as seen in the next section. Therefore, the
observed \km are neither sensitive to nuclear properties nor to
the \km N potential at densities higher than 0.5 $\rho_0$ in
strong contrast to the \kp. It is important to note that the total
\km yield is linked to the total \kp yield because when more
strange baryons are produced more secondary reactions take place
in which \km mesons are created. This relation has been shown
in~\cite{Oeschler:2000dp} and has been studied in detail by
applying the law of mass action in~\cite{Cleymans:2004bf}. There
it has been shown that the \km production via strangeness exchange
seems to be dominant also at AGS energies.

\subsection{Influence of the \km N potential}

The nuclear matter calculations presented in section II have shown
that in cold matter \km mesons can hardly be considered as quasi
particles because the width of their spectral function is too
large. Calculation at finite temperature, however,
indicate that with increasing temperature the \km regains its
quasi-particle properties. Dealing with reactions in which inverse
slope parameters of more than 80 MeV are observed  we here treat
the \km meson as a quasi particle. Whereas the theories differ
little on the increase of the \kp mass as a function of the
density, there are substantial differences between the different
theories for the \km pole mass at zero temperature as a function
of the density, as shown in section II. These differences are due to details of the
calculation, due to a different pion dressing and due to different
channels which have been included. A similar uncertainty is also
expected for the \km masses at finite temperature.

\begin{figure}
\epsfig{figure=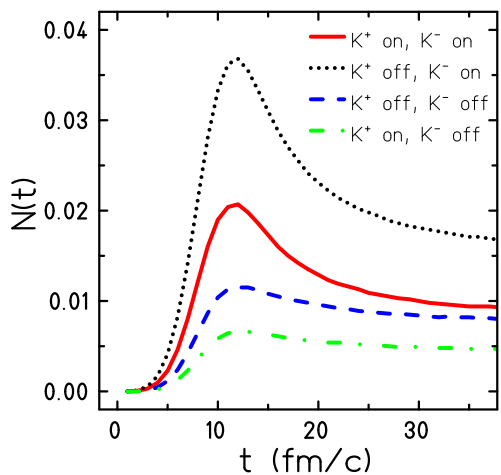,width=0.4\textwidth}
 \caption{
Time evolution of the \km  yield in central Au+Au collisions at
1.5 \AGeV for different options of the KN-potentials.}
\label{potential}
\end{figure}

In the IQMD approach the mass, or more precise $\omega({\bf k} =
0)$, is given by the mean field calculation  of
Ref.~\cite{Schaffner:1994bx,Schaffner:1996kv} (see
Eq.~\ref{schaf}). It is expected that the attractive \km N
interaction increases the \km yield in heavy-ion reactions as
compared to an inactive \km N interaction because the \km becomes
"lighter" in matter ($\omega({\bf k}=0)<m_0$). The situation is becoming more complex if we
include both, the attractive \km N and the repulsive \kp N
interaction, as seen in Fig.~\ref{potential}. Then, we have almost
the same \km multiplicity as if none is active (dashed and solid
lines in Fig.~\ref{potential}). The gain of the yield by a smaller
\km mass (dotted line) is almost completely compensated by the
loss of the number of $\Lambda$ (dash-dotted line) due to a
smaller production cross section if the \kp becomes heavier. This
might appear astonishing as the density dependence of the \kp N
potential is rather weak, while that of the \km N interaction is
rather strong and less cancellation is expected. The cancellation
occurs, however, because the \kp production occurs at high densities
(Fig.~\ref{AuAu_CC_kp}.) whereas the surviving \km are created at
densities at or below $\rho_0$, as shown before.

\subsection{Uncertainties in the input quantities}

Besides the little-known density dependence of the  \km N
potential a second, even less controlled source of uncertainty is
the lack of knowledge of the essential elementary cross sections.
E.g.~the cross section of the $\Delta\Lambda(\Sigma)\to \N\N\km$
channel is not known at all. Depending on the life time of the
$\Delta$ in matter it may be  as important
as the $\pi\Lambda(\Sigma)$ channel (see
\figref{KM_time_channels}) in heavy-ion collisions. It is also nontrivial to extrapolate
the measured free \km N $\to \Lambda(\Sigma) +\pi$ cross section
to a nuclear environment. Calculations have shown that the
$\Lambda (1405)$ resonance disappears in matter
\cite{Koch:1994mj,Waas:1996xh,Waas:1996fy,Waas:1997pe}
but the details, especially the density dependence of the
disappearance, depend on the \km N interaction. As can be seen
from Fig.~\ref{KM_survival}, there is a broad distribution of
densities, which ranges from almost zero to 1.5 (2) $\rho_0$ in C+C (Au+Au)
collisions, to which
the emitted \km mesons are sensitive. Detailed information on the
properties of the $\Lambda (1405)$ at these densities and hence
its role for the \km production and annihilation cross sections
is not available. Experiments like pA $\to $ X \km could in
principle elucidate this question but the few available data are
not at all sufficient. Therefore, the two essential input
quantities, the \km N potential and the \km N $ \to \Lambda \pi$
cross section, are much less known than the corresponding
quantities for the \kp.

Besides these uncertainties directly related to the production of the
\km the dominance of the hyperon induced channels link the production
of \km directly to the yield of the hyperons. Since these hyperons are
dominantly produced together with a positive
kaon, there is a direct link between
the yield of \kp and \km. Thus, all uncertainties related to the production
of \kp - like the unknown cross section of the production via the
N$\Delta$ channel, see  Fig.~\ref{xsections-NDN} - are inherited by the
\km. This effects the absolute yield of the \km but does not effect
the relative \kp/\km ratio.


\subsection{\km multiplicity}

\begin{figure}
\epsfig{figure=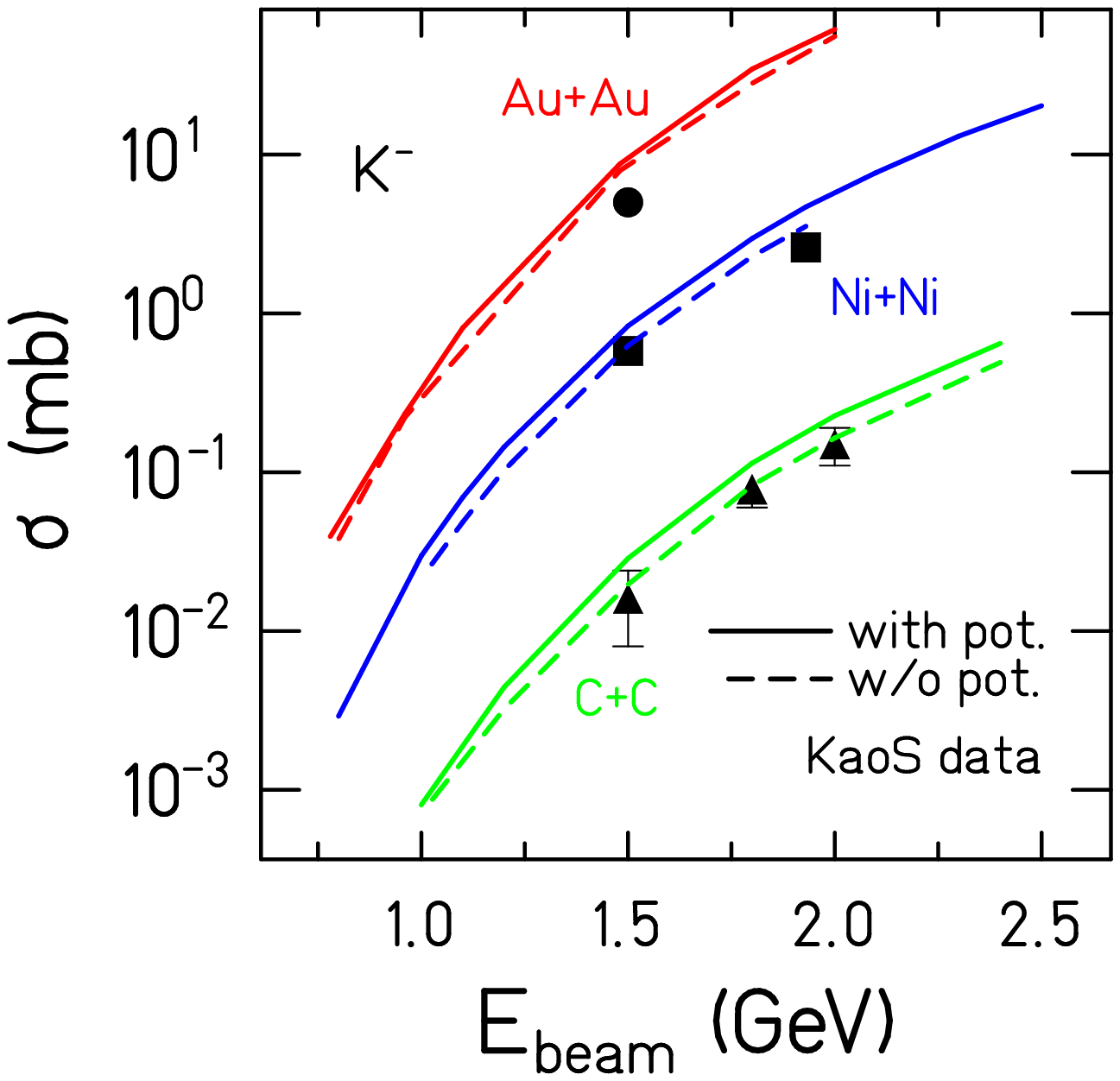,width=0.32\textwidth}
\epsfig{figure=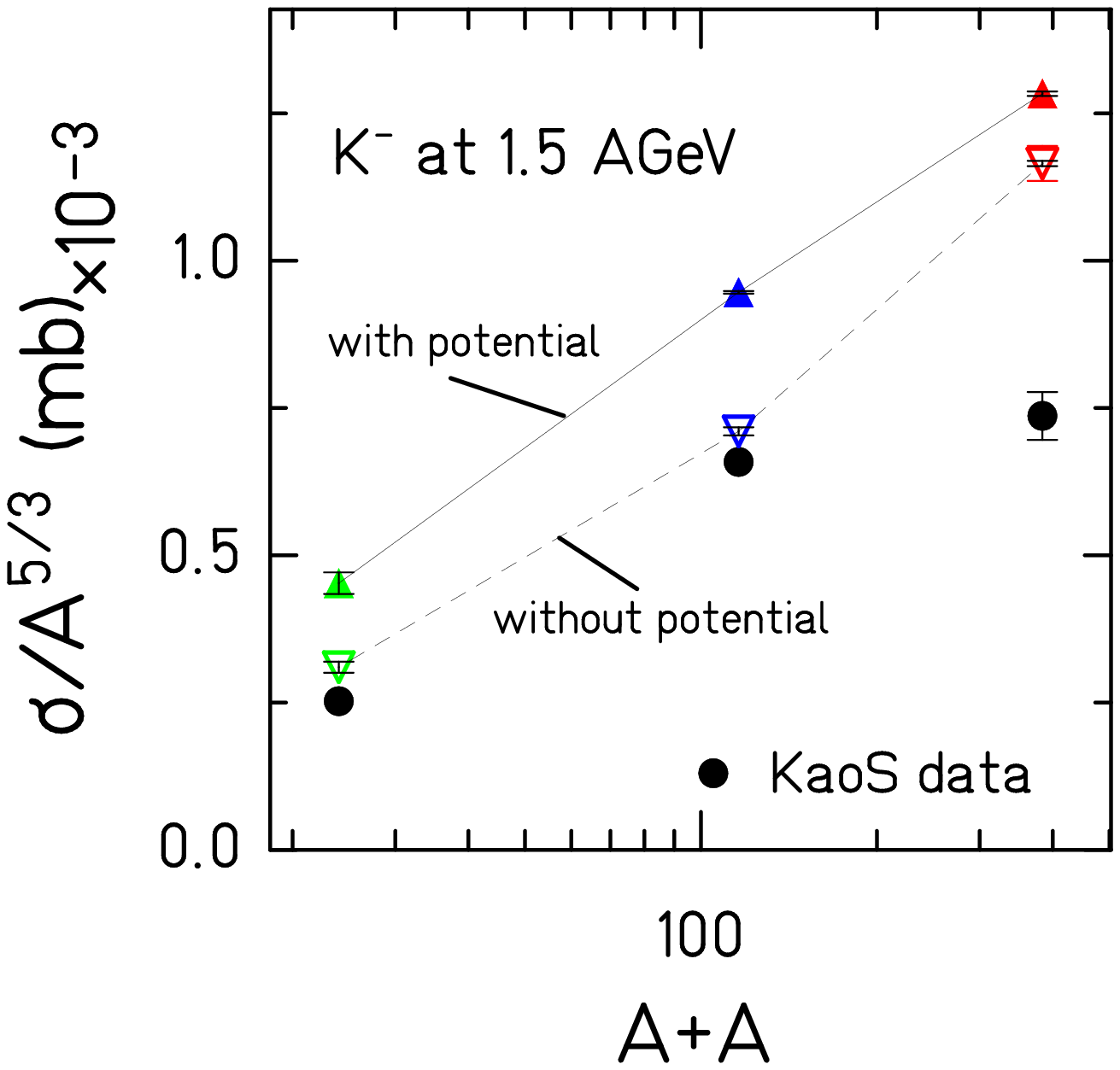,width=0.32\textwidth}
\epsfig{file=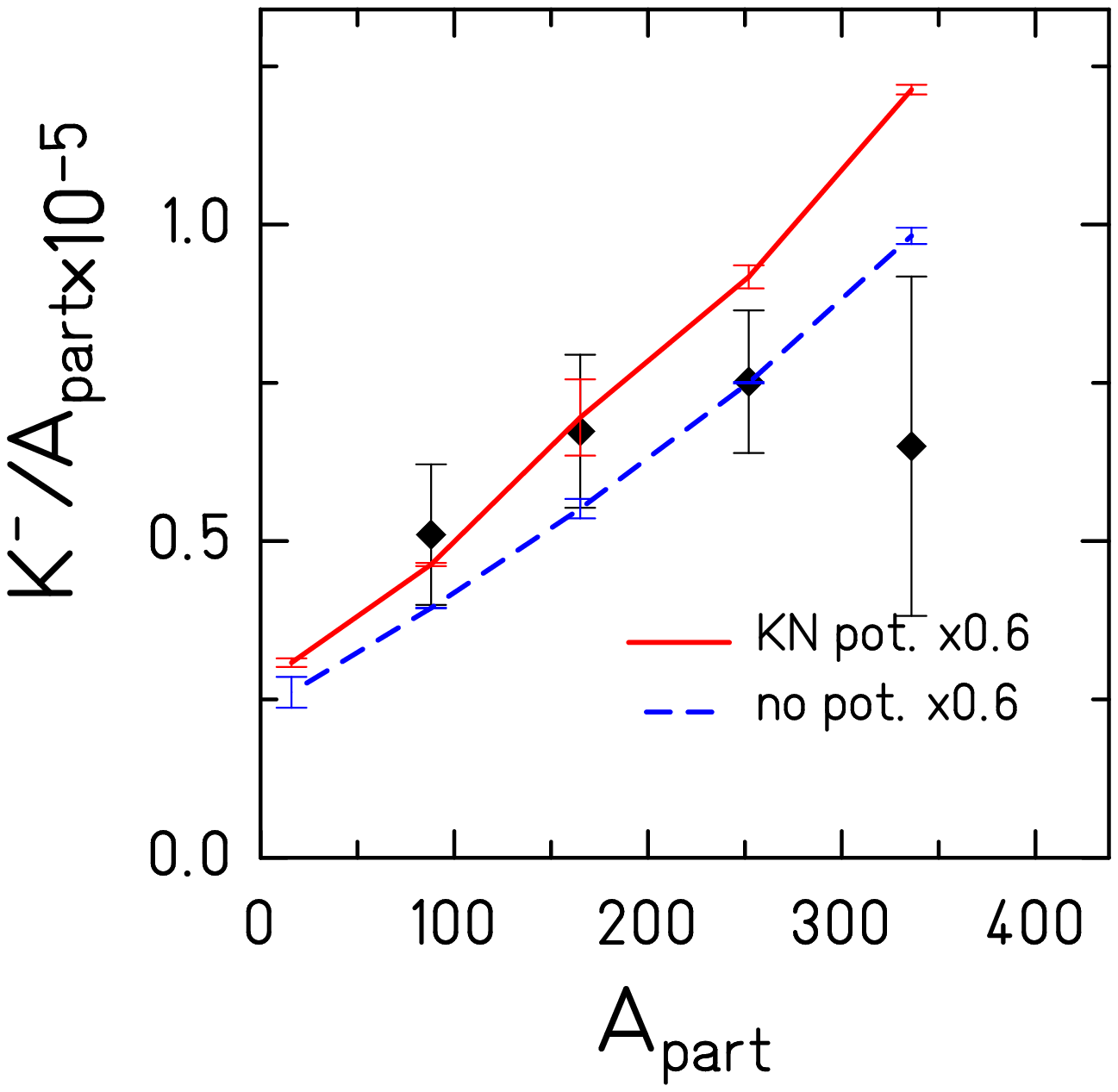,width=0.32\textwidth} \caption{Left:
Excitation function of the inclusive \km cross section with and
without the KN potential. The lines refer to IQMD calculations and
the symbols to the experimental results of the KaoS
Collaboration~\cite{Forster:2007qk}. Middle: \km cross section
divided by $A^{5/3}$ as a function of the system size in theory
and experiment. The line are to guide the eye.
Right: \km/\Apart  versus \Apart for \mbox{Au+Au}
collisions  at 1.5 \AGeV (right) as compared with IQMD
calculations. The IQMD yields are multiplied with the factor which
is indicated in the figure.} \label{km_exc}
\end{figure}

The excitation function of the \km cross section in heavy-ion
reactions is shown in Fig.~\ref{km_exc} (left), including and
excluding the KN interaction and compared to the results of the
KaoS collaboration~\cite{Forster:2007qk}. As already seen in
Fig.~\ref{potential}, for Au+Au reactions the combined influence
of both KN potentials is small. It increases for lighter
systems because there the density at the production point of the
\km is higher (see \Figref{KM_survival}) and that of the \kp
lower. Then the \kp N potential hardly lowers the \kp yield
but the stronger density dependence of the \km mass makes it more
probable that a \km can be produced. The form of the excitation
function is well reproduced but the calculations have the tendency
to over-predict slightly the yield. This tendency is also visible
in the middle part of Fig.~\ref{km_exc} which shows the
theoretical and experimental inclusive cross section divided by
$A^{5/3}$ as a function of the system size for 1.5 \AGeV. The
experimental trend that the \km yield increases stronger than the
system size is reproduced by the calculations. This
tendency is also visible in the centrality dependence
of the \km yield in Au+Au reactions at 1.5 \AGeV (right).

\subsection{Are data compatible with a strong \km potential?}

One of the essential questions in  \km physics is the depth of the
\km N interaction. As discussed in Section II many efforts have
been made to determine this potential but a definite answer is not
at hand yet. Therefore, it is interesting to explore how the
results of simulations of heavy-ion reactions depend on the
potential depth and whether the existing data limit the strength.
For this purpose we simulate the reaction - keeping the (well
established) \kp N potential at its standard value - with a
variety of \km N potentials. They are obtained by multiplying the
potential constants by a factor $\alpha$ as it is shown in
Eq.~\ref{kpotfactor}.

\begin{figure}
\epsfig{figure=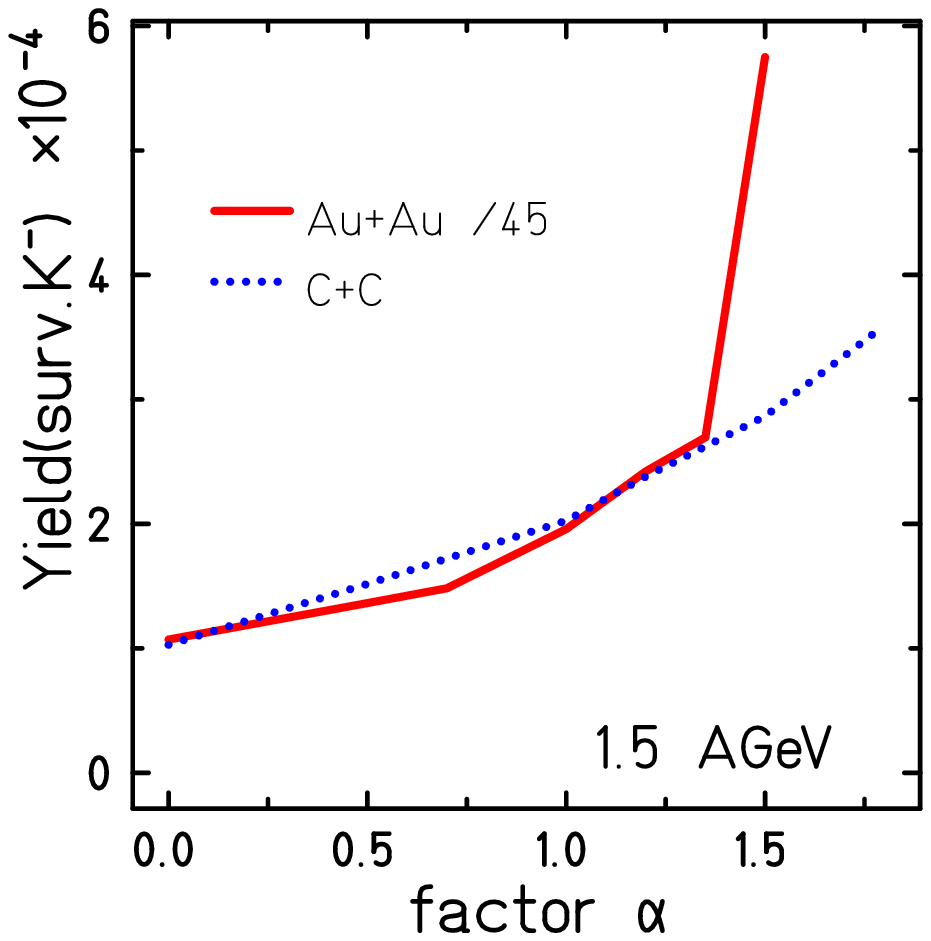,width=0.35\textwidth}
\epsfig{figure=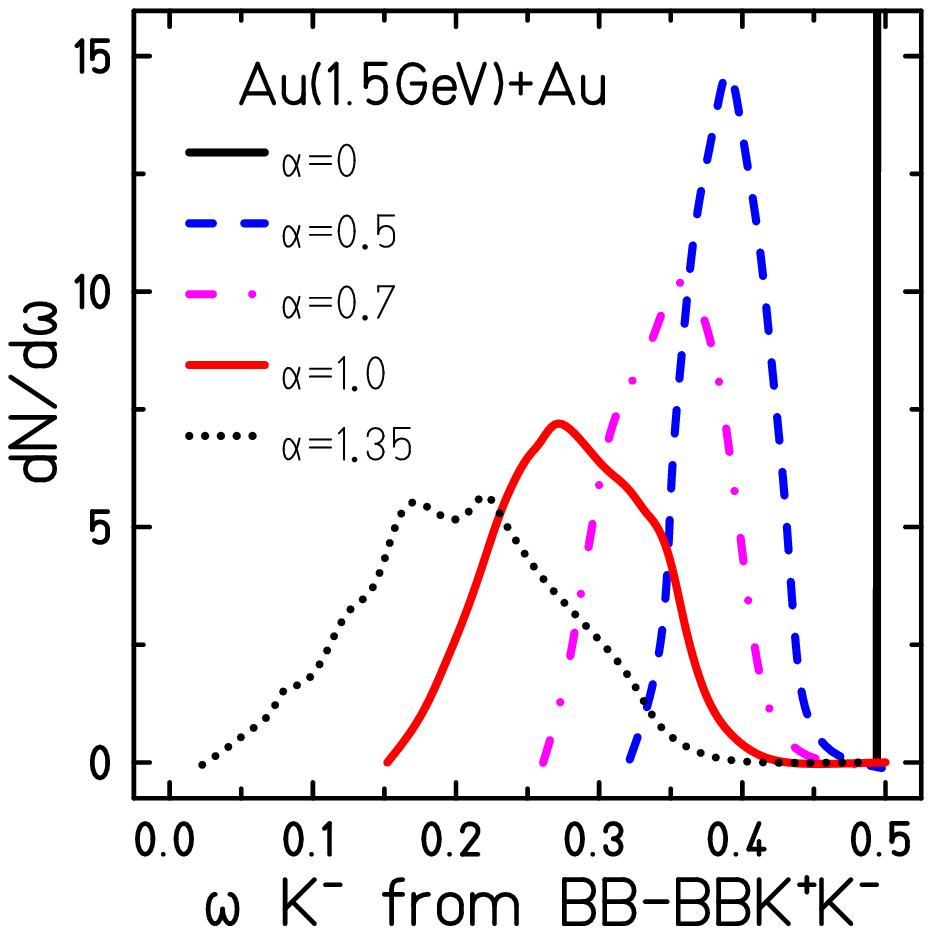,width=0.35\textwidth}
\caption{Left: \km multiplicity as a function of the strength of
the \km N potential. Right: $\omega(\bf{k}=0)$ distribution as a
function of the strength of the \km N potential.} \label{depth}
\end{figure}

In Fig.~\ref{depth}, left,  the calculated \km yield in Au+Au and
C+C systems for different values of $\alpha$ is presented. Below
$\alpha$ = 1.25 the yield increases moderately and in the same way
for Au+Au as for C+C. Above $\alpha = 1.25$ we observe a very
sudden and strong increase of the yield in Au+Au reactions whereas
the increase in the C+C system is continuous. Figure~\ref{depth},
right, shows the origin of the different behavior of the Au+Au and
C+C systems above $\alpha$ = 1.25. It displays the
$\omega({\bf k}=0)$ distribution as a function of the strength of
the potential. Already in our standard parameterization ($\alpha
=1$) we observe in Au+Au reactions small values of
$\omega({\bf k}=0)$. This fraction increases with the strength of
the potential and a larger and larger fraction of the \km has a
very small mass. Above $\alpha$ =1.25 we find that in some of the
production reactions $\omega({\bf k}=0)=0$. This means a very low
threshold in this reaction and hence a large yield. In C+C systems
where the density is lower $\omega({\bf k}=0)$ remains large.

Thus the measurable ratio of the \km yield in Au+Au and C+C systems is
relatively independent of acceptance cuts and
detector response, but is quite sensitive to the depth of the
potential. The fact that the yield ratio is reproduced in our
simulations (see \figref{km_exc}) therefore allows for the conclusion that $\alpha$ has to
be smaller than 1.25. The strong \km potentials which has been
discussed in section II and which are a prerequisite for the
existence of a \km condensate, are not compatible with heavy ion
data.

\subsection{\km spectra}

\begin{figure}
\epsfig{figure=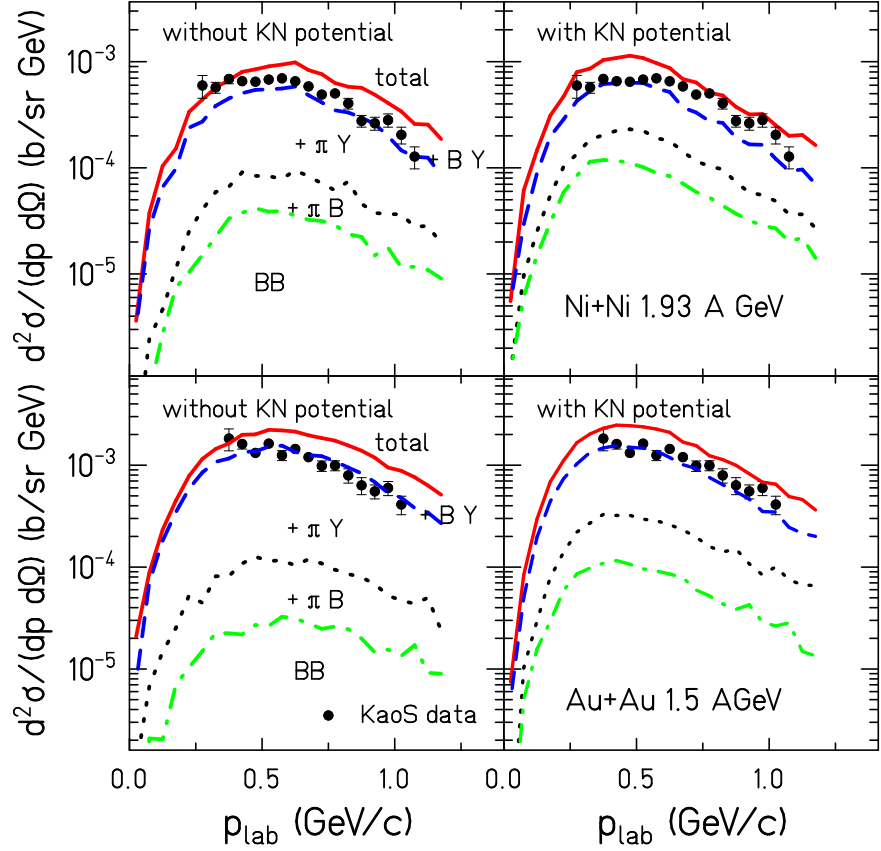,width=0.8\textwidth}
\caption{Contribution of various channels to the production of \km
in inclusive Ni+Ni (top) and Au+Au (bottom) collisions at
$\theta_{\rm lab} = 40^\circ$ with (right) and without (left) KN
potential. The respective contributions are cumulative. Hence each line
represents the sum of the channels below.  }
\label{spectra_potential}
\end{figure}

The \km spectrum at production is a convolution of the momentum
distributions of the $\Lambda (\Sigma)$ and of the $\pi$ or
baryon, respectively, weighted with the production cross section.
Figure~\ref{spectra_potential} shows the various channels which
contribute to the \km spectra. The sum is compared to the
experimental data of the KaoS Collaboration~\cite{Forster:2007qk}.
As the \km momentum distribution in the different channels does
not differ substantially, the form of the spectra does not allow
for conclusions on the production mechanism. On the left hand side
we have switched off both KN potentials in the calculation. As
expected from \Figref{potential}, the KN potential changes the
yield only marginally. The calculation reproduces quite well the
form of the spectra but over-predicts slightly the yield, as
already seen in \Figref{km_exc}.

\begin{figure}[htb]
\epsfig{file=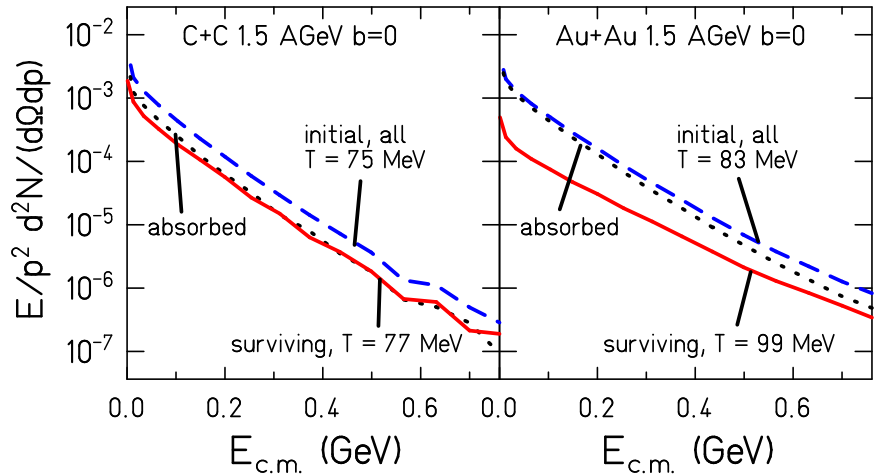,width=0.8\textwidth}
\caption{Spectral form of \km in central
collision of C+C (left) and Au+Au (right, yields are divided by 100)
at 1.5 \AGeV.} \label{km-spectra-effects-auc}
\end{figure}

The slope of the surviving \km at production is around 8 MeV higher
for Au+Au than for the C+C reactions at 1.5 \AGeV as shown in
\Figref{km-spectra-effects-auc}. This is due to the large fraction
of $N\Delta$ collisions in the strangeness-production process in
Au+Au which produces $\Lambda$ with a higher energy. After
production the shape of the \km spectra may be modified in several
ways:

(i) The absorption cross section is strongly momentum dependent.
Therefore, the slope of the surviving \km differ if reabsorption
becomes important. The influence of this effect is still small
for the C+C system but it becomes an important contribution to the
change of the slope in Au+Au reactions. Figure
\ref{km-spectra-effects-auc} illustrates the absorption process
showing the \km spectra at the initial production, the distribution
of the absorbed ones and that of the surviving ones. The
corresponding slopes are given in the figure.

\begin{figure}[htb]
\epsfig{file=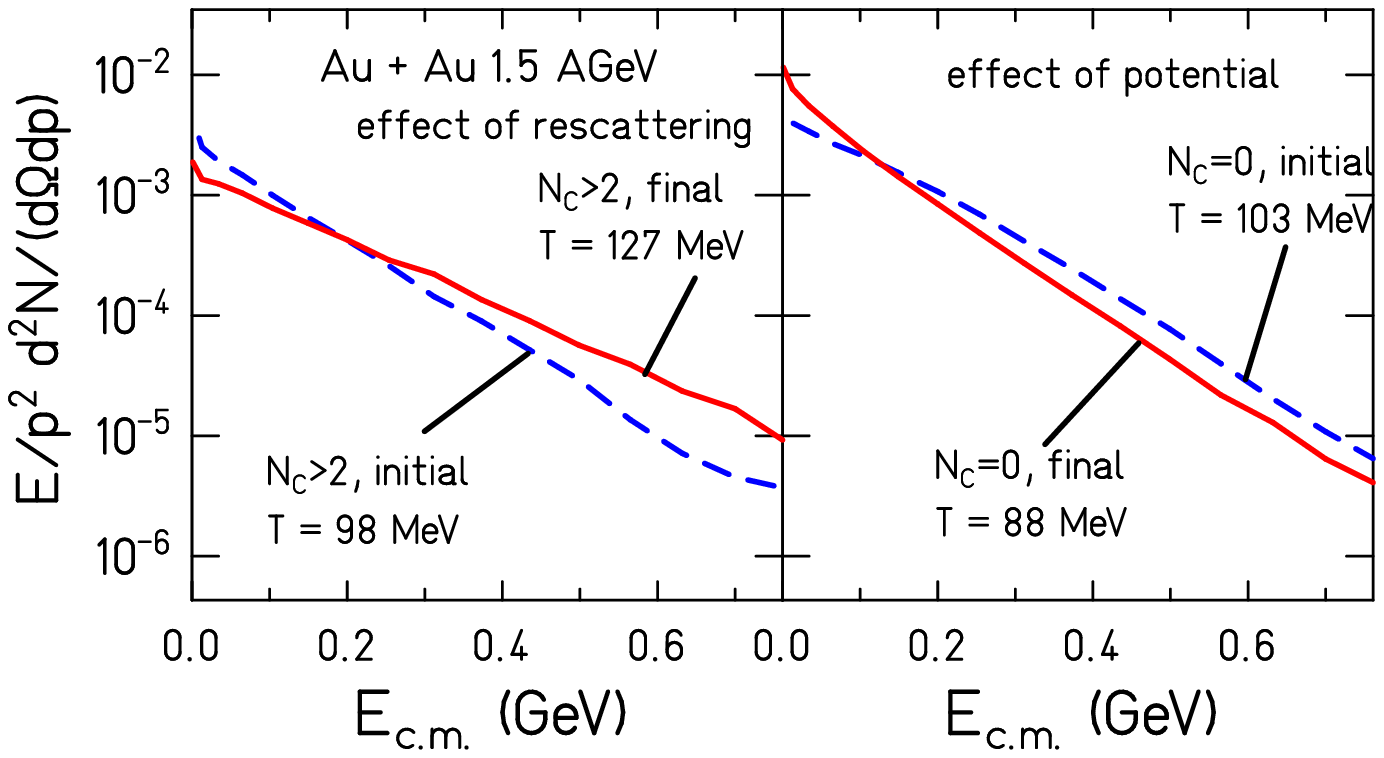,width=0.8\textwidth}
\epsfig{file=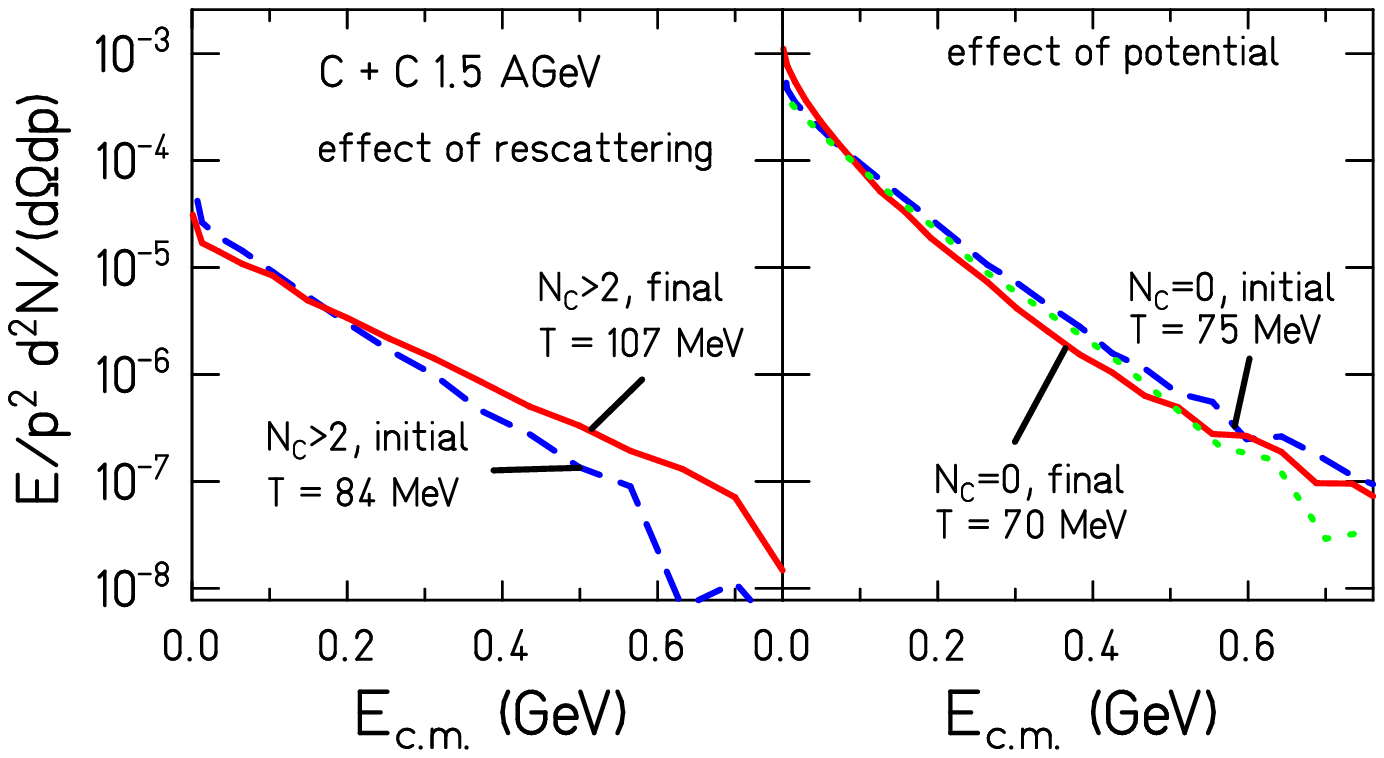,width=0.8\textwidth}
\caption{Influence of elastic rescattering and of the KN potential
on the \km spectrum for central Au+Au (top) and C+C (bottom)
collisions at 1.5 $A$ GeV. Left: Influence of the rescattering of
\km by selecting kaons which have scattered twice or more, showing
their initial and final distribution. Right: Influence of the KN
potential on the spectral shape demonstrated by selecting kaons
which never rescattered ($N_C$ = 0) and comparing the initial and
final spectra.} \label{km-au-spectra-effects}
\end{figure}

(ii) Elastic rescattering of the \km with the surrounding nucleons,
which have a larger inverse slope parameter, increases the \km
slope. This is seen on the left hand side of
\Figref{km-au-spectra-effects} which shows a more than twenty
percent increase of the slope of those \km which had at least two
rescattering collisions, independent of the system.

\begin{figure}[htb]
\epsfig{file=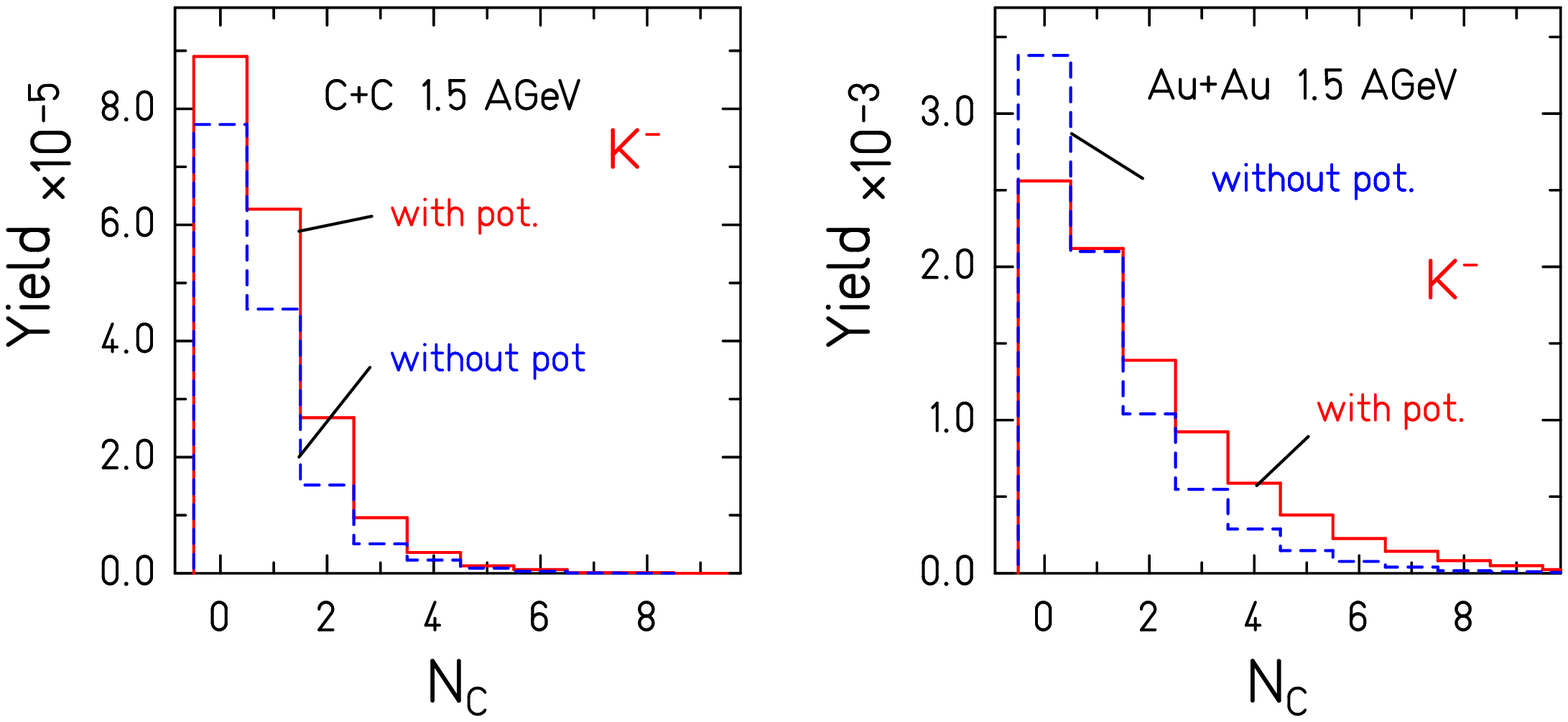,width=0.8\textwidth}
\caption{Distribution of the number of elastic collisions of \km in
IQMD simulations of C+C and Au+Au collisions.} \label{NC_AuAu_CC}
\end{figure}

(iii) The potential interaction reduces the \km momentum (opposite
to the \kp case) when it approaches the surface of the system
because energy is needed to bring the \km mass to its free value
and this energy is partially taken from the kinetic energy of the
\km . This leads to an enhancement at small momenta in the c.m.
system and to a change of the high-momentum slope, as demonstrated
on the right hand side of \Figref{km-au-spectra-effects}, where we
have selected \km which do not rescatter in order to separate this
change from that due to rescattering.
Because the distribution of the number of elastic rescatterings is different for
the different systems, shown in \Figref{NC_AuAu_CC}, the change of
the slope of the \km spectra due to rescattering is different in
Au+Au as compared to C+C collisions.

\begin{figure}[htb]
\epsfig{file=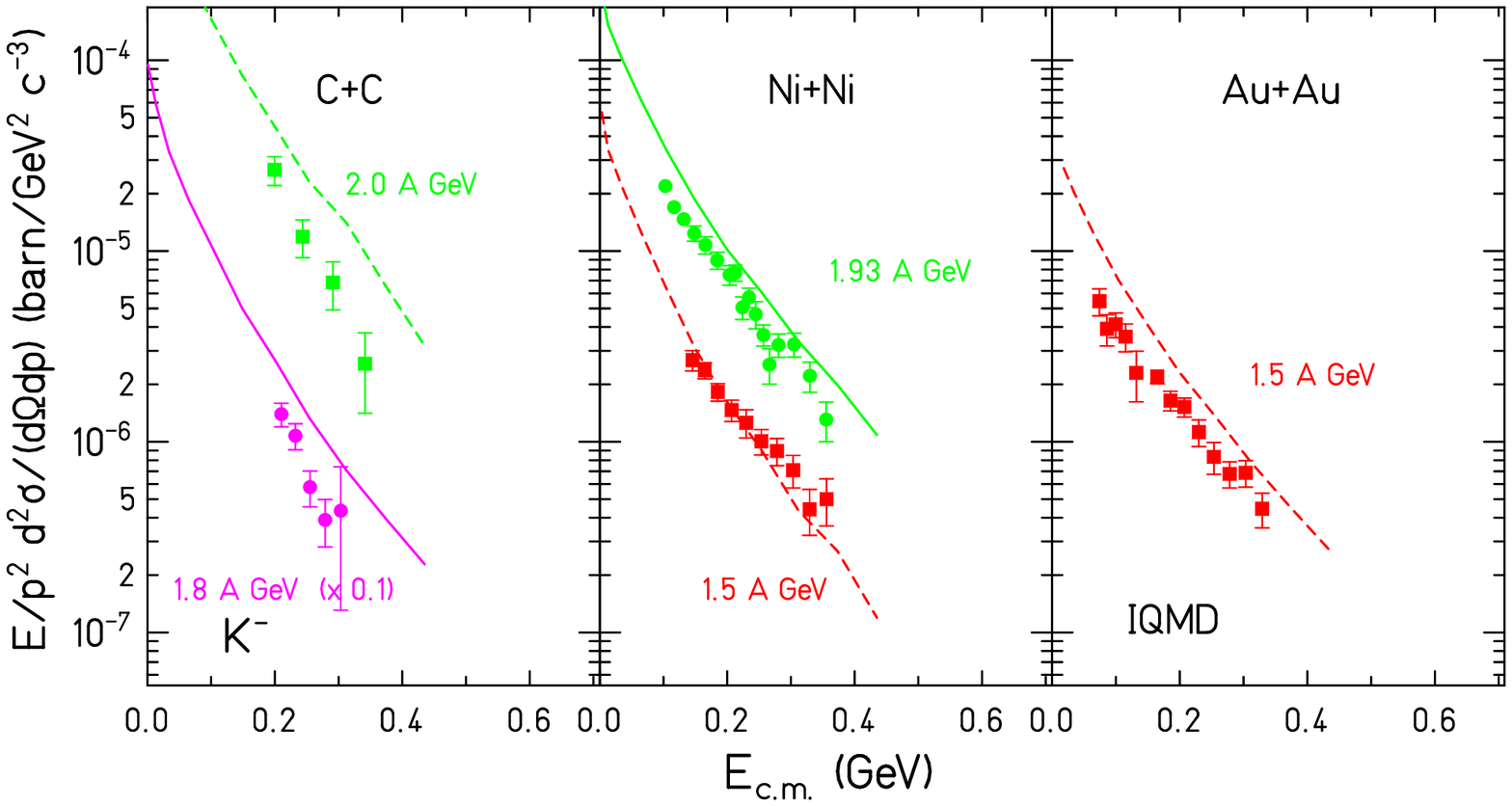,width=1.0\textwidth}\\
\epsfig{file=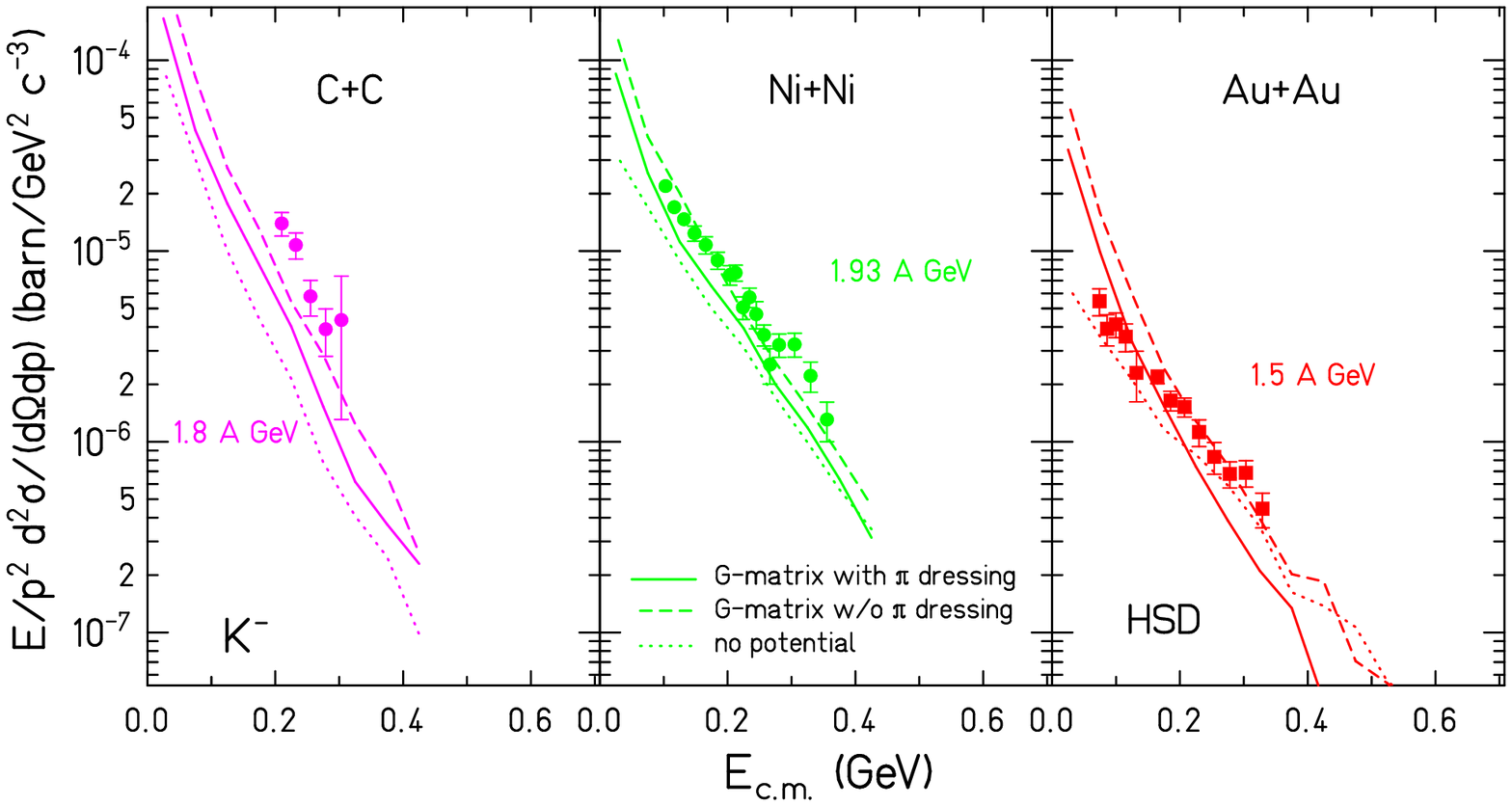,width=1.0\textwidth}\\
 \caption{Measured (symbols) and calculated (lines) inclusive
 invariant cross sections of \km mesons at mid-rapidity
 as a function of the kinetic energy $E_{\rm c.m.}$
 for Au+Au (right), Ni+Ni (middle) and C+C (left) reactions at various beam energies.
 On the top we display IQMD at the bottom HSD calculations. The mid-rapidity condition is a selection
 of $\theta_{\rm c.m.} = 90^\circ \pm 10^\circ$ both for the data and the calculations.
 The data are from the KaoS Collaboration~\cite{Forster:2007qk}.}
 \label{midrap_KM}
 \end{figure}

Figure \ref{midrap_KM} (top) compares the spectra, which have been
measured by the KaoS Collaboration~\cite{Forster:2007qk}, with
IQMD (upper part) and HSD calculations \cite{Cassing:2003vz}
(lower part). Both theories reproduce the slope of all three
systems quite well although they use quite a different potential:
IQMD employs a mean field potential (Eq.~\ref{schaf}) whereas in
HSD different versions of a G-matrix approach are used (for
details see \cite{Cassing:2003vz}). Both approaches use also a
different description of the in-medium modification of the \km
cross section. IQMD (HSD) calculation have the tendency to
overpredict (underpredict) slightly the yield. Calculations using
the same potential and the same \km cross section are not
available.
\begin{figure}
\epsfig{file=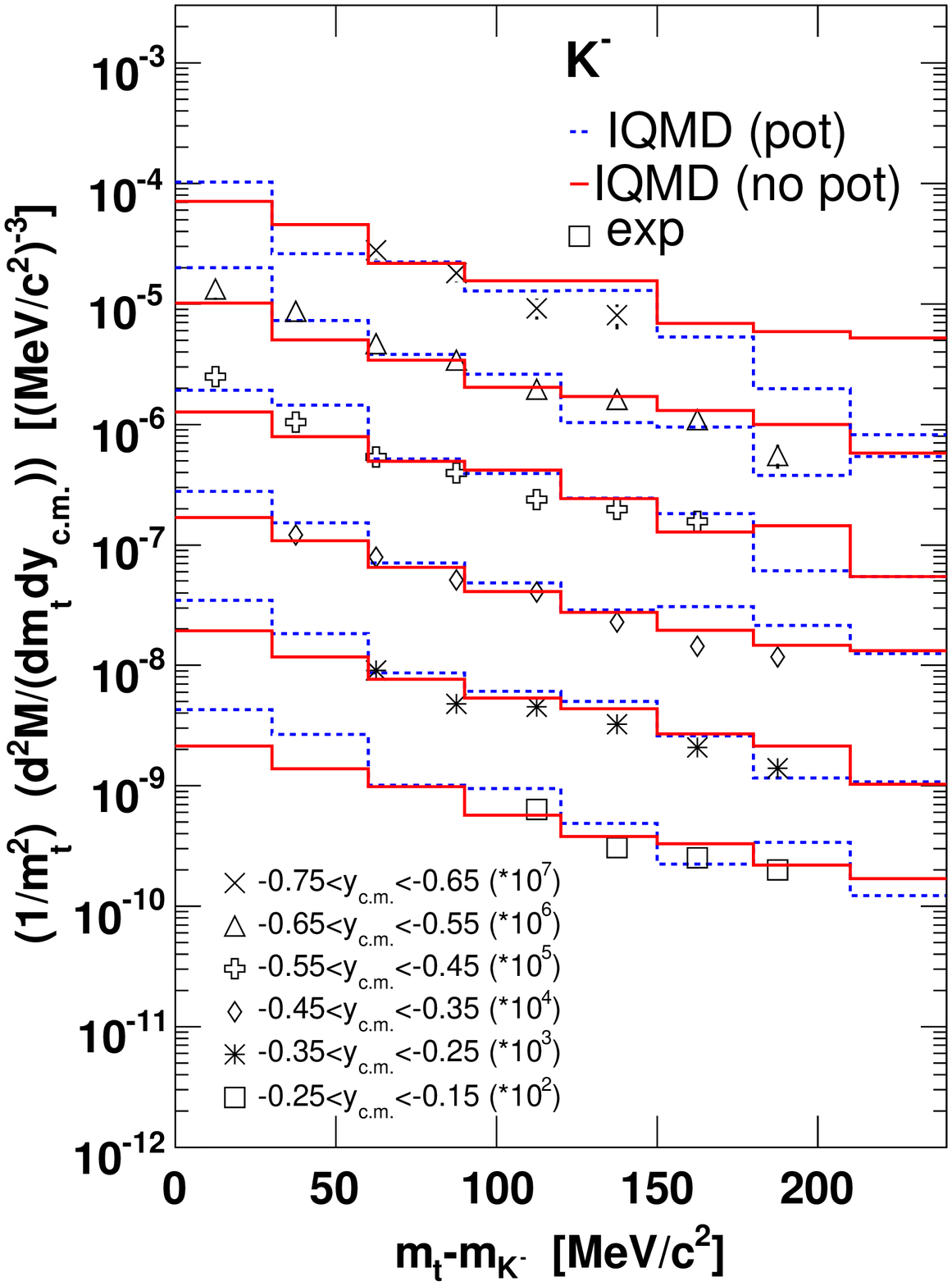,width=0.4\textwidth}
 \caption{\km transverse mass spectrum for semi-central
collisions of  \mbox{Ar+KCl} at $1.75$ \AGeV measured by the HADES
Collaboration~\cite{Agakishiev:2009ar} as compared with IQMD
calculations with active and inactive KN potential. The spectra
are normalized to $K^0$ spectra.} \label{spec_hades}
\end{figure}

The HADES Collaboration has recently measured \km spectra as a
function of $m_t-m_0$ for different rapidity intervals for the
reaction  Ar+KCl at 1.75 \AGeV \cite{Agakishiev:2009ar}. Their
results are compared  with IQMD calculations with and without KN
potential in \figref{spec_hades}. As the other HADES spectra in
this report, this spectrum is normalized to the K$^0$ yield
because an absolute normalization is not available. All \km
spectra have an exponential form at high momenta if plotted as
$(E/{p^2}) ({\rm d}\sigma/{\rm d} p)$ or as $(1/m_t^2){\rm
d}^2M/({\rm d}m_t {\rm d}y$).

As shown in \figref{spec_hades}, the inverse slope parameter of
the \km changes if a KN potential is employed, as could already be
inferred from \figref{km-au-spectra-effects}. The HADES
collaboration has parameterized the measured inverse slope
parameter $T$ as a function of the rapidity by $T(y) =
T_0/\cosh(y)$ with $T_0 = 69\pm2$. Such a form is expected if the
emission is isotropic in the center-of-mass system. IQMD
calculation, including (excluding) \km N potential yields 80 (62)
MeV.

\begin{figure}
\epsfig{file=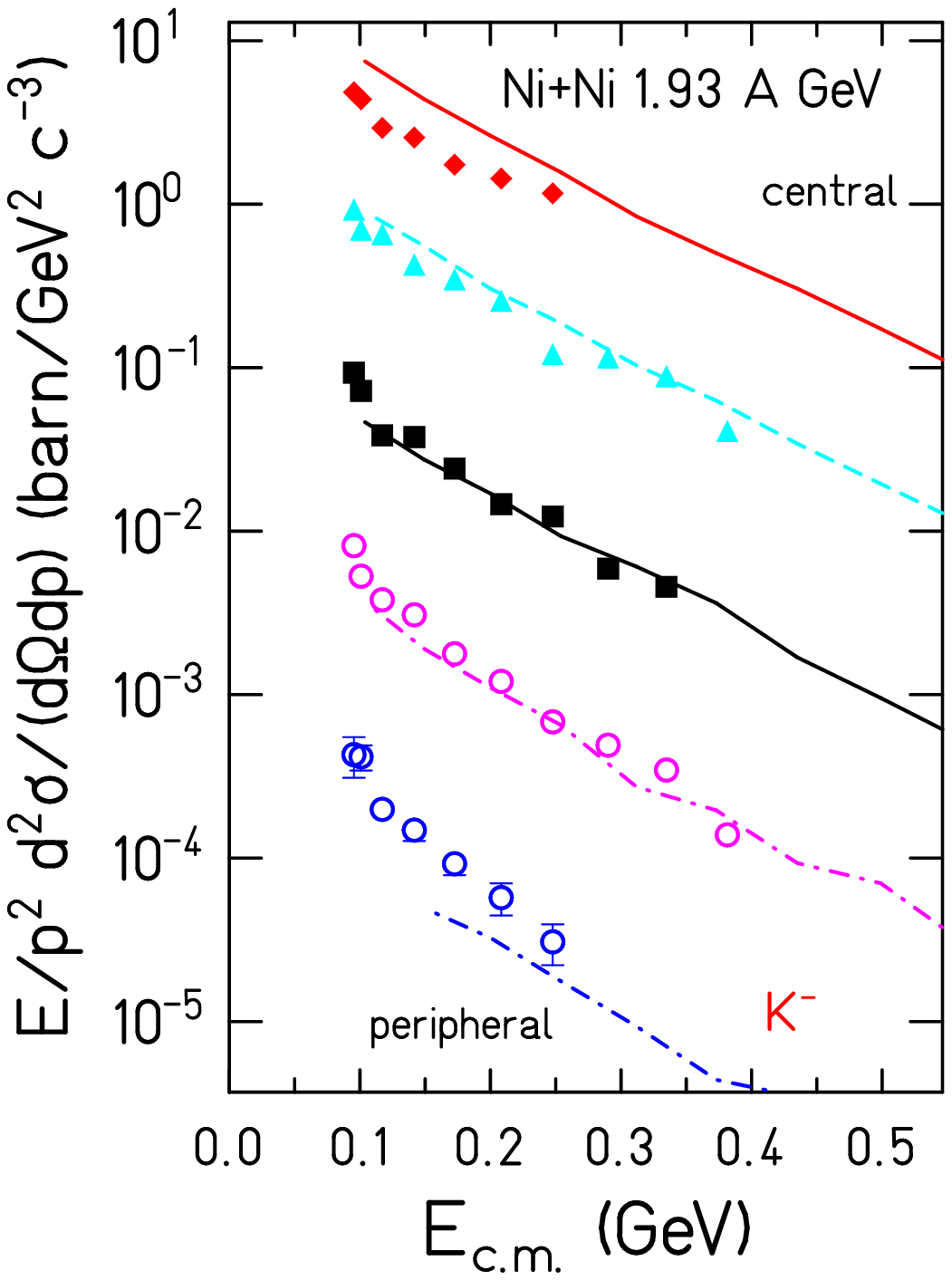,width=0.45\textwidth}
\epsfig{file=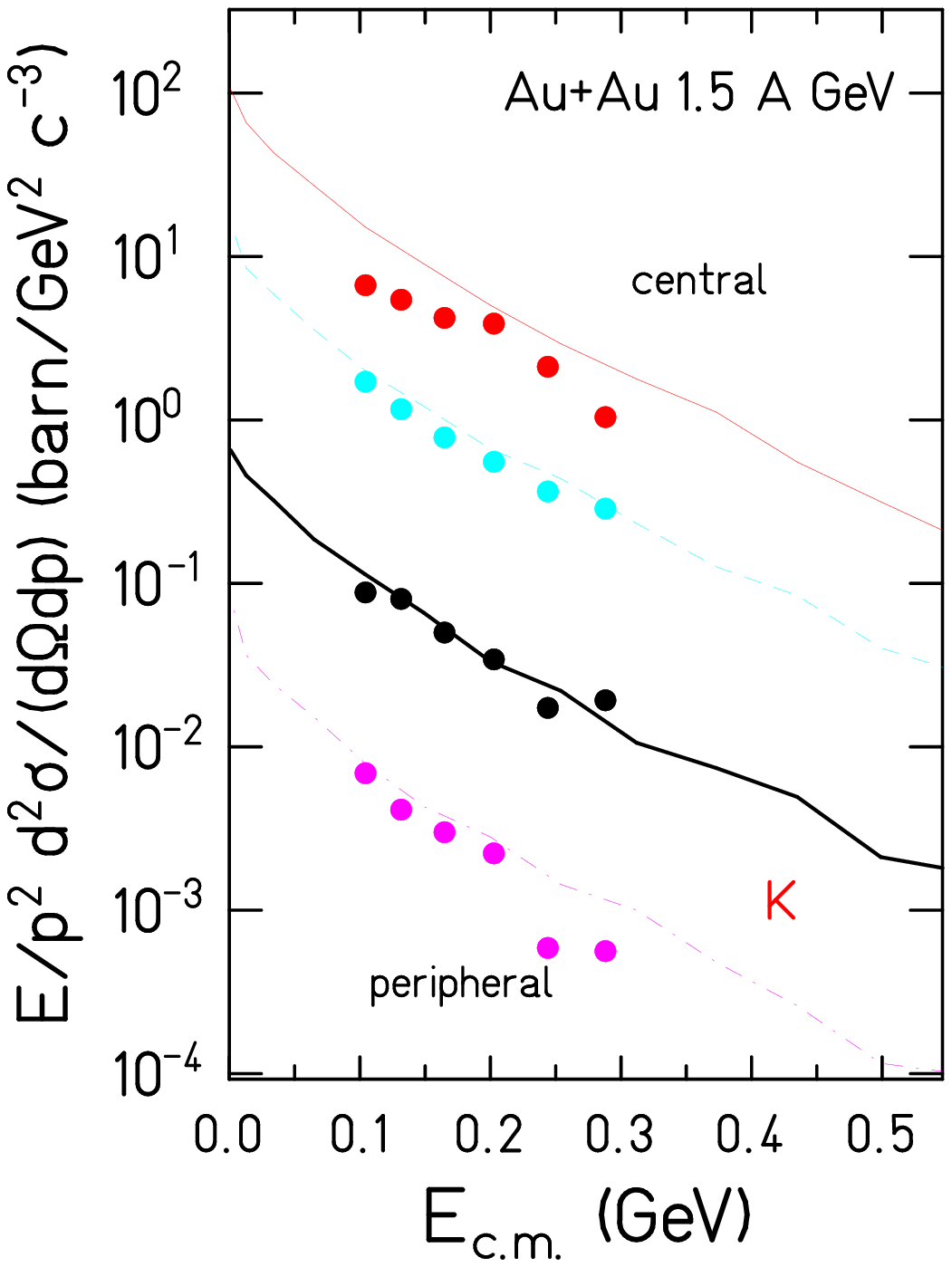,width=0.45\textwidth} \caption{
Centrality selected energy spectra for \mbox{Ni+Ni} at
$1.93$~$A$~GeV and Au+Au at 1.5 \AGeV measured at $\theta_{\rm
lab} = 40^\circ$  by the KaoS Collaboration~\cite{Forster:2007qk}
as compared with IQMD calculations.} \label{spec_cent_km}
\end{figure}

The KaoS Collaboration has measured centrality-selected spectra
for the Ni+Ni and Au+Au systems~\cite{Forster:2007qk}, which are
displayed and compared to IQMD calculations in
\Figref{spec_cent_km}. Except for the peripheral Ni+Ni data the
slopes are reproduced but the calculation overpredicts the data
for the most central collisions. For mid-central events the
calculation reproduce the data quantitatively.

Similarly as has been observed for the system-size dependence, due
to rescattering and absorption the theoretical inverse slope
parameters of the surviving \km mesons vary for heavy systems with
centrality as well. This is summarized  in \Figref{dyn-km-apart}
for Au+Au at 1.5 \AGeV and compared to the KaoS
data~\cite{Forster:2007qk}. The slope of the \km spectra at the
moment of their production is determined by the elementary
production process and quite independent of the centrality.
Selecting only those \km mesons which survive the heavy-ion
reaction, an inverse slope parameter which increases with
centrality, can be observed (\Figref{km-spectra-effects-auc}) due
to the momentum dependence of the absorption cross section.
Elastic \km collisions increase the slope even further until
finally the potential interaction with the nucleons leads to a
reduction. The final slope is in good agreement with experiment.

\begin{figure}[hbt]
\epsfig{file=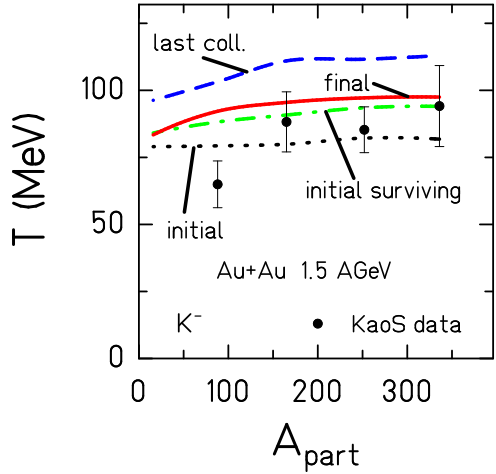,width=0.5\textwidth}
\caption{Centrality dependence of the inverse slope parameters of
the \km spectra showing the influence of the different effects and
compared to the data of the KaoS
Collaboration~\cite{Forster:2007qk}. The final curve includes in addition
the \km N potential interaction after the last collision.} \Label{dyn-km-apart}
\end{figure}

\subsection{Rapidity distribution}

\begin{figure}[htb]
\epsfig{figure=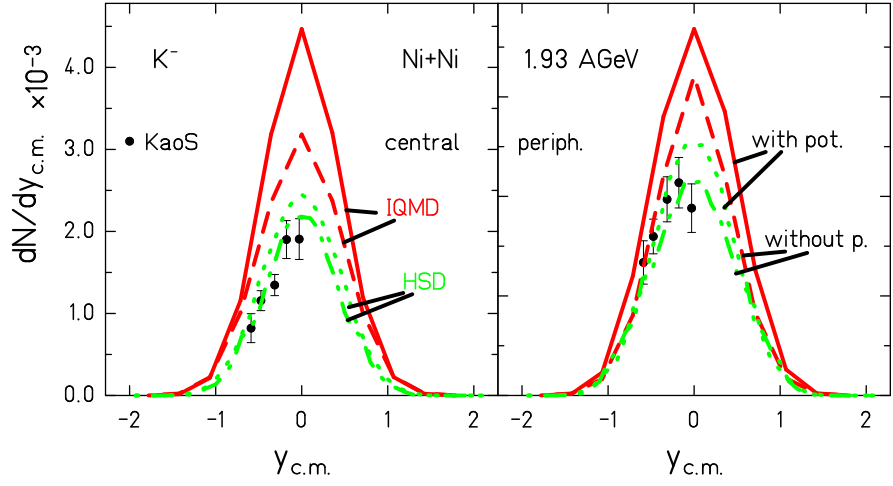,width=0.8\textwidth}
\caption{The rapidity distribution of \km mesons  for central
($b\leq 4.5$~fm, left) and peripheral ($b=4.5 - 7.5$~fm, right)
Ni+Ni reactions at 1.93 \AGeV. The results from the KaoS
Collaboration~\protect\cite{Menzel:2000vv} are compared with IQMD
(red solid and dashed lines) and HSD \cite{Cassing:2003vz} (green
dotted and dashed-dotted) calculations. The solid and dotted lines
correspond to the calculations including the $K^+$ and $K^-$
potential, while the dashes and dash-dotted lines show the results
without KN potential. Both calculations use different \km potentials (see text).}
\label{ele_yNi19-CP}
\end{figure}

As for the \kp mesons, the measured spectra are extrapolated to
small $p_t$ momenta 
in order to obtain rapidity density distributions.
The result for 1.93 \AGeV Ni+Ni is shown in
\figref{ele_yNi19-CP} and compared with theory, separately for
central and peripheral collisions. Both theories, HSD as well as
IQMD, show that the \km N potential has no influence on the form of
the distribution but it enhances the total yield considerably. The
HSD \cite{Cassing:2003vz} calculation reproduces the 1.93 \AGeV
Ni+Ni data whereas the IQMD
calculation 
overestimates the yield by a roughly a factor of 1.5.

\subsection{Polar distributions}

\begin{figure}[htb]
\epsfig{file=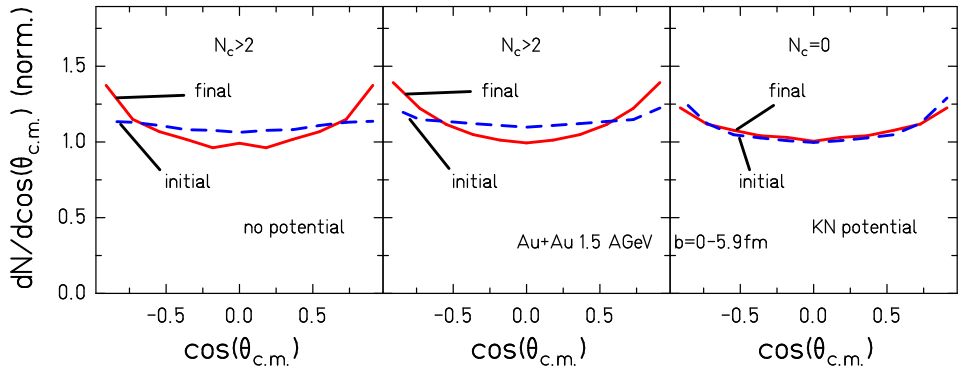,width=.85\textwidth}
\caption{Influence of rescattering and of the KN potential on the
polar distribution of surviving \km for Au+Au reaction at 1.5 $A$ GeV. The
distribution marked 'initial' is that of surviving \km.}
\label{KM_theta}
\end{figure}

Similar as for the \kp mesons, the polar distribution of the \km
is rather flat at the moment of their production   if averaged
over their momenta. 
The distribution is also flat if one selects only those \km mesons
which survive. Thus, the observed anisotropy is due to the
interactions of the \km with the system. Elastic rescattering of
the surviving \km brings the polar distribution closer to that of
the scattering partners and therefore leads to an enhancement in
forward and in backward direction. It is independent of the
strength of the KN potential as can be seen comparing the left and
the middle parts of \figref{KM_theta} which show calculations for
$b < 5.9$ fm Au+Au reactions at 1.5 \AGeV. Choosing the surviving
\km which did not rescatter ( right panel), one selects those with the shortest
path through the matter. This path depends on the geometry and
therefore, by selecting those particles, one creates an anisotropy
already at production. Again, the potential has no influence on
the distribution.

\begin{figure}[htb]
\epsfig{file=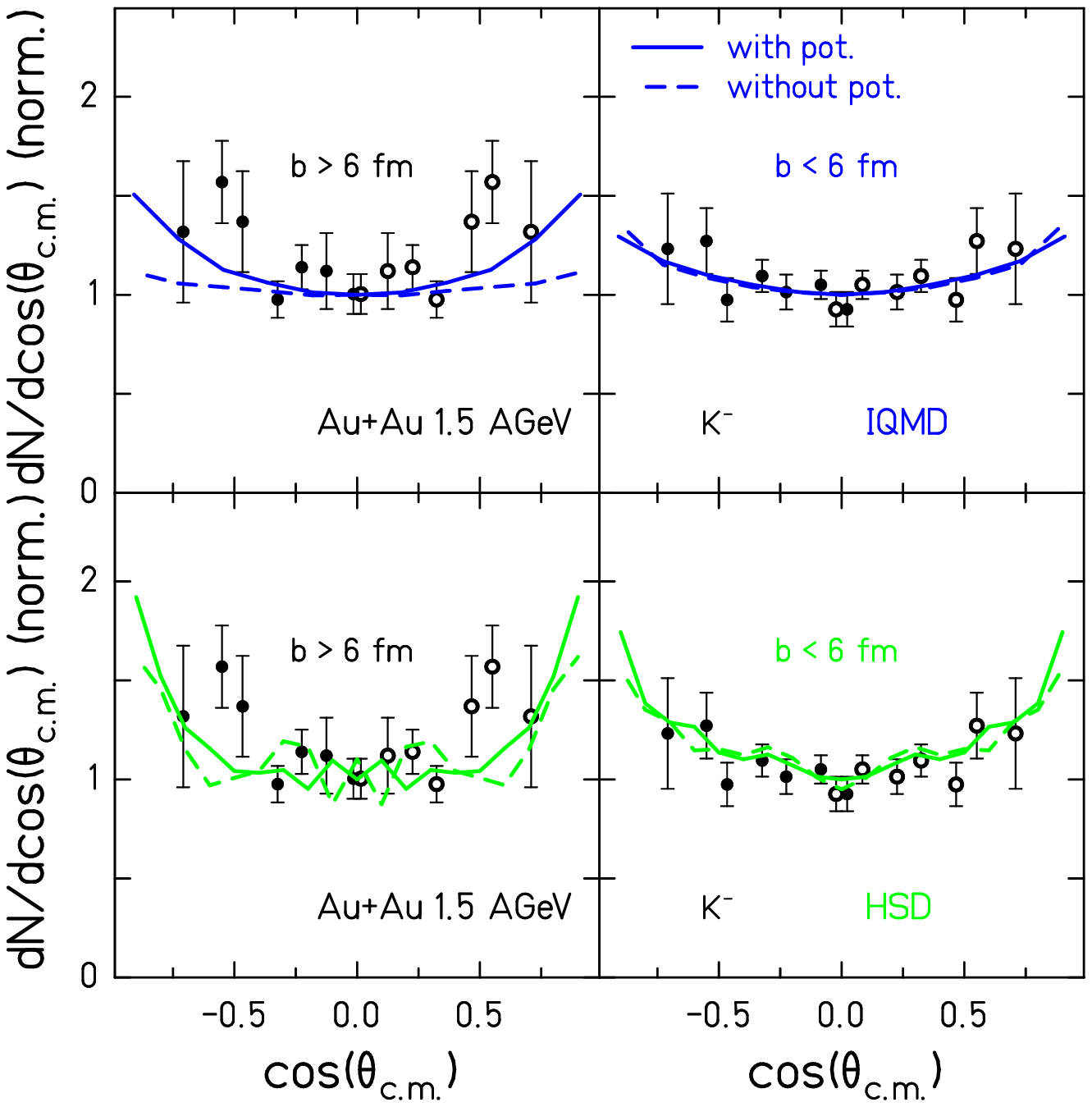,width=0.65\textwidth}
\epsfig{file=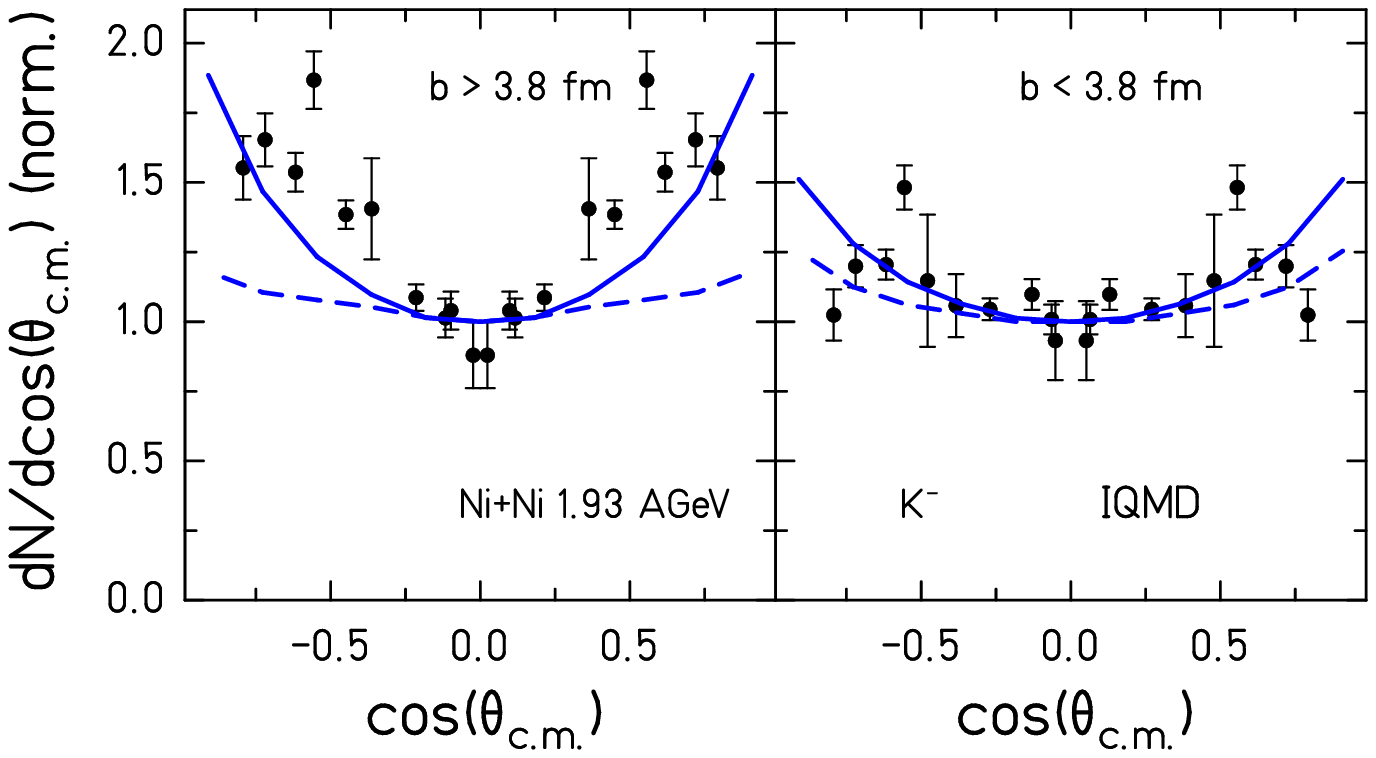,width=0.65\textwidth}
 \caption{The measured polar angle distribution of \km mesons for Au+Au at 1.5 $A$ GeV
(upper part) and for Ni+Ni at 1.93 $A$ GeV (lower part)
 for two impact parameter ranges are
compared to IQMD and HSD \cite{Cassing:2003vz} (only for Au+Au)  calculations.
The \km N potential in both approaches is different (see text).
The data are from the KaoS Collaboration~\cite{Forster:2007qk}. The
solid (dashed) lines refer to calculations with (without) KN
potential. The distributions are normalized to unity for $\cos
\theta_{\rm c.m.} = 0 $.} \label{angdis_data_calc}
\end{figure}

Figure~\ref{angdis_data_calc} confronts IQMD and HSD
\cite{Cassing:2003vz} calculations with experiment. On the left
(right) hand side, we display the results for peripheral
(semi-central) collisions. The upper figure shows Au+Au
collisions at 1.5 \AGeV, the lower figure Ni+Ni collisions at 1.93
\AGeV. For central collisions, data and theory agree and both
theoretical approaches confirm that the KN potential is without
any influence on the distribution for heavy systems, as discussed
before. For the lighter Ni+Ni system we observe a small influence
of the potential in central collisions. For peripheral reactions
the potential is more influential but the present error bars are
too large in order to draw firm conclusion from the comparison of
theory and experiment, may be with the exception of the Ni data.

\subsection{Azimuthal distribution}

\subsubsection{In-plane flow $p_x(y)$ and $v_1$}

\begin{figure}[hbt]
\epsfig{file=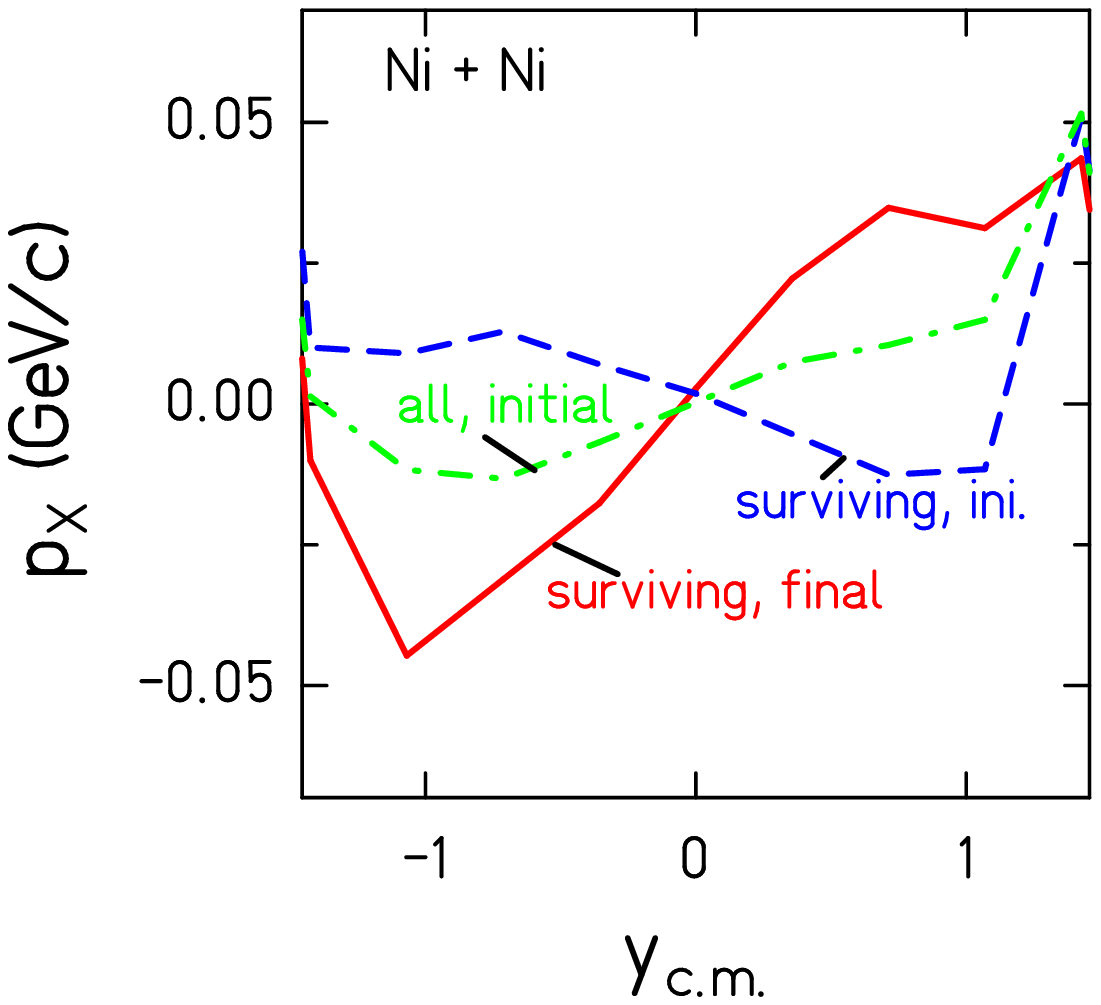,width=0.4\textwidth}
\epsfig{file=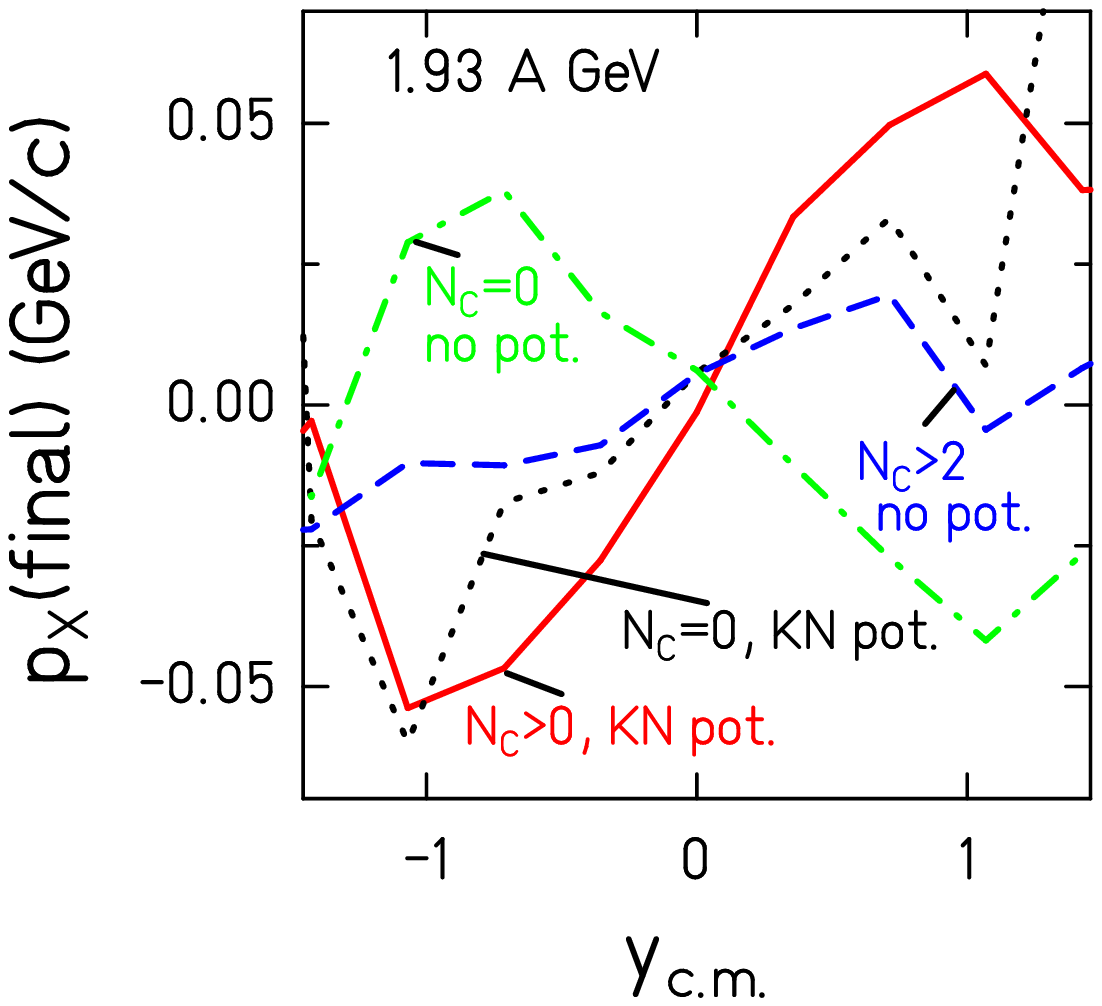,width=0.4\textwidth}
\caption{In-plane flow $<p_x>(y_{\rm c.m.})$ of \km in Ni+Ni
reactions at 1.93 $A$ GeV. Left: Influence of the absorption cross
section on the in-plane flow. Right: Influence of potential and
rescattering collisions on the in-plane flow.}
\label{kmpx}
\end{figure}

The in-plane flow of \km mesons is, similar to the inverse slope
parameter of the spectrum, a sensitive probe for the different
processes which the \km suffer from creation to observation. The
left hand side of \Figref{kmpx} displays the in-plane flow as a
function of the rapidity $y_{\rm c.m.}$. Already at production a
small in-plane flow can be observed. Absorption modifies this flow
substantially. Those \km which flow with the nucleons have, due to
geometry, a much higher probability to get absorbed. Therefore,
the surviving \km mesons show initially an anti-flow. Subsequent
elastic collisions bring the distribution closer to that of the
baryons.  In contradiction to the repulsive \kp N interaction, the
attractive \km N interaction pulls the \km also towards the
baryons. The influence of the rescattering collisions is less
important than that of the potential, see \Figref{kmpx}, right.

\begin{figure}[hbt]
\epsfig{file=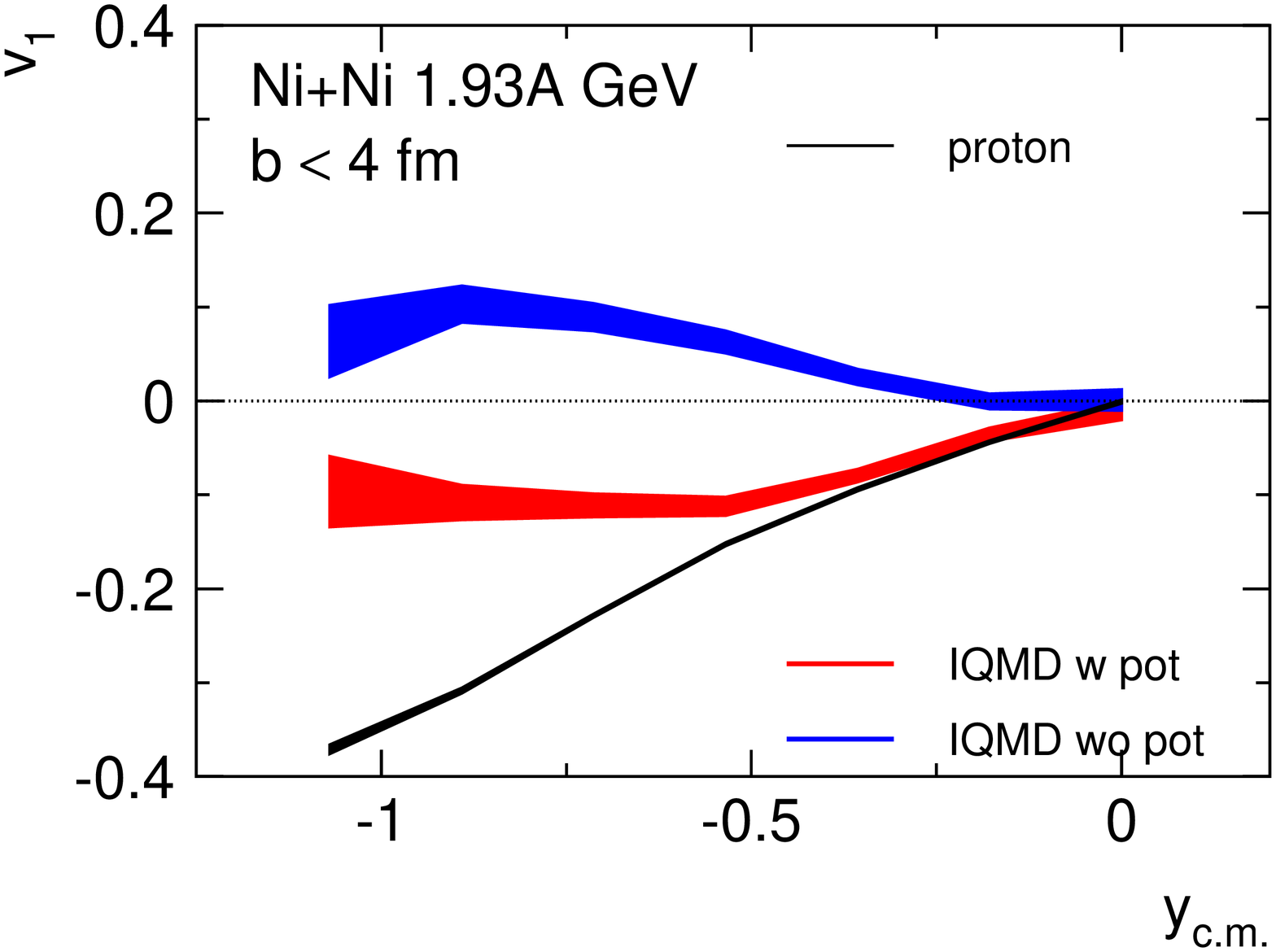,width=0.4\textwidth}
\epsfig{file=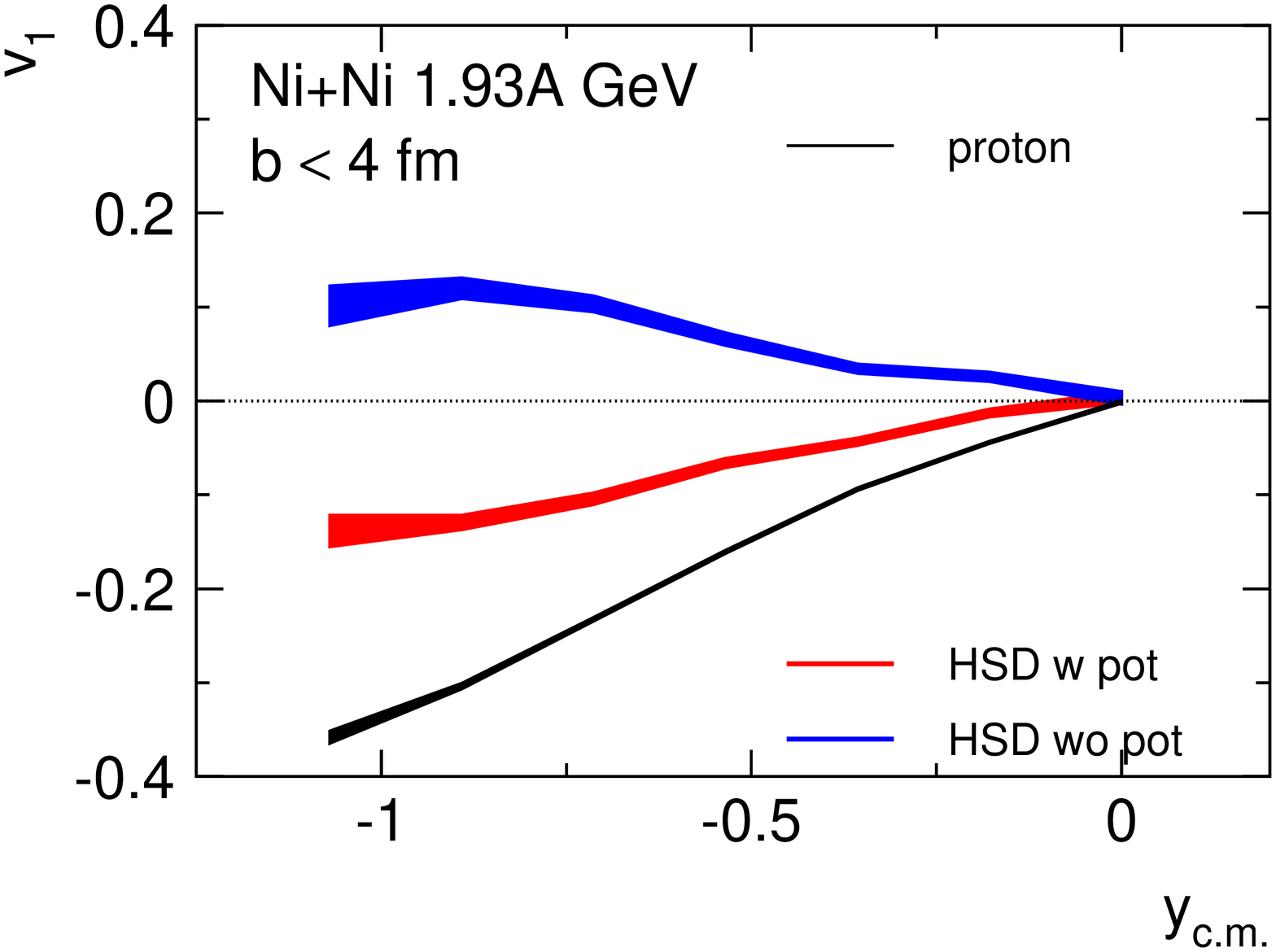,width=0.4\textwidth}
\caption{IQMD (left)  and HSD (right) predictions of $v_1$ as a function of the rapidity
for central Ni+Ni collisions at 1.93 \AGeV with and without
KN-potential. The results of IQMD predictions are filtered for the
acceptance of the new time-of-flight barrel of the FOPI detector.}
\label{v1-km-fopi}
\end{figure}

In Fig.~\ref{v1-km-fopi}, IQMD and HSD predictions for $v_1$ as a
function of rapidity are shown for Ni+Ni collisions at 1.93 \AGeV.
The calculations are filtered by the FOPI acceptance. Without the
\km N potential, $v_1$ (defined in
Eq.~\ref{fourier}) of the \km is positive and opposite to that of
the nucleons.  The attractive $\overline{\rm{K}}$N potential
pushes the \km closer to the nucleons. Consequently the flow of
the \km changes sign and becomes negative.

\subsubsection{Out-of-plane flow $v_2$}

\begin{figure}[htb]
\epsfig{file=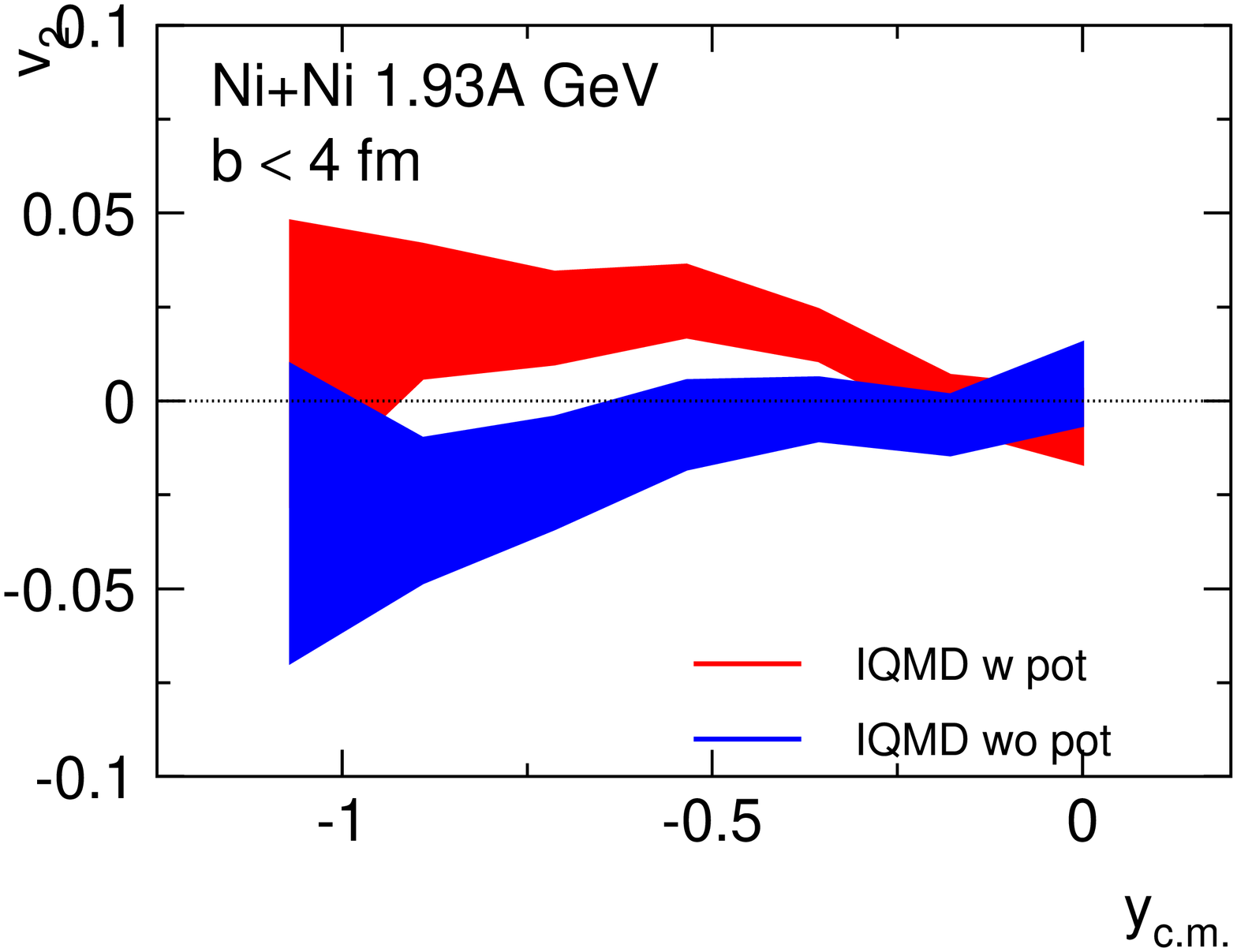,width=0.4\textwidth}
\epsfig{file=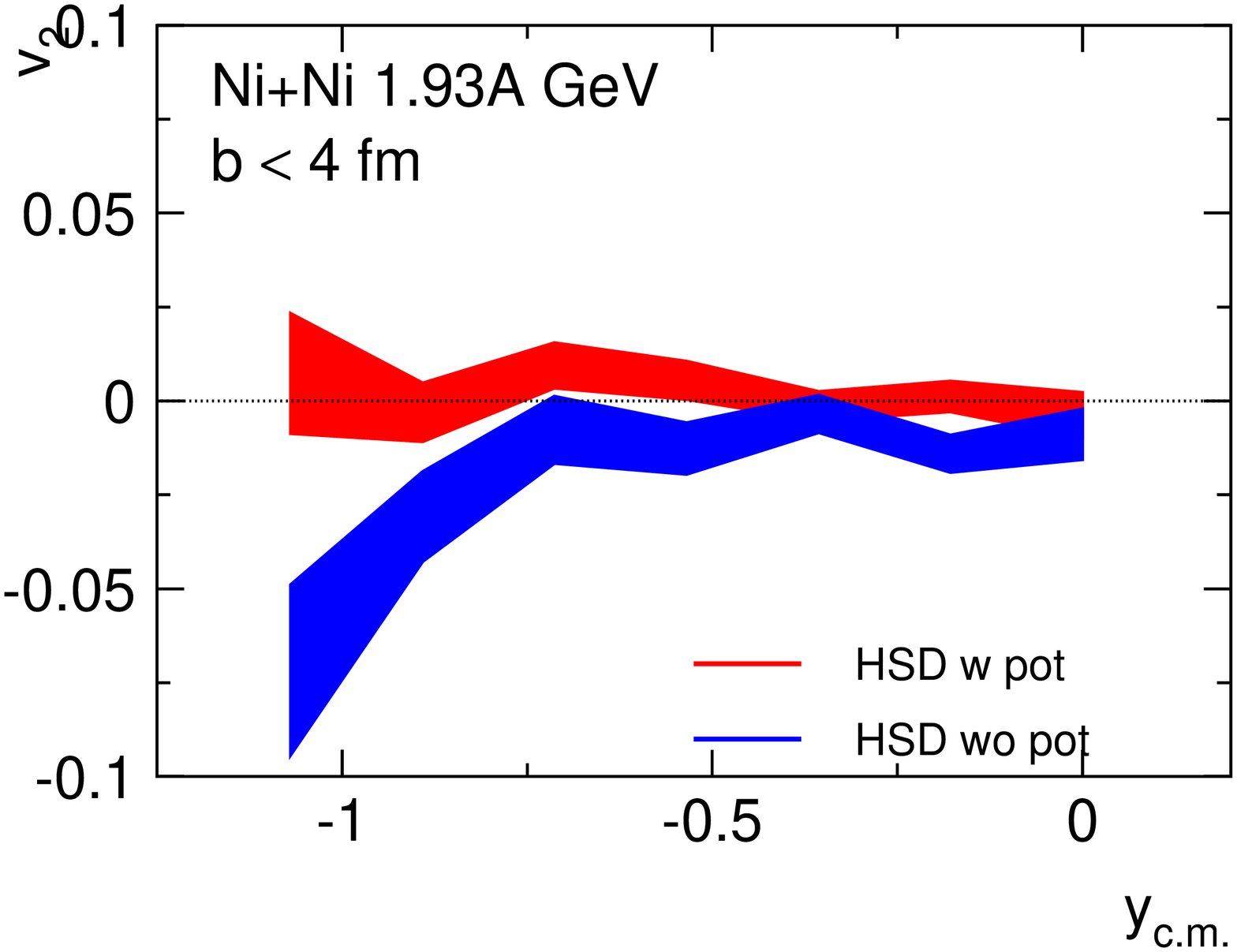,width=0.4\textwidth}
\caption{IQMD and HSD predictions for $v_2$ as a function of rapidity for
central collisions of Ni+Ni at 1.93 \AGeV. The results are filtered with
the FOPI acceptance filter.}
\label{v21-fopi}
\end{figure}

IQMD and HSD predictions for $v_2$ (defined in Eq.~\ref{fourier}) as a
function of rapidity for Ni+Ni collisions at 1.93 \AGeV within the
FOPI acceptance are displayed in \figref{v21-fopi}. Error bars for
the model predictions are rather large, but there is a tendency of
an in-plane enhancement of the elliptic flow , $v_2 >0$, at large rapidities
which is reduced if an  attractive $\overline{\rm{K}}$N potential.
Closer to mid-rapidity there is no influence of the potential on
the elliptic flow.

\begin{figure}[hbt]
\epsfig{file=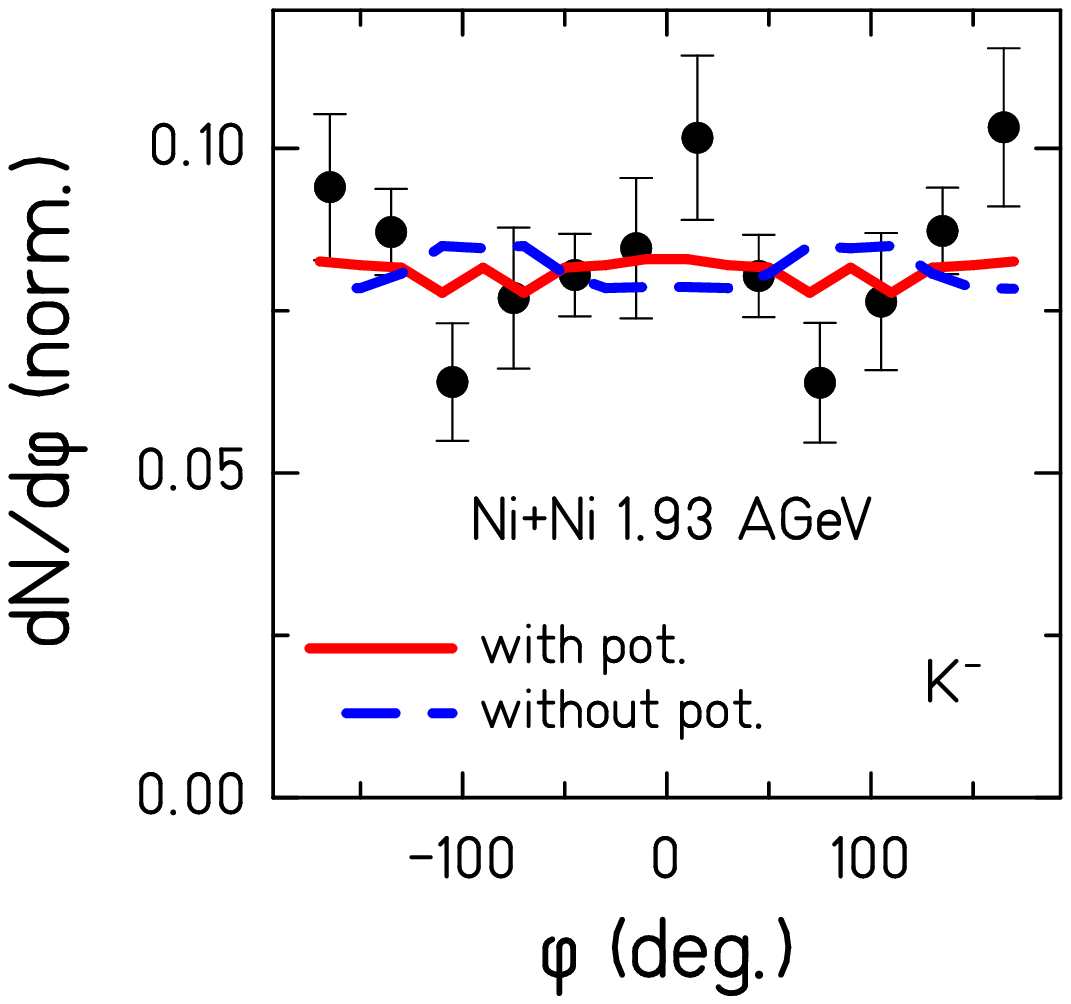,width=0.55\textwidth}
\caption{Azimuthal distribution of d$n$/d$\phi$ for Ni+Ni
reactions at 1.93 \AGeV from the KaoS
Collaboration~\cite{Uhlig:2004ue} as compared with IQMD
calculations with and without KN potential. The data are obtained
for impact parameters between 3.8 fm and 6.5 fm, at rapidities of
0.3 $< y/y_{\rm beam}$ $<$ 0.7 and for momenta between 0.2 and 0.8
GeV/$c$.} \Label{ni-v2}
\end{figure}

The azimuthal distribution of d$n$/d$\phi$ has been measured by
the KaoS Collaboration~\cite{Uhlig:2004ue} and is presented in
\figref{ni-v2} in comparison with IQMD calculations. The
calculations show no visible influence of the $\overline{\rm{K}}$N
potential and the error bars of the data are too large to verify
whether the distribution is as flat as theory predicts or whether
a small in-plane enhancement $(v_2>0)$ is observed.

\begin{figure}[hbt]
\epsfig{file=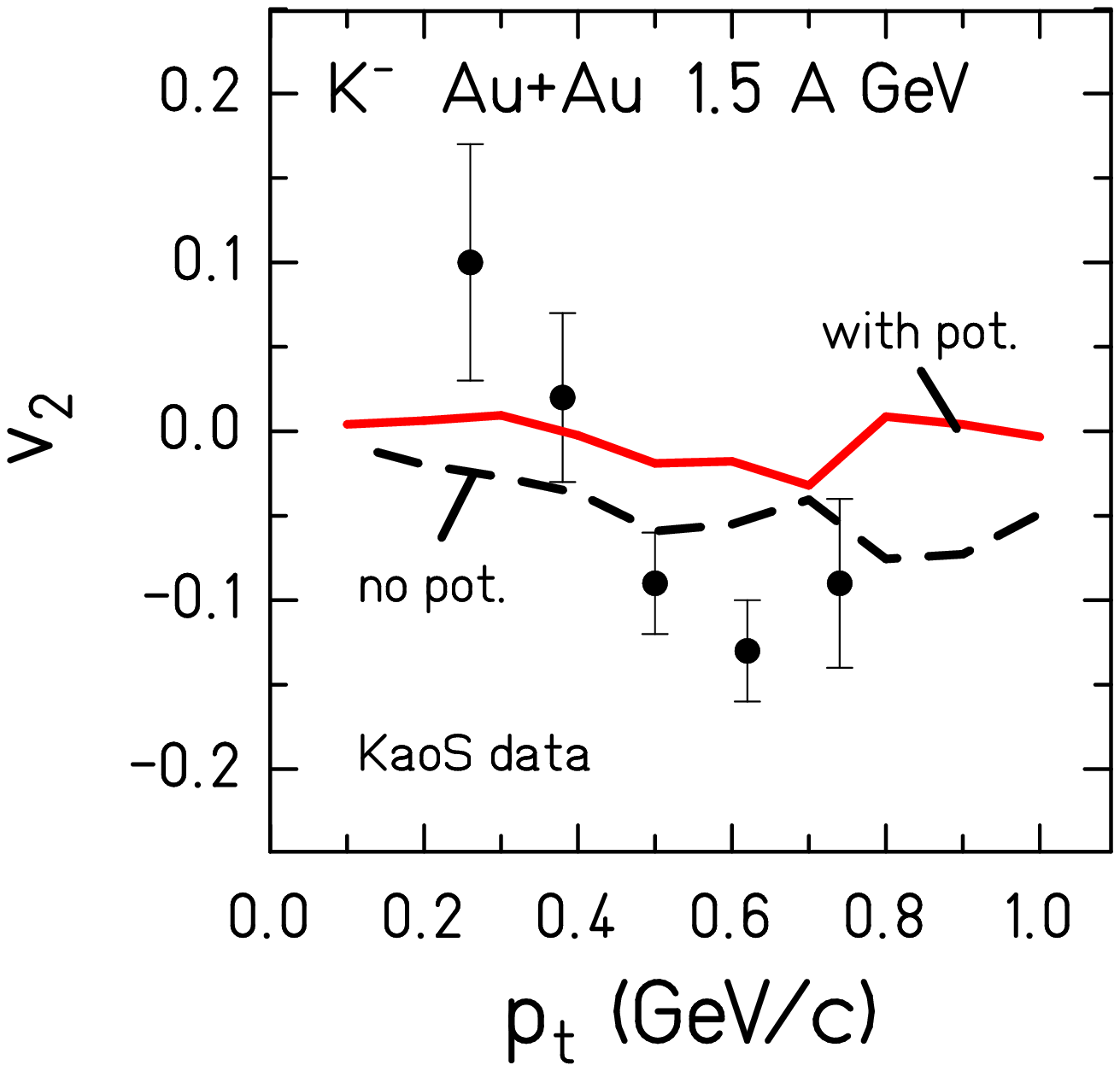,width=0.49\textwidth}
\caption{
Experimental $v_2(p_t)$ distribution measured by the KaoS
Collaboration~\cite{Ploskon:2005qr} for non-central Au+Au reactions at 1.5
\AGeV as compared with IQMD calculations with and without KN
potential.} \Label{au15-v2-pt-km}
\end{figure}

The differential $v_2(p_t)$ for the heavier Au+Au system at 1.5
\AGeV, measured by the KaoS Collaboration~\cite{Ploskon:2005qr}
and presented in \figref{au15-v2-pt-km}, shows a strong $p_t$
dependence which is not reproduced by the model. Non-central
collisions are selected by excluding the 20\% most central
collisions. This corresponds to impact parameters larger than 6.4
fm. The difference between the calculations including and
excluding a \km N potential is small. The reason for this
small effect is likely the late emission of the \km. Then the
spectators are gone and cannot absorb the \km emitted in the
reaction plane. A relation between emission time and out-of-plane
enhancement can be seen from Fig.~\ref{au15-km-phi-discussion}.
Early (0-8 fm/$c$) emitted \km  exhibit indeed an out-of-plane
enhancement evidencing the absorption in the spectator matter. In
the model calculations the bulk of the \km are emitted after 8
fm/$c$ which results in a rather flat distribution. One might then
speculate whether the observed positive $v_2$ values for low-$p_t$
\km indicate an earlier emission than in the model calculations.
HSD including off-shell propagation of \km and a G-Matrix approach
for the \km spectral function \cite{Cassing:2003vz} predicts a
stronger transverse momentum dependence for $v_2$. The details of
the distributions in particular the change of sign at low $p_t$ is
not described, however.

In conclusion, $v_2$ appears to be rather insensitive to the KN
potential and does not change much as a function of the rapidity. The
in-plane flow $<p_x(y)>$ or $v_1(y)$ is, on the contrary, quite sensitive
to the potential but also to the reabsorption cross section.

\begin{figure}[hbt]
\epsfig{file=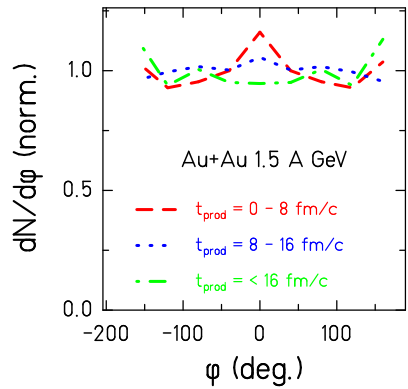,width=0.4\textwidth} \caption{
Azimuthal distribution selected according to emission time}
\Label{au15-km-phi-discussion}
\end{figure}

\clearpage
\section{Discussion and Summary}
\setcounter{figure}{0}

\subsection{Differences and similarities of \kp and \km emission}

\label{differences}

\begin{figure}[bht]
\epsfig{file=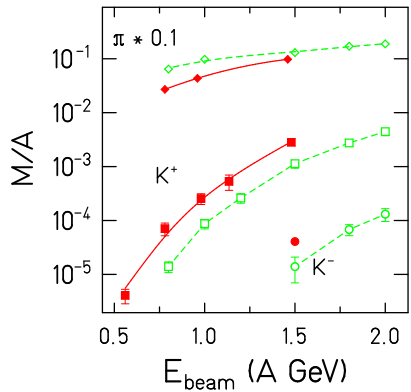,width=.4\textwidth}
 \caption{Multiplicity  per $A$ for \kp, \km and pions (sum of all three
 species). The dashed (green) lines connect the results for C+C, the full (red) lines those of
 Au+Au. Data are from Ref.~\cite{Forster:2007qk}.}
\label{pikpkm-ebeam}
\end{figure}

Having demonstrated the quite different production mechanisms for
the anti-strange (Section IV) and strange (Section V) mesons we
now confront the meson observables directly. In
Fig.~\ref{pikpkm-ebeam} we display the experimental excitation
function of the multiplicity per projectile nucleon for \kp and
\km as well as that of $\pi$ observed in  inclusive collisions of
C+C (dashed lines) and Au+Au (full lines). For both, the \kp and
\km mesons, the multiplicity per projectile nucleon is larger for
the heavy system than for the light one while for pions the
situation is just opposite. Pions can be absorbed in large
systems, while the production of \kp increases with density and
hence the yield rises for heavier systems where higher densities
are reached. In spite of strong absorption, the \km yield in the
heavy system is also higher than in the light one. This is again
an experimental verification that the \kp and \km production are
tied together because the \km are produced in a
strangeness-exchange reaction with the hyperon which has been
created together with the \kp.

\begin{figure}[bht]
\epsfig{file=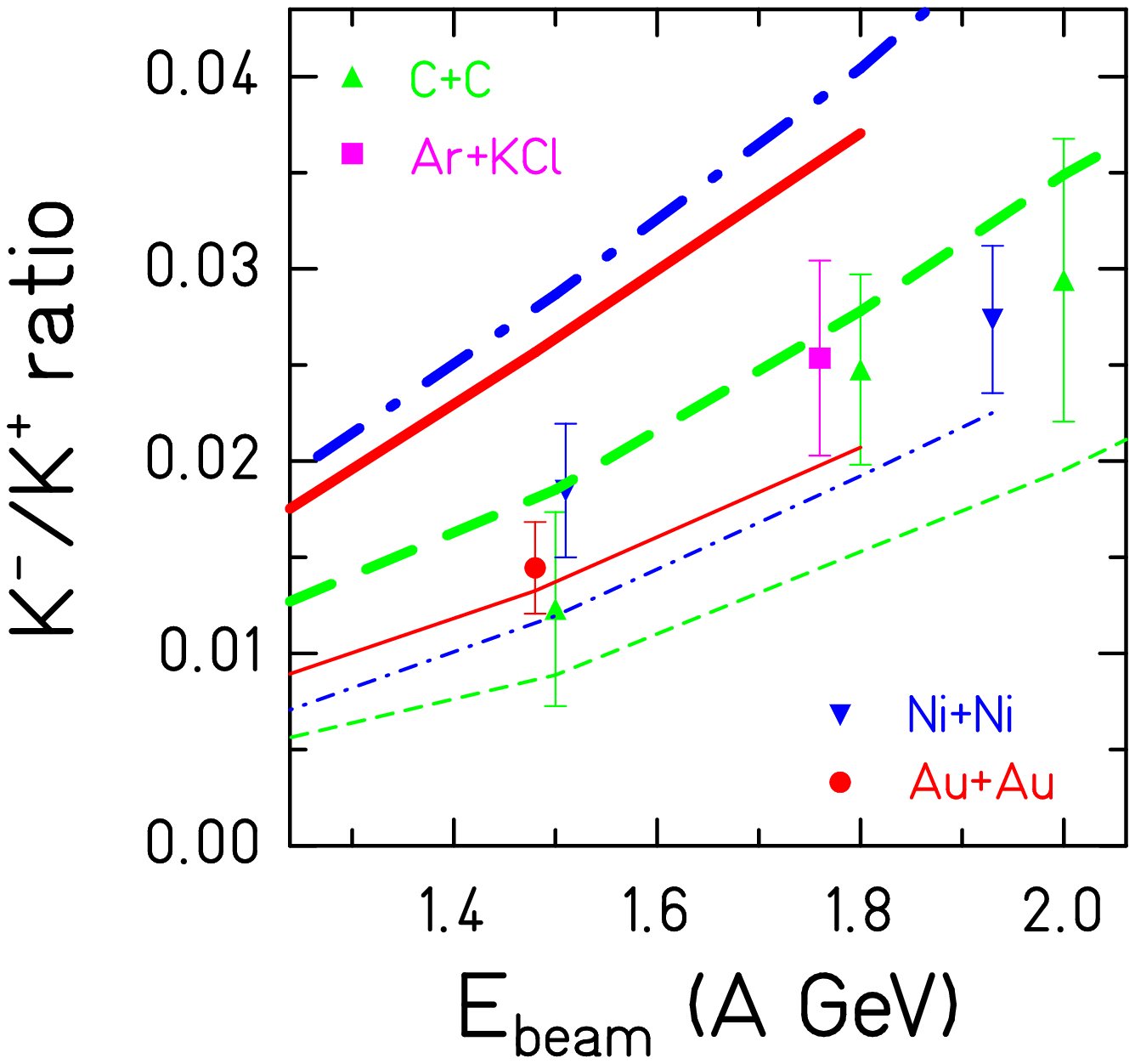,width=.4\textwidth}
\epsfig{file=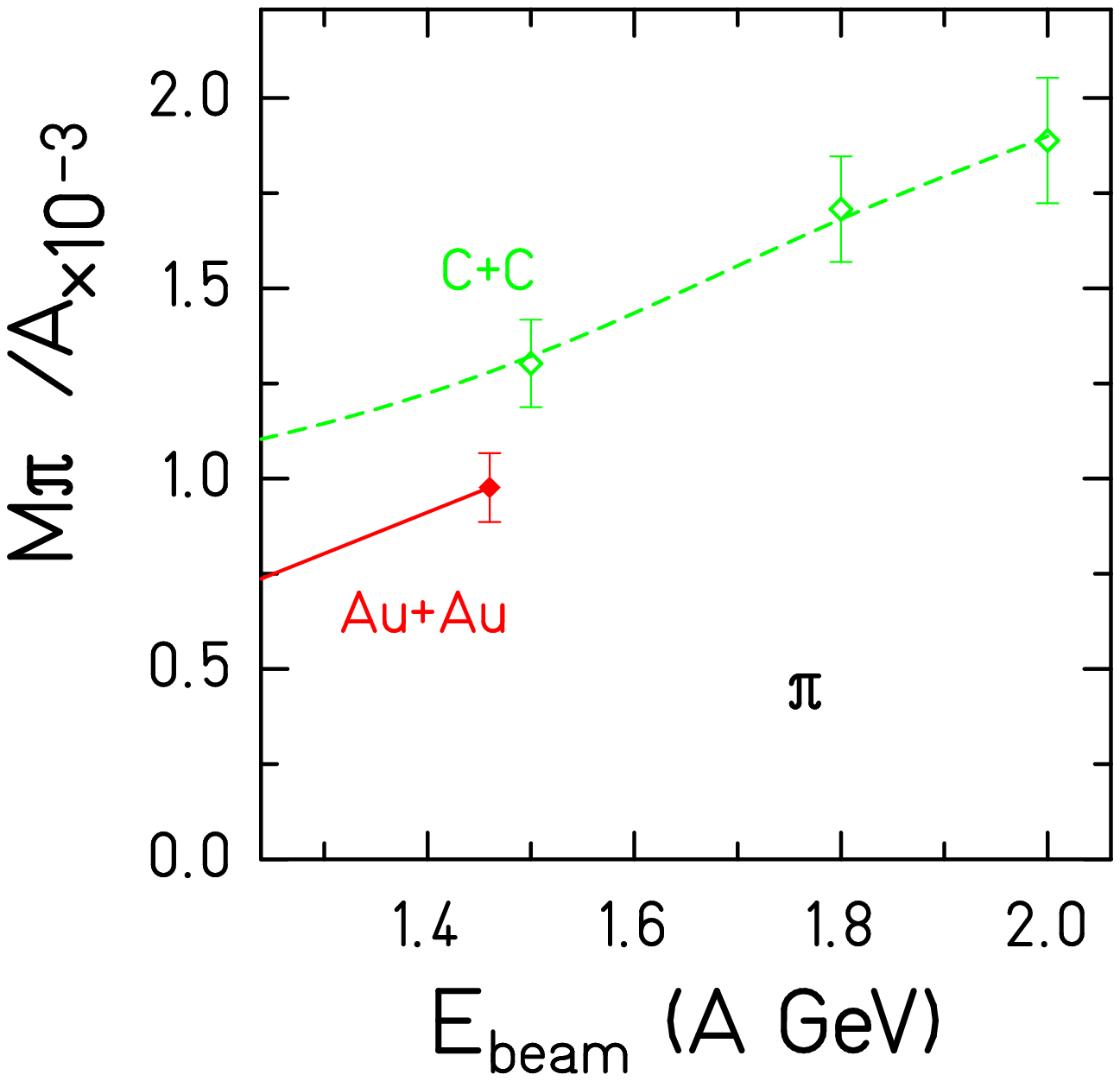,width=.4\textwidth}
 \caption{Left: Excitation function of the \km/\kp ratio for inclusive collisions of various
systems together with IQMD calculations. The experimental data are
from the KaoS Collaboration~\cite{Forster:2007qk} and the Ar+KCl
result from the HADES Collaboration~\cite{Schmah:2009eh} with
statistical and systematic errors added in quadrature. Thick
(thin) lines mark calculations including (excluding) the KN
potential. The full, dashed and dashed-dotted lines refer to
Au+Au, Ni+Ni and C+C, respectively. Right: Excitation function of
$\pi$ per projectile nucleon.} \label{kpkm-ebeam}
\end{figure}
The rise of the \kp and \km yields as a function of $E_{\rm beam}$
reflects a typical threshold behavior. Yet, in contrast to the
production in elementary processes, the shape of these two
excitation functions are very similar (see Fig.~\ref{KMelem}).
Again, this testifies how closely both are connected. Consequently
we see only the weak rise of the \km/\kp ratio with $E_{\rm beam}$
shown in Fig.~\ref{kpkm-ebeam}. Even more interesting, the rise is
about the same for all systems. Also IQMD calculation exhibit a
rather soft rise and a rather small variation for the different
collisions systems. Including the KN potential (thick lines ) the
calculations over-predict this ratio due to a over-prediction of
the \km yield, whereas without KN potential (thin lines) theory
under-predicts this ratio.  As shown in
Ref.~\cite{Oeschler:2000dp,Cleymans:2004bf}, the yield of pions
determines the  \km/\kp ratio, if the strangeness-exchange channel
reaches equilibrium. Indeed, the pion excitation function rises in
a very similar way, as demonstrated in the right panel. Although
in the IQMD only about 60\% of the \km are produced via the
$\Y+\pi=\km+N$ channels and the rest dominated ${\rm B} \Y$ (see
Fig.~\ref{KM_time_channels}) this trend is not modified.
Nevertheless it should be noted that for the 40\% remaining
channels there is a strong contribution of $\Delta \Y$ reactions.
The reacting $\Delta$ stem mostly from ${\rm N}\pi \to \Delta$
reactions and thus also reflect a dependence on the pion yield.

\begin{figure}[bht]
\epsfig{file=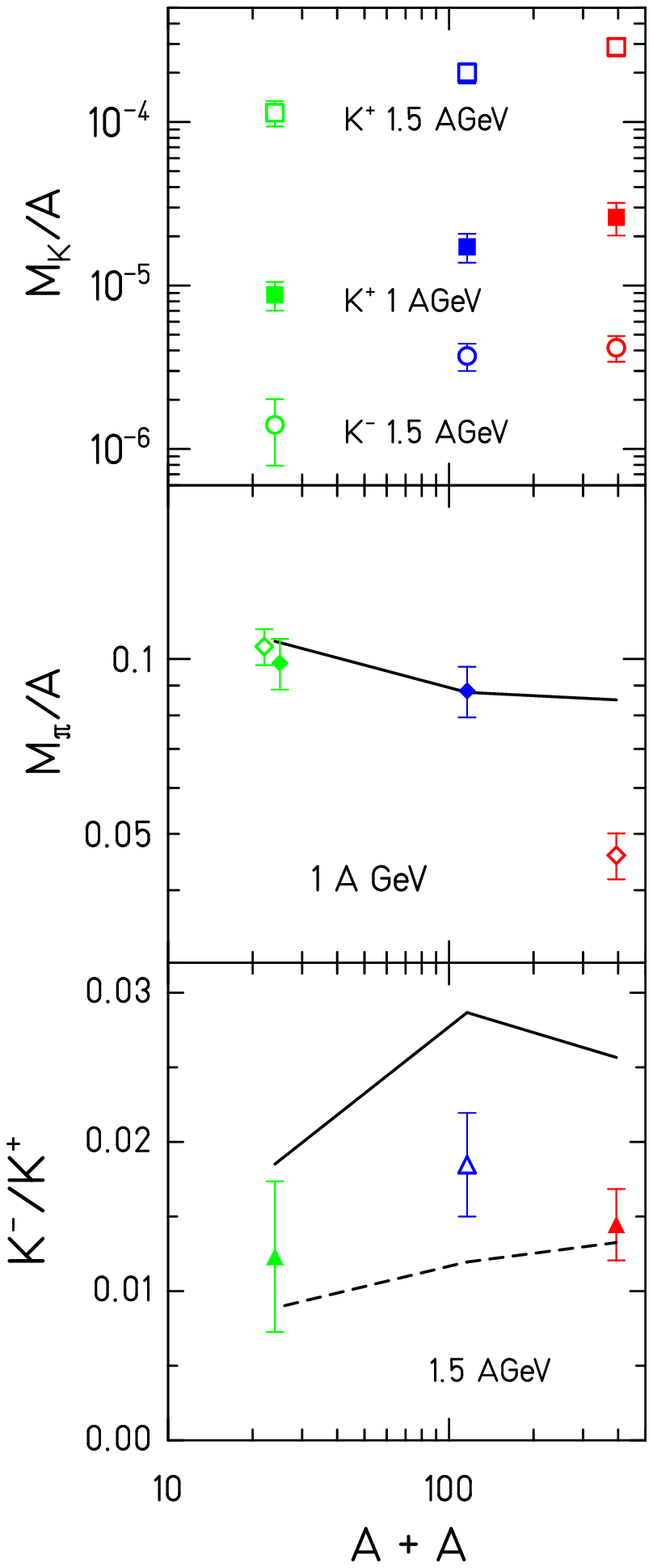,width=.4\textwidth}
\epsfig{file=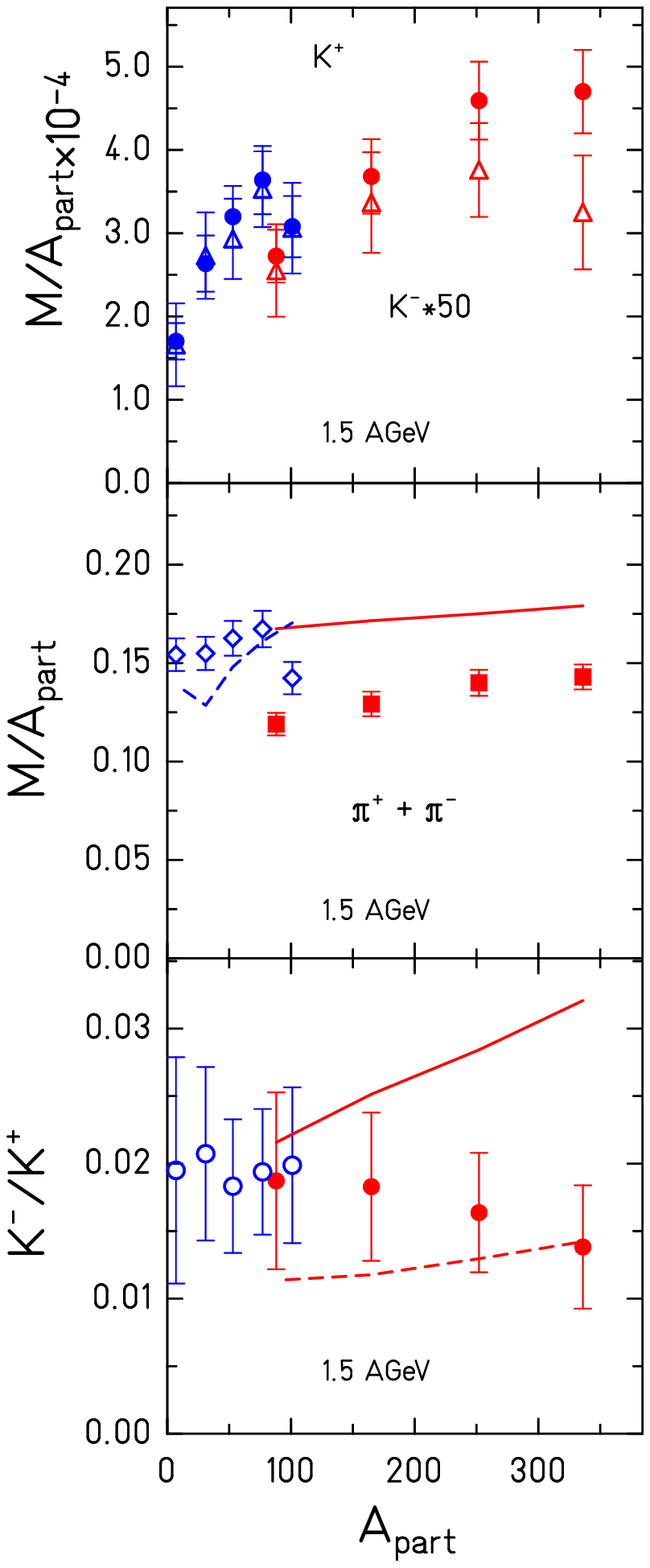,width=.4\textwidth}
\caption{Dependence of the multiplicities of \kp, \km, $\pi$
multiplicity per participating nucleon and of the \km/\kp ratio on
the system size (for inclusive symmetric reactions at the same
beam energy), left, and as a function of the participant number,
right. The green symbols refer to C+C collisions, the green ones
to Ni+Ni and the red ones to Au+Au. IQMD calculations including
(excluding) the KN potential are given by full (dashed) lines. The
data are from
Ref.~\cite{Averbeck:2000sn,Reisdorf:2006ie,Forster:2007qk,Sturm:2000dm}.}
 \label{kpkm-ebeam-A-Apart}
\end{figure}

The multiplicities per projectile or  target nucleon $A$, $M/A$,
and  per participating nucleons, $M/A_{\rm part}$, are summarized
in \figref{kpkm-ebeam-A-Apart}. On the left hand side, we display
these multiplicities for inclusive symmetric reactions as a
function of the system size ($A+A$), on the right hand side as a
function of the (geometrically determined) participant number,
\Apart (taking the values determined by the KaoS Collaboration).
As in the previous figures, the Au+Au data are marked by red
symbols, Ni+Ni data by blue symbols and C+C by green ones. Both
kaon multiplicities rise with system size as well as with
centrality. At the same beam energy (1.5 \AGeV) the increase with
$A$ is similar for both kaon species
(Fig.~\ref{kpkm-ebeam-A-Apart}, top left). IQMD calculations
confirm these findings as we have shown in earlier sections. As a
consequence, the ratio \kp/\km is about constant, both as a
function of $A+A$ as well as with \Apart, as seen in the bottom
panels.  For pions clearly a different behavior is seen. While the
multiplicities per $A$ for inclusive collisions decreases with
$A$, the centrality dependence is roughly constant as has been
found already at the BEVALAC~\cite{Harris:1981tb}. The system-size
dependence can be understood by absorption. At these low incident
energies, the multiplicity per $A$ in heavy-ion collisions is much
lower than that in elementary NN collisions~\cite{Muntz:1995am}.
With increasing beam energy, this difference vanishes and finally
at SPS energies the pion multiplicities per $A$ in heavy-ion
collisions exceeds those observed in elementary NN collisions.

It has been observed by the KaoS collaboration that the slopes of
the \kp and \km spectra show a very systematic behavior: Fitting
the mid-rapidity spectra by a function of the form $E {\rm d}^3
\sigma/{\rm d} p^3 \propto \exp({-E/T})$,  the observed inverse
slope parameters of the \kp spectra are always about 15-20 MeV
higher  than those of the \km \cite{Forster:2007qk} independent of
the system size and independently of the beam energy. This
seemingly simple correlation has, however, a quite complicated
origin, if one follows the IQMD calculations.

The elementary production of \kp mesons follows the three-body
phase-space and the energy of the \kp therefore depends (apart
from the beam energy) on the relative contribution of the
(on-the-average more energetic) $\Delta$ N collisions. Even though
the three-body phase space distribution does not have an
exponential slope, after averaging over the different channels and
different c.m.-energies the spectra have an exponential form, $E
{\rm d^3}N/{\rm d}p^3 \propto \exp({-E/T})$
(\figref{spectra-effects}). Immediately after production in
central Au+Au collisions the inverse slope parameter is $T =  79
(65)$ MeV for central Au+Au (C+C) reactions at 1.5 \AGeV (see
\figref{spectra-effects},\figref{CC_slopes}) . As seen there, the
KN potential does not change the slope considerable. Rescattering
on the contrary changes the slope considerably but for C+C
collisions the number of rescatterings is small and therefore we
observe a final slope of about 80 MeV. In the Au+Au case, on the
contrary, rescattering is very important and increases $T$ to 98
MeV and the strong repulsive \kp N potential shifts it further up
to 110 MeV.

For the \km the inverse slope parameter at production (averaged
over all production channels) in central Au+Au collisions at 1.5
\AGeV is 83 (75) MeV for Au+Au (C+C)
(\figref{km-spectra-effects-auc}) and therefore even higher than
that for the \kp. For C+C  the influence of absorption on the
spectra is negligible and also the potential interaction is weak.
It lowers slightly the inverse slope parameter whereas elastic
rescattering has the opposite effect (see
\figref{km-au-spectra-effects}). At the end, all effects nearly
compensate and the final slope is close to that at production. For
Au+Au the situation is different. The strongly momentum-dependent
rescattering cross section has the consequence that those \km
mesons, which are registered finally in the detector, have an
initial inverse slope parameter of 99 MeV. Rescattering increases
this value but the attractive \km N interaction reduces it again
and both effects together cancel each other resulting in an
inverse slope parameter of 98 MeV. For \km mesons the changes of
inverse slope parameter due to rescattering and due to the
potential compensate partially, whereas for the \kp they add and
the final inverse slope parameter of \kp is higher than that of
the final \km slope, as observed experimentally.

\begin{figure}[hbt]
\epsfig{file=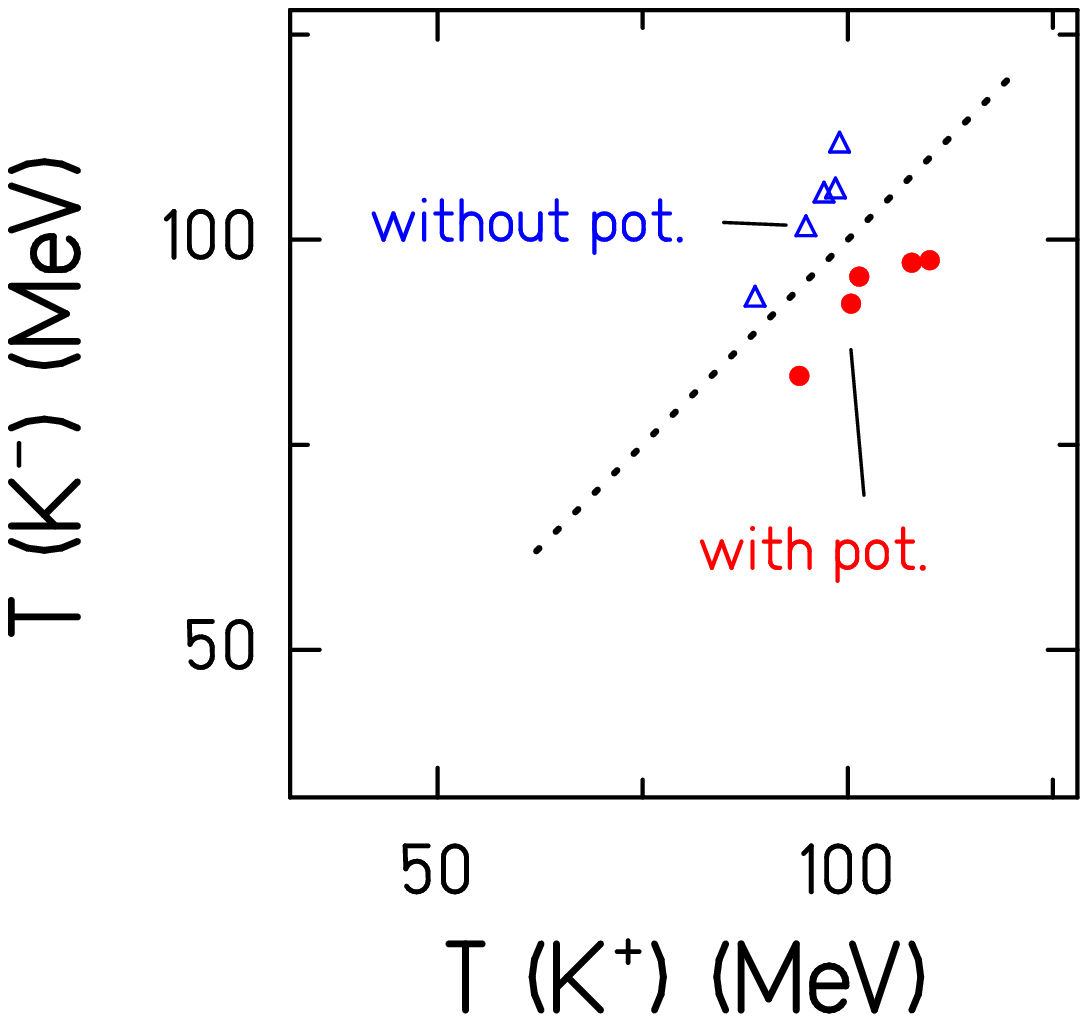,width=.32\textwidth}
\epsfig{file=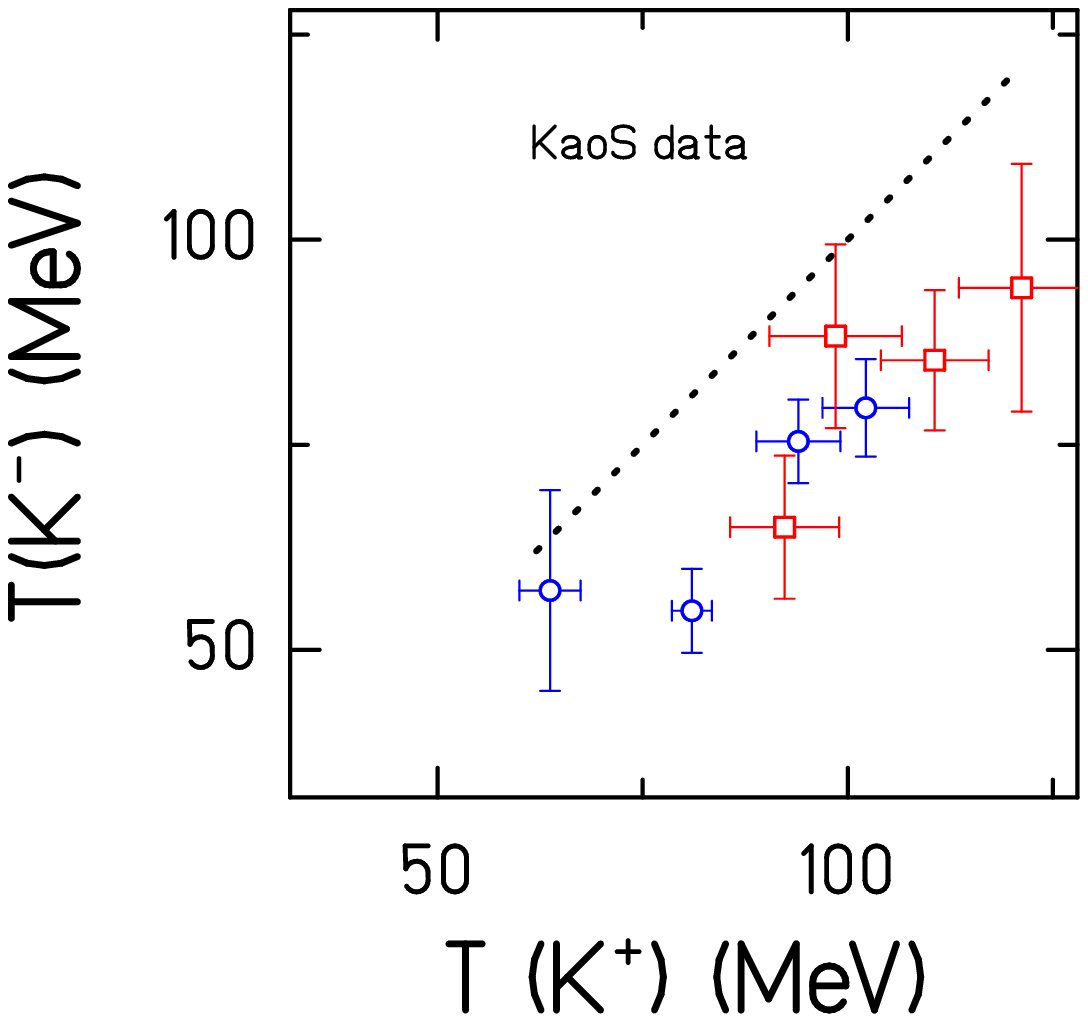,width=.32\textwidth}
\epsfig{file=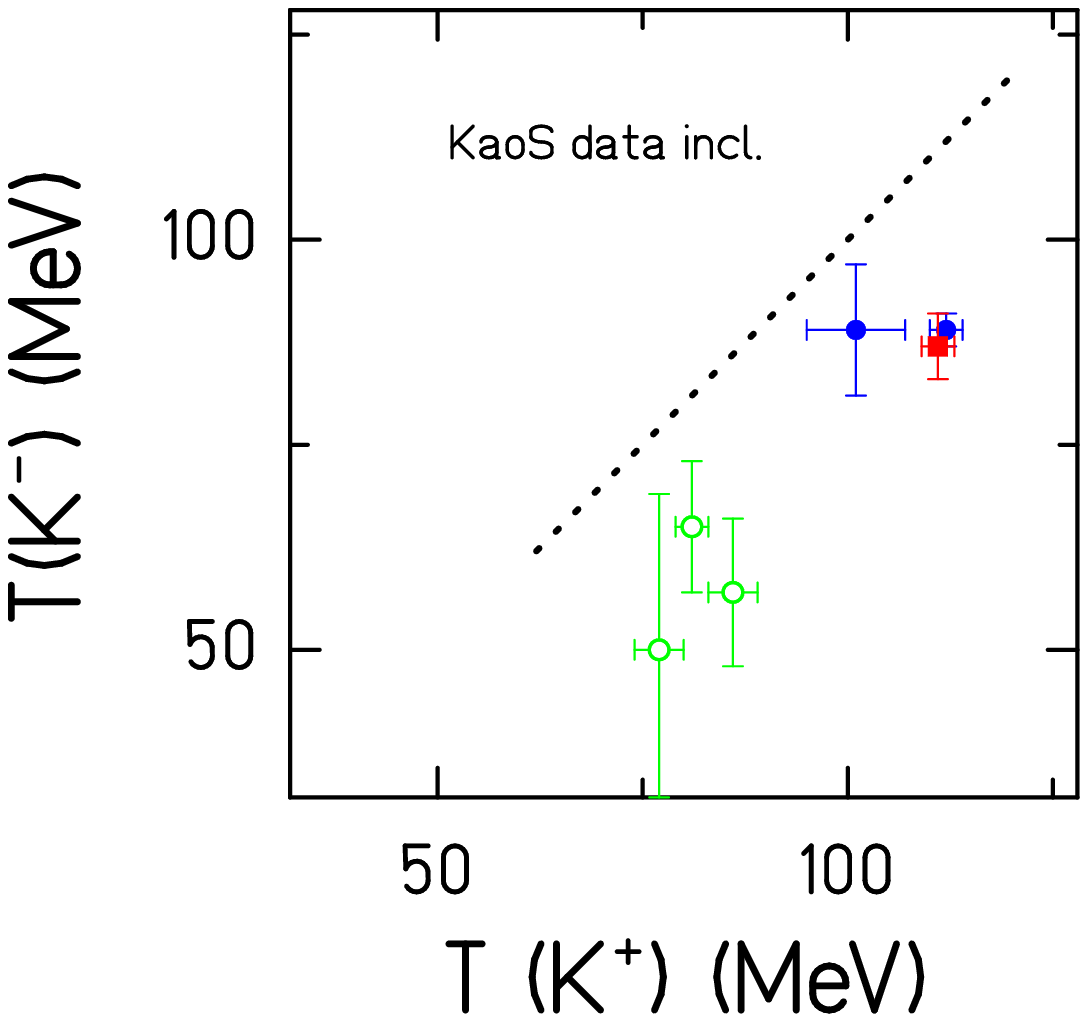,width=0.32\textwidth}
\caption{Relation between the inverse slope parameter $T(\km)$
versus $T(\kp)$. The color code is green for C+C, blue for Ni+Ni
and red for Au+Au. Left: IQMD calculations for Au+Au collisions at
1.5~\AGeV for different centralities with and without KN
potential. Middle: Experimental results for Au+Au collisions at
1.5~\AGeV and for Ni+Ni at 1.93~\AGeV for various centralities.
Right: Data from inclusive collisions of C+C, Ni+Ni and Au+Au at
several incident energies. The experimental results are from
Ref.~\cite{Forster:2007qk}}. \label{KPKM_slopes}
\end{figure}

The left part of Fig.~\ref{KPKM_slopes} shows this influence of
the KN potential on the inverse slope parameter $T$ of \kp and of
\km in IQMD calculations for Au+Au at various centralities at the
end of the reaction. The KN potential is the primary reason that
$T$(\kp ) $>$ $T$(\km ).  Whereas the \km N interaction lowers
the inverse slope parameter of the \km the \kp N interaction
increase the inverse slope parameter of the \kp. The IQMD results including KN potential are
confronted to experimental results from the KaoS Collaboration for Au+Au at 1.5
\AGeV ~\cite{Forster:2007qk} in the middle and right panel of
Fig.~\ref{KPKM_slopes}. The middle part shows the experimental centrality
dependence of the inverse slope parameter for Au+Au and Ni+Ni and the right part the data
obtained for inclusive reactions  at various incident energies.
Thus in experiments we see as well $T$(\kp ) $>$ $T$(\km ) another
piece of evidence for the importance of rescattering and
of the KN potential for the dynamics of strange mesons in matter.

It has been argued \cite{Cleymans:1998yb} that the multiplicities
of all hadrons observed at SIS energies fit very well into the
systematics which is expected in statistical model calculations.
For this analysis, only the multiplicities of \km and \kp have
been used. We have already seen that the different slopes of the
\kp and \km spectra do not support the assumption that the system
has reached a global equilibrium assuming a unique freeze out.
They depend in a complicated way on the dynamics of the system. It
is nevertheless interesting to see how these slopes fit into those
observed for other particles.
This point is approached in two steps. First, the more
general question of global equilibration is discussed.
\begin{figure}[bht]
\epsfig{file=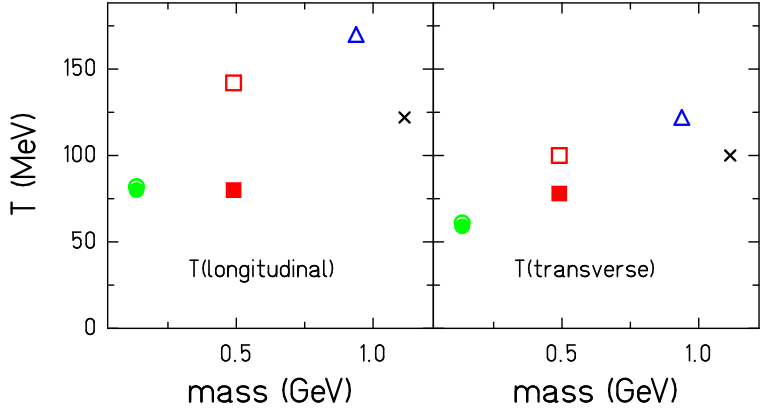,width=0.69\textwidth} \caption{
Transverse and longitudinal temperatures for different hadrons in
simulations of $b$=0 Au+Au reactions at 1.5 \AGeV. We have
converted the calculated $<y^2_t>$ with help of Eq.~(\ref{meansq})
into a temperature. \kp and \km are marked by open (full) squares. } \label{rap2}
\end{figure}
If the particles were distributed according to phase space \be
\frac{E{\rm d}N}{4\pi p^2{\rm d}p}=\frac{{\rm d}N}{{\rm d}y\ m_t
{\rm d}m_t\ {\rm d}\phi}= \frac{V}{(2\pi)^3}E\exp({-E/T}) \ee
 we expect for the rapidity
distribution \cite{Schnedermann:1993ws}
\be \frac{{\rm d}n}{{\rm d}y}=
\frac{V}{(2\pi)^2}T^3(\frac{m^2}{T^2}+\frac{m}{T}\frac{2}{\cosh{y}}+
\frac{2}{(\cosh{y})^2}) \exp{-\frac{m}{T}\cosh{y}}
\label{meansq}\ee where $m_t = \sqrt{p_t^2+m^2}$ is the transverse
energy with respect to the direction chosen to calculate $<y^2>$.
The averaged squared rapidity $<y^2>$ depends on the mass of the
particle $m$ and on the temperature of the system $T$. It turns
out that $<y^2>$ is quite different when comparing the
longitudinal and the transverse direction and also different for
the various particles. With help of Eq.~(\ref{meansq}) we determine
the inverse slope parameters $T_{long/trans}$ from $<y^2_{trans/long}>$.
They are shown in Fig.~\ref{rap2}, left in longitudinal direction and right in
transverse direction.
The values of the inverse slope parameter in longitudinal direction
exceed those in transverse direction and the values are different for all particles. A
global equilibrium is not established in heavy-ion reactions in
this energy domain, even in the most central reactions of the
largest system. This is not too astonishing because the reaction
is fast and a communication between particles in different space
regions is almost impossible. In both figures, two particles, the
\km and the $\Lambda$ do not follow the general trends. This again
is not astonishing from the discussion before: Different particles
see different regions of the phase space and separate from
the system at different times. Now, the $\pi$, the \kp mesons and
possibly the protons lie on a rising line. This trend could be
interpreted as resulting from a kinetic freeze out at
around $T$ = 60 - 70 MeV and a radial flow component. In most data
such a trend is seen and generally interpreted as collective flow.
However, the slope of $\Lambda$ and $\km$ cannot be interpreted in
such a model.

The question whether the system approaches a local equilibrium will be addressed now.
Rescattering brings the slope of particles close to the local
temperature. The \kp mesons and to a much smaller extent also the
\km mesons have elastic collisions with the surrounding nucleons
(only 17\% of the \kp and 30\% of the \km do not rescatter in
central Au+Au reactions at 1.5 \AGeV). It is interesting to
investigate whether there are sufficient collisions to bring the
\kp and the \km mesons  to a local equilibrium with their
environment.
\begin{figure}[tbh]
\epsfig{file=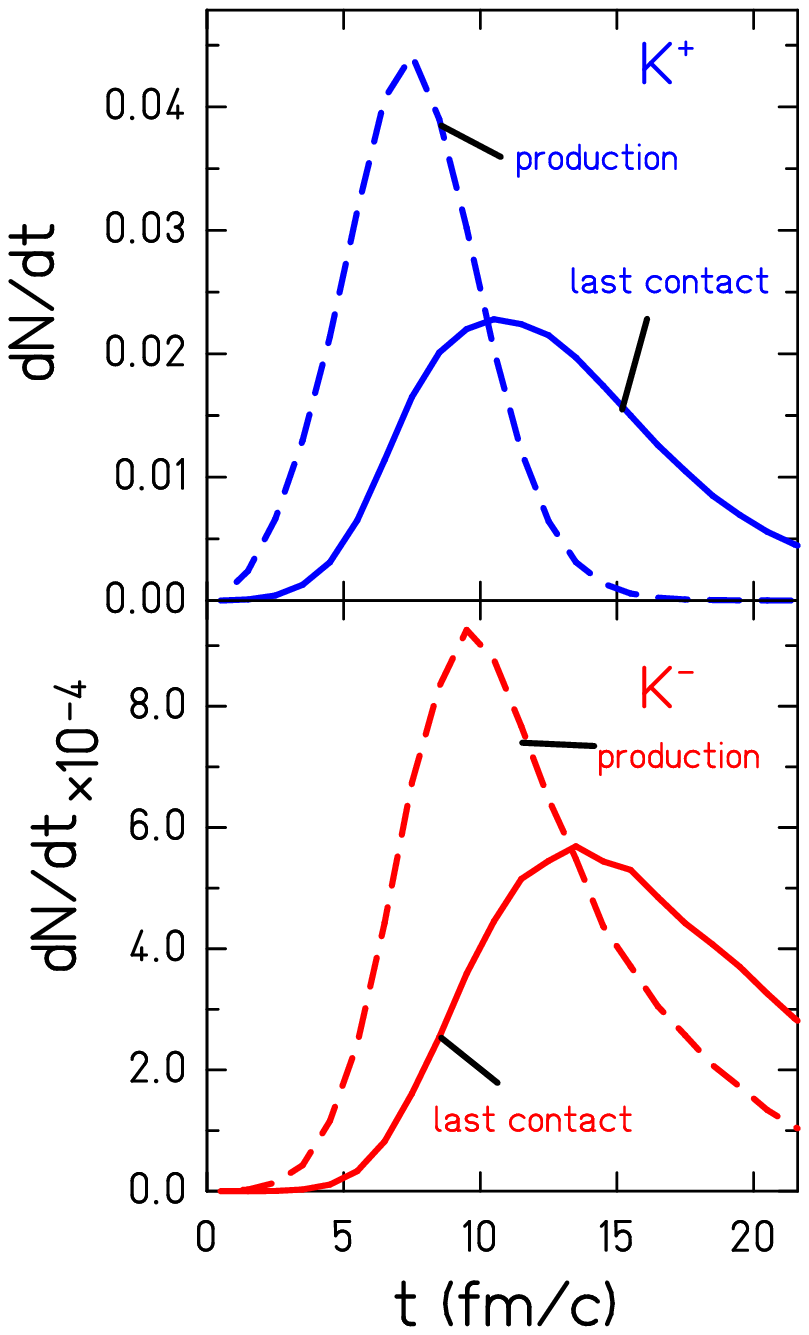,width=0.4\textwidth}
\epsfig{file=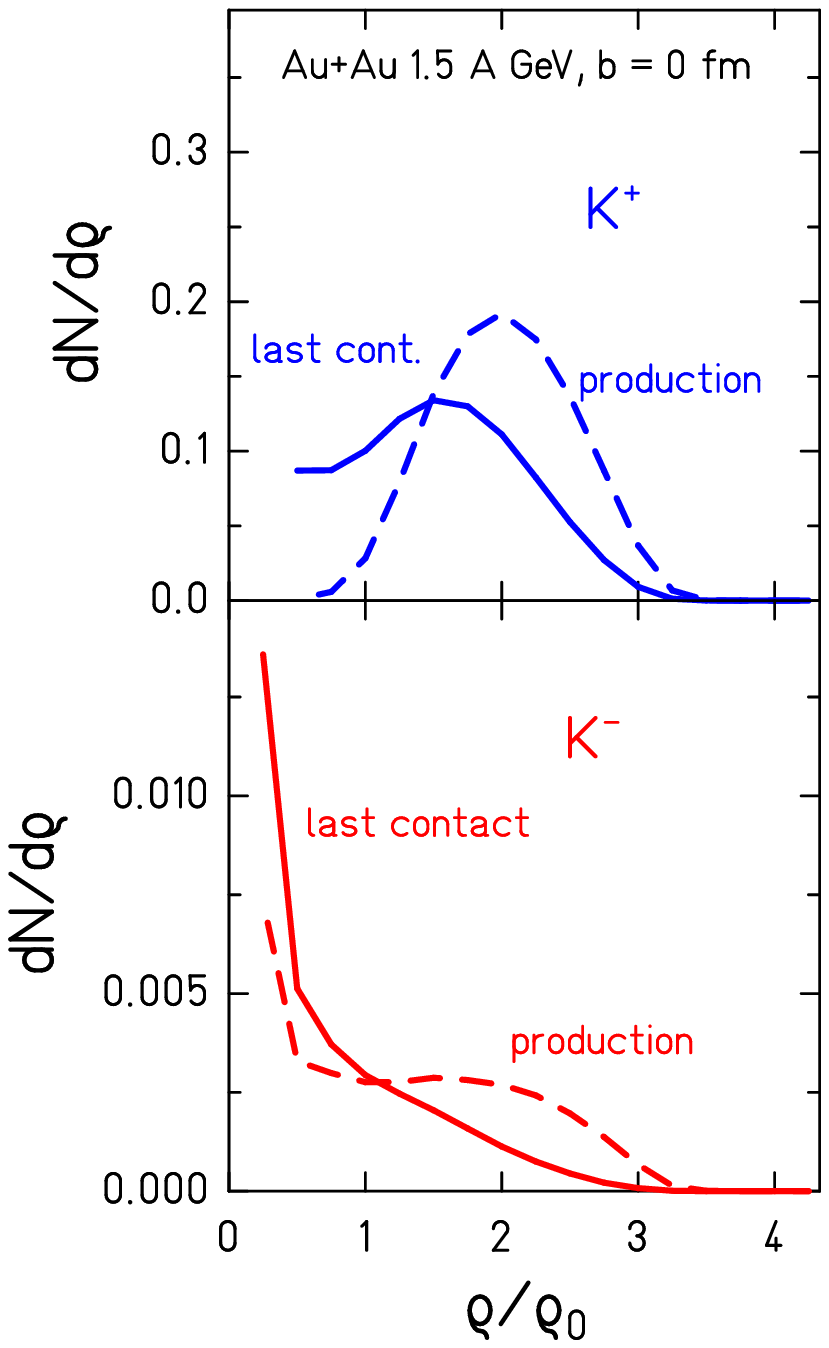,width=0.4\textwidth}
\caption{
Left: Time
profiles for production and last contact for \kp (top) and for \km
(bottom) mesons in central Au+Au collisions at 1.5 \AGeV. Right:
The corresponding distributions of the densities at the production point evidencing a huge
difference for the two mesons. } \Label{density-time}
\end{figure}
\begin{figure}[tbh]
\epsfig{file=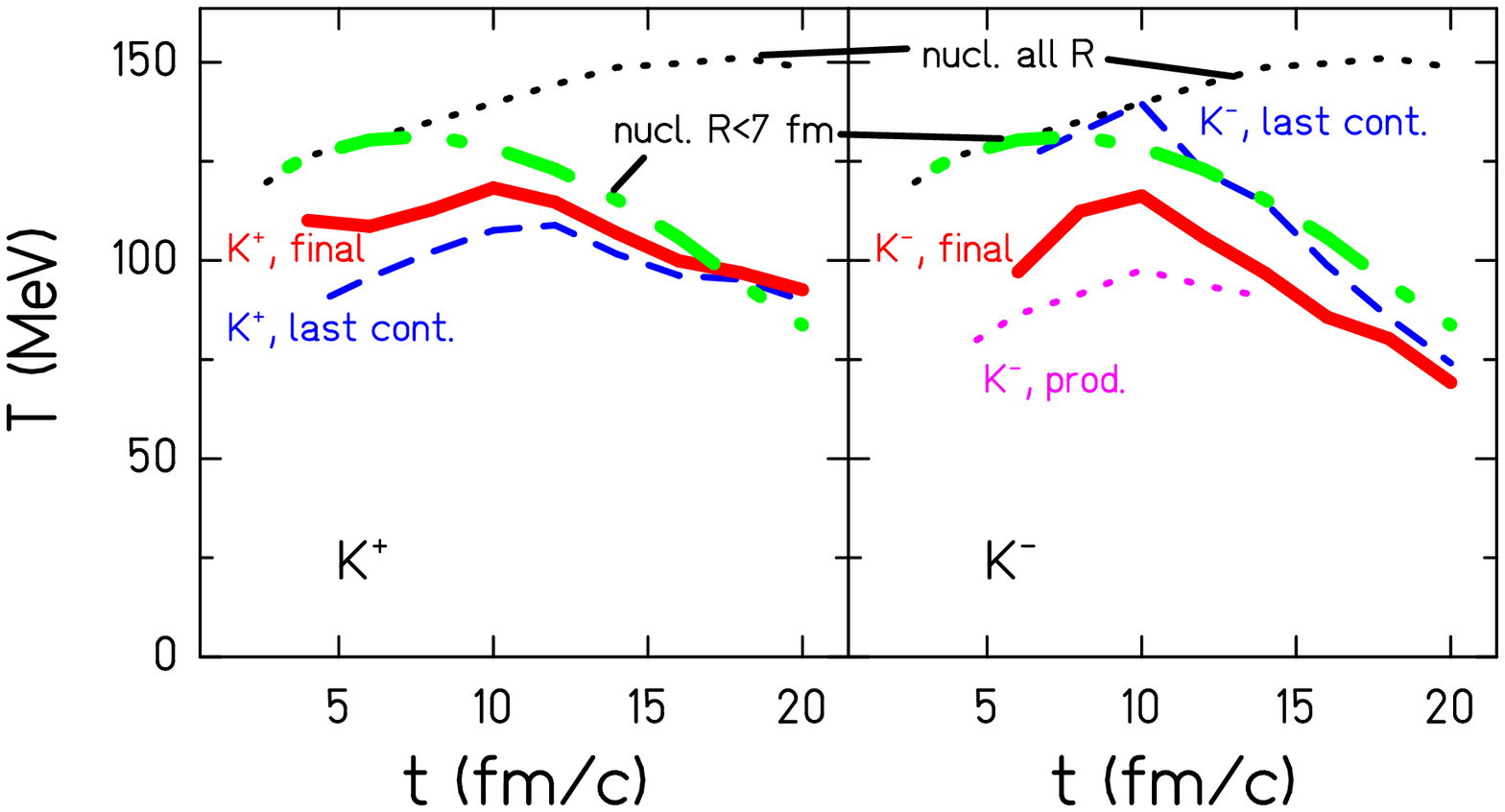,width=0.8\textwidth} \caption{The
inverse slope parameters obtained in IQMD calculations as a
function of time of last contact for \kp (left) and for \km
(right).} \Label{KPKM_slopes_time}
\end{figure}

In order to address this question we first study when the kaons
have their last contact with the system. In \figref{density-time}
(left) we see that the distribution of the last contact time is
rather wide with an average value of around 12 fm/$c$ for the \kp
mesons and 15 fm/$c$ for the \km mesons. Figure~\ref{density-time}
(right) shows the density distribution of the system at the time
when the kaons had their last contact (solid lines). In the case
of \kp mesons, the average density is around 1.5 $\rho_0$ whereas,
as expected, for \km mesons the density is much lower.

Now we study the question whether the kaons are in equilibrium
with their environment after their last contact.
Figure~\ref{KPKM_slopes_time} shows the inverse slope parameter
$T$ of the spectra  as a function of time (i) for nucleons whose
distance from the center of the reaction is closer than 7 fm
(dashed-dotted) at that time (and which are therefore in the
vicinity of the \kp) and (ii) for all nucleons (dotted). These
slopes are confronted with the slopes of the spectra of those
kaons which have their last contact at time $t$. We display the
slopes which the kaons have immediately after the last contact as
well as that of the final spectra of those kaons which had their
last contact at $t$. Both differ due to the potential interaction
between kaons and nucleons.
These two kaon slopes differ only for early emission where the
density and hence the KN potential is still strong. Yet, they
become very similar at later times. The later the \kp mesons are
emitted the closer is their (last contact) inverse  slope
parameter to that of the surrounding protons (dash-dotted line) as
shown in \figref{KPKM_slopes_time} left.  The majority of \kp
mesons is emitted earlier, as can be seen from
\Figref{density-time}, where the \kp slope differs from the proton
slope.

The \km mesons
come closer to a local equilibrium as can be seen in
\figref{KPKM_slopes_time}, right. From 10 fm/$c$ on, the slopes of
\km mesons (at last contact) and of protons are identical. Because
most of the \km mesons are emitted after 10 fm/$c$ (see
\figref{density-time}) their slope is a measure of the local
temperature and therefore, even a larger \km N elastic scattering cross
section would not change the slope of the kaons. In other words,
the slopes are insensitive to rescattering cross section as soon
as they are larger than a minimal value. The finally observed slope
(full line) is, however, different, because the KN potential
changes the slope considerably after the last contact.

\subsection{Sensitivity of the observables on properties of the system}

So far we have shown how the strange meson spectra are modified by
the interaction with the medium. Now we would like to address the
question: what information do strange mesons carry on properties
of the medium?  Figure~\ref{density-time} (left) shows the time
profile for the production and the last collision. The surviving \km  mesons are
produced about 4 fm/$c$ later than the \kp mesons. This is
understandable because  the $\Lambda$ (together with the \kp) has to be
produced before it can create a \km via strangeness exchange.

 A drastic difference between the two species
 is seen when studying the densities at which they have their last contact
 as shown in \figref{density-time}.
The results for \kp mesons are shown in top
 panel, those for \km in the bottom panel.
 The \kp production occurs at around twice
 normal nuclear matter density and since their number does not
 change, the \kp multiplicity measures the system properties at that density.
KN rescattering changes their momentum distribution but
their last contact  is still well above $\rho_0$.  The \kp
 momentum distribution is therefore sensitive to system properties at that density.
This makes them an ideal probe for the high-density phase of the reaction.
In particular, this fact allows for studying  the nuclear equation of state, as will be
discussed in the next section.

 The \km mesons, in contrast, are produced at lower densities and
 their number is changing due to the \km N $\to \Lambda \pi$ channel.
 They cannot serve  as a probe for high densities but may be
 sensitive to system properties like the KN potentials at densities
 below $\rho$ = 0.5 $\rho_0$.  They are,
however, useful to study the absorption cross section $\km  \N
\to \Y  \pi $ in the medium because its momentum dependence considerably
influences the slope of the spectra. It may be
possible that without this absorption the local equilibrium,
discussed in Section VI.A, is not obtained. One can envisage that
high-precision measurement allow to study the behavior of the
$\Lambda (1405)$ in the medium, a question which is highly debated
(see Section~\ref{theory}).

Thus, the information contained in the spectra of \kp
and \km mesons is sensitive to quite different phases and phase
space regions of the reaction.

Since the elementary production cross section for \kp and \km
mesons have been measured (with the exception of those with a
$\Delta$ in the entrance channel) heavy-ion collisions can help to
identify the influence of the medium on the \kp and \km
observables. As has been discussed in detail in the previous
chapters this interaction with the medium is a very complicated
process and very different for \kp and \km even if the $\Delta$
cross sections were known. This interaction depends on the
potential interaction between the kaons and the hadronic matter as
well as on in-medium modifications of the involved cross section
due to, for example, the dissolution of the $\Lambda(1405)$ in
nuclear matter, as predicted by theory. Presently, it is fair to
say that transport calculations which have been developed before
the large body of kaon data became available do describe the
ensemble of the data (with the exception of the $p_t$ dependence
of $v_2$).

We have seen that almost all of the observables are sensitive to
more than one of these unknown or only vaguely known quantities,
however with different sensitivity. For this reason we have
summarized in Tab.~\ref{sens} to which parameters (KN potential,
elastic rescattering, \km absorption cross section) the different
\kp and \km observables are sensitive to and where the main
uncertainties are. The most promising observable for further
studies is  the kaon spectra at low momentum. It depends  on the
KN potential but is relatively independent of the rescattering
cross section. Precision experiments should be able to determine
these potentials experimentally. Especially the experimental
determination of the K$^0$N potential,
which is not perturbed by Coulomb interactions, should be within reach. \\

\begin{table}
\begin{tabular}{|l|c|c|c|l|}
\hline
observable & KN pot & resc & abs & comments\\
\hline
\kp yield & + & -& -& uncertainties in input \\
\km yield & - & + & +&many uncertainties in input, including those of \kp \\
\km yield (C/Au)& + & - & -& limits the strength of the \km potential \\
\kp/\km production ratio & - & + & + & same, only the input of \kp cancels\\
\kp, \km  difference in slopes & + & + & + & mostly potential , also caused by rescattering \kp, \\
                    &   &   &  & absorption of \km \\
\kp, \km ratio  & + & + & + & also caused by rescattering \kp, \\
                    &   &   &  & absorption of \km\\
K$^{0,+,-}$ at low momenta & ++ & + &-&difficult to measure\\
$v_2$ of \kp & + & + & - & contribution of rescattering dilutes the signal\\
$v_2$ of \km & - & + & - & governed by emission time of \km \\
$v_{1,2}(p_t)$ of \kp & + & + & - & sensitive to potential in HSD \\
$v_{1,2}(p_t)$ of \km & + & - & + & sensitive to potential in HSD \\
angular distr. of \kp & - & + & -& not sensitive to KN potential\\
angular distr. of \km & - & + & -& not sensitive to KN potential\\
\hline
\end{tabular}
\caption{Summary of sensitivities of various kaon observables to the different aspects of kaon physics.} \label{sens}
\end{table}

\subsection{The nuclear equation of state (EoS)}
\label{eos}

The understanding of the global properties of infinite nuclear
matter is certainly one of the most challenging questions in
nuclear physics. Especially the change in the behavior of matter
with density is a key property which is relevant also outside of
the nuclear physics domain. It plays a major role in the
understanding of astrophysical phenomena, e.g.~the structure of
neutron stars. Indeed, these are expected to have densities around
2 $\rho_0$ to 5 $\rho_0$ just those reached in relativistic
heavy-ion collisions. \cite{Aichelin:2008jn}

Theoretical approaches like the Br\"uckner-Hartree-Fock approach
\cite{schuck} are available for densities moderately above and
below normal nuclear matter density. This has recently been
reviewed by Baldo and Maieron \cite{Baldo:2007wm}. It agrees well
with other approaches like variational calculations and Greens
function Monte-Carlo calculations \cite{Fuchs:2005yn} up to
densities around $2\rho_0$. At higher densities (for $a/r_0 \approx
1$), where $a$ is the range of the short-range repulsive nuclear
force and $r_0$ is the internuclear distance, other many body
diagrams become important. At $\rho_0$ this value is $a/r_0 \approx 0.3$
and for densities reached in heavy-ion collision, as discussed
here, this approach therefore comes to the limit of its validity.
Of course, by introducing phenomenological three-body potentials
one may extend the validity of these approaches but they limit
also their predictive power.

Many attempts have been made to determine the EoS from experiments
using heavy nuclei. However, the extraction of this relation is
hampered by several facts:

1. The nuclear binding energies are about -8 $A$ MeV while the one
in the Weizs\"acker mass formula for infinite matter is -16 $A$
MeV. This difference shows that the properties of nuclei are
strongly influenced by surface effect and also by the Coulomb
force.

2. In heavy-ion reactions nuclei become compressed {\it and} excited.
Excitations and compression cannot be
varied separately. Excited nuclear matter consists also of baryonic
resonances and of mesons. This means that only a hadronic equation
of state at $T >0$ is accessible via experiment. Transport models do
describe these excitations, yet they also indicate that in these
collisions equilibrium - at least a global one - is not
reached. The EoS has therefore to be extracted indirectly.

Three experimental observables have been studied to extract
information on the nuclear EoS: (i) the strength distribution of
giant isoscalar monopole resonances \cite{Youngblood:1999zza}, (ii) the
in-plane side-wards flow of nucleons in semi-central heavy-ion
reactions at energies between 100 $A$ MeV and 400 $A$ MeV
\cite{Stoecker:1986ci} and (iii) the production of $K^+$ mesons in heavy-ion
reactions at energies around 1 $A$ GeV suggested in
Ref.~\cite{Aichelin:1986ss} and recently studied in
Ref.~\cite{Hartnack:2005tr}.

(i) The study of monopole vibrations \cite{Youngblood:1999zza} has been
very successful. Giant monopole resonances are sensitive to the
energy which is necessary to change the density of a cold nucleus
close to the equilibrium density $\rho_0$ which limits the
information to a very small density range. The vibration frequency
depends directly on the force which counteracts any deviation
from the equilibrium and therefore to the potential energy. These
calculations allow for the determination of the compression
modulus
$$K = 
-V\frac{{\rm d}p}{{\rm d}V}= 9 \rho^2 \frac{{\rm
d}^2E/A(\rho,T)}{({\rm d}\rho)^2} |_{\rho=\rho_0}$$ which measures
the curvature of $E/A(\rho,T)$ at the equilibrium point. The
values found for the volume compressibility in different
non-relativistic and relativistic approaches are around $K = 240$
MeV \cite{Piekarewicz:2003br,Agrawal:2003xb,
Colo:2003zm,Colo:2004mj,Vretenar:2003qm}, close to what has been
dubbed ``soft equation of state''. Very recently this value has
been questioned because the influence of the surface
compressibility may have been underestimated \cite{Sharma:2008uy}. This
may cause an uncertainty of 30\%.

(ii) In semi-central heavy-ion collisions the overlap zone of
projectile and target becomes considerably compressed. The
transverse pressure on the baryons outside of the interaction
region causes an in-plane flow which is expected to be proportional
to the transverse pressure. This effect can only be observed if
the beam energy is large as compared to the Fermi energy, i.e.~at
well above 100 $A$ MeV.  The in-plane flow is one of the few observables
which depend on the density profile at the surface of the
nuclei during the interaction being directly proportional
to the density gradient. It's sensitivity to the EOS is weak.
As shown recently~\cite{Andronic:2004cp} the tiny variation of
the in-plane flow with the EOS cannot be consistently calculated in the
transport theories,
putting doubts on previous conclusions~\cite{Danielewicz:2002pu}.

(iii) The most promising method for the study of nuclear matter
properties at high densities is the  kaon production in - mainly
central - heavy-ion collisions.
Here, the yield of \kp depends sensitively on the density and the
energetic condition during the interaction.

As mentioned before, \kp production  below the NN threshold
requires several (inelastic) collisions as the energy in first-chance
collisions is not sufficient. Due to strangeness conservation
together with a \kp also a $\Lambda$ has to be created. Hence the
threshold is 671 MeV in the center of mass. The most effective way to accumulate energy
is the conversion of a nucleon into a $\Delta$ and to produce in a
subsequent collision a \kp meson via $\Delta \N \rightarrow \N
{\rm K^+} \Lambda$. Two effects influence the yield of produced
\kp with the density reached in the collision and give thus access
to the stiffness of the EoS. For a soft EoS  less energy is
needed to compress matter and hence (i) more energy is available
for the \kp production and (ii) the density which can be reached
in these reactions is higher. The latter effect is the most
important one as a higher density leads to a smaller mean free
path and therefore the probability that the $\Delta$ has a
collision before it decays is higher. Hence, the \kp yield
increases.

\begin{figure}[htb]
\epsfig{file=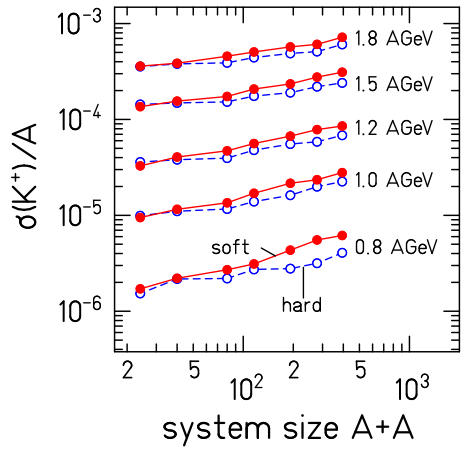,width=7.5cm}
  \caption{Cross section of \kp mesons per mass number $A$ as a function of system size $A+A$
  for several beam energies demonstrating the influence of the stiffness of the EoS on the \kp yield.}
  \label{alpha_gamma}
\end{figure}

The influence of the density reached in these collisions is
illustrated in Fig.~\ref{alpha_gamma}
showing the yield of \kp per mass number $A$ as a function of the
system size $A+A$ for different beam energies. The full (red) line
represents the calculations with the standard soft ($K$ = 200 MeV)
EoS, whereas the dashed (blue) line is the result for a hard ($K$
= 380 MeV) EoS. A hard  EoS requires more energy to compress
hadronic matter and therefore the kaon yield is lower for heavy
systems. Lighter systems do not stop that much and therefore their density
during the interaction depends only little on the stiffness of the EoS.  The
difference in yield for the heavy system is strongest at the
lowest beam energy. At beam energies $> 2~A$ GeV \kp production
by first-chance NN collisions dominates and the sensitivity on the
EoS is lost.

\begin{figure}[htb]
\vspace*{-1cm}
 \epsfig{file=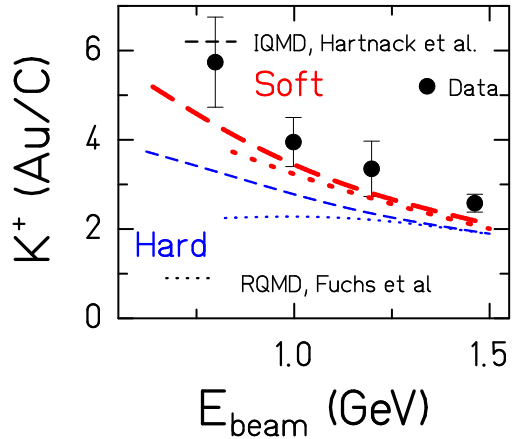,width=9cm}
  \caption{Comparison of the experimental \kp excitation functions \cite{Sturm:2000dm} of the
  double ratio $(M_{\kp}/A)_{\rm Au+Au} / (M_{\kp}/A)_{\rm C+C}$ (the \kp multiplicities per mass number $A$) obtained
  in \mbox{Au+Au} divided by the one in \mbox{C+C} with
  RQMD~\cite{Fuchs:2000kp} (dotted) and with
  IQMD calculations~\cite{Hartnack:2005tr} (dashed). We compare the results of a
  soft (bold red) with a hard (thin blue) EoS.}
  \label{EOS}
\end{figure}

It was found that the most sensitive observable for the stiffness
of EoS so far is the double ratio~\cite{Sturm:2000dm} of the \kp
production in a heavy and a light system
 \be
 (M_{\kp}/A)_{\rm Au+Au} / (M_{\kp}/A)_{\rm C+C}.
\ee This ratio is plotted in Fig.~\ref{EOS} as a function of the
beam energy for a soft (bold red) and a hard (thin blue) EoS together
with the data form the KaoS Collaboration~\cite{Sturm:2000dm}. The
calculations by Fuchs et al.~\cite{Fuchs:1998yy,Fuchs:2000kp,Fuchs:2001gv} are also shown as dotted
lines. This figure elucidates again that the sensitivity on the
EoS is highest at the lowest beam energies and demonstrates that
only a soft EoS is compatible with the experimental data of the
KaoS Collaboration~\cite{Sturm:2000dm,Fuchs:2000kp}.

\begin{figure}[htb]
\vspace*{-.5cm} \epsfig{file=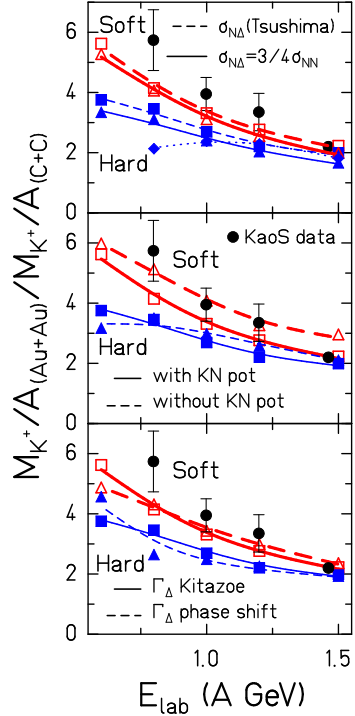,width=6.8cm}
\vspace*{-.3cm}
 \caption{Comparison of the measured excitation
function of the ratio of the $K^+$ multiplicities per mass number
$A$ obtained in Au+Au and in C+C reactions
(Ref.~\cite{Sturm:2000dm}) with various assumptions on the physical input.
The use of a
hard EoS is denoted by thin (blue) lines, a soft EoS by thick
(red) lines. The energies of the calculations are given by the symbols, the
lines are drawn to guide the eye. On top, two different versions
of the $N\Delta \rightarrow K^+\Lambda N$ cross sections are used.
One is based on isospin arguments \cite{Randrup:1980qd}, the other is
determined by a relativistic tree level calculation \cite{Tsushima:1994rj,Tsushima:1998jz}.
Middle: IQMD calculations with and without $KN$ potential are
compared. Bottom: The influence of different options for the life
time of  $\Delta$ in matter.} \label{ratio}
\end{figure}

Using the ratio of yields many uncertainties in the theoretical approaches
cancel because they act in the same way in both systems, as shown in Ref.~\cite{Hartnack:2005tr}.
Figure~\ref{ratio} displays some examples. From top to bottom we
show calculations with different $\N\Delta \to \kp \N\N$ cross
sections, with and without KN potential and for different life
times of the $\Delta$ resonance. None of these uncertainties is able
to weaken our previous conclusion that only a soft EoS is
compatible with the observed excitation function of the \kp yield.
\begin{figure}[htb]
 \epsfig{file=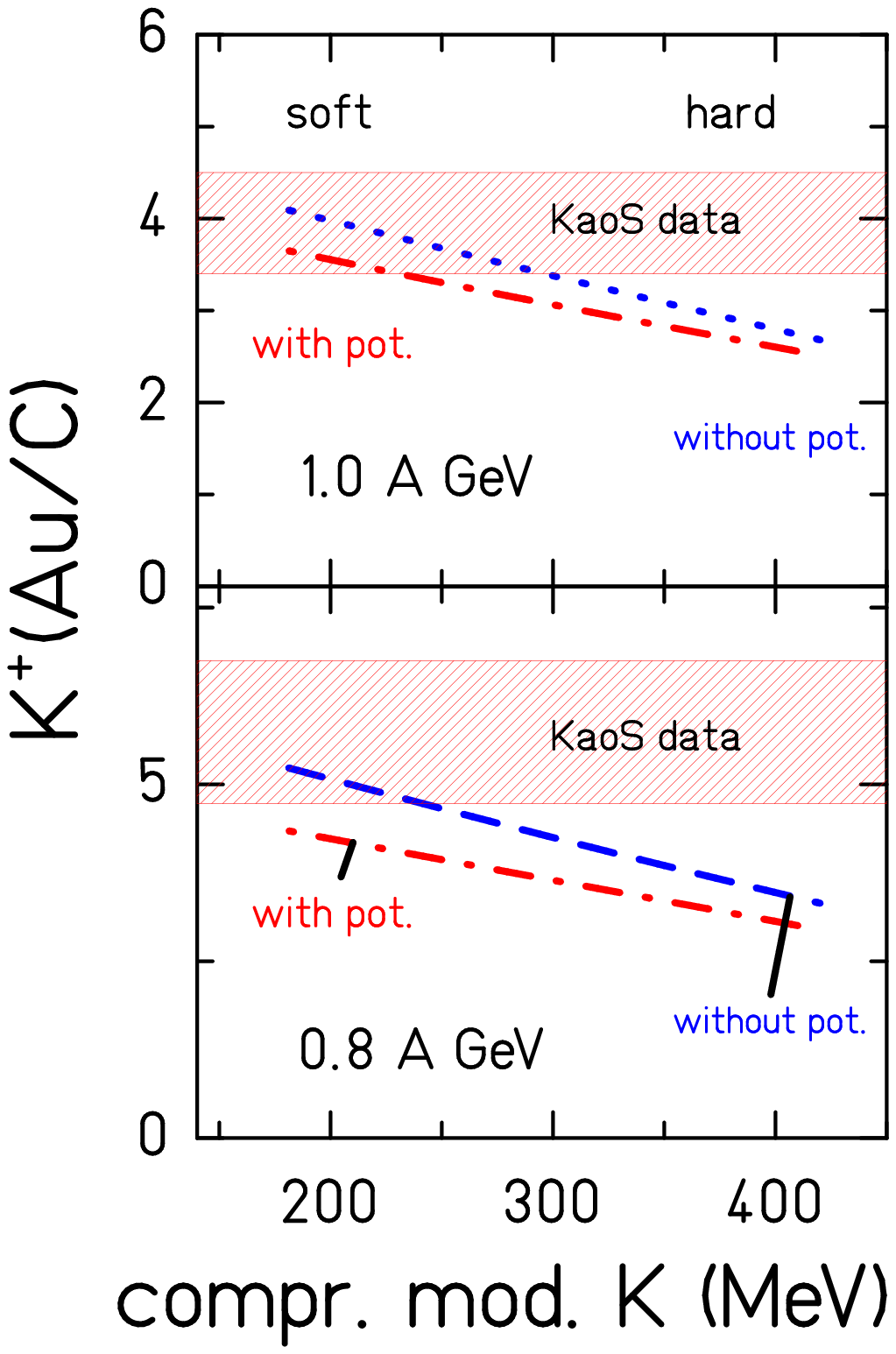,width=9cm}
  \caption{The double ratio
  $[M/A({\rm \mbox{Au+Au}})] / [M/A({\rm \mbox{C+C}})]$
calculated within the IQMD model (with and without KN potential)
as a function of $K$ at both 0.8 (top) and at 1.0 \AGeV (bottom).
The experimental values are given as band and allow to estimate
upper limits for the compressibility modulus $K$ as described in
the text.}
  \label{IQMD_ratio}
\end{figure}

\begin{figure}[hbt]
\vspace*{-.5cm} 
\epsfig{file=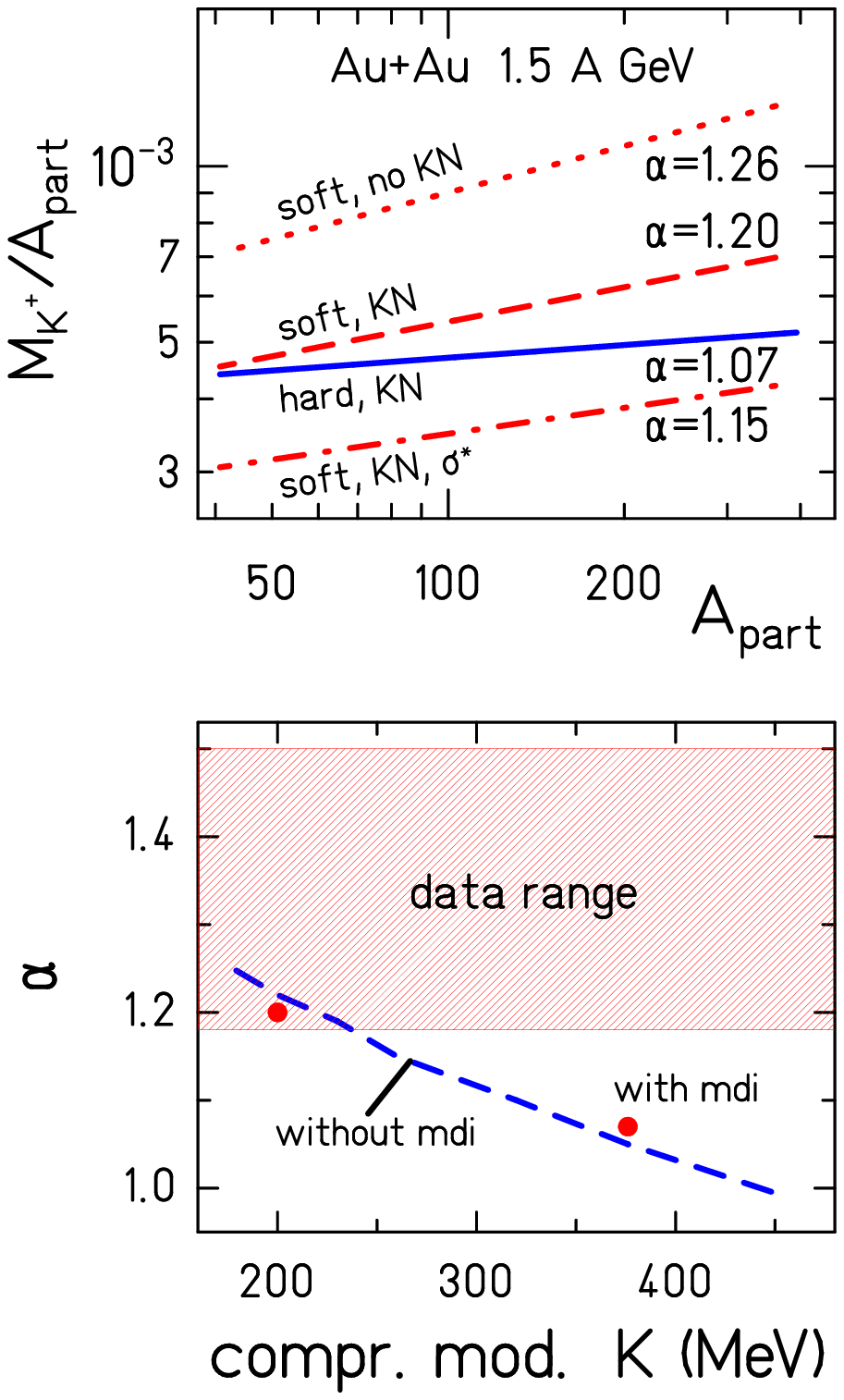,width=6.8cm} \vspace*{-.3cm}
\caption{Dependence of the \kp scaling on the nuclear equation of
state.  We present this dependence in form of $M_{{\rm
K}^+}=A_{{\rm part}}^\alpha$. On the top the dependence of
$M_{{\rm K}^+}/A_{{\rm part}}$ as a function of $ A_{{\rm part}}$
is shown for different options: A ``hard'' EoS with KN potential
(solid line), the other three lines show a ``soft'' EoS, without
KN potential and $\sigma({\rm N}\Delta)$ from
Tsushima~\cite{Tsushima:1994rj,Tsushima:1998jz} (dotted line),
with KN potential and the same parametrization of the cross
section (dashed line) and with KN potential and $\sigma({\rm
N}\Delta) = 3/4 \sigma({\rm NN})$. On the bottom the exponent
$\alpha$ is shown as a function of the compression modulus $K$ for
calculations with momentum-dependent interactions (mdi) and for
static interactions ($t_4=0$, dashed line, see Eq.~\ref{Vij}).}
\label{part}
\end{figure}

For a more quantitative extraction of the compression modules $K$,
Fig.~\ref{IQMD_ratio} exhibits the double ratio obtained in IQMD
calculations as a function of $K$ for the two options, with and
without the KN potential. On top (bottom) the results for a beam
energy of 1 (0.8) \AGeV are shown. Again, a compression modulus of
less than 250 MeV is required to reproduce the data. Instead of
varying the size of the system one can also vary the centrality
and this will give an independent observable. Because more central
collisions yield a higher compression we expect that the \kp yield
per participant as a function of the centrality depends as well on
the EoS. This is indeed observed in the simulations and displayed
in Fig.~\ref{part}, top, showing the dependence of the \kp
multiplicity per participant as a function of the participant
number \Apart for a hard EoS as well as for a soft EoS  with various options: Different
N$\Delta\to$ \kp YN cross sections with and without KN potentials.
These choices modify only little the slope of the curves. In
contrast, the calculation with a hard EoS exhibits a drastic
decrease of the slope demonstrating the sensitivity on the
stiffness of the EoS. It shows as well that the momentum
dependence of the interaction (mdi) does not influence the result.

On a logarithmic scale $M/A_{\rm part}$ exhibits an almost linear
increase with \Apart. The curves can therefore be characterized by
$M/A_{\rm part} \propto A_{{\rm part}}^\alpha$ with a
characteristic slope parameter $\alpha$. The bottom part of
Fig.~\ref{part} shows for the standard N$\Delta\to$ \kp NN cross
section and including the KN potential the slope parameter
$\alpha$ as a function of the compression modulus $K$. Also for
this observable the data of the KaoS
Collaboration~\cite{Forster:2007qk} are only compatible with
values of $K$ around 200 MeV.
Thus two rather independent observables point towards a rather low
compression modulus.

\section{Summary}

This report is a comprehensive analysis of  strangeness production
in p+A and A+A collisions by comparing the recent
available experimental data on the production of three kaon
species, \kp , K$^0$, \km , and of $\Lambda$ hyperons with results from one and in some cases from
two transport theories. With the exception of $v_1(p_t)$ of \kp, $v_2(p_t)$ of \km
mesons the multitude of the available experimental data are simultaneously reproduced by the two
transport approaches. Consequently, one can conclude  that the physics of strangeness
production close to threshold in p+A and A+A collisions is
qualitatively and for most of the observables also quantitatively
understood. This is a remarkable result in view of the fact that experimental data for
several of the essential elementary cross sections are not available and that hence
this input to the transport models is based on theory only.

This agreement establishes the following scenario for strangeness
production in heavy ion reactions at these sub-threshold energies: Close to threshold  the
most important process in which strange particles are produced is
the associate strangeness production via the $\Delta$ N $\to$ Y
K$^{+/0}$N channel. In this reaction a sufficient $\sqrt{s}$
value can be more easily obtained than in NN collisions. Hence,
the creation of \kp mesons occurs preferably via
multi-step interactions. The production of \kp in nucleus nucleus
collisions  is therefore right from the beginning different than that in
elementary collisions.

In contrast, \km are produced dominantly in secondary collisions
where a strange baryon transfers its strange quark to a meson,
i.e.~via strangeness exchange. \km production is therefore closely
connected to the \kp production as strange baryons are created
together with \kp. This is again in contrary to elementary
reactions in which this strangeness-exchange channel is absent and
where \km are exclusively produced in N N $\to$
 N N \km \kp  reactions. This channel contributes only very moderately
to the  \km yield in heavy-ion reactions at these energies.

The \kp rescatter after being produced. Rescattering becomes more
important if the nuclei are larger.  Already in C+C reactions a
\kp rescatters on the average more than once. This rescattering is
visible in the spectra. At production, \kp mesons have in the
average a much smaller kinetic energy than the nucleons of the
surrounding matter. Rescattering  increases therefore the inverse
slope parameter of the \kp spectra. It causes also finite $v_1$
and $v_2$ values. The latter, however, are strongly modified by
the potential interaction between \kp and the surrounding
nucleons. This potential interaction modifies also the spectra,
mainly at low transverse momentum but also, in a very moderate
way,  at higher momenta where it is responsible for a further
increase of  the inverse slope parameter. Only for the small
fraction of \kp  which stay in collisional contact with the
nuclear system until very late times the slope of the \kp and that
of the nucleons come close to each other (but, as said, this \kp
slope is still modified by the \kp nucleus potential interaction
which acts when the \kp leave the nuclear environment). Generally
the inverse slope parameter of the \kp spectra is smaller than
that of the surrounding nucleons and much smaller than that of all
nucleons. (Nucleons do not come to a global thermal equilibrium.)
Most of the \kp are emitted early and are not in equilibrium with
their environment. It remains for the moment an open question,
which deserves a further investigation, why nevertheless the
multiplicity of strange particles is well described in statistical
models.

$\Lambda$'s are produced in association with \kp mesons.  They serve
as a cross check for the \kp production as far as their
multiplicity is concerned. They rescatter with the nuclear
environment with a  large but not very well known cross section which modifies the
slope considerably. Rescattering is also the reason for
the considerable in-plane flow which is as large as that for the
surrounding nucleons. For the $\Lambda$ the two available data
sets cannot be reconciled if compared with IQMD and HSD calculations. So
further investigations are necessary.

The \km meson is more difficult to describe than \kp and
$\Lambda$. First of all, in cold nuclear matter it is far away
from being a quasi particle, i.e. from a particle with a small width.
The only available calculation
indicates that for the temperatures which are of importance for
heavy-ion collisions it becomes close to a quasi-particle. This
is understandable because at high temperatures  the Pauli blocking of
the intermediate states is less important. Second, it is produced
in secondary collisions. For one channel, Y $\pi \to$ \km N, the
cross section can be obtained by detailed balance from the
measured inverse cross section provided that it is not modified in
the medium. This is a very strong assumption because it was
predicted that the $\Lambda  (1405) $ resonance disappears already
at low densities. For the other cross section,  Y N $\to$ \km N N
little information is available. Third, there is a strong debate
of the strength of the \km nucleus potential where
measurements of kaonic atoms and those from kaonic clusters are
difficult to reconcile.

The cross section of the reaction \km N $\to \Lambda \pi $  is
rather large, hence in heavy systems a large number of the
produced \km becomes absorbed before escaping from the expanding
system whereas in light systems like C+C this is of less
importance. Due to the large momentum dependence of this cross
section reabsorption modifies  strongly the slope of the surviving
\km. This explains the large difference of the slope of heavy
systems as compared to that of light ones. As a consequence of the
large reabsorption cross section most of the surviving \km are
produced close to the nuclear surface. Consequently, both, the
yield of \km and their spectral shape, are not clean observables
for the K nucleus potential.

As for the \kp , the \km nucleus interaction modifies the slope at
low transverse momentum considerably but only slightly the slope
of the \km spectra at intermediate and large transverse momentum.
Because the \km potential is attractive the change of the momentum
spectrum at low transverse momentum has an opposite curvature than
that for the \kp which are repelled by a repulsive \kp nucleus
potential. Experimentally the low transverse momentum region is
not easily accessible but this is the only observable which allows
to determine the K-nucleus potential.

The comparison between calculations and data show clearly that a
\km condensate is not within reach in the beam energy domain of
our studies. It would be signaled by a drastic change of the ratio of \km multiplicities
in a light and a heavy system as a function of the beam energy because
in light and heavy systems quite different densities are reached.
This has experimentally not been observed.

Using the free cross section and a standard mean field approach
for the \km nucleus interaction a reasonable description of all
\km observables can be obtained. The little known cross sections
limit the  predictive power of  transport calculations for \km observables.

To make progress in understanding the strength of the \kp and
\km nucleus interaction in the nuclear medium
it is necessary to reach a consistent
description of all sensitive observables in a heavy ion reaction, i.e.
directed flow \cite{Crochet:2000fz},
spectra at low transverse momenta \cite{Agakishiev:2010zw},
ratios of \km and \kp
spectra \cite{Wisniewski:2001dk}.
None of the observables is exclusively sensitive to the depth
of the \kp and \km nucleus as a function of density.
Also K$^0$ spectra at very low momenta, as measured by the HADES collaboration,
\cite{Agakishiev:2010zw} do not depend exclusively on the
\kp nucleon potential but in addition on the \kp nucleon rescattering
cross section. Hence, spectra of
different projectile target combinations are necessary to
establish this potential unambiguously.

Because \kp mesons are created during the dense phase of the
collision, several \kp observables turn out to be sensitive to the
hadronic equation of state. These data together with the model
calculations clearly point towards a rather soft equation of
state. Only values of the compression modulus $K$ at nuclear
ground state density below 250 MeV are compatible with the
heavy-ion collision data which are sensitive to densities of
around twice normal nuclear matter density. These values agree
with the one found in the analysis of giant monopole vibrations
which test, however, only  densities very close to the ground
state density.

In summary, employing the experimental free scattering cross
section and theoretical model predictions for cross sections
were experimental data are not available, theory agrees with experiment if
a repulsive \kp and an attractive \km nucleon potential in medium are
employed. Transport calculations can reveal the complicated processes which are at
the origin of multiplicities, spectral slopes and
angular distributions of kaons observed in proton-nucleus and heavy ion reactions.

Acknowledgement: We would like to thank Dr.~Ch.~Fuchs.
Dr.~L.~Tolos and Dr.~H.~Wolter for a careful reading of the
manuscript as well as Dr.~A.~F\"orster and Dr.~A.~Schmah for
discussions. This work was supported by the Helmholtz
International Center for FAIR within the framework of the LOEWE
program (Landesoffensive zur Entwicklung
Wissenschaftlich-\"Okonomischer Exzellenz) launched by the State
of Hesse.


\begin{thebibliography}{99}






\bibitem{Harris:1981tb}
J.~W.~Harris {\it et al.}
Phys.\ Rev.\ Lett.\  {\bf 47}, 229 (1981)
\bibitem{Schnetzer:1989vy}
S.~Schnetzer {\it et al.}
Phys.\ Rev.\  C {\bf 40}, 640 (1989)
  [Erratum-ibid.\  C {\bf 41}, 1320 (1990)]
\bibitem{Hartnack:1997ez}
C.~Hartnack, R.~K.~Puri, J.~Aichelin, J.~Konopka, S.~A.~Bass, H.~St\"ocker and W.~Greiner
Eur.\ Phys.\ J.\  A {\bf 1}, 151 (1998)
  [arXiv:nucl-th/9811015]
\bibitem{Cassing:1999es}
  W.~Cassing and E.~L.~Bratkovskaya,
  Phys.\ Rept.\  {\bf 308}, 65 (1999).
\bibitem{Fuchs:2003pc}
C.~Fuchs
Prog.\ Part.\ Nucl.\ Phys.\  {\bf 53}, 113 (2004)
  [arXiv:nucl-th/0312052]
\bibitem{Dohrmann:2006dx}
F.~Dohrmann
Int.\ J.\ Mod.\ Phys.\  E {\bf 15}, 761 (2006)

\bibitem{Kaiser:1995cy}
N.~Kaiser, P.~B.~Siegel and W.~Weise
Phys.\ Lett.\  B {\bf 362}, 23 (1995)
  [arXiv:nucl-th/9507036]
\bibitem{Kaiser:1996js}
N.~Kaiser, T.~Waas and W.~Weise
Nucl.\ Phys.\  A {\bf 612}, 297 (1997)
  [arXiv:hep-ph/9607459]
\bibitem{Oller:2000ma}
J.~A.~Oller, E.~Oset and A.~Ramos
Prog.\ Part.\ Nucl.\ Phys.\  {\bf 45}, 157 (2000)
  [arXiv:hep-ph/0002193]
\bibitem{SchaffnerBielich:2000jy}
J.~Schaffner-Bielich
J.\ Phys.\ G {\bf 27}, 337 (2001)
  [arXiv:nucl-th/0009083]
\bibitem{Tolos:2005jg}
L.~Tolos, D.~Cabrera, A.~Ramos and A.~Polls
Phys.\ Lett.\  B {\bf 632}, 219 (2006)
  [arXiv:hep-ph/0503009]
\bibitem{Nekipelov:2002sd}
M.~Nekipelov {\it et al.}
Phys.\ Lett.\  B {\bf 540}, 207 (2002)
  [arXiv:nucl-ex/0202021]
\bibitem{Dover:1971hr}
C.~B.~Dover, J.~Huefner and R.~H.~Lemmer
Annals Phys.\  {\bf 66}, 248 (1971)
\bibitem{Lutz:1997wt}
M.~Lutz
Phys.\ Lett.\  B {\bf 426}, 12 (1998)
  [arXiv:nucl-th/9709073]
\bibitem{Lutz:1994cf}
M.~Lutz, A.~Steiner and W.~Weise
Nucl.\ Phys.\  A {\bf 574}, 755 (1994)
\bibitem{Martin:1980qe}
A.~D.~Martin
Nucl.\ Phys.\  B {\bf 179}, 33 (1981)
\bibitem{Dover:1982zh}
C.~B.~Dover and G.~E.~Walker
Phys.\ Rept.\  {\bf 89}, 1 (1982)
\bibitem{luhab} M. Lutz, nucl-th/0212021
\bibitem{Baca:2000ic}
A.~Baca, C.~Garcia-Recio and J.~Nieves
Nucl.\ Phys.\  A {\bf 673}, 335 (2000)
  [arXiv:nucl-th/0001060]
\bibitem{Koch:1994mj}
V.~Koch
Phys.\ Lett.\  B {\bf 337}, 7 (1994)
  [arXiv:nucl-th/9406030]
\bibitem{Waas:1996xh}
T.~Waas, N.~Kaiser and W.~Weise
Phys.\ Lett.\  B {\bf 365}, 12 (1996)
\bibitem{Waas:1996fy}
T.~Waas, N.~Kaiser and W.~Weise
Phys.\ Lett.\  B {\bf 379}, 34 (1996)
\bibitem{Waas:1997pe}
T.~Waas and W.~Weise
Nucl.\ Phys.\  A {\bf 625}, 287 (1997)
\bibitem{Kaplan:1986yq}
D.~B.~Kaplan and A.~E.~Nelson
Phys.\ Lett.\  B {\bf 175}, 57 (1986)
\bibitem{Nelson:1987dg}
  A.~E.~Nelson and D.~B.~Kaplan,
  Phys.\ Lett.\  B {\bf 192}, 193 (1987).
\bibitem{Ko:1996yy}
  C.~M.~Ko and G.~Q.~Li,
  J.\ Phys.\ G {\bf 22} (1996) 1673
  [arXiv:nucl-th/9611027].
\bibitem{Lee:1994my}
  C.~H.~Lee, H.~Jung, D.~P.~Min and M.~Rho,
  Phys.\ Lett.\  B {\bf 326} (1994) 14
  [arXiv:hep-ph/9401245].
\bibitem{Kaiser:1995eg}
  N.~Kaiser, P.~B.~Siegel and W.~Weise,
  Nucl.\ Phys.\  A {\bf 594}, 325 (1995)
  [arXiv:nucl-th/9505043].
\bibitem{Dong:1994zs}
  S.~J.~Dong and K.~F.~Liu,
  Nucl.\ Phys.\ Proc.\ Suppl.\  {\bf 42}, 322 (1995)
  [arXiv:hep-lat/9412059].
\bibitem{Fukugita:1994ba}
  M.~Fukugita, Y.~Kuramashi, M.~Okawa and A.~Ukawa,
  Phys.\ Rev.\  D {\bf 51} (1995) 5319
  [arXiv:hep-lat/9408002].
\bibitem{bhaduri} R.~K.~Bhaduri, {\it Models of the Nucleon}
(Addison-Wesley, Reading, MA, 1988).
\bibitem{Korpa:2004ae}
  C.~L.~Korpa and M.~F.~M.~Lutz,
  Acta Phys.\ Hung.\  A {\bf 22} (2005) 21
  [arXiv:nucl-th/0404088].
\bibitem{Oset:1997it}
  E.~Oset and A.~Ramos,
  Nucl.\ Phys.\  A {\bf 635} (1998) 99
  [arXiv:nucl-th/9711022].
\bibitem{Ramos:1999ku}
  A.~Ramos and E.~Oset,
  Nucl.\ Phys.\  A {\bf 671}, 481 (2000)
  [arXiv:nucl-th/9906016].
\bibitem{Tolos:2002ud}
  L.~Tolos, A.~Ramos and A.~Polls,
  Phys.\ Rev.\  C {\bf 65}, 054907 (2002)
  [arXiv:nucl-th/0202057].
\bibitem{Friedman:1993cu}
  E.~Friedman, A.~Gal and C.~J.~Batty,
  Phys.\ Lett.\  B {\bf 308}, 6 (1993).
\bibitem{Oset:2000eg}
  E.~Oset and A.~Ramos,
  Nucl.\ Phys.\  A {\bf 679}, 616 (2001)
  [arXiv:nucl-th/0005046].
\bibitem{Rios:2005ps}
  A.~Rios, A.~Polls, A.~Ramos and I.~Vidana,
  Phys.\ Rev.\  C {\bf 72}, 024316 (2005)
  [arXiv:nucl-th/0503074].
\bibitem{Tolos:2000fj}
  L.~Tolos, A.~Ramos, A.~Polls and T.~T.~S.~Kuo,
  Nucl.\ Phys.\  A {\bf 690}, 547 (2001)
  [arXiv:nucl-th/0007042].
\bibitem{Lutz:2001dq}
  M.~F.~M.~Lutz and C.~L.~Korpa,
  Nucl.\ Phys.\  A {\bf 700}, 309 (2002)
  [arXiv:nucl-th/0105067].
\bibitem{Tolos:2006ny}
  L.~Tolos, A.~Ramos and E.~Oset,
  Phys.\ Rev.\  C {\bf 74}, 015203 (2006)
  [arXiv:nucl-th/0603033].
\bibitem{MuellerGroeling:1990cw}
  A.~Mueller- Groeling, K.~Holinde and J.~Speth,
  Nucl.\ Phys.\  A {\bf 513}, 557 (1990).
\bibitem{Friedman:1994hx}
  E.~Friedman, A.~Gal and C.~J.~Batty,
  Nucl.\ Phys.\  A {\bf 579}, 518 (1994).
\bibitem{Friedman:1998xa}
  E.~Friedman, A.~Gal and J.~Mares,
  Phys.\ Rev.\  C {\bf 60}, 024314 (1999)
  [arXiv:nucl-th/9804072].
\bibitem{Cieply:2001yg}
  A.~Cieply, E.~Friedman, A.~Gal and J.~Mares,
  Nucl.\ Phys.\  A {\bf 696}, 173 (2001)
  [arXiv:nucl-th/0104087].
\bibitem{Barnea:2006kv}
  N.~Barnea and E.~Friedman,
  Phys.\ Rev.\  C {\bf 75}, 022202 (2007)
  [arXiv:nucl-th/0611020].
\bibitem{Akaishi:2002bg}
  Y.~Akaishi and T.~Yamazaki,
  Phys.\ Rev.\  C {\bf 65}, 044005 (2002).
\bibitem{Suzuki:2005bj}
  T.~Suzuki {\it et al.},
  Nucl.\ Phys.\  A {\bf 754}, 375 (2005)
  [arXiv:nucl-ex/0501013].
\bibitem{Agnello:2006jt}
  M.~Agnello {\it et al.}  [FINUDA Collaboration],
  Nucl.\ Phys.\  A {\bf 775}, 35 (2006).
\bibitem{Oset:2005sn}
  E.~Oset and H.~Toki,
  Phys.\ Rev.\  C {\bf 74}, 015207 (2006)
  [arXiv:nucl-th/0509048].
\bibitem{Yamazaki:2002uh}
  T.~Yamazaki and Y.~Akaishi,
  Phys.\ Lett.\  B {\bf 535}, 70 (2002).
\bibitem{Dote:2003ac}
  A.~Dote, H.~Horiuchi, Y.~Akaishi and T.~Yamazaki,
  Phys.\ Rev.\  C {\bf 70}, 044313 (2004)
  [arXiv:nucl-th/0309062].
\bibitem{Ramos:2007vg}
  A.~Ramos, V.~K.~Magas, E.~Oset and H.~Toki,
  arXiv:nucl-th/0702019.
\bibitem{Magas:2008bp}
  V.~K.~Magas, E.~Oset and A.~Ramos,
  Phys.\ Rev.\  C {\bf 77} (2008) 065210
  [arXiv:0801.4504 [nucl-th]].
\bibitem{Magas:2009qd}
  V.~K.~Magas, E.~Oset and A.~Ramos,
  arXiv:0901.1086 [nucl-th].
\bibitem{Schulze:1995jx}
  H.~J.~Schulze, A.~Lejeune, J.~Cugnon, M.~Baldo and U.~Lombardo,
  Phys.\ Lett.\  B {\bf 355}, 21 (1995).
\bibitem{Schulze:1998jf}
  H.~J.~Schulze, M.~Baldo, U.~Lombardo, J.~Cugnon and A.~Lejeune,
  Phys.\ Rev.\  C {\bf 57}, 704 (1998).
\bibitem{Vidana:1999jm}
  I.~Vidana, A.~Polls, A.~Ramos, M.~Hjorth-Jensen and V.~G.~J.~Stoks,
  Phys.\ Rev.\  C {\bf 61}, 025802 (2000)
  [arXiv:nucl-th/9909019].
\bibitem{Rijken:1998yy}
  T.~A.~Rijken, V.~G.~J.~Stoks and Y.~Yamamoto,
  Phys.\ Rev.\  C {\bf 59}, 21 (1999)
  [arXiv:nucl-th/9807082].


\bibitem{Hartnack:1989sd}
  C.~Hartnack {\it et al.},
  Nucl.\ Phys.\  A {\bf 495}, 303C (1989).
\bibitem{hart}
Ch. Hartnack,
\newblock PhD thesis, GSI-Report 93-5 (1993)
\bibitem{chhab} Christoph Hartnack, nucl-th/0507002
\bibitem{Bass:1995pj}
  S.~A.~Bass, C.~Hartnack, H.~St\"ocker and W.~Greiner,
  Phys.\ Rev.\  C {\bf 51}, 3343 (1995)
  [arXiv:nucl-th/9501002].
\bibitem{Stoecker:1986ci}
  H.~St\"ocker and W.~Greiner,
  Phys.\ Rept.\  {\bf 137}, 277 (1986).
\bibitem{Cassing:1990dr}
  W.~Cassing, V.~Metag, U.~Mosel and K.~Niita,
  Phys.\ Rept.\  {\bf 188}, 363 (1990).
\bibitem{Aichelin:1991xy}
  J.~Aichelin,
  Phys.\ Rept.\  {\bf 202}, 233 (1991).
\bibitem{Bass:1998ca}
  S.~A.~Bass {\it et al.},
  Prog.\ Part.\ Nucl.\ Phys.\  {\bf 41}, 255 (1998)
  [Prog.\ Part.\ Nucl.\ Phys.\  {\bf 41}, 225 (1998)]
  [arXiv:nucl-th/9803035].
\bibitem{Cassing:1996xx}
  W.~Cassing, E.~L.~Bratkovskaya, U.~Mosel, S.~Teis and A.~Sibirtsev,
  Nucl.\ Phys.\  A {\bf 614}, 415 (1997)
  [arXiv:nucl-th/9609050].
\bibitem{Cassing:2003vz}
  W.~Cassing, L.~Tolos, E.~L.~Bratkovskaya and A.~Ramos,
  Nucl.\ Phys.\  A {\bf 727}, 59 (2003)
  [arXiv:nucl-th/0304006].
\bibitem{Mishra:2004te}
  A.~Mishra, E.~L.~Bratkovskaya, J.~Schaffner-Bielich, S.~Schramm and H.~St\"ocker,
  Phys.\ Rev.\  C {\bf 70}, 044904 (2004)
  [arXiv:nucl-th/0402062].
\bibitem{Jeukenne:1976uy}
  J.~P.~Jeukenne, A.~Lejeune and C.~Mahaux,
  Phys.\ Rept.\  {\bf 25}, 83 (1976).
\bibitem{Arnold:1982rf}
  L.~G.~Arnold {\it et al.},
  Phys.\ Rev.\  C {\bf 25}, 936 (1982).
\bibitem{pa67}
G. Passatore,
\newblock Nucl.~Phys. {\bf A95}, 694 (1967).
\bibitem{Hama:1990vr}
  S.~Hama, B.~C.~Clark, E.~D.~Cooper, H.~S.~Sherif and R.~L.~Mercer,
  Phys.\ Rev.\  C {\bf 41}, 2737 (1990).
\bibitem{Aichelin:1987ti}
  J.~Aichelin, A.~Rosenhauer, G.~Peilert, H.~St\"ocker and W.~Greiner,
  Phys.\ Rev.\ Lett.\  {\bf 58}, 1926 (1987).
\bibitem{Bertsch:1988ik}
  G.~F.~Bertsch and S.~Das Gupta,
  Phys.\ Rept.\  {\bf 160}, 189 (1988).
\bibitem{Hartnack:1994zz}
  C.~Hartnack and J.~Aichelin,
  Phys.\ Rev.\  C {\bf 49}, 2801 (1994).
\bibitem{Li:2008gp}
  B.~A.~Li, L.~W.~Chen and C.~M.~Ko,
  Phys.\ Rept.\  {\bf 464}, 113 (2008)
  [arXiv:0804.3580 [nucl-th]].
\bibitem{Hartnack:1994bs}
  C.~Hartnack, J.~Aichelin, H.~St\"ocker and W.~Greiner,
  Mod.\ Phys.\ Lett.\  A {\bf 9}, 1151 (1994).
\bibitem{Stoicea:2004kp}
  G.~Stoicea {\it et al.}  [FOPI Collaboration],
  Phys.\ Rev.\ Lett.\  {\bf 92}, 072303 (2004)
  [arXiv:nucl-ex/0401041].
\bibitem{Andronic:2004cp}
  A.~Andronic {\it et al.}  [FOPI Collaboration],
  Phys.\ Lett.\  B {\bf 612}, 173 (2005)
  [arXiv:nucl-ex/0411024].
\bibitem{Schaffner:1996kv}
  J.~Schaffner, J.~Bondorf and I.~N.~Mishustin,
  Nucl.\ Phys.\  A {\bf 625}, 325 (1997)
  [arXiv:nucl-th/9607058].
\bibitem{Tolos:2008di}
L.~Tolos, D.~Cabrera and A.~Ramos
Phys.\ Rev.\  C {\bf 78}, 045205 (2008)
  [arXiv:0807.2947 [nucl-th]]
\bibitem{Danielewicz:1991dh}
  P.~Danielewicz and G.~F.~Bertsch,
  Nucl.\ Phys.\  A {\bf 533}, 712 (1991).
\bibitem{Bass:2001up}
  S.~A.~Bass,
  J.\ Phys.\ G {\bf 28}, 1543 (2002)
  [arXiv:nucl-th/0112046].
\bibitem{bleicher} M. Bleicher, UFTP Frankfurt, private communication.
\bibitem{Hartnack:1993bq}
  G.~Hartnack, L.~Sehn, J.~Jaenicke, H.~St\"ocker and J.~Aichelin,
  Nucl.\ Phys.\  A {\bf 580}, 643 (1994).
\bibitem{OZI}
  B.R. Martin, G. Shaw; "Particle physics", John Wiley and Sons, Chichester (England) 2nd ed. (1997),
  Chapt. 6.1.1 Charmonium
\bibitem{Hogan:1968zz}
  W.~J.~Hogan, P.~A.~Piroue and A.~J.~S.~Smith,
  Phys.\ Rev.\  {\bf 166}, 1472 (1968).
\bibitem{Sibirtsev:1995xb}
  A.~Sibirtsev,
  Phys.\ Lett.\  B {\bf 359}, 29 (1995).
\bibitem{Tsushima:1998jz}
  K.~Tsushima, A.~Sibirtsev and A.~W.~Thomas,
  Phys.\ Rev.\  C {\bf 59}, 369 (1999)
  [Erratum-ibid.\  C {\bf 61}, 029903 (2000)]
  [arXiv:nucl-th/9801063].
 \bibitem{Balewski:1998pd}
  J.~T.~Balewski {\it et al.},
  Phys.\ Lett.\  B {\bf 420}, 211 (1998)
  [arXiv:nucl-ex/9803003].
\bibitem{Tsushima:1994rj}
  K.~Tsushima, S.~W.~Huang and A.~Faessler,
  Phys.\ Lett.\  B {\bf 337}, 245 (1994)
  [arXiv:nucl-th/9407021].
\bibitem{Randrup:1980qd}
  J.~Randrup and C.~M.~Ko,
  Nucl.\ Phys.\  A {\bf 343}, 519 (1980)
  [Erratum-ibid.\  A {\bf 411}, 537 (1983)].
\bibitem{Sibirtsev:1996rh}
  A.~A.~Sibirtsev, W.~Cassing and C.~M.~Ko,
  Z.\ Phys.\  A {\bf 358}, 101 (1997)
  [arXiv:nucl-th/9612040].
\bibitem{totkm}
C. Amsler {\it et al.} (Particle Data Group), Phys. Lett. B {\bf
667}, 1 (2008) and 2009 partial update for the 2010 edition and
references therein
\bibitem{Wagner:2000ak}
  A.~Wagner {\it et al.},
  Phys.\ Rev.\ Lett.\  {\bf 85}, 18 (2000)
  [arXiv:nucl-ex/0005004].
\bibitem{Averbeck:2000sn}
  R.~Averbeck, R.~Holzmann, V.~Metag and R.~S.~Simon,
  Phys.\ Rev.\  C {\bf 67}, 024903 (2003)
  [arXiv:nucl-ex/0012007].
\bibitem{Reisdorf:2006ie}
  W.~Reisdorf {\it et al.}  [FOPI Collaboration],
  Nucl.\ Phys.\  A {\bf 781}, 459 (2007)
  [arXiv:nucl-ex/0610025].
\bibitem{Tschuck} T.  Schuck, Diplomarbeit, University of Frankfurt, 2003.
\bibitem{Oeschler:2002ch}
  H.~Oeschler,
  J.\ Phys.\ G {\bf 28}, 1787 (2002)
  [arXiv:nucl-ex/0202003].
\bibitem{Sturm:2000dm}
  C.~Sturm {\it et al.}  [KAOS Collaboration],
  Phys.\ Rev.\ Lett.\  {\bf 86}, 39 (2001)
  [arXiv:nucl-ex/0011001]; C. Sturm, PhD thesis, TH Darmstadt, 2001.
\bibitem{Hong:1997mr}
  B.~Hong {\it et al.}  [FOPI Collaboration],
  Phys.\ Rev.\  C {\bf 57}, 244 (1998)
  [Erratum-ibid.\  C {\bf 58}, 603 (1998\ PHRVA,C58,603.1998)]
  [arXiv:nucl-ex/9707001].


\bibitem{Buescher:2001qu}
  M.~B\"uscher {\it et al.},
  Phys.\ Rev.\  C {\bf 65}, 014603 (2002)
  [arXiv:nucl-ex/0107011].
\bibitem{Buescher:2002np}
  M.~B\"uscher and M.~Nekipelov  [ANKE collaboration],
  arXiv:nucl-ex/0207002.
\bibitem{Anke_priv}
M. B\"uscher, private communication.
\bibitem{Scheinast:2005xs}
  W.~Scheinast {\it et al.}  [KaoS Collaboration],
  Phys.\ Rev.\ Lett.\  {\bf 96} (2006) 072301
  [arXiv:nucl-ex/0512028].
\bibitem{Kolomeitsev:2004np}
  E.~E.~Kolomeitsev {\it et al.},
  J.\ Phys.\ G {\bf 31}, S741 (2005)
  [arXiv:nucl-th/0412037].
\bibitem{Forster:2007qk}
  A.~F\"orster {\it et al.},
  Phys.\ Rev.\  C {\bf 75}, 024906 (2007)
  [arXiv:nucl-ex/0701014].
\bibitem{Merschmeyer:2007zz}
  M.~Merschmeyer {\it et al.},
  Phys.\ Rev.\  C {\bf 76}, 024906 (2007).
\bibitem{Schmah:2009eh}
  A.~Schmah and L.~Fabbietti,
  arXiv:0911.0300 [nucl-ex].
\bibitem{Agakishiev:2010zw}
  G.~Agakishiev {\it et al.},
  Phys.\ Rev.\  C {\bf 82} (2010) 044907
  [arXiv:1004.3881 [nucl-ex]].



\bibitem{Agakishiev:2010rs}
  G.~Agakishiev {\it et al.}  [HADES Collaboration],
  arXiv:1010.1675 [nucl-ex].

\bibitem{Wisniewski:2001dk}
  K.~Wisniewski {\it et al.} [ FOPI Collaboration ],
  Eur.\ Phys.\ J.\  {\bf A9 } (2000)  515-519.
  [nucl-ex/0101009].
\cite{Best:1997ua}
\bibitem{Best:1997ua}
  D.~Best {\it et al.}  [FOPI Collaboration],
  Nucl.\ Phys.\  A {\bf 625}, 307 (1997)
  [arXiv:nucl-ex/9704005].
\bibitem{Forster:2003vc}
  A.~F\"orster {\it et al.}  [KaoS Collaboration],
  Phys.\ Rev.\ Lett.\  {\bf 91}, 152301 (2003)
  [arXiv:nucl-ex/0307017].
\bibitem{Workshop2000} J. Ritman, private communication
\bibitem{Crochet:2000fz}
  P.~Crochet {\it et al.}  [FOPI Collaboration],
  Phys.\ Lett.\  B {\bf 486}, 6 (2000)
  [arXiv:nucl-ex/0006004].
\bibitem{Uhlig:2004ue}
  F.~Uhlig {\it et al.},
  Phys.\ Rev.\ Lett.\  {\bf 95}, 012301 (2005)
  [arXiv:nucl-ex/0411021].
\bibitem{Shin:1998hm}
  Y.~Shin {\it et al.}  [KaoS Collaboration],
  Phys.\ Rev.\ Lett.\  {\bf 81}, 1576 (1998)
  [arXiv:nucl-ex/9807003].
\bibitem{Ploskon:2005qr}
  M.~Ploskon  [KaoS Collaboration],
  Nucl.\ Phys.\  A {\bf 749}, 170 (2005) and PhD Thesis,
  University of Frankfurt, 2005 .
\bibitem{Ritman:1995tn}
  J.~L.~Ritman {\it et al.}  [FOPI Collaboration],
  Z.\ Phys.\  A {\bf 352}, 355 (1995)
  [arXiv:nucl-ex/9506002].
\bibitem{Alexander:1969cx}
  G.~Alexander, U.~Karshon, A.~Shapira, G.~Yekutieli, R.~Engelmann, H.~Filthuth and W.~Lughofer,
  Phys.\ Rev.\  {\bf 173}, 1452 (1968).
\bibitem{Anderson:1975rh}
  K.~J.~Anderson, N.~M.~Gelfand, J.~Keren and T.~H.~Tan,
  Phys.\ Rev.\  D {\bf 11}, 473 (1975)
  [Erratum-ibid.\  D {\bf 11}, 3359 (1975)].
\bibitem{David:1998qu}
  C.~David, C.~Hartnack and J.~Aichelin,
  Nucl.\ Phys.\  A {\bf 650}, 358 (1999)
  [arXiv:nucl-th/9805017].
\bibitem{Chung:2001je}
  P.~Chung {\it et al.},
  Phys.\ Rev.\ Lett.\  {\bf 86}, 2533 (2001)
  [arXiv:nucl-ex/0101002].

\bibitem{Oeschler:2000dp}
  H.~Oeschler,
  J.\ Phys.\ G {\bf 27}, 257 (2001)
  [arXiv:nucl-ex/0011007].


\bibitem{Cleymans:2004bf}
  J.~Cleymans, A.~F\"orster, H.~Oeschler, K.~Redlich and F.~Uhlig,
  Phys.\ Lett.\  B {\bf 603}, 146 (2004)
  [arXiv:hep-ph/0406108].

\bibitem{Schaffner:1994bx}
  J.~Schaffner, A.~Gal, I.~N.~Mishustin, H.~St\"ocker and W.~Greiner,
  Phys.\ Lett.\  B {\bf 334}, 268 (1994).
\bibitem{SchaffnerBielich:1999cp}
  J.~Schaffner-Bielich, V.~Koch and M.~Effenberger,
  Nucl.\ Phys.\  A {\bf 669}, 153 (2000)
  [arXiv:nucl-th/9907095].
\bibitem{Menzel:2000vv}
  M.~Menzel {\it et al.}  [KaoS Collaboration],
  Phys.\ Lett.\  B {\bf 495}, 26 (2000)
  [arXiv:nucl-ex/0010013]; M. Menzel,
Ph.D. Thesis, Universit\"at Marburg, 2000.

\bibitem{Muntz:1995am}
  C.~M\"untz {\it et al.} [KaoS Collaboration],
  Z.\ Phys.\  A {\bf 352}, 175 (1995);
  C. M\"untz {\em et al.,}  Z.\ Phys.\ A {\bf 357}, 399 (1997).
\bibitem{Cleymans:1998yb}
  J.~Cleymans, H.~Oeschler and K.~Redlich,
  Phys.\ Rev.\  C {\bf 59}, 1663 (1999)
  [arXiv:nucl-th/9809027].
\bibitem{Agakishiev:2009ar}
  G.~Agakishiev {\it et al.}  [HADES Collaboration],
  Phys.\ Rev.\  C {\bf 80}, 025209 (2009)
  [arXiv:0902.3487 [nucl-ex]],  A. Schmah, Ph.D.thesis, Darmstadt University of Technology, 2008.
\bibitem{Schnedermann:1993ws}
  E.~Schnedermann, J.~Sollfrank and U.~W.~Heinz,
  Phys.\ Rev.\  C {\bf 48} (1993) 2462
  [arXiv:nucl-th/9307020].

\bibitem{Aichelin:2008jn}
  J.~Aichelin and J.~Schaffner-Bielich,
Relativistic Heavy Ion Physics (Landolt-B\"ornstein: Numerical Data and Functional Relationships in Science and Technology - New Series / Elementary Particles, Nuclei and Atoms), Springer, Heidelberg, 2010
arXiv:0812.1341 [nucl-th].
\bibitem{schuck}
P. Ring and P. Schuck, The Nuclear Many Body Problem, Springer, Heidelberg
\bibitem{Baldo:2007wm}
  M.~Baldo and C.~Maieron,
  J.\ Phys.\ G {\bf 34}, R243 (2007)
  [arXiv:nucl-th/0703004].
\bibitem{Fuchs:2005yn}
  C.~Fuchs and H.~H.~Wolter,
  Eur.\ Phys.\ J.\  A {\bf 30} (2006) 5
  [arXiv:nucl-th/0511070].
\bibitem{Youngblood:1999zza}
  D.~H.~Youngblood, H.~L.~Clark and Y.~W.~Lui,
  Phys.\ Rev.\ Lett.\  {\bf 82}, 691 (1999).
\bibitem{Aichelin:1986ss}
  J.~Aichelin and C.~M.~Ko,
  Phys.\ Rev.\ Lett.\  {\bf 55}, 2661 (1985).
\bibitem{Hartnack:2005tr}
  C.~Hartnack, H.~Oeschler and J.~Aichelin,
  Phys.\ Rev.\ Lett.\  {\bf 96}, 012302 (2006)
  [arXiv:nucl-th/0506087].
\bibitem{Piekarewicz:2003br}
  J.~Piekarewicz,
  Phys.\ Rev.\  C {\bf 69}, 041301 (2004)
  [arXiv:nucl-th/0312020] and references therein.
\bibitem{Agrawal:2003xb}
  B.~K.~Agrawal, S.~Shlomo and V.~K.~Au,
  Phys.\ Rev.\  C {\bf 68}, 031304 (2003)
  [arXiv:nucl-th/0308042].
\bibitem{Colo:2003zm}
  G.~Colo and G.~N.~Van,
  arXiv:nucl-th/0309002.
\bibitem{Colo:2004mj}
  G.~Colo, N.~Van Giai, J.~Meyer, K.~Bennaceur and P.~Bonche,
  Phys.\ Rev.\  C {\bf 70}, 024307 (2004)
  [arXiv:nucl-th/0403086].
\bibitem{Vretenar:2003qm}
  D.~Vretenar, T.~Niksic and P.~Ring,
  Phys.\ Rev.\  C {\bf 68}, 024310 (2003)
  [arXiv:nucl-th/0302070].
\bibitem{Sharma:2008uy}
  M.~M.~Sharma,
  Nucl.\ Phys.\  A {\bf 816}, 65 (2009)
  [arXiv:0811.2729 [nucl-th]].
\bibitem{Danielewicz:2002pu}
  P.~Danielewicz, R.~Lacey and W.~G.~Lynch,
  Science {\bf 298}, 1592 (2002)
  [arXiv:nucl-th/0208016].
\bibitem{Fuchs:1998yy}
  C.~Fuchs, D.~S.~Kosov, A.~Faessler, Z.~S.~Wang and T.~Waindzoch,
  Phys.\ Lett.\  B {\bf 434}, 245 (1998)
  [arXiv:nucl-th/9801048].
\bibitem{Fuchs:2000kp}
  C.~Fuchs, A.~Faessler, E.~Zabrodin and Y.~M.~Zheng,
  Phys.\ Rev.\ Lett.\  {\bf 86}, 1974 (2001)
  [arXiv:nucl-th/0011102].
\bibitem{Fuchs:2001gv}
  C.~Fuchs, A.~Faessler, S.~El-Basaouny and E.~Zabrodin,
  J.\ Phys.\ G {\bf 28}, 1615 (2002)
  [arXiv:nucl-th/0103036].


\end{thebibliography}
\end{document}